Hadron Hall Users' Association

# Write-ups for workshop on the project for the extended hadron experimental facility of J-PARC

Partial collection of LOIs at the extended hadron hall and the related topics

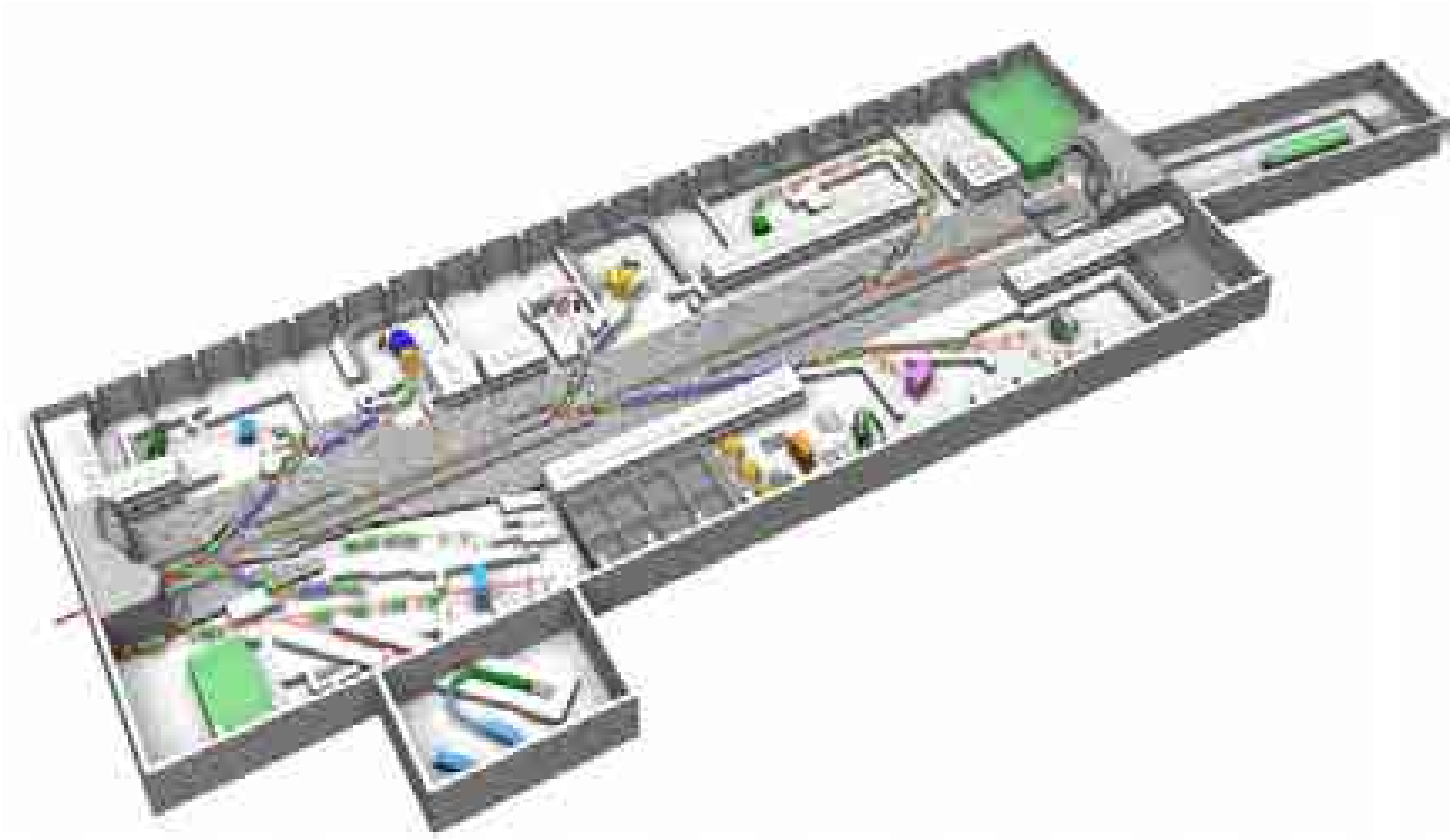

Hadron Hall Users' Association (HUA)

2019 June

# Preface

This write-up document is a summary of "International workshop on the project for the extended hadron experimental facility" which was held from March 26 to 28, 2018 at KEK Tokai Campus. This is the second of the series of workshops organized by J-PARC Hadron Hall Users' Association (HUA), the first one was held in March 2016, to discuss physics and experimental ideas at new beam lines or upgraded beam lines in the extended hadron experimental facility (HEF) of J-PARC. This time we call for contributions to work for Letter Of Intent (LOI) related to the proposed beam line facilities. In this workshop, we invited theorists as keynote talks to discuss what kinds of problems to be attacked at the extended HEF. We also invited speakers from GSI in Germany and Jefferson Laboratory (JLab) in USA, which play complimentary roles to J-PARC HEF. A lot of proposals and experimental plans which should be carried out at the extended HEF were presented and discussed intensively, together with those planned at other related facilities such as ELPH-Tohoku, Mainz, GSI, JLab, and so on so forth.

In 2015, Committee for the study of the extension of the Hadron Experimental Facility under HUA published so-called White Paper on HEF extension project in Japanese. In Part 1 of the White Paper, we introduce unanswered fundamental questions in nature: the origin of matter, how the matter is developed, and the mechanism of the relevant phenomena. To answer these questions, we describe roles of the project, overview of facilities, and how particle and nuclear physics attack the questions. In part 2, details of the beam lines and many experimental researches to be carried out were collected. The English translation of Part1 was published in 2016, as arXiv.1706.07916 [nucl-ex]. This write-up is expected to correspond to an updated version of Part.2 of the White Paper. We wish these documents push forward the project at HEF and its extension.

On behalf of
Hadron Hall Users' Association and
Organizing Committee of the Workshop

Toshiyuki TAKAHASHI, Chair of HUA in JFY2018

# Contents

Nuclear physics experiments and their results at the present Hadron Hall

**T. Takahashi**

KEK-IPNS/J-PARC Center

The status of the present Hadron Hall and results of the conducted experiments are briefly summarized. Some of the experiments which will be carried out in near future at the present Hadron Hall are also shown.

# 1 Beam lines and the achieved performance of the present facility

Three secondary beam lines are in operation at the present Hadron Hall (Fig.1). K1.8 and K1.8BR beam lines in the north area share the upstream part of the beam line. K1.8BR started its operation in February 2009, which can deliver mass-separated secondary beams in the range from 0.7 to 1.1 GeV/c with a single electrostatic (ES) separator. In the area, Cylindrical Detector System (CDS) and neutron TOF counter of 17 m flight length are equipped. K1.8 has double-stage ES separators and swerves mass-separated beams of $K^-$ (1.5 – 2.0 GeV/$c$) and pions (1.2 – 2.0 GeV/$c$). It started operation in October 2009. In an early stage of the operation, SKS was equipped and it was replaced by KURAMA in 2016. Experiments on strangeness nuclear physics have been and will be conducted at both beam lines. KL is a neutral kaon beam line for kaon rare decay experiment (KOTO).

In south area of Hadron Hall and South Experimental Hall, new primary beam lines, high-p and COMET, are under cons traction. High-p beam line can deliver primary proton beam of 30 GeV for hadron nuclear physics experiment with a large dipole magnet (FM). It is also planned to deliver high momentum unseparated secondary beams. K1.1 for low-momomentum mass-separated particle is planned to be construct. After the construction, K1.1 will be operated in time shared with high-p beam line. SKS was moved to K1.1 are for future strangeness nuclear physics experiments.

In the SX operation from January 2018, 51kW beam power with 2.0 sec beam spill length in 5.2 sec. duration was achieved. Intensity and purity of $K^-$ beam at 51 kW ( 5.4E+13 ppp) are as follows,

- K1.8
  - 325k/spill 82.5% (1.8 GeV/$c$, E07)
  - 700k/spill 44% (1.8 GeV/$c$, E05-pilot)
- K1.8BR
  - 400k/spill 20% (1.1 GeV/$c$)
  - 9k/spill 3% (0.7 GeV/$c$)
- K1.1 (design)
  - 180k/spill (1.1 GeV/$c$)

Beam power of 51kW is limited by the present production target. New production target which is allowable to 90kW will be fabricated in 2018 and installed in 2019.

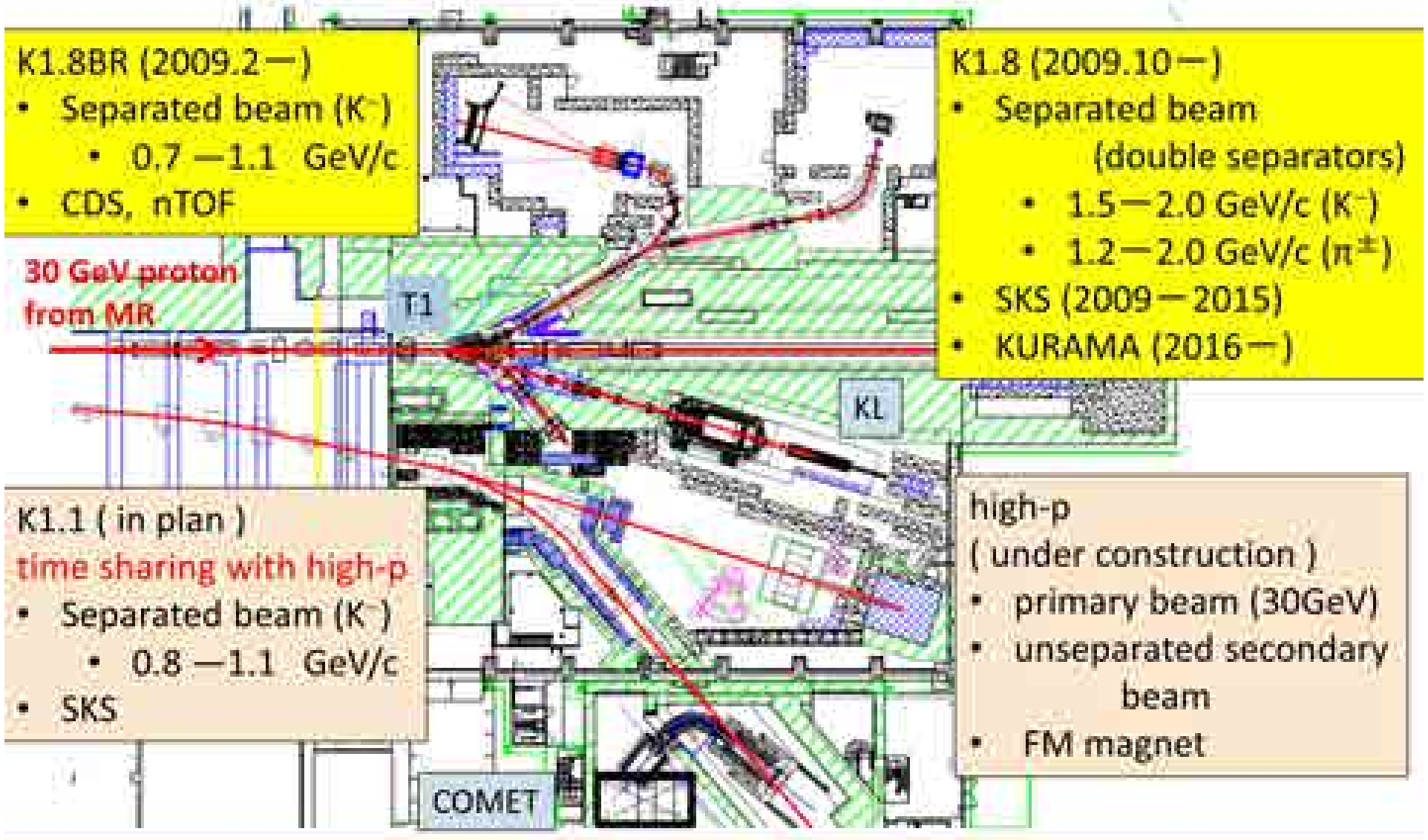

Figure 1: Beam lines of the present and near future Hadron Hall.

## 2 Physics results of the conducted experiments

Almost 10 years has passed since the start of the beam operation, although long term shutdown of the operation happened ( ~1 year for East Japan Earthquake and ∼2 years for the radiation incident at Hadron Hall). The conducted experiments and physics results are briefly summarised.

**E19: Search for $\Theta^+$ via the $\pi^- + p \to K^- + X$ reaction (K1.8)**
A pentaquark, $\Theta^+$, was searched for via the $\pi^- + p \to K^- + X$ reaction on a liquid $H_2$ target at 2.0 GeV/$c$ with the missing-mass resolution of 2.0 MeV (FWHM) No peak structure was observed[1, 2]. Upper limit of the production cross section was obtained to be 0.28$\mu$b/sr (90% C.L.) Assuming spin and parity of $\Theta^+$, upper limits of the widths were also obtained to be 0.36 MeV and 1.9 MeV for $1/2^+$ and $1/2^-$, respectively.

**E10: Neutron-rich $\Lambda$ hypernucleus, ${}^{6}_{\Lambda}$H (K1.8)**
Neutron-rich $\Lambda$ hypernuclei are important and interesting in studying $\Lambda$ potential in neutron-rich nuclear matter such as core of neutron stars. ${}^{6}_{\Lambda}$H, of which candidates had been reported by FINUDA experiment[3], was searched for using double charge exchange $(\pi^-, K^+)$ reaction on $^6$Li. No peak was observed. Upper limit of 1.2nb/sr (90% C.L.) was reported in the first paper[4]. Since no event was seen in the bound region and near threshold by the improved analysis, the upper limit reduced to be 0.56 nb/sr. Thus, not only bound states but also the resonance ones seems not to exist [5]. The existence of the ${}^{6}_{\Lambda}$H is now in discussion. Studies on neutron-rich $\Lambda$ hypernuclei should be continued in future at HIHR beam line.

**E13: $\gamma$-ray spectroscopy of ${}^{4}_{\Lambda}$He and ${}^{19}_{\Lambda}$F (K1.8)**
The $\gamma$-ray corresponding to $1^+ \to 0^+$ transition of ${}^{4}_{\Lambda}$He was measured with Hyperball-J detector in coincidence with the ${}^{4}_{\Lambda}$He production via the $(K^-, \pi^-)$ reaction on a liquid He target at 1.5 GeV/$c$[6]. The $1^+$-$0^+$ level spacing of ${}^{4}_{\Lambda}$He was determined to be 1.406±0.002(stat.)±0.002(syst.) MeV, which is much larger than that of mirror hypernucleus, ${}^{4}_{\Lambda}$H, (1.09±0.02 MeV). Thus a large charge symmetry breaking (CSB) was observed in the $1^+$-$0^+$ level spacing. Furthermore, $\Lambda$ binding energies are very similar for $1^+$ state, while they are very different for $0^+$ state, taking $0^+$ ground state mass from measured the emulsion data. Therefore CSB is spin-dependent, which suggests the effect of $\Lambda N$-$\Sigma N$ mixing and importance of three-body force via this mixing. The ground state mass of ${}^4$H was measured at MANI-C and was consistent with the emulsion data[7]. The level spacing of ${}^{4}_{\Lambda}$H will be measured near future at K1.1 beam line (E63 experiment). It is also important to measure ${}^{4}_{\Lambda}$ ground state mass by independent methods with emulsion, which can be done at HIHR beam line future.

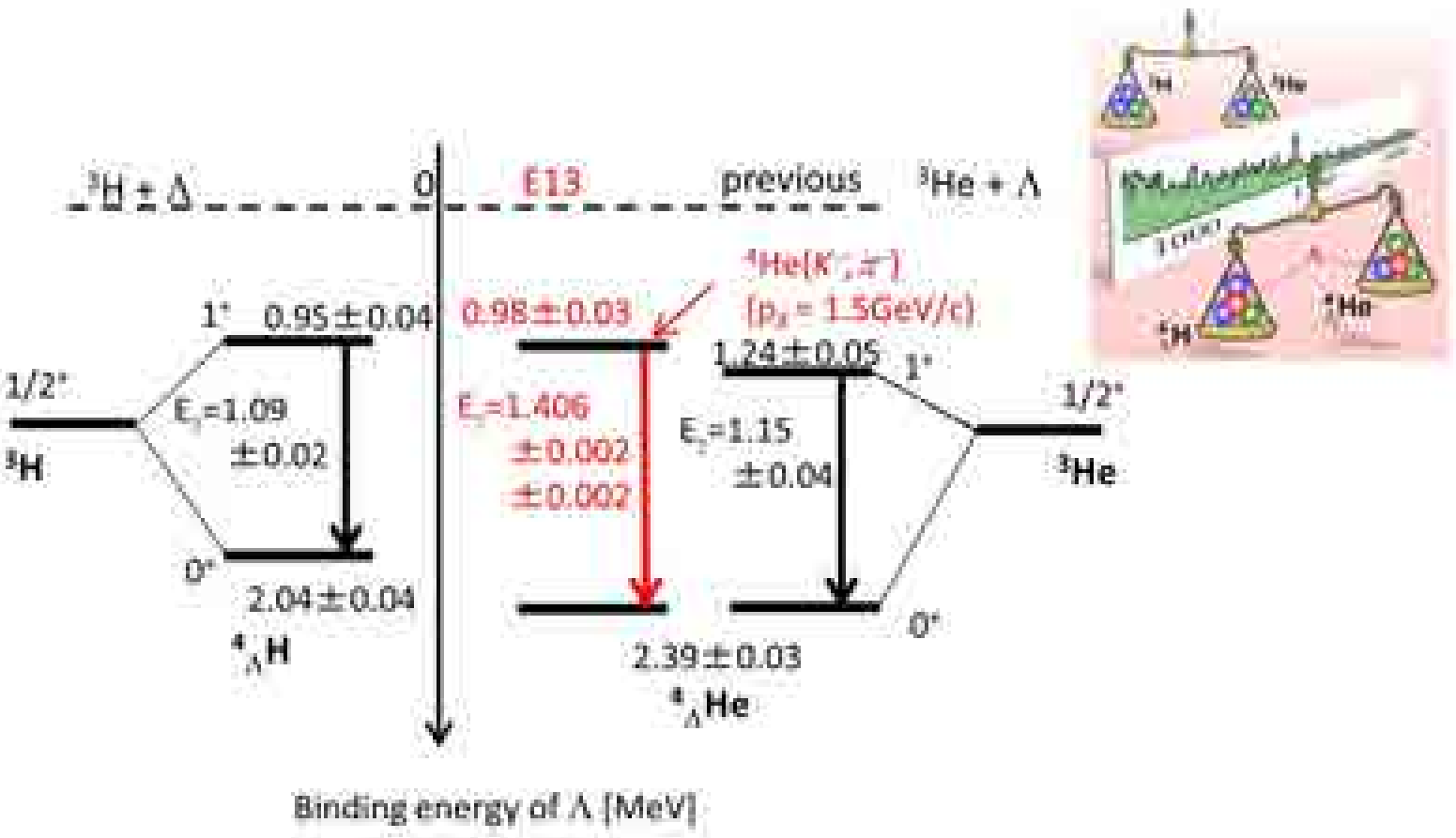

Figure 2: Level scheme of mirror hypernuclei ${}^{4}_{\Lambda}$H and ${}^{4}_{\Lambda}$He.

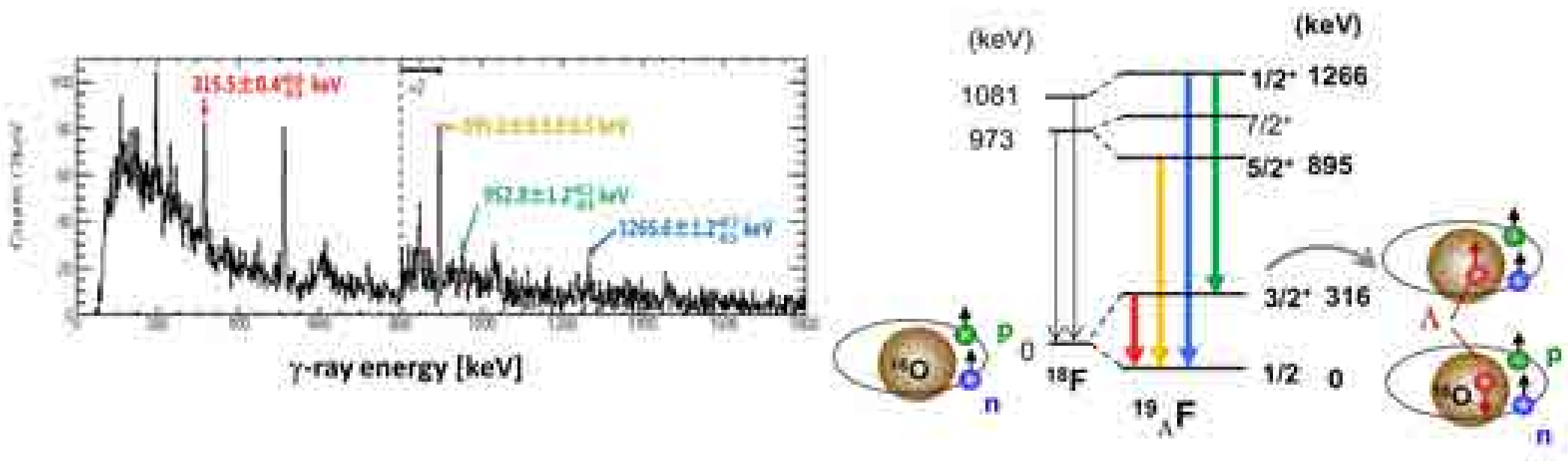

Figure 3: (Left)The *gamma*-ray spectrum for ${}^{19}_{\Lambda}$F. (Rgiht) The reconstructed level scheme.

The level structure of *sd*-shell hypernucleus, ${}^{19}_{\Lambda}$F, was determined for the first time by measuring $\gamma$-rays in coincidence with the hypernuclear production via the $K^-, \pi^-)$ reaction on liquid $CF_4$ target at 1.8 GeV/$c$[8]. Four *gamma*-rays were observed and the level scheme was reconstructed as shown in Fig.3. The ground state doublet spacing, which was determined to be 315.5±0.4(stat.)$^{+0.6}_{-0.5}$(syst.), is well described by theoretical models based on the existing *s*- and *p*-shell hypernuclear data.

**E05: Spectroscopy of ${}^{12}_{\Xi}$Be via the ${}^{12}$C$(K^-, K^+)$ reaction (K1.8)**

A missing-mass spectrum for the ${}^{12}$C$(K^-, K^+)$ reaction at 1.8 GeV/$c$ was measured using SKS with the best resolution of 5.4 MeV (FHWM) and high statistics, so far. Although a peak structure was not observed, there is an excess of events in the bound region of $\Xi^-$, which strongly suggests the existence of bound states. Fitting of the spectrum assuming single or double peaks plus $\Xi$ quasi-free production and background gives relatively deep binding energy[9]. New measurement with much better resolution using S-2S spectrometer is planned in future (E70).

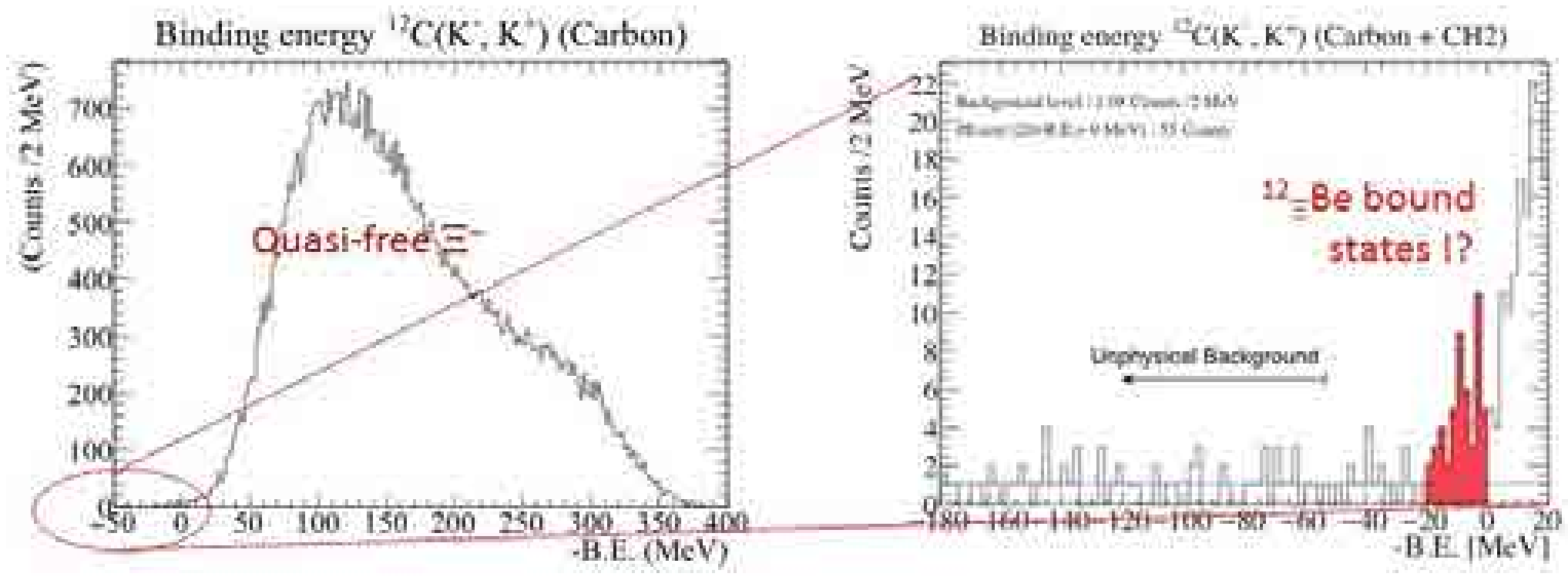

Figure 4: The missing-mass spectrum for $^{12}\mathrm{C}(K^-, K^+)$ reaction.

**E07: Study of double strangeness system by hybrid emulsion method (K1.8)**
Double strangeness nuclei, such as double-$\Lambda$ and $\Xi$ nuclei, are searched for in the emulsion by following $\Xi-$ produced via the $(K^-, K^+)$ reaction on a diamond target and identified by K1.8 beam and a newly installed KURAMA spectrometers. Beam exposure to total 118 emulsion modules were successfully completed by June 2017. Photographic development finished by February 2018. Scanning of the emulsion is in progress. Several candidate events for double strangeness nuclei were found, although their nuclear species were not yet uniquely identified. We expect 10,000 $\Xi^-$ stop events, 100 double strangeness nuclei, and 10 well-identified nuclei among them. X-rays were also measured for the first time to obtain $\Xi-$ potential by level shift and widths of $Xi$-atoms such as Br or Ag.

**E27: Search for $K^-pp$ bound state via the $\mathrm{d}(\pi^+, K^+)$ reaction at 1.69 GeV/$c$ (K1.8)**
In this experiment, $K-pp$ is produced via the $\Lambda(1405)$ doorway and searched for by missing-mass of the $\mathrm{d}(\pi^+, K^+)$ reaction. Two protons in final state were measured in coincidence with Range Counter Arrays to suppress huge background of quasi-free e $Y*$ production processes. A $K-pp$-like structure was observed in the missing-mass by selecting $\Sigma^0 p$ final state as shown in Fig.5 left. If this structure is true, it is deeply bound state. Obtained binding energy and width are 95 $^{+18}_{-17}$(stat.) $^{+30}_{-21}$(syst.) MeV and 162 $^{+87}_{-45}$(stat.) $^{+66}_{-78}$(syst.) MeV, respectively[10].

**E15: Search for $K^-pp$ bound state via the $K^-+^3\mathrm{He} \to n+\Lambda p$ (K1.8BR)**
Another experiment to search for $K^-pp$ bound state was carried out using 1 GeV/$c$ $K^-$ beam at K1.8BR beam line. In this experiment, $K^-pp$ is produced via the $^3\mathrm{He}(K^-, n)$ reaction with the forward neutron TOF counter of 17m flight path and decayed particles are measured by a large acceptance cylindrical detector system. Thus, both missing-mass and invariant-mass spectroscopy can be applied. Figure 5 right shows the $\Lambda p$ invariant-mass of $\Lambda p + n$ final state, where $n$ is identified by the missing-mass of the reaction. The position of the peak above the $K^-+p+p$ threshold depends on the missing neutron angle, which indicates quasi-free process. On the other hand, a peak below the threshold with no position dependence on the angle is a signal of the bound state. However this peak shows a shallow bound state of ~40 MeV binding energy and ~70 MeV width, much different result with that of E27. Further studies both experimentally and theoretically might be necessary.

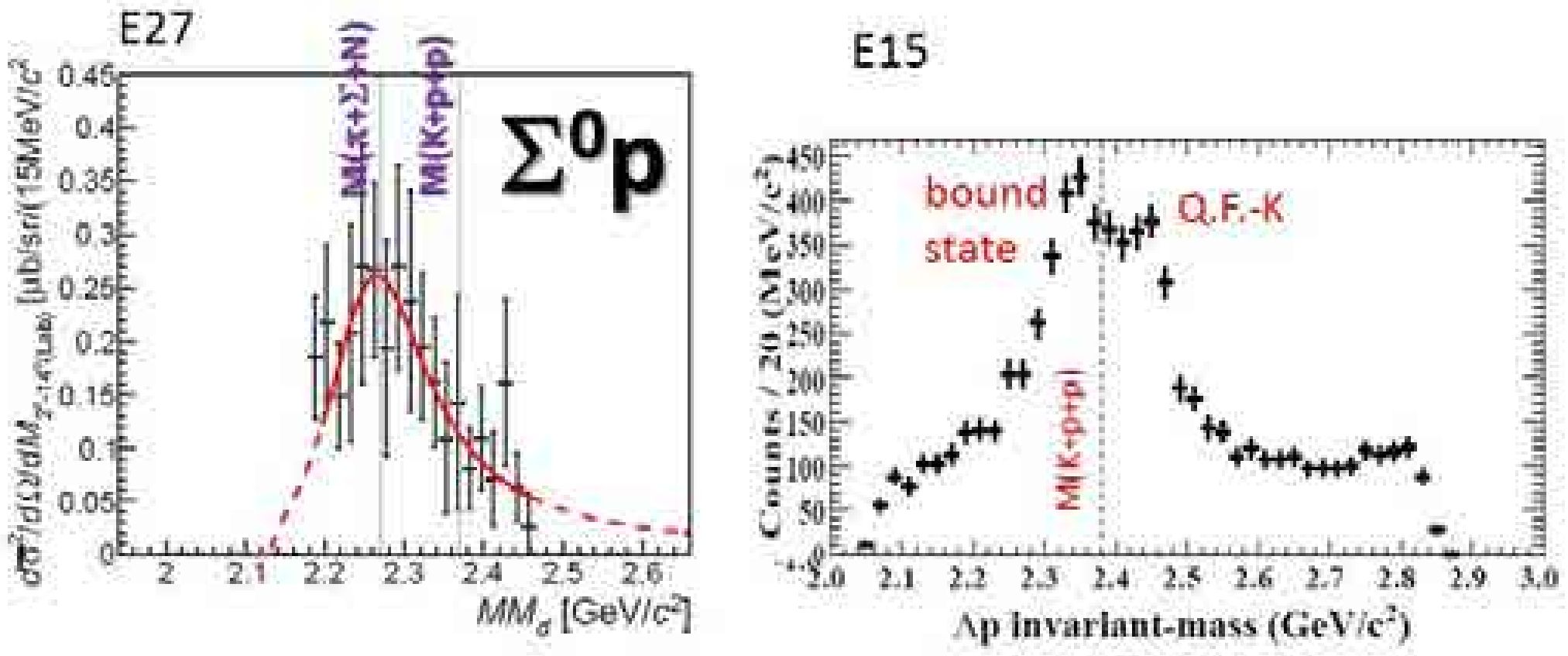

Figure 5: (Left) A missing-mass spectrum for the d$(\pi^+, K^+)$ after selecting the $\Sigma^0 p$ final state obtained in E27. (Right) The $\Lambda p$ invariant-mass spectrum for the $K^- + {}^3\text{He} \rightarrow n + \Lambda p$ obtained in E15.

**E31: Study of $\Lambda(1405)$ in the $K^- d \rightarrow n\Sigma\pi$ reaction (K1.8BR)** The $\Lambda(1405)$ structure is in discussion whether 3 quarks baryon or baryon-meson bound state. It also has two poles of $\bar{K}N$ and $\pi\Sigma$. The structure of $\Lambda(1405)$ is studied in the $K^- d \rightarrow n\Sigma\pi$ reaction by separating the isospin channels of the final $\Sigma\pi$ states. Data-taking was completed in February 2018 and analysis is in progress.

## 3 Near future experiments at the present Hadron Hall

The understanding of the nuclear force both qualitatively and quantitatively is one of the important subjects in nuclear physics. Meson exchange model successfully describes medium to long-range attraction, but repulsive core is phenomenologically introduced in the short-range part. In the quark based picture, on the other hand, the short-range part could be explained naturally by Pauli effect and color-magnetic interaction between quarks. Those pictures will be confirmed by the following experiments.

**E40: $\Sigma^{\pm} p$ scattering (K1.8)**
The $\Sigma N$ interaction is studied by the direct scattering experiment using modern detectors with high-rate abilities to ac hive high statistics. A $\Sigma^+ p$ channel is a Pauli forbidden channel in the $SU(3)_f$ baryon-baryon interaction. A large repulsion in this channel is experimentally co firmed. $\Sigma^- p$ scattering and $\Sigma^- p \rightarrow \Lambda n$ conversion channels will be also measured. Commissioning just started form January 2018.

**E42: Search for $H$-dibaryon (K1.8)**
There is no repulsive core but attractive one in $SU(3)_f$ singlet channel due to no quark Pauli effect and strong color-magnetic interaction. Therefore *uuddss* six-quark state, $H$-dibaryon is expected in this channel as predicted by Jaffe]citeJaffe. Recent lattice QCD calculations also predict $H$-dibaryon near the threshold. Thus discovery of $H$ is the evidence of an existence of the attractive core and the measurement gives the strength of the $SU(3)_f$ singlet channel. E42 will search for $H$-dybaryon will be searched and determine the strength of the $SU(3)_f$ singlet channel using a large acceptance hyperon spectrometer based on a TPC under preparation.

Low-energy $\bar{K}N$ interaction is another subject in the present Hadron Hall. Two experiments on X-ray spectroscopy of kaonic-atoms are planned at K1.8BR. In E62, 3d→2p X-rays from kaonic helium 3 and 4 will be measured in ultra-high resolution with TES detector to conclude deep or shallow of K-nucleus potential problem. Data-taking is scheduled in June 2018. Another one, E57, aims to measure X-rays from kaonic deuterium to disentangle isospin I=0 and 1 components of $\bar{K}N$ interaction.

## References

[1] K. Shirotori *et al.*, Phys. Rev. Lett. **109**, 132002 (2012).

[2] M. Moritsu *et al.*, Phys. Rev. C**90**, 035205 (2014).

[3] M. Agnello *et al.*, FINUDA Collaboration, Phys. Rev. Lett. **108**, 042501 (2012).

[4] H. Sugimura *et al.*, Phys. Lett. B729, 39 (2014).

[5] R. Honda *et al.*, Phys. Rev. C**96**, 014005 (2017).

[6] T. O. Yamamoto *et al.*, Phys. Rev. Lett. **115**, 22501 (2015).

[7] A. Esser *et al.*, Phys. Rev. Lett. **114**, 232501 (2015).

[8] S. B. Yang *et al.*, Phys. Rev. Lett. bf 120, 132504 (2018).

[9] T. Nagae *et al.*, PoS, INPC2016, 038 (2017).

[10] Y. Ichikawa *et al.*, Prog. Theor. Exp. Phys. 2015,. 021D01 (2015).

[11] R. L. Jaffe, Phys. Rev. Lett. **38**, 195 (1977).

# Strangeness nuclear physics at $\overline{\text{P}}$ANDA in a nutshell

**Josef Pochodzalla[1,2], Patrick Achenbach[1,2], Sebastian Bleser[1], Michael Bölting[1], Falk Schupp[1], Marcell Steinen[1]**
**on behalf of the $\overline{\text{P}}$ANDA Collaboration**

[1]Helmholtz Institute Mainz, Johannes Gutenberg University, 55099 Mainz, Germany

[2]Institute for Nuclear Physics, Johannes Gutenberg University, 55099 Mainz, Germany

$\overline{\text{P}}$ANDA at FAIR will address the physics of strangeness in nuclei by several novel and unique measurements. These studies are only made possible by the one-of-a-kind combination of the stored antiproton beam at FAIR and the modular $\overline{\text{P}}$ANDA detector which will be complemented by a germanium detector array. The latter was made available by a cooperation with the NUSTAR collaboration.

- $\overline{\text{P}}$ANDA also offers the unique possibility to search for X-rays from very heavy hyperatoms such as e.g. $\Xi^-$-$^{208}$Pb. This will complement experiments at J-PARC which attempt to measure X-rays in light and (possibly) medium-heavy nuclei by the J-PARC E03 and E07 experiments. The measurement at $\overline{\text{P}}$ANDA will allow to constrain the real and imaginary potential of $\Xi^-$ hyperons in the neutron skin of the lead nucleus.
- $\overline{\text{P}}$ANDA will extend the studies on double hypernuclei by performing high resolution $\gamma$-spectroscopy of these nuclei for the first time. Thus, $\overline{\text{P}}$ANDA complements measurements of ground state masses of double hypernuclei in emulsions at J-PARC conducted by the E07 Collaboration [1, 2] or the production of excited resonant states in heavy ion reactions which may, for example, be explored in the future by the CBM Collaboration. Together, these measurements will provide a unrivaled information on the structure of double $\Lambda\Lambda$ hypernuclei.
- Furthermore, the exclusive production of hyperon-antihyperon pairs close to their production threshold in $\overline{\mathrm{p}}$ - nucleus collisions offers a unique and yet unexplored opportunity to elucidate the behaviour of antihyperons in cold nuclei which is intimately related to the short-range part of the baryonic interaction.

In all these studies, the production of low momentum $\overline{\Lambda}$ within nuclei or slow $\Xi^-$ hyperons which can eventually be stopped in a secondary target material play an essential role.

## 1 $\overline{\text{P}}$ANDA - a unique antihyperon - hyperon factory

A negatively charged, strongly interacting particle which has been stopped in a target may be captured into an outer atomic orbit thus forming an exotic atom. This atom will emit Auger electrons and characteristic X-rays whilst cascading down. The time to cascade down has been measured for $\Sigma^-$ hyperons in emulsions and was found to be $<10^{-12}$ s [3]. This is more than a factor of 100 smaller than the typical hyperon lifetime. If this strongly interacting, negative charged particle reaches an orbit which has a significant overlap with the nucleus, the intensity of the X-ray transition will be reduced by the strong nuclear capture and the energy levels are shifted and broadened with respect to the electromagnetic situation. The broadening usually terminates the atomic cascade and the hyperon may be captured by the strong interaction in the nucleus. As a consequence, the strong interaction shift and broadening is in most cases significant and directly measurable only for this last level. Higher transitions are much less affected by the strong interaction.

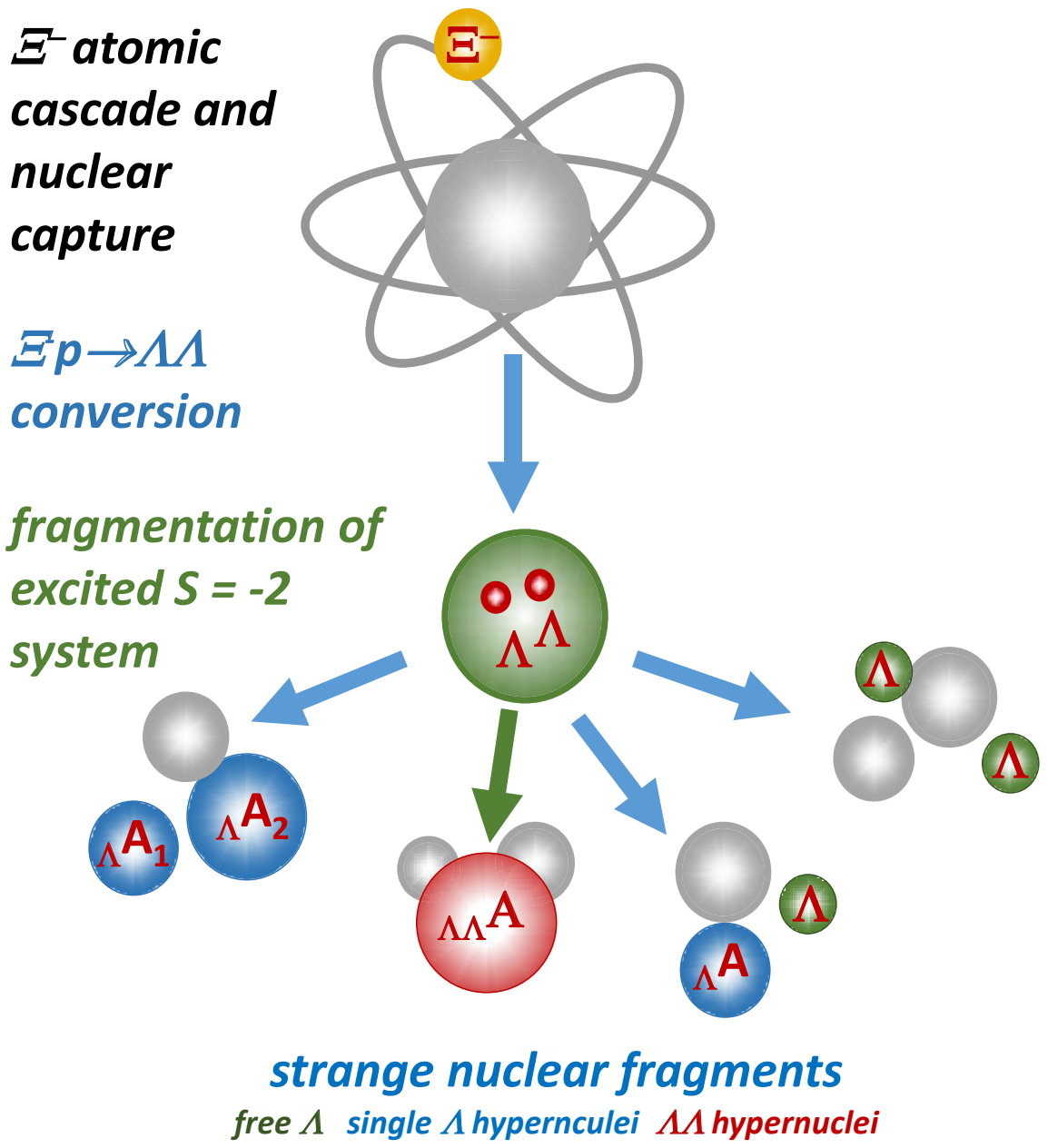

**Figure 1:** The capture of stopped $\Xi^-$ hyperon into an atomic orbit is the starting point for the formation of a $\Xi^-$ atom which itself may convert into an excited $\Lambda\Lambda$ hypernucleus system. Possible fragmentation channels after the capture and conversion of a stopped $\Xi^-$ hyperon are indicated in the bottom part.

In a second step, $\Xi^-$-hyperatoms can serve as a doorway to the production of $\Lambda\Lambda$-hypernuclei. Indeed, the simultaneous production and implantation of two individual $\Lambda$'s in a nucleus in two distinct processes is not feasible. However, $\Lambda\Lambda$-hypernuclei can be produced in a 'controlled' way by the conversion of a $\Xi^-$ hyperon within the nucleus into two $\Lambda$ hyperons via the reaction $\Xi^- p \to \Lambda\Lambda$. It is one of the lucky coincidences in physics that the conversion of a $\Xi^-$ and a proton into two $\Lambda$'s releases – ignoring binding energy effects – only 28 MeV of excitation energy. Because of this low value there is a significant chance of typically a few percent that both created $\Lambda$ hyperons stick to the same nucleus [4, 5, 6]. Of course, also events with two single hypernuclei or free $\Lambda$'s are produced in this deexcitation state (Fig. 1). Eventually, the resulting hyperfragments may be in excited states which de-excite via $\gamma$ emission.

Unfortunately, $\Xi^-$ hyperons produced in reactions with hadron beams typically have rather high momenta and the direct trapping of the $\Xi^-$ in the *same* nucleus where it has been produced is rather unlikely. Therefore, the capture of $\Xi^-$ hyperons usually proceeds in two (or even three) steps. Using energetic hadronic beams, a $\Xi^-$ (together with its associated antistrange particles) is produced in a primary target. In a second step this $\Xi^-$ is slowed down in a dense, solid material (e.g. a nuclear emulsion) and forms a $\Xi^-$ atom [8]. As a consequence, the principle design of any double hypernucleus experiment with $\Xi^-$ conversion is determined by the interplay between the finite $\Xi^-$ mean lifetime of $\tau_{\Xi^-}$=1.64·10$^{-10}$ s and the time it takes to stop a hyperon inside the detection system.

As representative examples for the $\Lambda\Lambda$-hypernucleus and $\Xi^-$-hyperatom production process in $\overline{\text{P}}$ANDA, Fig. 2 shows the probability for stopping, decay and hadronic interactions of $\Xi^-$ hyperons in boron and lead absorbers. The dashed lines show the case when the $\Xi^-$ hyperons are produced inside the absorber whereas the solid lines assume a point of creation 10 mm apart from the absorber. The first situation is relevant for e.g. $\Xi^-$ production in $(K^-,K^+)$ double strangeness exchange reactions, while the latter case describes a more realistic configuration for the $\overline{\text{P}}$ANDA experiment. Obviously only $\Xi^-$ hyperons with momenta less than about 800 $\mathrm{MeV}/c$ have a significant chance for being stopped (green solid line). The required thickness of the moderator material is typically in the range of 10 mm (see upper scales in Fig. 2). The difference between the dashed and solid green lines reflects the decay in flight of very slow $\Xi^-$ hyperons and underlines the need to position the absorbing secondary target material as close as possible to the $\Xi^-$ production point.

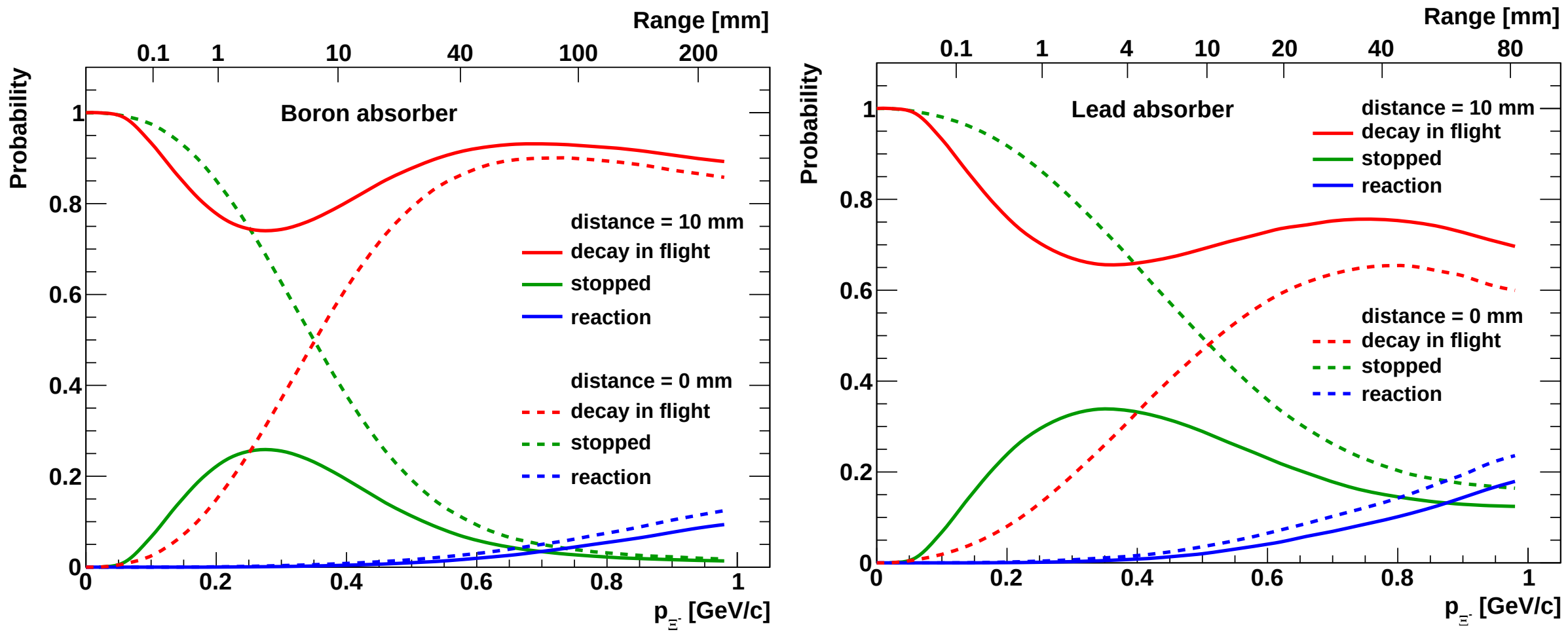

**Figure 2:** Probability for stopping (green lines), decay (red lines) and nuclear reaction (blue lines) for $\Xi^-$ hyperons produced inside boron (left) and lead (right). The $\overline{\text{P}}$ANDA experiment will employ B and Pb as secondary target materials for the $\Lambda\Lambda$-hypernucleus and $\Xi^-$-hyperatom studies, respectively. In case of the dashed lines the $\Xi^-$ hyperons were generated inside the absorber, the solid lines show the same quantities if the $\Xi^-$ hyperon is produced 10 mm apart from point of entrance into the absorber. The latter geometry describes the situation at $\overline{\text{P}}$ANDA (from Ref. [7]).

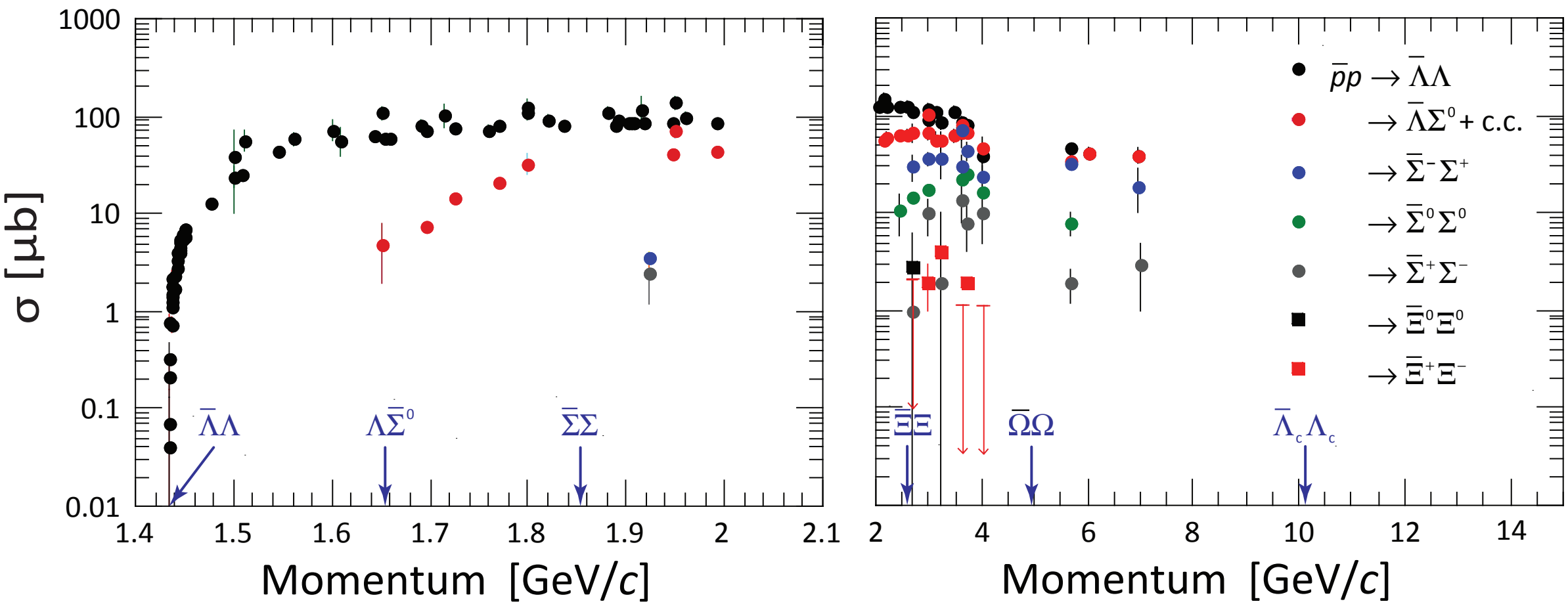

**Figure 3:** Total cross sections for $\overline{\text{p}}\text{p} \to \text{Y}\overline{\text{Y}}$ reactions in the momentum range of the HESR. The upper limits in red refer to the $\overline{\Xi}^+\Xi^-$-channel. The arrows pointing to the momentum axis indicate the threshold momenta for the different hyperon families (from Ref. [9]).

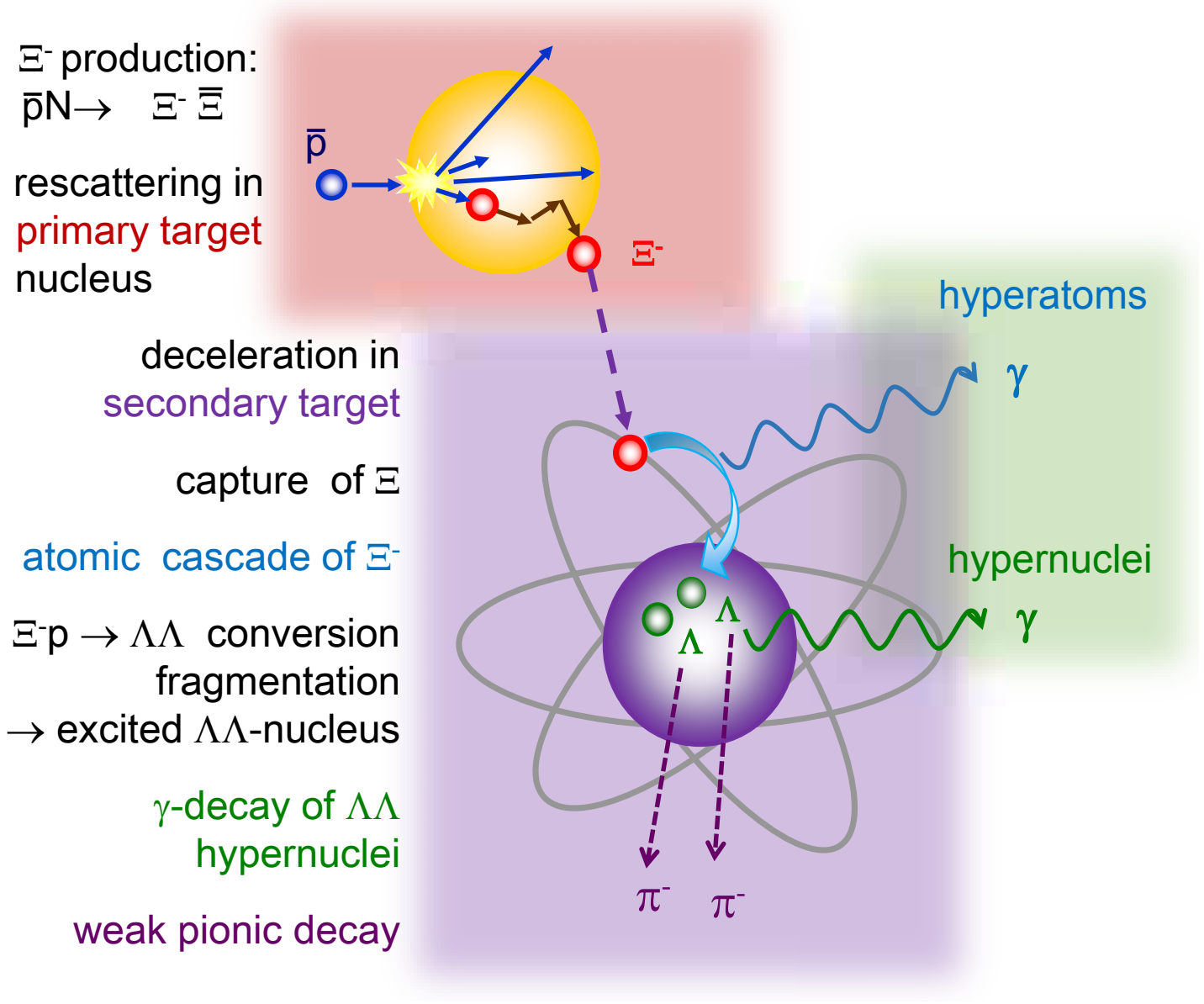

**Figure 4:** Production scheme of $\Xi^-$-hyperatoms and $\Lambda\Lambda$-hypernuclei at $\overline{\text{P}}$ANDA.

Using the $(K^-,K^+)$ double strangeness exchange reaction, $\Xi^-$ are produced with typical momenta of about 500 MeV/$c$ [10]. In only a few percent of the reactions the $\Xi^-$ hyperon is produced and captured in the same target nucleus. In most cases ($\sim$80%), the $\Xi^-$ escape from the initial target nucleus. Those $\Xi^-$ hyperons need to be stopped in a second step. The main advantage of the $(K^-,K^+)$ reaction is the fact that the outgoing $K^+$ can be used as a tag for this reaction. The main drawback, however, is the low kaon beam intensity because of its low lifetime and hence the need for thick secondary targets which is identical with or very close to the primary target (compare the green dashed and solid lines in Fig. 2).

At this point it is interesting to note that the $\overline{\Xi}$ was discovered in antiproton-proton interactions at 3 GeV/$c$ [11]. Because of the additional energy released in the annihilation process, $\Xi^-$ hyperons with relatively low laboratory momenta are produced in the $\overline{p}+p \to \Xi\overline{\Xi}$ reaction. The main advantage of antiprotons compared to the kaon induced production is the fact that the antiproton is stable and can thus be retained in a storage ring. Combined with the large cross sections for the production of associated hyperon-antihyperon pairs (see Fig. 3), this enables rather high luminosities even with very thin primary targets. Reactions close to the $\Xi\overline{\Xi}$ threshold also minimize the production of associated particles as well as the number of secondary particles produced in other nuclear reactions.

Even lower momentum $\Xi^-$ hyperons can be produced via the $\overline{\text{p}}\text{p} \to \Xi^-\overline{\Xi}^+$ or $\overline{\text{p}}\text{n} \to \Xi^-\overline{\Xi}^0$ reactions *within* a complex nucleus where the produced $\Xi^-$ can re-scatter and thus decrease its momentum [12] (see Fig. 4). After stopping in a secondary target, a bound $\Xi^-$ system may be formed, allowing the study of both, $\Xi^-$ hyperatoms and $\Lambda\Lambda$ hypernuclei. Also the $\overline{\Xi}$ can here be used as a tag of this reaction if it is not absorbed in the primary target nucleus. Unlike in the $(K^-,K^+)$ reaction, the primary and secondary targets can be separated by a few mm, thus allowing a wide range of materials for the secondary target.

## 2 $\Xi^-$ Hyperatoms at $\overline{\text{P}}$ANDA

Also $\Xi^-$ or $\Xi^0$ hyperons may play a crucial role in the interior of neutron stars. Unfortunately, the behaviour of $\Xi^-$ hyperons in nuclear matter is experimentally not well constrained. However, we have several tools at hand to approach this problem. Experimentally, the interaction of $\Xi^-$ hyperons in nuclei can be studied by $\Xi^-$-hypernuclei, $\Xi^-$-atoms, $\Xi^-$-proton scattering or $\Xi^-$-proton final state interaction in e.g. energetic heavy ion reactions and by $\Xi^-\overline{\Xi}^+$ production in $\overline{p}$+nucleus collisions (see Fig. 5). However, the present experimental knowledge on $\Xi^-$ hyperons in nuclear systems is rather disillusioning. At $\overline{\text{P}}$ANDA we will perform unique studies of heavy $\Xi^-$ atoms, thus probing the $\Xi^-$ interaction in a neutron-rich nuclear surrounding.

All negative particles that live long enough to be captured by an atomic nucleus may form an exotic atom. Compared to electrons, the higher mass of these charged particles ($\mu^-$, $\pi^-$, $K^-$, $\overline{\text{p}}$, $\Sigma^-$, $\Xi^-$, $\Omega^-$) causes a shrinkage of the radial spatial probability distribution which is inversely proportional to the reduced mass of the exotic system and the charge of the core nucleus. Because the wave function of the bound particle overlaps with the finite size nucleus, nuclear effects come into play. In case of hadrons, the strong interaction modifies the energy of such atomic levels as well as their lifetime.

At J-PARC, light and medium heavy $\Xi^-$-hypernuclei and $\Xi^-$-hyperatoms will be studied in the coming years. Because of the spatial separation between the primary and secondary targets, the $\overline{\text{P}}$ANDA experiment offers the unique possibility to study very heavy $\Xi^-$ atoms like $\Xi^-$-$^{208}$Pb and thus the interaction of $\Xi^-$ hyperons with the neutron skin. The left part of Fig. 6 shows a pictorial sketch of the relevant level scheme of the $\Xi^-$-$^{208}$Pb atom. The (n=10,l=9) level is the lowest level not yet fully absorbed by the strong interaction. Thus, the (n=10,l=9)$\rightarrow$(9,8) transition is the last transition which can be observed with a reasonable $\gamma$ branching ratio.

The right part of Fig. 6 summarizes in a more quantitative way some relevant properties of the $^{208}$Pb nucleus and of the (n=9,l=8) atomic $\Xi^-$ level which is the closest to the nuclear surface. These calculations were performed with a program provided by Eli Friedman [8, 13]. The neutron and proton distributions are described by a two-parameter Fermi distribution

$$\rho_{p,n}(r) = \frac{\rho_{0p,0n}}{1+\exp((r-R_{p,n})/a_{p,n})}. \tag{1}$$

For protons and neutrons the same diffuseness parameter $a_n$=$a_p$=0.475 fm was adopted. The proton radius parameter of $R_p$=6.654 fm was constrained by reproducing the experimental rms charge radius of $^{208}$Pb. The quantity most frequently used to define the relative radial extent of the neutrons with respect to the protons is the difference in rms radii. The lower part of the right panel in Fig. 6 shows the density distributions for protons and neutrons for $^{208}$Pb as a function of the radial distance from the center of the nucleus. For protons we extract from this distribution a density rms radius of $r_p$=5.4481 fm, close to the experimental rms radius of 5.5012 fm (see Ref. [14] for an updated list of experimental nuclear ground state charge radii). For neutrons a radius parameter of 6.966 fm was used as input, leading to a rms radius of $r_n$=5.677 fm. As expected, the difference of $r_n$-$r_p$= 0.229 fm agrees well with measured values of the neutron skin thickness of $^{208}$Pb.

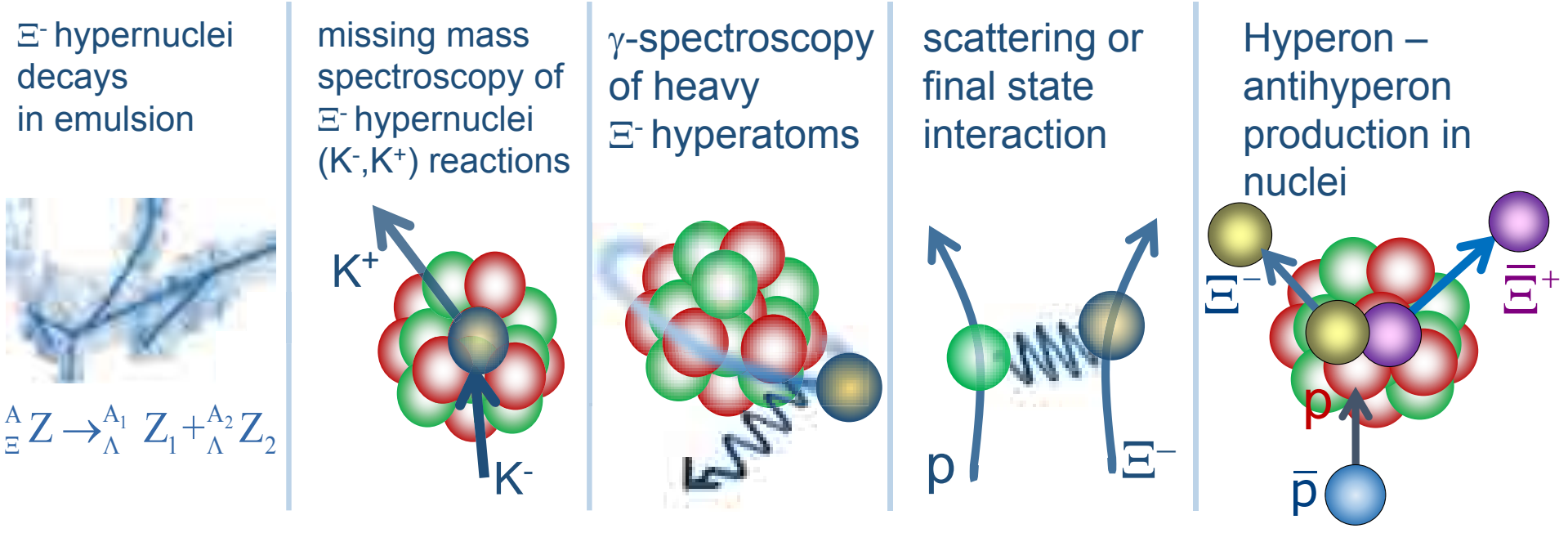

**Figure 5:** Possible processes to study the interaction of hyperons – here exemplarily illustrated for $\Xi^-$ hyperons – in nuclear interactions.

The blue curve in the top part of the right panel of Fig. 6 shows the squared wave function of $\Xi^-$ hyperons in the n=9 and l=8 state. The wave function peaks at 20.3 fm which coincides with the value of 20.326 fm expected from the Bohr radius $a_0 = \hbar^2/e^2\mu$ and the well known relation

$$\langle r \rangle = a_0[3n^2 - l(l+1)]/2Z. \qquad (2)$$

For the next higher state n=10, l=9 , which is relevant for determining the width via the relative intensities of subsequent transitions, the maximum lies at 25 fm. In such nuclei the orbiting hadron is closer to the nucleus than the innermost electrons. The system is therefore hydrogenic, with the electron cloud contributing an almost negligible screening effect, and the fundamental equations of the hydrogen atom are applicable.

The nuclear densities are an essential ingredient of the optical potential. As expected, the real part of the $\Xi^-$ potential shown in the center panel follows the nuclear density. The red curve

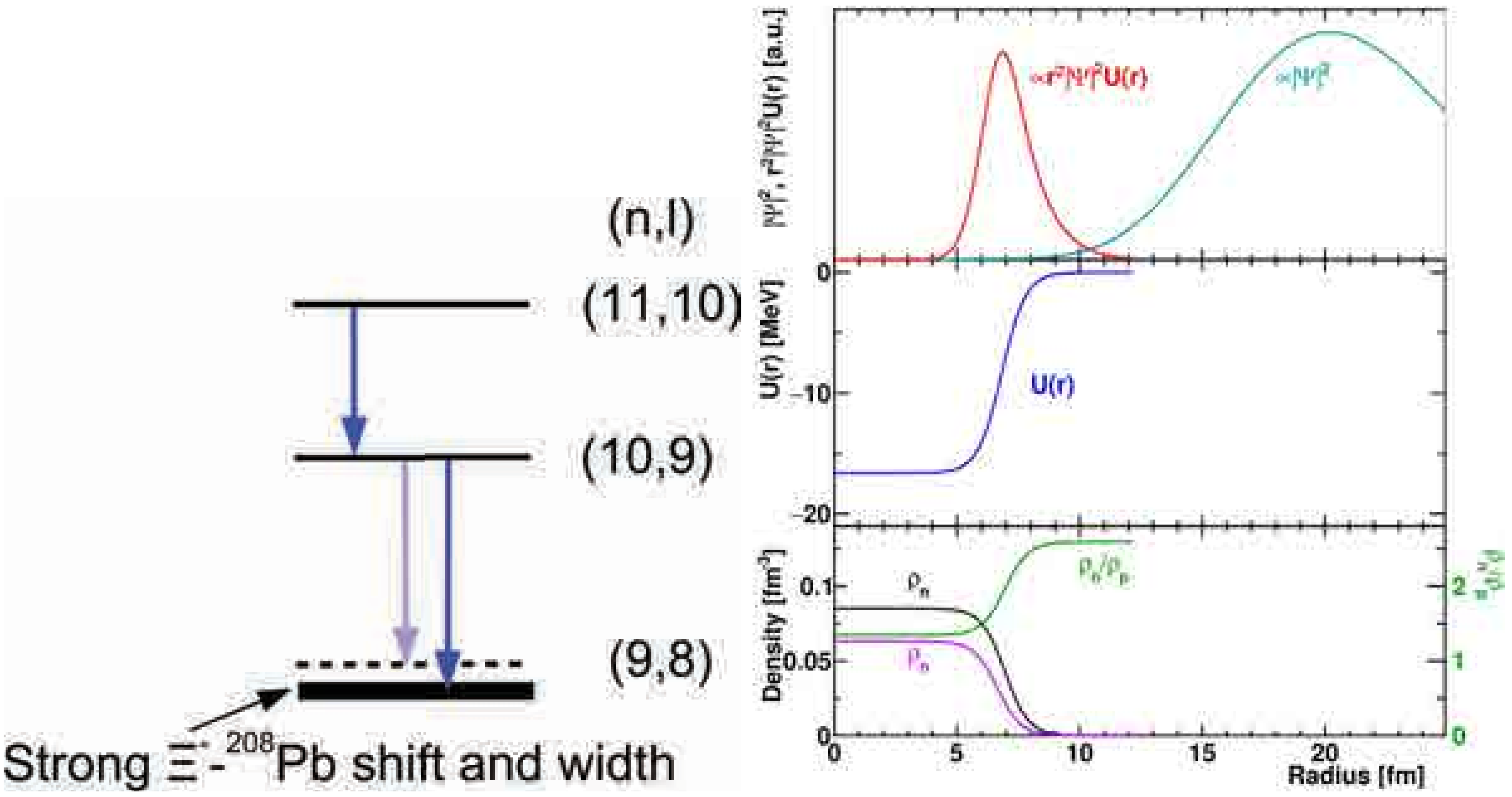

**Figure 6:** Left: If a strongly interacting, negatively charged particle reaches an orbit which has a significant overlap with the nucleus, the intensity of the X-ray transition will be reduced by the strong nuclear capture and the energy levels are shifted and broadened with respect to the electromagnetic situation marked by the dashed line. The broadening usually terminates the atomic cascade. Right: The lower part shows the density distribution for protons and neutrons for $^{208}$Pb. The green line is the neutron-to-proton density ratio. The real part of the nuclear potential for $\Xi^-$ hyperons is shown in the center panel. In the top panel the squared wave function of a $\Xi^-$ hyperon in the n=9 and l=8 state and their absorption probability in the nuclear periphery (arbitrary units) are displayed. The calculations were performed with a program provided by Eli Friedman [8, 13] (from Ref. [15]).

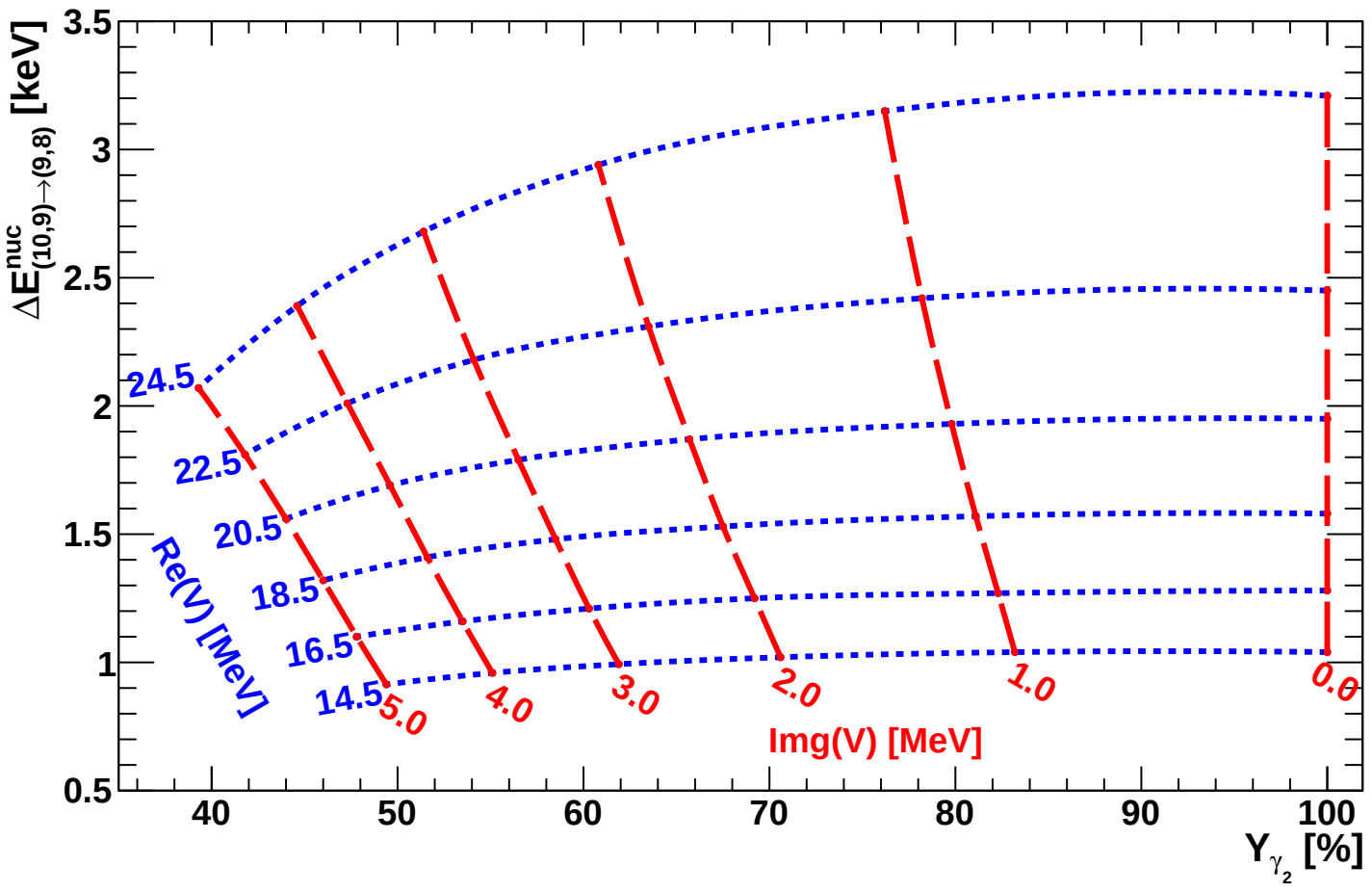

**Figure 7:** Mapping of the real and imaginary parts of the $\Xi^-$-$^{208}$Pb potential on the energy shift of the (n=10,l=9)→(9,8) transition and the relative yield $Y_{\gamma 2}$ of the (n=10,l=9)→(9,8) transition with respect to (11,10)→(10,9). The calculations were performed with a program provided by Eli Friedman [8, 13] (from Ref. [15])

in the upper panel shows the 'strong interaction strength' proportional to $r^2 ReU(r)|\Psi|^2$ which is relevant for the shift of the energy level. The average $\Xi^-$ radius weighted by this factor gives $\langle R_{\Delta E}\rangle$ = 7.16 fm and a weighted neutron-proton density ratio of $\langle \rho_n/\rho_p\rangle$ = 2.03. As expected, the n=10, l=9 state samples larger radii with $\langle R_{\Delta E}\rangle$ = 7.73 fm and , therefore, even more neutron rich matter with $\langle \rho_n/\rho_p\rangle$ = 2.23. Apparently, heavy $\Xi^-$-hyperatoms probe the nuclear periphery which is strongly influenced by the $\Xi^-$-neutron interaction. Complemented by the study of light $\Xi^-$-hyperatoms at J-PARC, $\overline{\text{P}}$ANDA will thus explore the isospin dependence of the $\Xi^-$-nucleus interaction.

While the measurement of the energy shift of a transition is in principle straightforward, the direct determination of the broadening of a state caused by the strong interaction – typically in the range of 1 keV – is in many cases hampered by the limited energy resolution of the used X-ray detectors. Fortunately, in some cases the measurement of relative X-ray yields of transitions preceding the last transition provides an indirect measurement of the width of the upper level (in Pb the (n=10,l=9) level) [16]. Shifts and widths (or relative yields) of the atomic levels caused by the strong interaction may then be expressed e.g. in terms of the real and imaginary parts of an optical potential as illustrated in Fig. 7 for the $\Xi^-$-$^{208}$Pb system.

# 3 $\Lambda\Lambda$ Hypernuclei

Because of the two-step production process, we can not perform spectroscopic studies of $\Lambda\Lambda$ systems via the analysis of two-body reactions as in single hypernuclei. Spectroscopic information on double hypernuclei can only be obtained via their decays. We can distinguish three different steps of the decay process (Fig. 8):

- Particle unstable states in the continuum may be studied in line with conventional two-particle correlation measurements (e.g. ref. [18]) in energetic heavy ion reactions. The large yield of hypernuclei expected to be produced by the CBM experiment at FAIR will allow to observe correlations between $\Lambda$-hypernuclei and $\Lambda$ hyperons or even between two light single hypernuclei.

- Double hypernuclei also offer a rich spectrum of particle-stable excited states [19]. The excitation spectrum of a double $\Lambda$ hypernucleus ${}^{\rm A}_{\Lambda\Lambda}$Z is strongly related to the spectrum of the conventional core nucleus ${}^{\rm A-2}$Z. Of course, the presence of the two $\Lambda$ hyperons will cause structural changes of the core nucleus which will be reflected in the excitation spectrum (see e.g. Ref. [20]). The $\overline{\rm P}$ANDA experiment aims to explore this part of the excitation spectrum for the first time. High-resolution $\gamma$-spectroscopy is essential to obtain relevant information on this region of the excitation spectrum [12].

- Once the ground state is reached, the weak decay of the hyperon(s) will initiate the emission of several particles. The determination of the ground state mass requires the knowledge of all masses of the various decay products and their kinetic energies. Hence unique identification of the final state particles and precise determination of their kinetic energies is mandatory. Nuclear emulsions are even today the only technique which meets all requirements for precision measurements of ground state masses.

Combining these three complementary methods we will have access to the complete level scheme of $\Lambda\Lambda$-hypernuclei in the future, which nicely demonstrates the necessity of cooperation between different scientific communities in this field. Complemented by hyperon-hyperon correlation studies in heavy ion collisions which will be discussed in the next section, these measurements will provide comprehensive information on the hyperon-hyperon interaction and on the role of $\Lambda\Lambda$ - $\Sigma\Sigma$ - $\Xi N$ mixing in nuclei [21].

In 2016 and 2017 the E07 experiment at J-PARC employed the (K$^-$,K$^+$) reaction to produce low momentum $\Xi^-$. These hyperons were subsequently stopped in nuclear emulsion plates. E07 is expected to provide precise information on the absolute ground state masses of a few ten new $\Lambda\Lambda$-hypernuclei [22]. It is however impossible to detect neutrons or $\gamma$-rays from intermediate particle unstable fragments with this method. As a consequence, the determination of the ground state mass of double hypernuclei is limited to light nuclei which decay exclusively into charged particles. Therefore, emulsion studies are limited to light nuclei with mass numbers A $\leq$ 12.

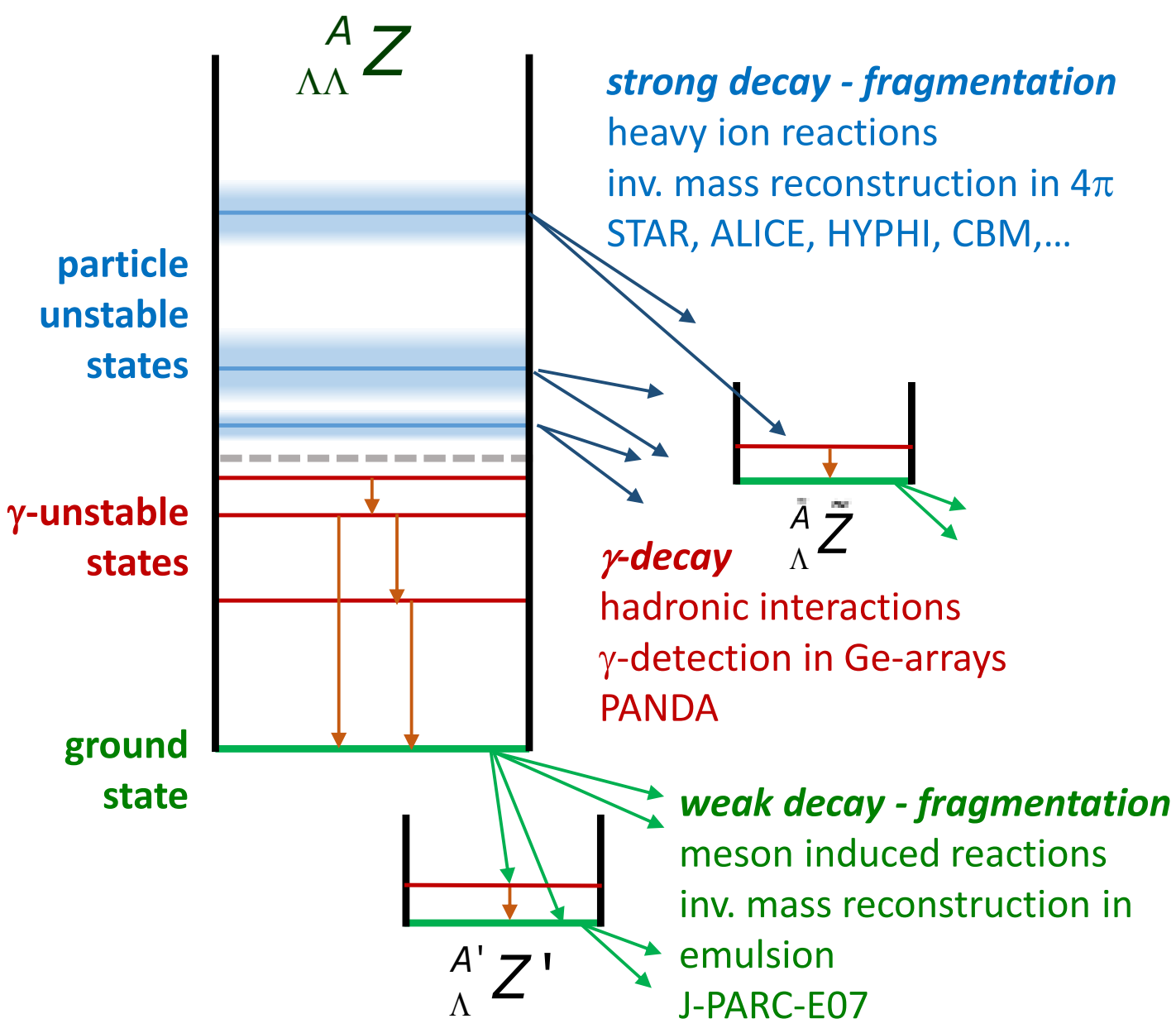

**Figure 8:** Various decays which allow to study the level scheme of $\Lambda\Lambda$-hypernuclei [17].

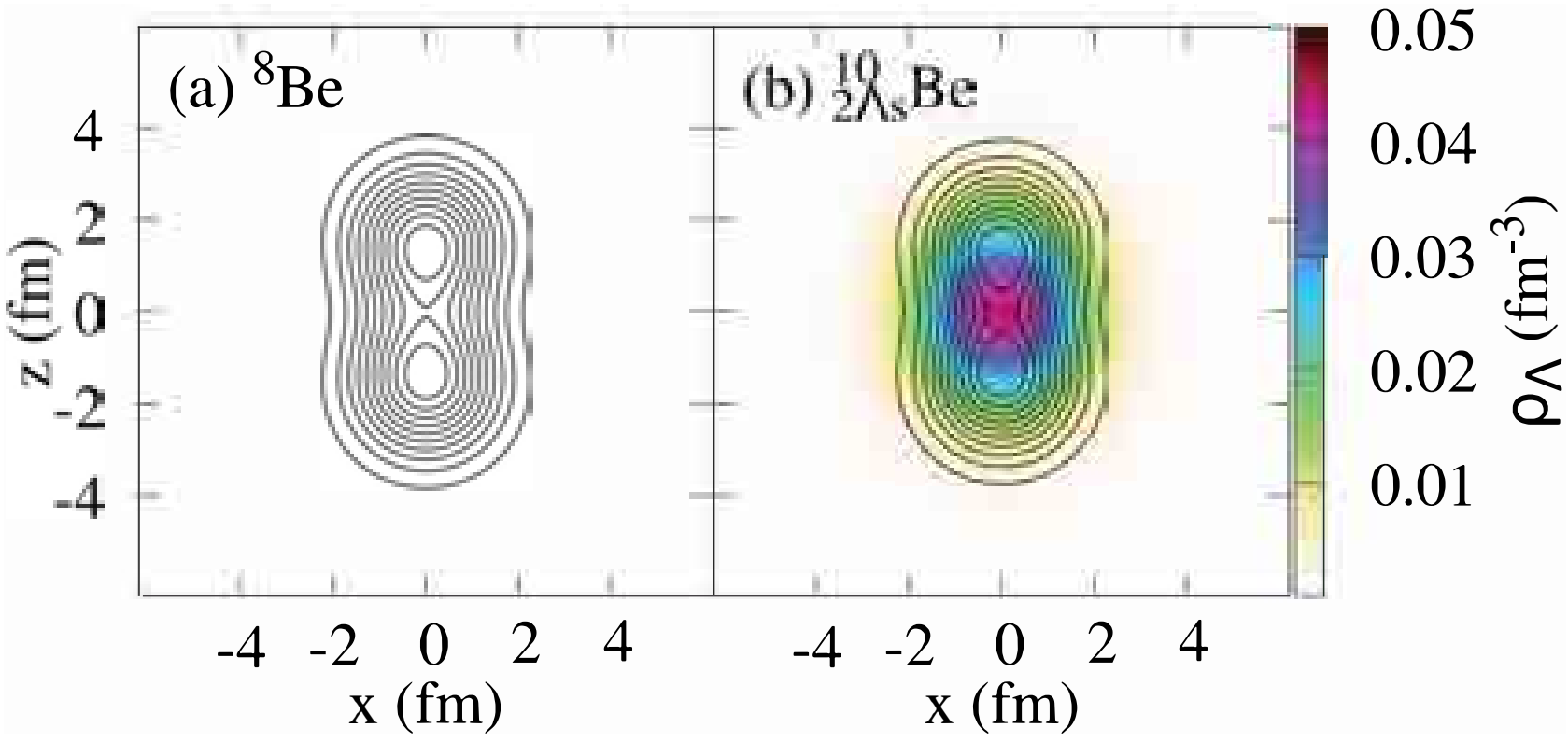

**Figure 9:** Neutron and $\Lambda$ hyperon density distributions of (a) $^8$Be and (b) $^{10}_{\Lambda\Lambda}$Be with both $\Lambda$'s in s orbit ($^{10}_{\Lambda\Lambda\text{-s}}$Be). The neutron densities $\rho_n$ are shown by black contour lines, which starts from 0.1 fm$^3$ with the increments by 0.1 fm$^3$. The $\Lambda$ hyperon densities $\rho_\Lambda$ are shown by color map (extracted from Ref. [23]).

As for conventional nuclei, high-resolution $\gamma$-spectroscopy of $\Lambda\Lambda$-hypernuclei will deepen our understanding of the structure of double hypernuclei. As an example, Fig.9 shows the neutron and $\Lambda$ hyperon density distribution of (a) $^8$Be and (b) $^{10}_{\Lambda\Lambda}$Be nuclei where the two $\Lambda$'s populate an s-state [23]. Similar to the $^7_\Lambda$Li single-$\Lambda$ hypernucleus, where the $^6$Li core shrinks when a $\Lambda$-particle is added [20], a sizable shrinkage of the $^8$Be core is predicted. In the mean-field calculations shown in Fig. 9, the distance between the two "$\alpha$ clusters" reduces from 2.8 fm in $^8$Be down to 2.6 fm in $^{10}_{\Lambda\Lambda}$Be with two $\Lambda$'s in the s-orbital.

At $\overline{\text{P}}$ANDA, $^{10}_{\Lambda\Lambda}$Be nuclei can be produced with large probability by using a secondary $^{10}$B target [24]. For $^{10}_{\Lambda\Lambda}$Be, the $^6_{\Lambda\Lambda}$He+$^4$He decay defines the lowest particle threshold at about 7.5 MeV above the ground state. As a consequence, the $2^+$ state at $\approx$2.86 MeV will be particle stable and the $^{10}_{\Lambda\Lambda}$Be(2.86 MeV;$2^+$) $\rightarrow$ $^{10}_{\Lambda\Lambda}$Be(0;$0^+$) E2-transition will be observable. The transition energy will be related to the effective moment of inertia. Of course, neither $^8$Be nor $^{10}_{\Lambda\Lambda}$Be are expected to be ideal rigid rotors with two $\alpha$-particles at a fixed distance [25]. Indeed, such effects as well as structural changes caused by the presence of the two $\Lambda$ hyperons will be reflected in the level scheme measured with high precision by $\overline{\text{P}}$ANDA.

# 4 Antihyperons in nuclear matter

Studying antibaryons in conventional nuclei provides a unique opportunity to elucidate strong in-medium effects in baryonic systems. Because of the strongly attractive antibaryon-nucleus potential, antibaryons embedded in nuclear matter might cause a local compression of rather cold baryonic matter [26, 27, 28, 29, 30, 31]. Thus, the study of the antibaryon-nucleus interaction can provide new information on the high density regime of the nuclear EOS. In turn, antibaryon-nucleus systems may serve as a laboratory for the antibaryon-baryon interaction. Considering the important role played e.g. by strange baryons and antibaryons for a quantitative interpretation of dense hadronic systems created in high-energy heavy-ion collisions, it is clearly necessary and beneficial to study antibaryons in nuclear systems under rather well defined conditions.

Unfortunately, experimental information on antibaryons in nuclei is rather scarce. Only for the antiproton its nuclear potential could be constrained by experimental studies. The (Schrödinger equivalent) antiproton potential at normal nuclear density turns out to be in the range of $U_{\overline{p}} \simeq -150$ MeV, i.e. a factor of approximately four weaker than naively expected from G-parity relations [32] (see Eq. 3 below). Gaitanos *et al.* [33] suggested that this discrepancy can be traced back to the missing energy dependence of the proton-nucleus optical potential in conventional relativistic mean-field models. The required energy and momentum dependence could be recovered by extending the relativistic hadrodynamics Lagrangian by non-linear derivative interactions [33, 34, 35, 36] thus also mimicking many-body forces [37]. This development is certainly promising and should be extended to the antihyperon sector in the future. An implementation of these potentials into transport models would be very beneficial for the experimental program at $\overline{\text{P}}$ANDA.

Considering the complete lack of data on the behaviour of antihyperons in nuclei, the antihyperon-meson couplings are usually obtained from the hyperon-meson coupling constants by applying the G-parity transformation:

$$g_{\sigma\overline{\mathrm{B}}} = \xi \cdot g_{\sigma\mathrm{B}}, \;\; g_{\omega\overline{\mathrm{B}}} = -\xi \cdot g_{\omega\mathrm{B}}, \;\; g_{\rho\overline{\mathrm{B}}} = \xi \cdot g_{\rho\mathrm{B}}. \tag{3}$$

Here, the scaling factor $\xi$ ensures, that the empirically determined antiproton-nucleus potential is reproduced (see e.g. Ref. [32]). It is clear though, that a relativistic mean-field model can not fully describe in-medium antibaryon interactions using the G-parity motivated coupling constants. Many-body effects and the presence of strong annihilation channels could cause large deviations from the simple G-parity relation. Therefore, G-parity motivated couplings can at most be a starting point to determine the antibaryon-meson couplings in a more rigourous way.

For illustration, we show in the right part of Fig. 10 the total potential acting on an extra baryon (a) and on an extra antibaryon (b) in an 1s state in an $^{16}$O nucleus [29, 30]. In the used relativistic mean-field model the hyperon-meson coupling constants were constrained by $\Lambda$ hypernuclei, $\Sigma$ atoms, and (K$^-$,K$^-$)$\Xi^-$ data. For the construction of the antibaryon-nucleus interaction via the G-parity symmetry the same scaling factor $\xi$=0.2 was assumed for all antibaryons. For all antihyperons rather deep potentials arise which differ strongly from the corresponding hyperon potentials.

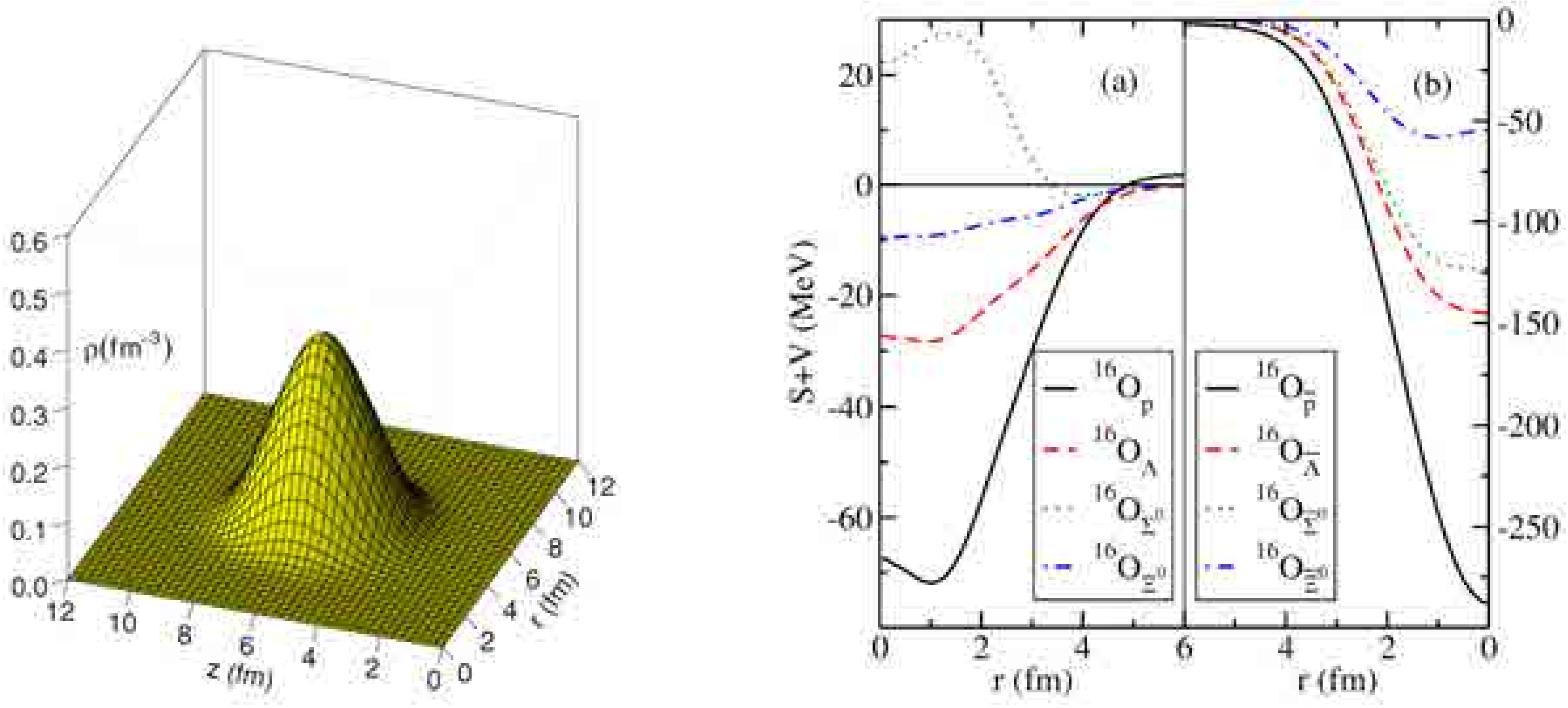

**Figure 10:** Left: 3D plot of nucleon density in the bound $\overline{\Lambda}$+$^{16}$O system [27]. Right: The baryon-nucleus (a) and antibaryon-nucleus (b) potentials in $^{16}$O, calculated dynamically in a relativistic mean-field model. Motivated by experimentally constrained antiproton-nucleus potentials, the scaling factor for the coupling constants was assumed to be $\xi$ = 0.2 for all antibaryons [29, 30].

Fig. 10 shows a 3D plot of the summed proton and neutron density of an $^{16}$O nucleus with an implanted $\overline{\Lambda}$, i.e. an $^{17}_{\overline{\Lambda}}$O hypernucleus [27]. A compression to more than twice the normal nuclear density is predicted. In these relativistic mean-field calculations, a large scaling factor of $\xi$=1 was applied. However, qualitatively similar compressional effects are also found with significantly smaller scaling factors in the case of antiprotons embedded in a nucleus [27].

Antihyperons annihilate quickly in nuclei and conventional spectroscopic studies of bound systems are not feasible. As a consequence, no experimental information on the nuclear potential of antihyperons exists so far. As suggested recently [38, 39, 40], quantitative information on the antihyperon potentials may be obtained via exclusive antihyperon-hyperon pair production close to threshold in antiproton-nucleus interactions. (left part of Fig. 11). Apart from Fermi motion, the initial transverse momentum in a $\overline{\mathrm{p}}$+A$\rightarrow$ $\mathrm{B}\overline{\mathrm{B}}$X reaction is 0. Therefore, the transverse momenta of the baryon and antibaryon should be opposite and equal at the point of their production inside the nucleus. Once these hyperons leave the nucleus and are detected, their asymptotic momentum distributions will reflect the depth of the respective potentials. A deep potential for one species could result in a momentum distribution of antihyperons which differs from that of the *coincident* hyperon.

The schematic calculations of Ref. [39, 40] revealed significant sensitivities of the transverse momentum asymmetry $\alpha_T$ to the depth of the antihyperon potential where $\alpha_T$ is defined in terms of the transverse momenta of the coincident particles

$$\alpha_T = \frac{p_T(\Lambda) - p_T(\overline{\Lambda})}{p_T(\Lambda) + p_T(\overline{\Lambda})} \tag{4}$$

In order to go beyond the schematic calculations presented in Refs. [39, 40] and to include simultaneously secondary deflection and absorption effects, we recently analyzed [38] more realistic calculations of this new observable with the Giessen Boltzmann-Uehling-Uhlenbeck transport model (GiBUU, Release 1.5) [41]. In order to explore the sensitivity of the transverse momentum asym-metry on the depth of the $\overline{\Lambda}$-potential calculations, the default antihyperon potentials [42] were scaled, leaving all other input parameters of the model unchanged.

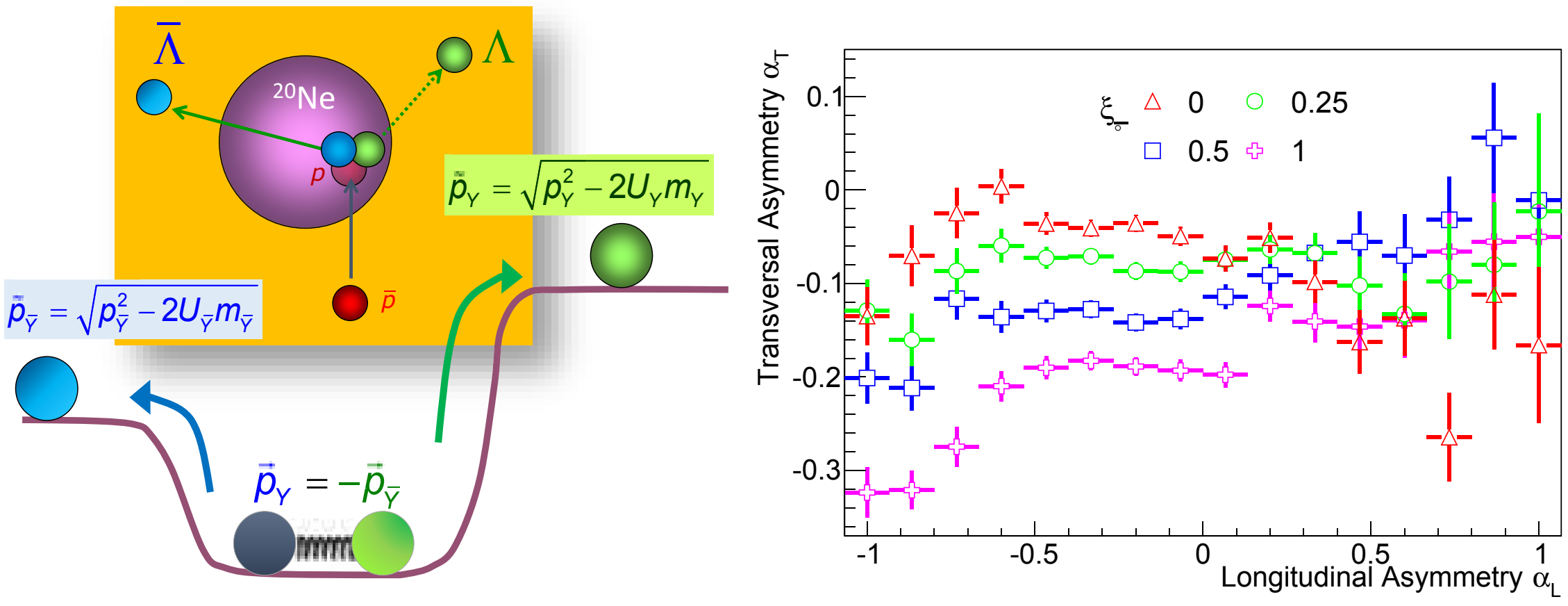

**Figure 11:** Left: Scheme of the reaction proposed to explore the relative nuclear potentials of baryons and antibaryons. Right: Average transverse momentum asymmetry as a function of the longitudinal momentum asymmetry for $\Lambda\overline{\Lambda}$-pairs produced exclusively in 0.85 GeV $\overline{\mathrm{p}}^{20}$Ne interactions [17, 38]. The different symbols show the GiBUU predictions for different scaling factors $\xi_{\overline{\Lambda}}$ of the antibaryon-potentials. Note that these results were obtained with the GiBUU release 1.5.

Besides hydrogen isotopes, also noble gases can be used by default as nuclear targets by the $\overline{\text{P}}$ANDA experiment. We, therefore, studied the exclusive reaction $\overline{\text{p}}$+$^{20}$Ne→ $\text{B}\overline{\text{B}}$X at several beam energies close the respective thresholds [17, 38]. The right panel of Fig. 11 shows the GiBUU prediction for the average transverse asymmetry $\alpha_T$ (Eq. 4) plotted as a function of the longitudinal momentum asymmetry $\alpha_L$ which is defined for each event as

$$\alpha_L = \frac{p_L(\Lambda) - p_L(\overline{\Lambda})}{p_L(\Lambda) + p_L(\overline{\Lambda})}. \tag{5}$$

The transverse asymmetry evaluated for $\Lambda\overline{\Lambda}$ pairs shows a remarkable sensitivity on the scaling factor $\xi_{\overline{\Lambda}}$ for the $\overline{\Lambda}$-potential. In Ref. [38] it was demonstrated that this sensitivity is strongly related to the rescattering process of the hyperons and antihyperons within the target nucleus.

$\Sigma^-\overline{\Lambda}$-momentum correlations also show a strong sensitivity to the depth of the antihyperon potential [38]. However, as already stressed in Ref. [42], the attractive $\Sigma^-$-potential adopted in the present GiBUU model is not compatible with experimental data. Clearly, the description of the $\Sigma^-$-potential in the transport model needs to be improved so that the predictions for $\Sigma^-\overline{\Lambda}$ pairs can be put on firm footing. Nevertheless, $\Sigma^-\overline{\Lambda}$ pairs are produced in $\overline{\text{p}}$-n interactions. As a consequence a comparison of $\Lambda\overline{\Lambda}$ and $\Sigma^-\overline{\Lambda}$-correlations in neutron-rich nuclei might help to explore the isospin dependence of the $\overline{\Lambda}$-potential in nuclear matter.

# 5 The HYPER-$\overline{\text{P}}$ANDA setup

In addition to the general purpose $\overline{\text{P}}$ANDA setup, the hyperatom and hypernuclear measurements require a dedicated primary target which is placed approximately 55 cm upstream of the nominal target position (Fig. 12). The main task of the primary target is the production of $\Xi^-$ hyperons.

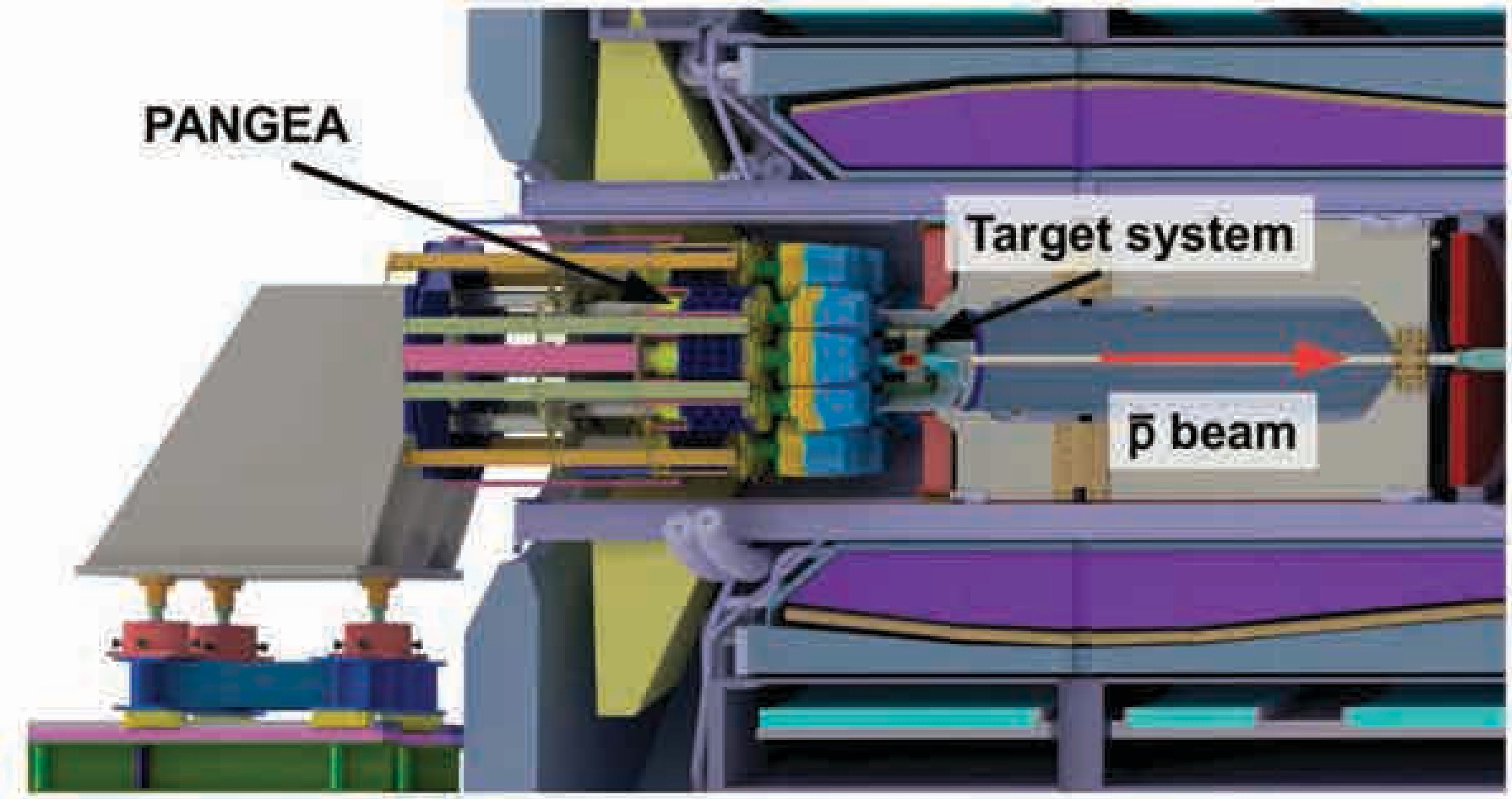

**Figure 12:** Integration of the hyperatom or hypernucleus setup into the $\overline{\text{P}}$ANDA detector.

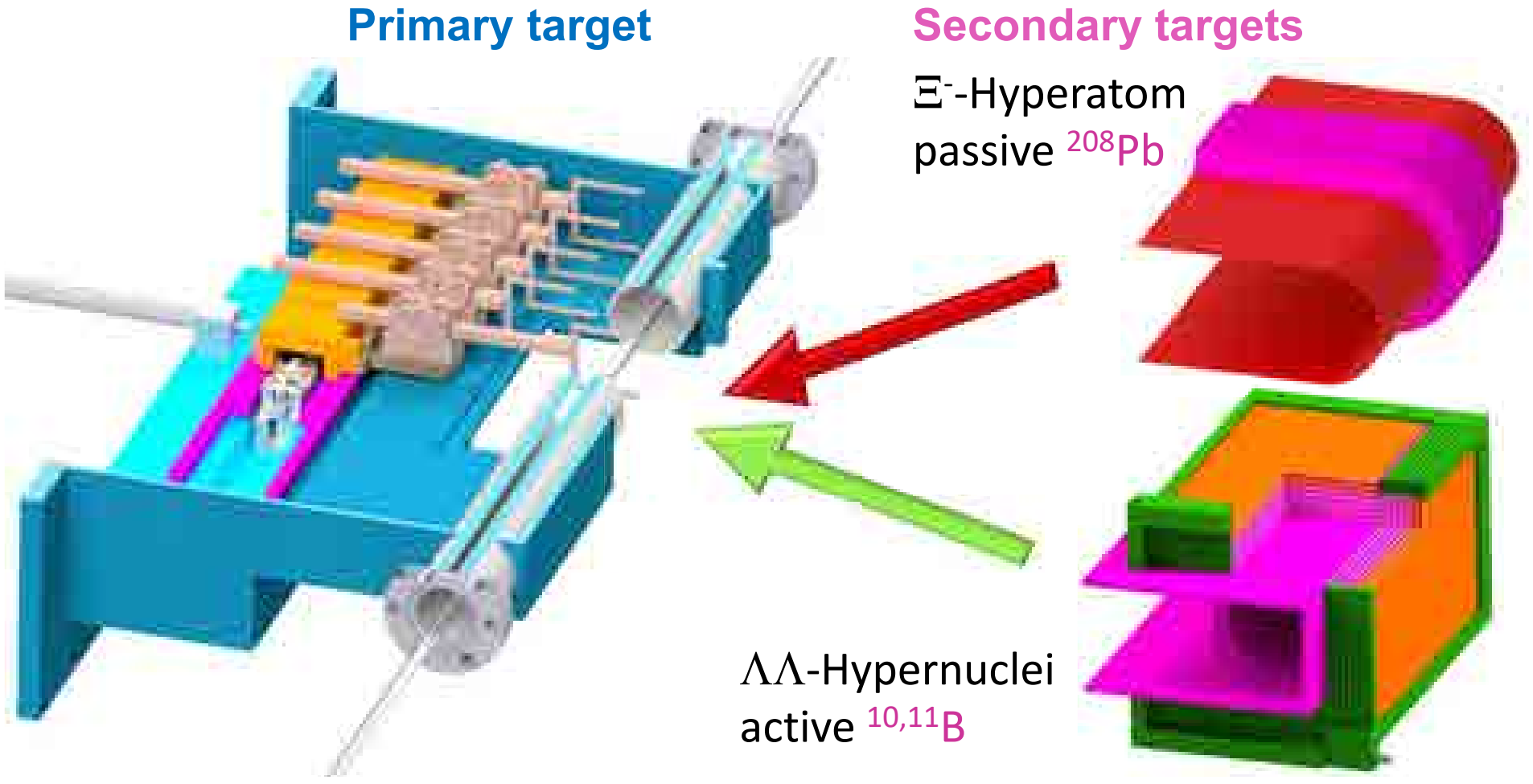

**Figure 13:** Target station for the primary target (left part) and secondary target modules surrounding the primary target. The secondary target modules differ for the $\Xi^-$-hyperatoms (top) and $\Lambda\Lambda$-hypernuclei studies at $\overline{\text{P}}$ANDA .

This internal target consists of a thin carbon filament which will be operated in the halo of the antiproton beam.

The secondary targets differ for the hyperatom and the hypernucleus experiments (Fig.13). However, the principle design of these secondary target systems is in both cases determined by the competition between the finite $\Xi^-$ mean lifetime of $\tau_{\Xi^-}$=1.64$\cdot$10$^{-10}$ s and the time it takes to stop a hyperon inside the secondary absorber (see Fig. 2). For the hypernucleus measurements, an active secondary target of silicon layers sandwiched between boron absorbers to stop the $\Xi^-$ hyperons and to detect pions from the weak decay of $\Lambda\Lambda$- and $\Lambda$-hypernuclei is used. Because of the short lifetime of the $\Xi^-$ hyperons and their brief stopping time in the secondary target, it is essential to place the secondary absorber as close as possible to the primary target to reach a maximum stopping probability. It was therefore decided to build the wall of the vacuum chamber in the region of the secondary target out of the secondary absorber material. Additional absorber material will be placed inside the vacuum chamber, thus forming a cylindrical beam pipe surrounding the primary target.

For the hyperatom studies this active secondary target will be replaced by a passive lead absorber. The geometry of the lead material is determined by the stopping distribution of $\Xi^-$ hyperons and the absorption of $\gamma$-rays from formed hyperatoms. The present design of the secondary target is optimized by means of $\Xi^-$ momentum distributions predicted by microscopic transport models.

To detect the $\gamma$-rays emitted from the produced hyperatoms or hypernuclei, the setup will be complemented by a high purity Germanium (HPGe) detectors array. This array is composed of 20 DEGAS triple detectors which are developed by the NUSTAR Collaboration [43, 44]. As shown in Fig. 12, these cluster detectors replace the backward end-cap of the $\overline{\text{P}}$ANDA calorimeter.

## 6 Summary

Antiproton beams at the High Energy Storage Ring (HESR) at FAIR facilitate the production of low energy hyperons and antihyperons with high production rates. By stopping these hyperons in secondary targets, the $\overline{\text{P}}$ANDA experiment will make absolutely unique contributions to the physics of strange nuclear systems with S = -2 by precision spectroscopic studies of heavy $\Xi^-$ atoms as well as $\Lambda\Lambda$-hypernuclei. In addition, by tagging on antihyperon–hyperon production in nuclei, antihyperons can be traced in cold nuclei for the first time.

## Acknowledgements

This work uses data collected within the framework of the PhD theses of Sebastian Bleser, Michael Bölting, Falk Schupp and Marcell Steinen at the Johannes Gutenberg University Mainz.

## References

[1] K. Imai, K. Nakazawa, and H. Tamura. *J-PARC E07 experiment: Systematic study of double-strangeness system with an emulsion-counter hybrid method*. eprint: `http://j-parc.jp/researcher/Hadron/en/pac_0606/pdf/p07-Nakazawa.pdf`.

[2] K. Nakazawa and J. Yoshida. "Double-$\Lambda$ hypernuclei at J-PARC - E07 experiment." In: *Chiral Symmetry in Hadrons and Nuclei*. WORLD SCIENTIFIC, 2014, pp. 132–136. eprint: `http://www.worldscientific.com/doi/pdf/10.1142/9789814618229_0025`.

[3] D. Tovee, D. Davis, J. Simonović, G. Bohm, J. Klabuhn, F. Wysotzki, M. Csejthey-Barth, J. Wickens, T. Cantwell, C. Ghógáin, A. Montwill, K. Garbowska-Pniewska, T. Pniewski, and J. Zakrzewski. "Some properties of the charged $\Sigma$ hyperons." In: *Nuclear Physics B* 33.2 (1971), pp. 493–504. ISSN: 0550-3213.

[4] T. Yamada and K. Ikeda. "Double-$\Lambda$, twin-$\Lambda$, and single-$\Lambda$ hypernuclear productions for stopped $\Xi^-$ particles in $^{12}$C." In: *Phys. Rev. C* 56 (1997), pp. 3216–3230.

[5] Y. Hirata, Y. Nara, A. Ohnishi, T. Harada, and J. Randrup. "Formation of twin and double $\Lambda$ hypernuclei from $\Xi^-$ absorption at rest on $^{12}$C." In: *Nuclear Physics A* 639.1 (1998), pp. 389c–392c. ISSN: 0375-9474.

[6] Y. Hirata, Y. Nara, A. Ohnishi, T. Harada, and J. Randrup. "Quantum Fluctuation Effects on Hyperfragment Formation from $\Xi^-$ Absorption at Rest on $^{12}$C." In: *Progress of Theoretical Physics* 102.1 (1999), p. 89. eprint: `/oup/backfile/Content_public/Journal/ptp/102/1/10.1143_PTP.102.89/3/102-1-89.pdf`.

[7] S. Bleser. "High resolution $\gamma$-spectroscopy of $\Lambda\Lambda$-hypernuclei at $\overline{\text{P}}$ANDA." PhD thesis. Johannes Gutenberg-Universität Mainz.

[8] C. J. Batty, E. Friedman, and A. Gal. "Experiments with $\Xi^-$ atoms." In: *Phys. Rev. C* 59 (1999), pp. 295–304.

[9] PANDA Collaboration and M.F.M. Lutz and B. Pire and O. Scholten and R. Timmermans. "Physics Performance Report for PANDA: Strong Interaction Studies with Antiprotons." In: (2009). arXiv: `0903.3905 [hep-ex]`.

[10] A. Ohnishi, Y. Hirata, Y. Nara, S. Shinmura, and Y. Akaishi. "Hyperon distribution and correlation in $(K^-, K^+)$ reactions." In: *Nuclear Physics A* 684.1 (2001), pp. 595–597. ISSN: 0375-9474.

[11] H. N. Brown, B. B. Culwick, W. B. Fowler, M. Gailloud, T. E. Kalogeropoulos, J. K. Kopp, R. M. Lea, R. I. Louttit, T. W. Morris, R. P. Shutt, A. M. Thorndike, M. S. Webster, C. Baltay, E. C. Fowler, J. Sandweiss, J. R. Sanford, and H. D. Taft. "Observation of Production of a $\Xi^-+\overline{\Xi}^+$ Pair." In: *Phys. Rev. Lett.* 8 (1962), pp. 255–257.

[12] J. Pochodzalla. "Future experiments on hypernuclei and hyperatoms." In: *Nuclear Instruments and Methods in Physics Research Section B: Beam Interactions with Materials and Atoms* 214 (2004). Low Energy Antiproton Physics (LEAP'03), pp. 149–152. ISSN: 0168-583X.

[13] E. Friedman. Private communication.

[14] I. Angeli and K. Marinova. "Table of experimental nuclear ground state charge radii: An update." In: *Atomic Data and Nuclear Data Tables* 99.1 (2013), pp. 69–95. ISSN: 0092-640X.

[15] M. Steinen. "Study of heavy $\Xi^-$ hyperatoms at $\overline{\text{P}}$ANDA." PhD thesis. Johannes Gutenberg-Universität Mainz.

[16] G. Backenstoss, A. Bamberger, T. Bunaciu, J. Egger, H. Koch, U. Lynen, H. Ritter, H. Schmitt, and A. Schwitter. "Strong interaction effects in antiprotonic atoms." In: *Physics Letters B* 41.4 (1972), pp. 552–556. ISSN: 0370-2693.

[17] T. P. C. B. Singh et al. "Study of doubly strange systems using stored antiprotons." In: *Nuclear Physics A* 954 (2016). Recent Progress in Strangeness and Charm Hadronic and Nuclear Physics, pp. 323–340. ISSN: 0375-9474.

[18] J. Pochodzalla et al. "Two-particle correlations at small relative momenta for $^{40}$Ar induced reactions on $^{197}$Au at E/A=60 MeV." In: *Phys. Rev. C* 35 (1987), pp. 1695–1719.

[19] E. Hiyama, M. Kamimura, T. Motoba, T. Yamada, and Y. Yamamoto. "Four-body cluster structure of $A$=7-10 double-$\Lambda$ hypernuclei." In: *Phys. Rev. C* 66 (2002), p. 024007.

[20] K. Tanida et al. "Measurement of the B(E2) of $^{7}_{\Lambda}$Li and Shrinkage of the Hypernuclear Size." In: *Phys. Rev. Lett.* 86 (2001), pp. 1982–1985.

[21] E. Hiyama, Y. Yamamoto, T. Motoba, and M. Kamimura. "Structure of $A = 7$ iso-triplet $\Lambda$ hypernuclei studied with the four-body cluster model." In: *Phys. Rev. C* 80 (2009), p. 054321.

[22] K. Nakazawa. "Double-$\Lambda$ Hypernuclei via the $\Xi^-$ Hyperon Capture at Rest Reaction in a Hybrid Emulsion." In: *Nuclear Physics A* 835.1 (2010), pp. 207–214. ISSN: 0375-9474.

[23] Y. Tanimura. "Clusterization and deformation of multi-$\Lambda$ hypernuclei within a relativistic mean-field model." In: *Phys. Rev. C* 99 (2019), p. 034324.

[24] A. S. Lorente, A. S. Botvina, and J. Pochodzalla. "Production of excited double hypernuclei via Fermi breakup of excited strange systems." In: *Physics Letters B* 697.3 (2011), pp. 222–228. ISSN: 0370-2693.

[25] E. Garrido, A. S. Jensen, and D. V. Fedorov. "Rotational bands in the continuum illustrated by $^{8}$Be results." In: *Phys. Rev. C* 88 (2013), p. 024001.

[26] T. Bürvenich and I.N. Mishustin and L.M. Satarov and J.A. Maruhn and H. Stöcker and W. Greiner. "Enhanced binding and cold compression of nuclei due to admixture of antibaryons." In: *Physics Letters B* 542.3 (2002), pp. 261–267. ISSN: 0370-2693.

[27] I. Mishustin, L. Satarov, T. Bürvenich, H. Stöcker, and W. Greiner. "Antibaryons bound in nuclei." In: *Phys. Rev. C* 71 (2005), p. 035201.

[28] A. B. Larionov, I. N. Mishustin, L. M. Satarov, and W. Greiner. "Possibility of cold nuclear compression in antiproton-nucleus collisions." In: *Phys. Rev. C* 82 (2010), p. 024602.

[29] J. Hrtánková and J. Mareš. "Antibaryon-nucleus bound states." In: *Journal of Physics: Conference Series* 599.1 (2015), p. 012007.

[30] J. Hrtánková and J. Mareš. "Antibaryon interactions with the nuclear medium." In: *PoS* INPC2016 (2017), p. 280. arXiv: 1710.02045 [nucl-th].

[31] J. Hrtánková and J. Mareš. "Interaction of antiprotons with nuclei." In: *JPS Conf. Proc.* 18 (2017), p. 011032. arXiv: 1711.10258 [nucl-th].

[32] A. B. Larionov, I. A. Pshenichnov, I. N. Mishustin, and W. Greiner. "Antiproton-nucleus collisions simulation within a kinetic approach with relativistic mean fields." In: *Phys. Rev. C* 80 (2009), p. 021601.

[33] T. Gaitanos, M. Kaskulov, and H. Lenske. "How deep is the antinucleon optical potential at FAIR energies." In: *Physics Letters B* 703.2 (2011), pp. 193–198. ISSN: 0370-2693.

[34] T. Gaitanos, M. Kaskulov, and U. Mosel. "Non-linear derivative interactions in relativistic hadrodynamics." In: *Nuclear Physics A* 828.1 (2009), pp. 9–28. ISSN: 0375-9474.

[35] T. Gaitanos and M. Kaskulov. "Momentum dependent mean-field dynamics of compressed nuclear matter and neutron stars." In: *Nuclear Physics A* 899.Supplement C (2013), pp. 133–169. ISSN: 0375-9474.

[36] T. Gaitanos and M. Kaskulov. "Toward relativistic mean-field description of $\overline{N}$ -nucleus reactions." In: *Nuclear Physics A* 940.Supplement C (2015), pp. 181–193. ISSN: 0375-9474.

[37] R. O. Gomes, V. Dexheimer, S. Schramm, and C. A. Z. Vasconcellos. "Many-body Forces in the Equation of State of Hyperonic Matter." In: *The Astrophysical Journal* 808.1 (2015), p. 8.

[38] A. S. Lorente, S. Bleser, M. Steinen, and J. Pochodzalla. "Antihyperon potentials in nuclei via exclusive antiproton-nucleus reactions." In: *Physics Letters B* 749 (2015), pp. 421–424. ISSN: 0370-2693.

[39] J. Pochodzalla. "Exploring the potential of antihyperons in nuclei with antiprotons." In: *Physics Letters B* 669.5 (2008), pp. 306–310. ISSN: 0370-2693.

[40] J. Pochodzalla. "Exploring the nuclear potential of antihyperons with antiprotons at PANDA." In: *Hyperfine Interactions* 194.1 (2009), p. 255. ISSN: 1572-9540.

[41] O. Buss, T. Gaitanos, K. Gallmeister, H. van Hees, M. Kaskulov, O. Lalakulich, A. Larionov, T. Leitner, J. Weil, and U. Mosel. "Transport-theoretical description of nuclear reactions." In: *Physics Reports* 512.1 - 2 (2012), pp. 1–124. ISSN: 0370-1573.

[42] A. B. Larionov, T. Gaitanos, and U. Mosel. "Kaon and hyperon production in antiproton-induced reactions on nuclei." In: *Phys. Rev. C* 85 (2012), p. 024614.

[43] DEGAS Collaboration. *Technical Report for the Design, Construction and Commissioning of the DESPEC Germanium Array Spectrometer (DEGAS).* https://edms.cern.ch/document/1813618/1.

[44] M. Doncel, B. Cederwall, A. Gadea, J. Gerl, I. Kojouharov, S. Martin, R. Palit, and B. Quintana. "Performance and imaging capabilities of the DEGAS high-resolution $\gamma$-ray detector array for the DESPEC experiment at FAIR." In: *Nuclear Instruments and Methods in Physics Research Section A: Accelerators, Spectrometers, Detectors and Associated Equipment* 873 (2017). Imaging 2016, pp. 36–38. ISSN: 0168-9002.

# High-intensity. high-resolution beam line

**Hiroyuki Noumi[1,2]**

[1]Reserach Center for Nuclear Physics, Osaka University

[2]Insitute of Particle and Nuclear Studies, High Energy Accelerator Research Organization (KEK)

Design performances of the high-momentum beam line for secondary beams and its extension in the extended hadron experimental facility are presented.

## 1 Beam Line

The high intensity, high resolution beam line (HIHR) ion-optically realizes a strongly correlated beam in position and momentum. Combining a properly-designed spectrometer to HIHR, which is the so-called dispersion matching ion-optical technique, as described below, one can carry out high-precision spectroscopy in nuclear physics. In particular, $\Lambda$-single particle energies in various $\Lambda$ hypernuclei can be measured at an energy resolution of as high as a few handred keV. As the dispersion matching beam line does not require a beam measurement, one can remove beam line detectors and reduce beam line materials that affect the energy resolution due to multiple scattering effects. HIHR provides a solution to utilize full-intensity beams potentially available at J-PARC, while available beam intensity would be limited a capability of detectors used as beam counters in usual beam lines.

The layout of HIHR and calculated beam envelope are shown in Figs. 1 and 2, respeectively. HIHR is composed of 4 sections as explained below. In the most upstream section, the secondary particles produced at the primary target are collected at the production angle of 6 degrees. The maximum momentum of the secondary beam is designed to be 2 GeV/c. This specification is the same as that of the K1.8 beam line [1]. The beam line layout of the extraction part would be essentially the same as that of K1.8. The secondary beam is focused vertically at the intermediate focal (IF) point. The secondary beam image is redefined by the IF slit placed at the IF point.

In the second section after the IF point, a particle beam is separated from the other particle beams. In this section, by using two sets of a pair of quadrupole magnets (Q-doublet), a so-called point-to-point ion-optics in the vertical direction is realized between the IF point and the MS point at the end of this section. An electrostatic separator (ESS) having a pair of parallel plate electrodes of 10 cm gap and 4.5 m long is placed between the Q-doublets. Charged particles are kicked vertically by an electrostatic field produced in the ESS. One can compensate the vertical kick by the pitching magnets placed just before and after the ESS. While the kicked angle by the ESS is proportional to the inverse of the particle velocity, the kicked angle by the magnets is proportional to the particle momentum, Namely, a particle beam compensated its kicked angle keeps the beam level and the other particle beams not fully compensated their kicked angles are focused at the different positions in vertical at the MS point. A mass slit (MS) is placed at the MS point so as to select a particle beam and block the other unwanted particle beams.

In the third section after the MS point, the secondary beam is focused achromatically in the horizontal and vertical directions at the HS point. The beam profile is defined again by the beam slits placed at the HS point.

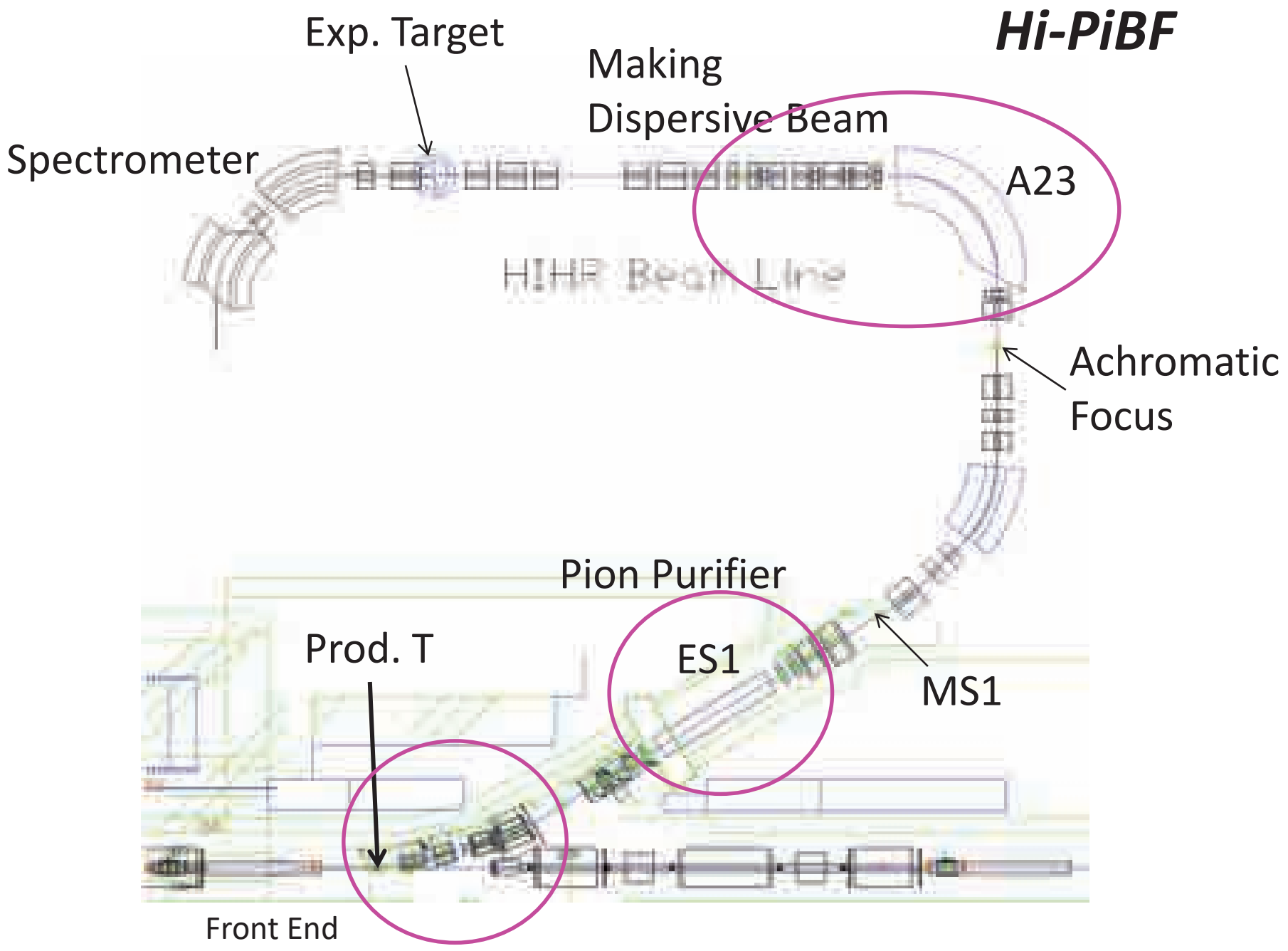

Figure 1: Layout of the HIHR beam line

In the final section after the HS point, by placing a large bending magnet and a several quadrupole magnets, the beam is focused with a magnification of 0.5 and a dispersion of -10 [cm/%] at the experimental target. Here, ion-optical aberrations to the second order are eliminated by three sextupole magnets.

Figure 3 shows expected beam intensities in the case that a 30-GeV, 50-kW primary proton beam is irradiated on a 60-mm thick platinum target. One expects $2.8\times10^8$ positive pions at 1.2 GeV/c every spill.

## 2 Dispersion matching ion optics

A spectrometer matched to HIHR comprises the QSQDMD configuration, where D, Q, S, and M stand for a dipole, a quadrupole, a sextupole, and a multipole magnet, respectively. Applying a dispersion matching ion optical technique [2], a $\Lambda$ hypernuclear excitation energy spectrum can be measured in the $(\pi^+, K^+)$ reaction by measuring a scattered particle $(K^+)$ position distribution at the final focal plane of the QSQDMD spectrometer. A conceptual illustration how to realize a momentum dispersion matching ion optics is shown in Fig. 4(right). Ion optical conditions for the dispersion matching are described below. The followings are ion-optical transformation of the initial coordinate ($x_0$, $\theta_0$, $\delta_0$) to the final one ($x_f$, $\theta_f$, $\delta_f$) by transfer matrices to the first order, where $x_0(x_f)$, $\theta_0(\theta_f)$, and $\delta_0(\delta_f)$ represent position, angle, and momentum difference with respect to the central orbit at the initial (final) point.

$$\begin{pmatrix} x_f \\ \theta_f \\ \delta_f \end{pmatrix} = \begin{pmatrix} s_{11} & s_{12} & s_{16} \\ s_{21} & s_{22} & s_{26} \\ 0 & 0 & 1 \end{pmatrix} \begin{pmatrix} T & 0 & 0 \\ 0 & \theta/\theta_1+1 & 0 \\ 0 & 0 & (K\theta+DQ)/\delta_0+C \end{pmatrix} \begin{pmatrix} b_{11} & b_{12} & b_{16} \\ b_{21} & b_{22} & b_{26} \\ 0 & 0 & 1 \end{pmatrix} \begin{pmatrix} x_0 \\ \theta_0 \\ \delta_0 \end{pmatrix}, \quad (1)$$

$$\theta_1 = b_{21}x_0 + b_{22}\theta_0 + b_{26}\delta_0, \quad (2)$$

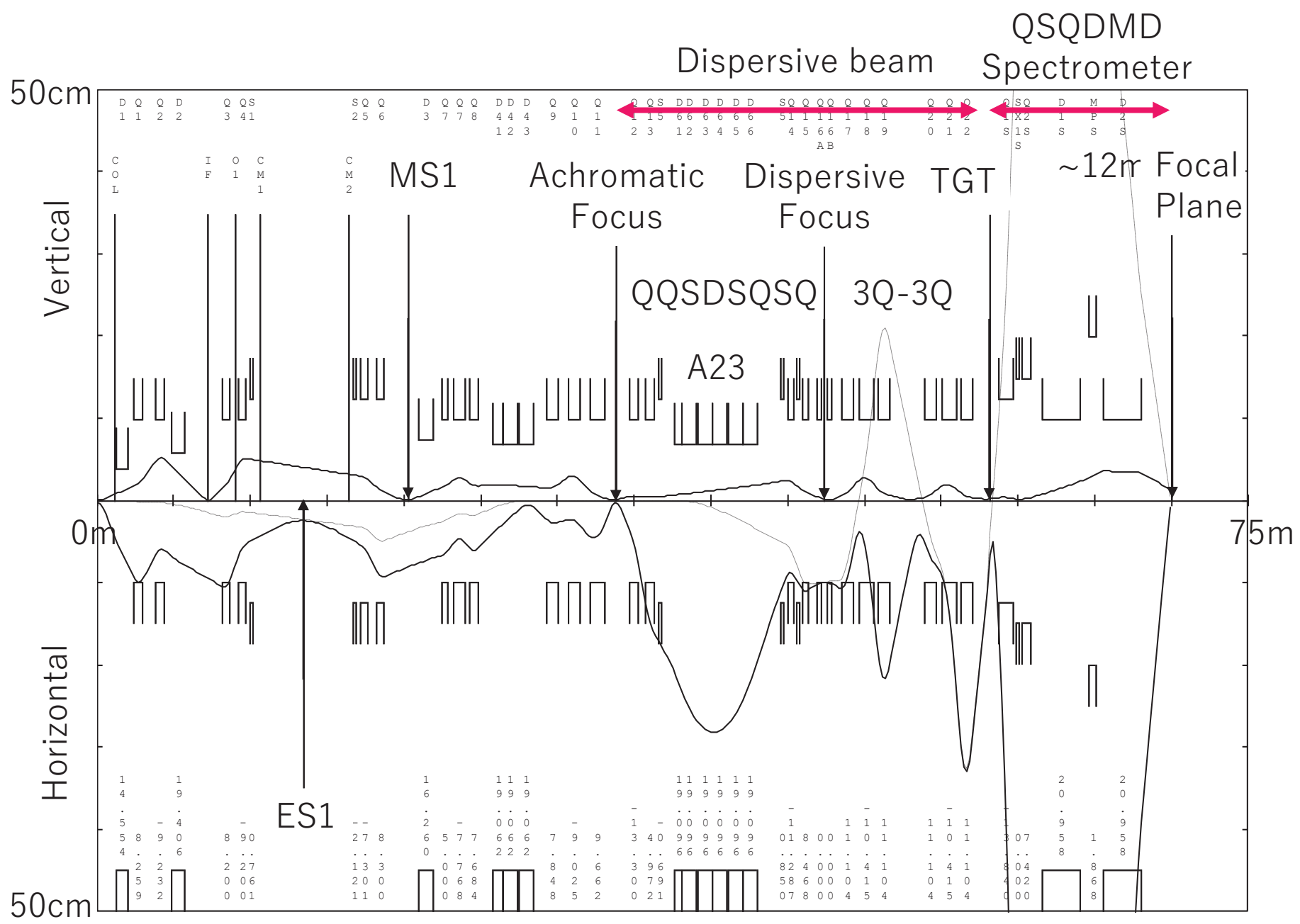

Figure 2: Beam envelope of HIHR calculated with dispersion matching optics to the first order.

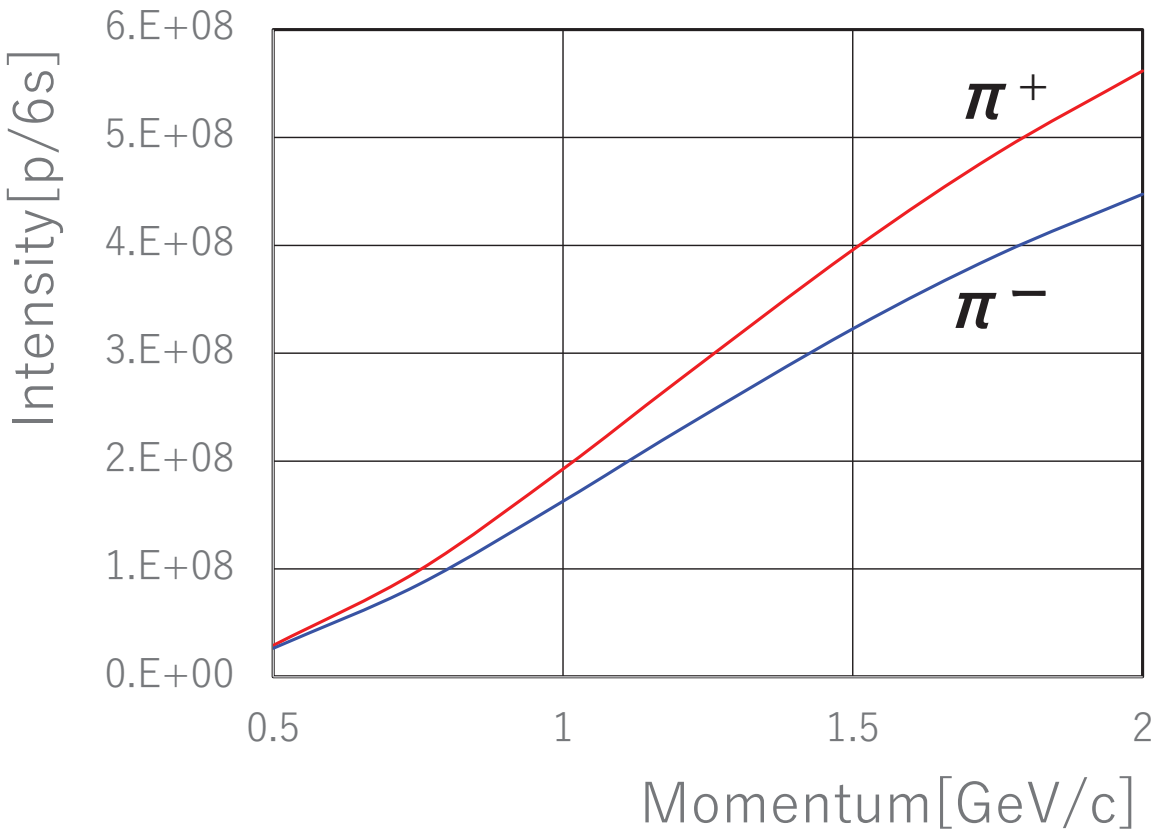

Figure 3: Expected beam intensities as a function of beam momentum in the case that a 30-GeV, 50-kW primary proton beam is irradiated on a 60-mm thick platinum target, calculated by using the so-called Sanford-Wang formula.

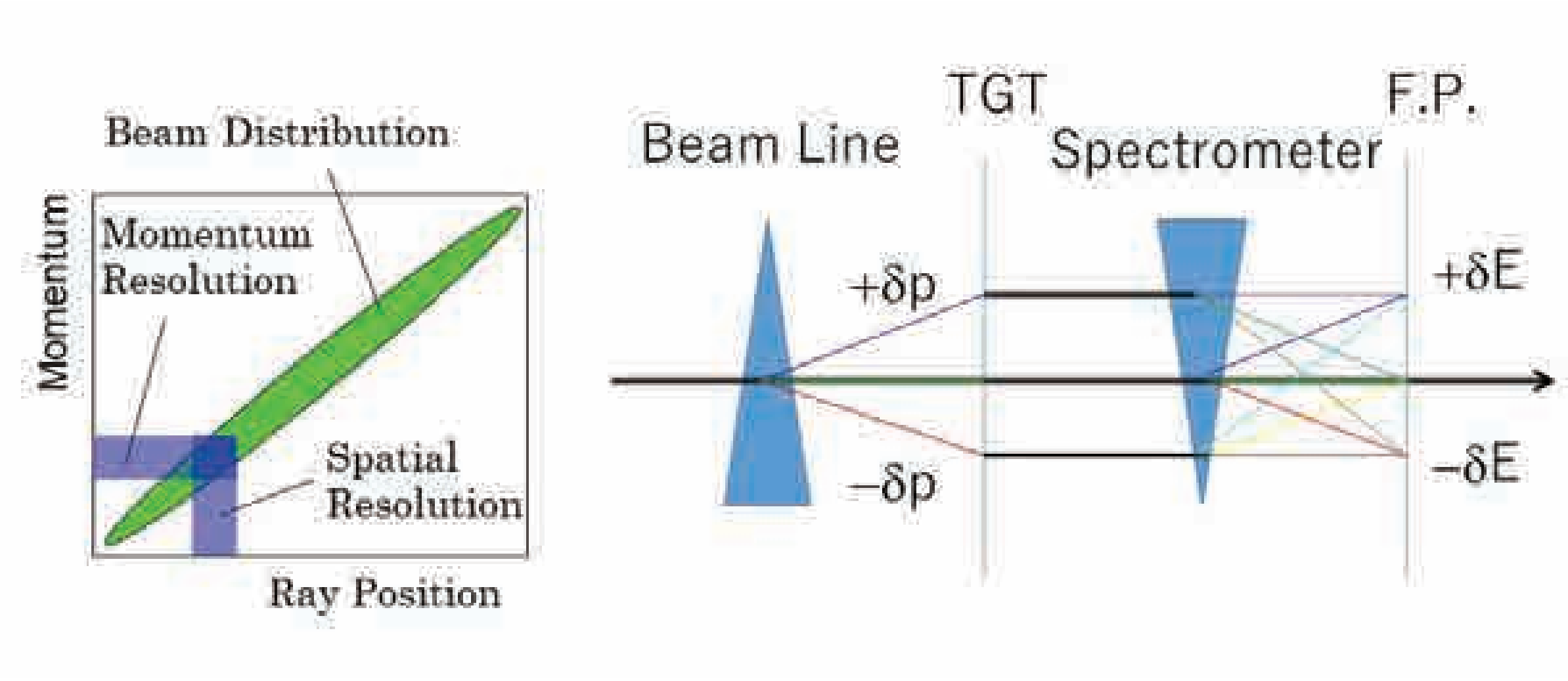

Figure 4: Conceptual illustrations of the beam with correlation of momentum to position (left) and momentum dispersion matching ion optics (right).

$$K = (\partial p_{scat}/\partial\theta)(1/p_{scat}), \tag{3}$$
$$C = (\partial p_{scat}/\partial p_{beam})(p_{beam}/p_{scat}), \tag{4}$$
$$D = (\partial p_{scat}/\partial Q)(1/p_{scat}). \tag{5}$$

Here, $T$, $\theta$, and $Q$ are respectively a cosine of an angle between a beam direction and a normal to a target plane, a scattering angle, and an excitation energy. For a value of $Q$, $K$ and $C$ are respectively a derivative of the scattered momentum ($p_{scat}$) to the scattering angle($\theta$) and that to the beam momentum ($p_{beam}$). $D$ is a derivative of $p_{scat}$ to $Q$. Then, the position $x_f$ at the final focal plane can be expressed as

$$x_f = (\partial x_f/\partial x_0)x_0 + (\partial x_f/\partial\theta_0)\theta_0 + (\partial x_f/\partial\delta_0)\delta_0 + (\partial x_f/\partial\theta)\theta + s_{16}DQ. \tag{6}$$

The matching conditions are obtained as follows.

$$(\partial x_f/\partial x_0) = s_{11}b_{11}T + s_{12}b_{21} \to \text{minimize}, \tag{7}$$
$$(\partial x_f/\partial\theta_0) = s_{11}b_{12}T + s_{12}b_{22} \to 0, \tag{8}$$
$$(\partial x_f/\partial\delta_0) = s_{11}b_{16}T + s_{12}b_{26} + s_{16}C \to 0, \tag{9}$$
$$(\partial x_f/\partial\theta) = s_{12} + s_{16}K \to 0. \tag{10}$$

Once reaction kinematics and a scattering angle is fixed, field gradients of relevant quadrupole magnets are to be adjusted. There are 6 and 3 tunable parameters in the beam line and the spectrometer, respectively. In the case that the scattering angle is set at zero, the number of tunable parameters are reduced to be 3. When the matching conditions are satisfied, a shift of $x_f$ due to the excitation energy $Q$ is determined by $s_{16}DQ$ in the last term in Eq. 6. The shift is proportional to $D$. An energy resolution is estimated as $(\partial x_f/\partial x_0)x_0$, which depends linearly on $x_0$, a beam size at the primary target.

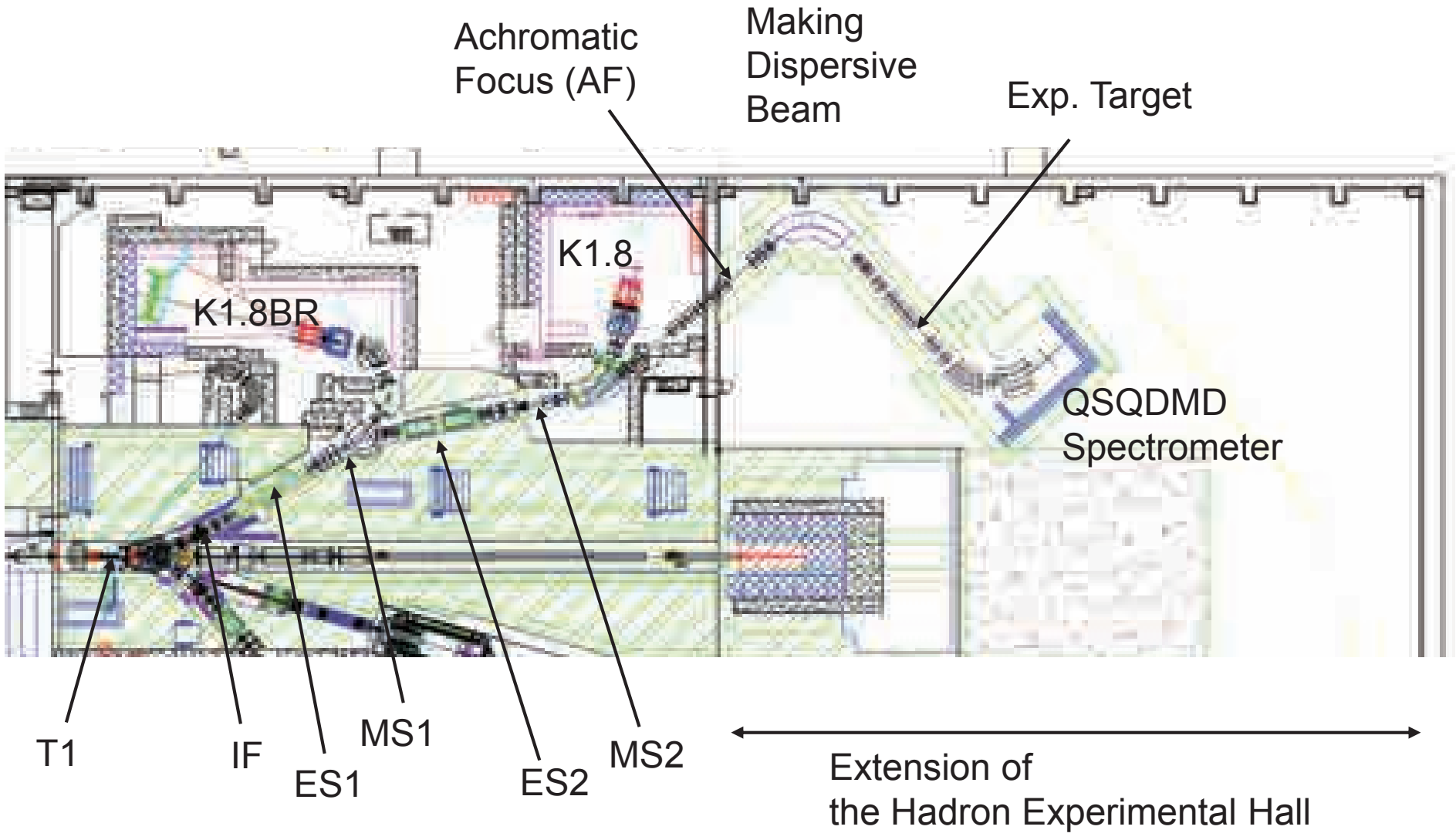

Figure 5: Layout of optional HIHR branched from K1.8

# Appendix A Optional layout of HIHR

An optical consideration has been made as an optional layout of HIHR, where HIHR is branched from the existing K1.8 beam line, as shown in Fig. 5. The current K1.8 bema line originates from the T1 target. K1.8 has been designed to deliver purified, intense kaon beams up to 2 GeV/c, having an IF slits to re-define horizontal aperture and vatical image of the scondary beam produced at T1, double-state mass separation with two electrostatic separators of 6 m long and correspoding mass slits (MS1, MS2). After MS2, a beam spectrometer with a QQDQQ magnet configuration is placed to analyze beam particle momenta. The present option of HIHR shares the K1.8 beam line elements up to the first “QQ” of the beam spectromter. Replacing the last “DQQ” to a different “DQQ”, the new beam line is extended to realize a dispersive beam at the exprimental target, employing another 1 dipole, 12 quadropole, and 3 sxtupole magnets are employed. The beam envelope calculated to the first order optics is shown in Fig. 6. An achromatic beam focus (AF) is realized before making the dispersive beam, as described in Section 1. A large dispersion of -10 [cm/%] with a magnification of -1 at the experimental target is realized to the first order. However, expected momentum resolution is found to be $\Delta p/p$=0.02%. Second order abberations have not yet been fully elliminated and the expected. Expected beam intensities for pions are about 80% reduced than those expected in HIHR presented in Fig.3. Since the total beam line length is as long as 74.2 m, the hadron experimntal hall should be extended in ordr to accommodate the extended part of the beam line and the QSQDMD spectrometer. It is noted that the primary beam size focused on T1 must be changed to 1 mm$\times$2,5 mm($1\sigma$) from current typical size of 2.5 mm$\times$1 mm($1\sigma$) in order to operate the beam line as an HIHR mode. Since the beam cross sectioon is kept, an energy deposit to the T1 target does not change very much. However, it should be evaluated if the thermal stress due to a distribution change of the energy deposit does not damage T1.

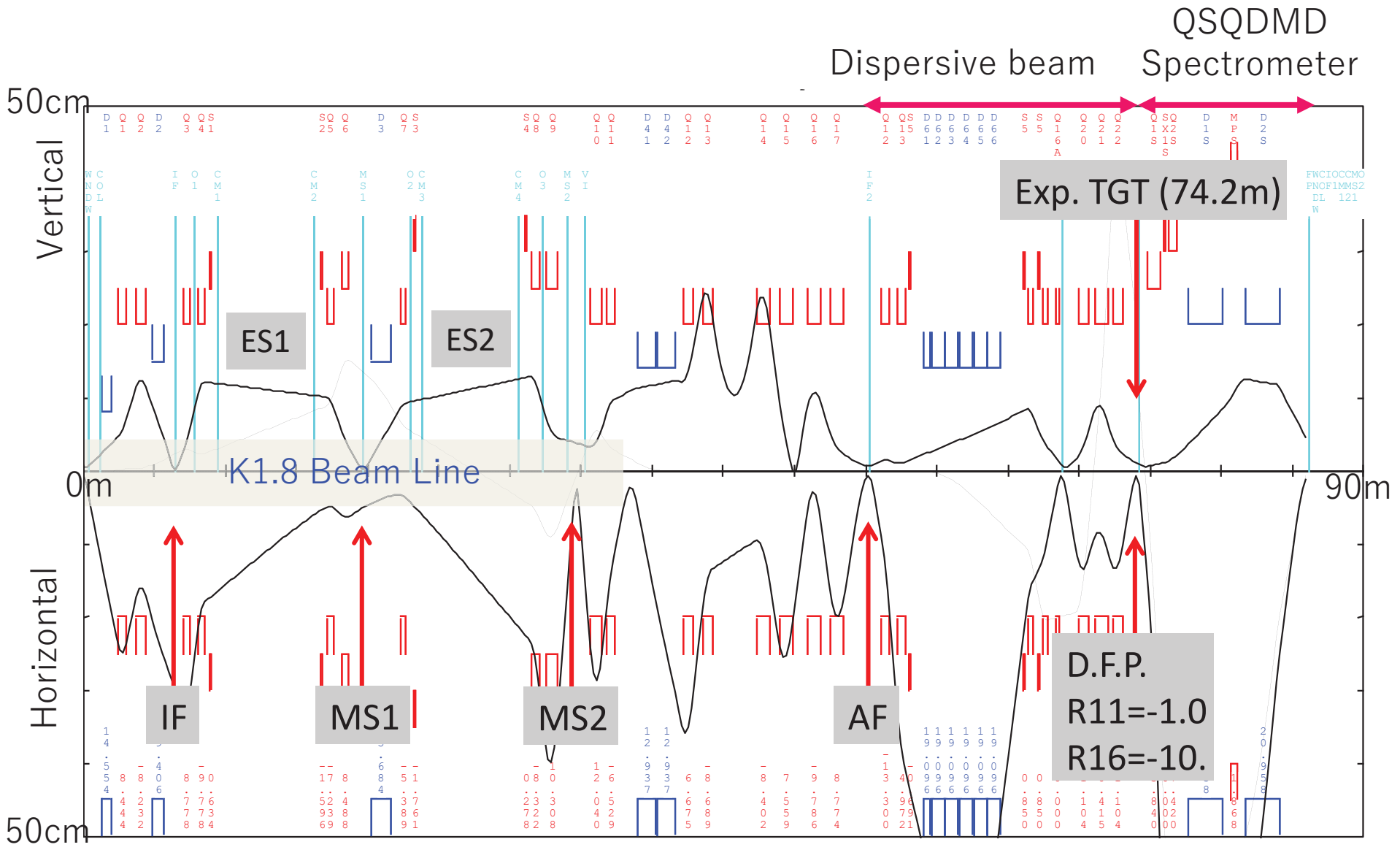

Figure 6: Beam envelope of optional HIHR branched from K1.8, calculated with dispersion matching optics to the first order.

# References

[1] K. Agari *et al.*, Prog. Theor. Exp. Phys. 2012, 02B009(2012).

[2] Y. Fujita *et al.*, Nucl. Instr. Meth. **B126**, 274(1997).

# K1.1 and K1.1BR beam lines in the Extended Hadron Experimental Facility

**Mifuyu Ukai and Toshiyuki Takahashi**

IPNS, KEK

## 1 Introduction

K1.1 and K1.1BR beam lines can supply secondary hadron beams up to 1.2 GeV/$c$. In particular, high-intensity and high-purity kaon beams are available by the 1.9 m-long separators (Wien filter type). Figure 1 shows the schematic view of the K1.1 and K1.1BR beam lines in the Extended Hadron Experimental Facility. The present K1.1 beam line elements will be moved with several modifications.

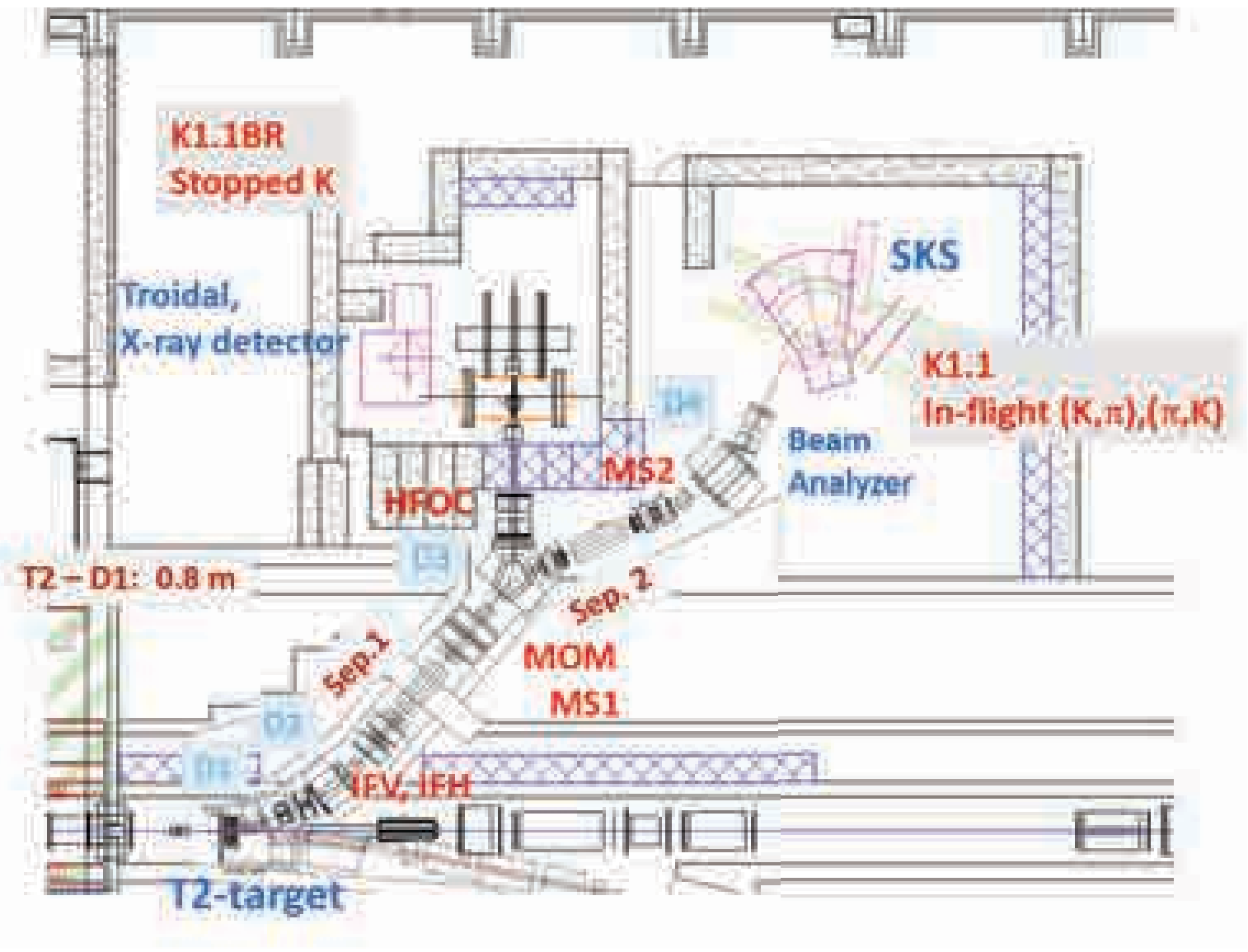

Figure 1: Schematic view of the K1,1 and K1.1BR beam lines in the Extended Hadron Experimental Facility. Sep. is the Wien filter type separators. IFV, MS1,and MS2 are vertical slits and IFH, MOM and HFOC are the horizontal slits.

## 2 K1.1 beam line

At K1.1, high purity and high intensity secondary beam is available up to 1.2 GeV/$c$ by the two stage separators. The beam line length is 28 m from the primary target to the final focus point. This beam line is optimum for the $S = -1$ sector study such as $\Lambda$ and $\Sigma$ hyperons and hypernuclear productions via the $(K^-, \pi)$ and $(\pi, K^+)$ reactions.

The expected $K^-$ intensities and purities for the beam momentum of 1.1 GeV/$c$ calculated by Decay-TURTLE[1] for several conditions are summarized in Table 1. The primary proton intensity (proton per pulse =ppp) is assumed to be $50 \times 10^{12}$ (50% loss in T2). The $K^-$ yield at the primary target (T2) was estimated by Sangford-Wang formulae [2, 3] with kinematical reflection factor. In this simulation, contributions from so-called cloud pions due to $K^0$ decay are not considered. As shown in the table, > 300 k/spill $K^-$ will be available for the reasonable purity.

Table 1: Expected $K^-$ yield and purity ($K^-$/total) at K1.1 for the beam momentum of 1.1 GeV/$c$ with $50 \times 10^{12}$ ppp on 50 % primary target loss

| Slit opening | Acceptance | ESS 70 kV/cm | ESS 60 kV/cm |
|---|---|---|---|
| IFV:±1.5 mm, MS1/MS2:±1.0 mm | 1.20 msr | 220k (98%) | 221k (97%) |
| IFV:±3.0 mm, MS1/MS2:±1.0 mm | 1.33 msr | 244k (98%) | 245k (93%) |
| IFV:±1.5 mm, MS1/MS2:±1.5 mm | 1.70 msr | 313k (74%) | 313k (19%) |
| IFV:±3.0 mm, MS1/MS2:±1.5 mm | 2.15 msr | 395k (33%) | 396k (11%) |
| IFV:±1.5 mm, MS1/MS2:±2.0 mm | 2.36 msr | 357k (13%) | 357k (3.9%) |
| IFV:±3.0 mm, MS1/MS2:±2.0 mm | 2.80 msr | 515k (5.2%) | 396k (1.5%) |

A momentum resolution of a K1.1 beam analyzer with a DQQ configulation is expected to be $\Delta p/p = 1.0 \times 10^{-3}$(FWHM) in the first order optics calculation.

Figure 2 shows the simulated beam profile at the final focus point with the slit conditions of IFV/MS1/MS2: ± 1.5 mm. As shown in the figure, the beam profile is obtained to be 3.0 cm × 1.0 cm.

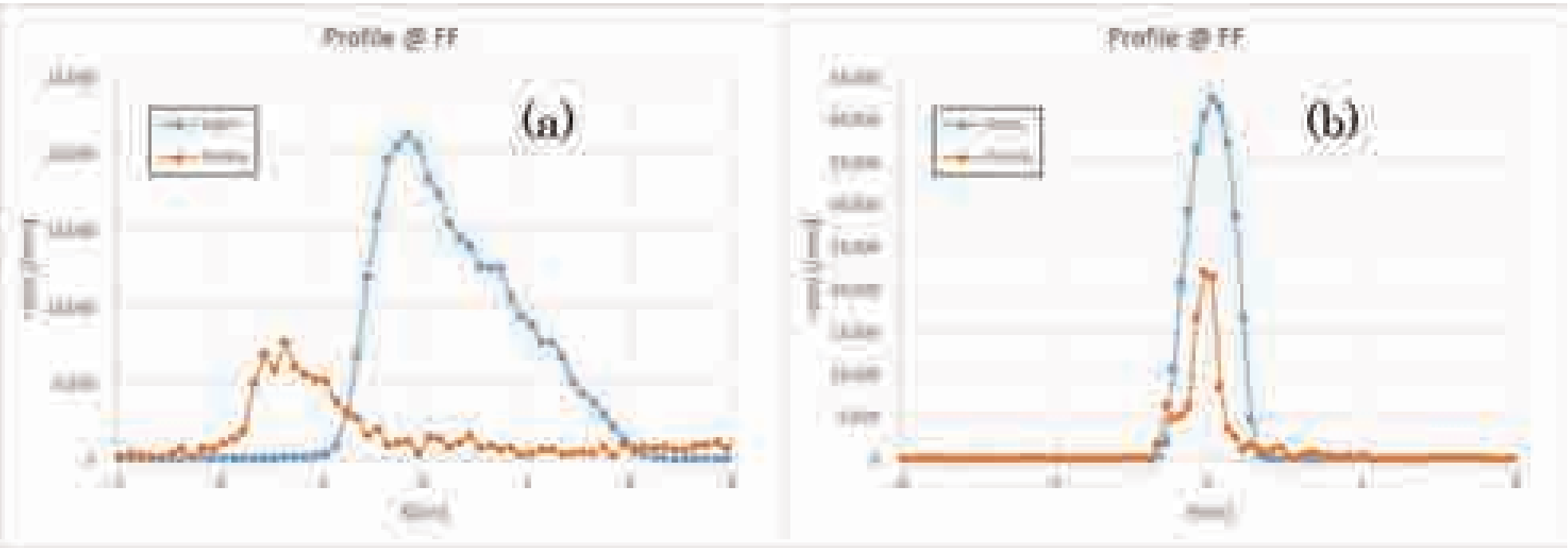

Figure 2: Simulated beam profiles at final focus point at K1.1 beam line for (a) horizontal and (b) vertical, respectively. The slit conditions are IFV/MS1/MS2: ± 1.5 mm.

## 3 K1.1BR beam line

The feature of the K1.1BR beam line is its short beam line length as 21m from the primary target to the final focus point. This beam line is optimum for experiments using the stopped $K^{\pm}$ reactionsca and decays.

The expected $K^{\pm}$ intensities and purities for the beam momentum of 0.8 GeV/$c$ for various conditions are summarized in The primary proton intensity (proton per pulse =ppp) is assumed to be $50 \times 10^{12}$ (50% loss in T2). The $K$ yield at the primary target (T2) was estimated by Sangford-Wang formulae [2, 3] with kinematical reflection factor. In this simulation, contributions from so-called cloud pions due to $K^0$ decay are not considered. As shown in the table, ∼180 k/spill for $K^+$ and ∼110 k/spill for $K^-$ are available for kaon/all purity of > 50%.

Table 2: Expected $K$ yield (/spill) and purity ($K$/total) at K1.1BR for the beam momentum of 0.8 GeV/$c$ with $50 \times 10^{12}$ ppp on 50 % primary target loss. In this calculation, separator setting is 60 kV/cm and HFOC width is ± 10 mm.

| Slit opening | Acceptance | $K^+$ | $K^-$ |
|---|---|---|---|
| IFV:±1.5 mm, MS1/MS2:±1.0 mm | 1.40 msr | 135k (90%) | 81k (87%) |
| IFV:±3.0 mm, MS1/MS2:±1.0 mm | 1.67 msr | 161k (53%) | 97k (45%) |
| IFV:±1.5 mm, MS1/MS2:±1.5 mm | 1.90 msr | 183k (69%) | 110k (61%) |
| IFV:±3.0 mm, MS1/MS2:±1.5 mm | 2.63 msr | 253k (39%) | 153k (31%) |
| IFV:±1.5 mm, MS1/MS2:±2.0 mm | 2.12 msr | 205k (47%) | 113k (39%) |
| IFV:±3.0 mm, MS1/MS2:±2.0 mm | 3.30 msr | 318k (31%) | 192k (24%) |

Figure 2 shows the simulated beam profile at the final focus point with the slit conditions of IFV/MS1/MS2: ± 1.5 mm. As shown in the figure, the beam profile is obtained to be 5.0 cm × 1.5 cm.

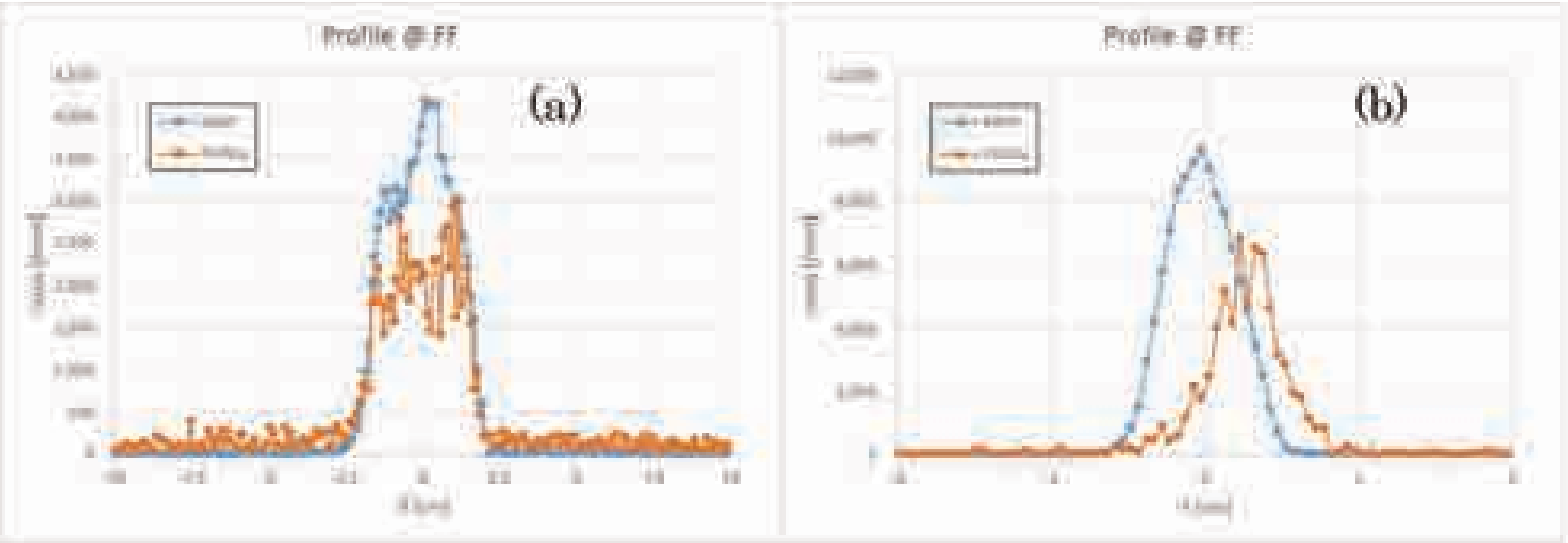

Figure 3: Simulated beam profiles at final focus point at K1.1 beam line for (a) horizontal and (b) vertical, respectively. The slit conditions are IFV/MS1/MS2: ± 1.5 mm.

# References

[1] PSI Graphic Turtle Framework by U. Rohrer based on a CERN-SLAC-FERMILAB version by K.L. Brown et al. http://aea.web.psi.ch/Urs_Rohrer/MyWeb/turtle.htm

[2] Sanford J. R., Wang C. L.. BNL-AGS internal report 11279, 1967.

[3] Sanford J. R., Wang C. L.. BNL-AGS internal report 11479, 1967.

# Experimental proposal of hyperon proton scattering experiment at the K1.1 beam line

**Koji Miwa**[1]

[1]Tohoku University

Investigation of two-body hyperon-nucleon (YN) interaction from scattering experiment is crucially important to construct realistic theoretical frameworks for YN interaction. Experiment to accumulate comprehensive data including spin observables for wide beam energy range ($0.2 \sim 1.2$ GeV/$c$) should be performed with high intensity $K^-$ beam at the K1.1 beamline. For this experiment, we think the liquid hydrogen target should be activated to detect low energy recoil protons and to select Yp scattering event without background. Therefore we would like to develop liquid $H_2$ TPC based on the knowledge of liquid Ar TPC and $H_2$ gas TPC. After this development, we want to perform Yp scattering experiment with the $(K^-, \pi^{\pm})$ reaction.

## 1 Physics motivation

We want to change the direction to derive the YN interaction to the same direction of normal NN interaction. It means that the YN interactions should be understood from the Yp scattering experiment. The observation of neutron stars with two solar mass [1] demanded especially the hypernuclear physicists to reconsider the role of hyperons in the neutron stars, because the two solar mass neutron stars can not be supported with hyperons due to the softness of equation of state. In order to support the two solar mass neutron stars, more repulsive forces are essential. Such candidates are repulsive force in YNN three body interactions. Now the experiments to measure the energy level of heavy $\Lambda$ hypernuclei to exclude the effect of the three-body YNN interaction are proposed [2]. For such a study, the reliable information of YN two body interactions is indispensable. So far, the theoretical frame works of YN interaction have been constructed to explain the experimental data of $\Lambda$ hypernuclei and the two-body YN interaction and the three-body YNN interaction could not be well separated. Now we really need to investigate the YN interaction from Yp scattering experiment. We think the purpose of the Yp scattering experiments are the followings.

- Understanding of new phenomena appeared in two body baryon interaction by introducing strange quark
- Data accumulation for better interaction model
- Direct information for YNN three-body interaction

The first topics is related to the origin of nuclear force such as short range core and spin-orbit interaction. The second and third topics are necessity of comprehensive data taking of YN channels. We need the Yp scattering data of the differential cross section including the spin observables. We also need to consider the feasibility of $\Lambda$d and $\Lambda^3He$ scattering experiments to obtain the direct information for YNN three-body interaction.

## 2 Liquid hydrogen TPC as a modern $H_2$ bubble chamber

Bubble chamber was a powerful tool to measure Yp scattering in low energy region. Actually almost all data of Yp scatterings were measured with a hydrogen bubble chamber[3]. The advantage of the bubble chamber was an imaging detector with $4\pi$ acceptance. Therefore the scattering events can be identified without background. The density of liquid hydrogen is low, so trajectory of low energy particle can be measured and total cross section can be measured. We want to make the $LH_2$ target active to enable us to accumulate comprehensive YN scattering data from low energy to high energy upto $\sim$1.2 GeV/$c$. For example, the ranges for 10 MeV proton in $LH_2$, $LD_2$ and CH (scintillation fiber) are 7, 6 and 1 mm, respectively. The 10 MeV proton can be measured in $LH_2$ and $LD_2$. This means that we can measure the YN scattering data for low energy region. Even for the high energy region, the recoil proton has a very low energy at the backward scattering angle. Because such a low energy proton can not escape from $LH_2$ target, we could not measure such proton and this sets the angular acceptance for E40 which is the proposed $\Sigma p$ scattering experiment at the K1.8 beam line [8]. If we can use an active $LH_2$ target, all angular ranges can be covered and total cross sections can be obtained. Therefore we want to exchange the hydrogen bubble chamber to modern active detector and we are considering the possibility of liquid $H_2$ TPC. The $H_2$ gas TPC [4] and liquid Ar TPC [5][6] are already established by other experimental groups. By adding these knowhow, we want to achieve the development of $LH_2$ TPC.

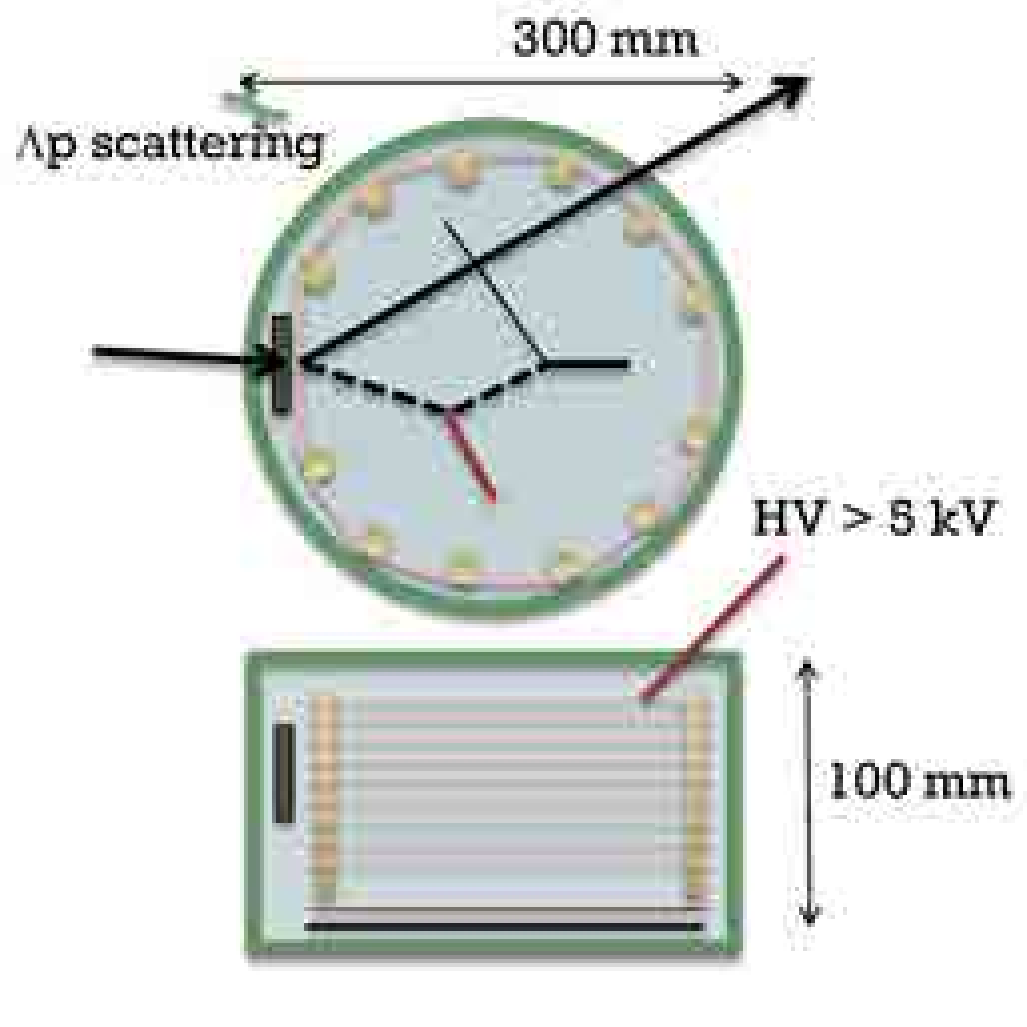

Figure 1: Conceptual drawing of $LH_2$ TPC. The upper figure shows the top view and the bottom figure shows the side view. The ionized electrons and ions are moved to the vertical direction toward the electric field.

The requirements for $LH_2$ TPC are summarized as followings. For tracking, position should be measured for each 1 mm in XY plane to detect a low energy proton of 10 MeV or less. For the drift velocity, it should be as fast as possible. In the liquid Ar TCP case, it might be $\sim$1.6 mm/$\mu s$. The drift velocity in $LH_2$ might be the same order and the drift time for 100 mm drift length, which is a typical target height, is 60 $\mu$s. This is too long to perform experiment with high intensity beam. Therefore we have to consider the solution. Finally, $LH_2$ should be exchanged to $LD_2$ or $L^3He$ to studyf $\Lambda$d scattering and $\Lambda^3$He scattering. For the $LH_2$ TPC, there are many items to be solved or to be studied.

## 3 Yp scattering experiment with $K^-$beam

In order to accumulate the comprehensive data of Yp scattering, the high intensity hyperon beams such as $\Lambda$ and $\Sigma^{\pm}$ is necessary. The polarization of the hyperon beams should be controlled. From these points of view, $K^-$ beam of 1.1 GeV/$c$ is quite suitable because the hyperon production cross section is the largest (~2.5 mb) and the hyperons are produced with a high polarization [7]. Although the Yp scattering experiment is difficult, there are big advantages to measure the spin observables. We can control the polarized hyperon beam and unpolarized hyperon beam by detecting only the $\pi$ particles scattered to the one direction or by detecting the $\pi$ particles scattered to the both directions. Then spin direction of hyperons can be identified from the direction of emitted proton in hyperon decay. Therefore it is possible to measure the polarization in scattering with unpolarized beam and Analyzing power with polarized beam. To realize this experiment, we need wide acceptance spectrometer which has the almost the same acceptance for left scattered and right scattered $\pi^{\pm}$. The $LH_2$ TPC should be installed inside of the spectrometer. We need to detect the $(K^-, \pi^{\pm})$ reaction up to ~90 degree which corresponds to the hyperon beam momentum from 0.2 GeV/$c$ to 1.2 GeV/$c$. Then we want to take data of differential cross sections and spin observables of $\Lambda$p, $\Sigma^{\pm}$p scatterings simultaneously. In the experimental points of view, there are many difficulties to be overcome. For example, the trigger rate must be huge, so we have to consider a clever trigger system and also fast DAQ system. Next, the yield estimation is shown, but this is quite preliminary value. We assume the $K^-$ beam intensity of 500 k/spill and $LH_2$ target thickness is 20 cm. When we assume a typical efficiencies and acceptance, the tagged $\Sigma$ beam is ~40 /spill. This value is about one fifth of the tagged $\Sigma$ in E40 [8] where 20 M/spill $\pi$ beam and 30 cm long $LH_2$ target are used and 3 weeks beam time is requested to accumulate 10,000 scattering events. Therefore roughly speaking, 3 month beam time is necessary to achieve the same level with E40. Although it needs rather long beam time, we can take the following data in the same setup. We summarize the possible measurements.

- Differential cross sections of Yp channels ($0.2 < p_{hyperon} < 1.2$ GeV/$c$)
  $\Sigma^+$ p, $\Sigma^-$p and $\Lambda$p elastic scatterings,
  $\Sigma^- p \to \Lambda n$ inelastic scattering
- Spin observables
  $\Sigma^+$p scattering : Polarization, Analyzing power
  $\Lambda$p scattering : Polarization, Analyzing power
  $\Sigma^- p \to \Lambda n$ scattering : Polarization
- $\Lambda$NN interaction
  $\Lambda$d scattering and $\Lambda^3 He$ scattering

## 4 Summary

Investigation of YN interaction from scattering experiment is crucially important. Comprehensive data taking of Yp scattering including spin observables should be performed at the extended hadron hall by using high intensity $K^-$ beam. In this measurement, $LH_2$ target should be activated to measure low energy proton's trajectory. We proposed the possibility of liquid $H_2$ TPC. This enables us not only to measure cross section at low energy region of a few hundred MeV/$c$ but also to measure total cross section at higher momentum region because the all scattering angles can be covered. We also proposed Yp scattering experiment with the $(K^{-}, \pi^{\pm})$ reactions at 1.1 GeV/$c$ to get the largest production cross section and high polarization of proposed hyperons. The detailed simulation and development of $LH_2$ TPC should be performed.

## References

[1] P.B. Demorest *et al.* Nature **467**,1081 (2010).

[2] S.N. Nakamura *et al.* Talk in this workshop.

[3] G. Alexander *et al.* Phys. Rev. **173**,1452 (1968).

[4] J. Egger *et al.* Eur. Phys. J. **A 50**,163 (2014).

[5] M. Antonello *et al.* Adv. High. Energy. Phys. **2013**,260820 (2013).

[6] C. Anderson *et al.* Phys. Rev. Lett. **108**,161802 (2012).

[7] B. Conforto *et al.* Nucl. Phys. **B 105**,189 (1976).

[8] K. Miwa *et al.* JPS Conf. Proc. **17**,012004 (2017).

# Λ binding energy measurement for various hypernuclei at HIHR

**Satoshi N. Nakamura**

Department of Physics, Graduate School of Science, Tohoku University

Study of Λ hypernuclei at HIHR will open a door to new physics. Especially it is essential for systematic study of 3-body repulsive force which is the key to solve the hyperon puzzle. Best mass resolution of reaction spectroscopy of Λ hypernuclei and reasonable beamtime will enable us to measure $B_\Lambda$ for wide mass range of Λ hypernuclei at HIHR.

## 1 Introduction

Understanding of many-body systems whose dynamics is governed by the strong interaction is one of major goals of nuclear physics. Baryons, nuclei and neutrons stars have different size scale from 1 fm to 10 km by order of $10^{19}$ order of magnitudes but we would like to understand their nature based on the same framework and detailed knowledge on the baryonic interaction is the key for it.

So far, the baryonic interaction has been studied on precious but limited number of hyperon-nucleon scattering and indirectly from the binding information of hypernuclei. The baryonic interaction potential models were constructed based onf $SU(3)_f$ symmetry and reproduced reasonably existing experimental data.

However, recent finding of massive neutron stars with two solar mass [1, 2] reveals our understanding of baryonic interaction is not enough. Simple discussion of Fermi energy of neutrons results in natural inclusion of Λ particle in the deep inside of a neutron star. However, inclusion of Λ hyperon makes the neutron stars' equation of the state (EoS) too soft to support two solar mass and thus it is called as the "hyperon puzzle" that so far established baryonic potentials cannot explain the existence of massive neutron stars. One of promising explanations is inclusion of many-body repulsive forces such as 3-body force which is known to be quite important even for explanation of masses of normal nuclei [3, 4, 5]. Additional repulsive force makes the NS EoS stiffer or even prevents appearance of Λs in NS core and thus the hyperon puzzle can be solved if this scenario is right.

Figure 1 shows expected effects of 3-body repulsive hyperon forces [4, 11]. The differences of calculated Λ's binding energy with and without many-body repulsive force are $< 1$ MeV for medium to heavy hypernuclei, but precision of existing experimental data is not enough to constraint the model. Therefore, experiments for various hypernuclei with 100-keV accuracy are longed for.

By using (e,e′K$^+$) reaction, precise spectroscopy of ${}^{40,48}_{\Lambda}$K are proposed [9] at Jefferson Lab but study of a few kinds of hypernuclei is not enought and thus, a systematic study of Λ hypernuclei in wide mass range is essential. Study of tri-axially deformed $^{25}$Mg with Λ can be studied the $^{25}$Mg($\pi^+$,K$^+$)${}^{25}_{\Lambda}$Mg reaction [8].

Those systematic study of Λ hypernuclei can be carried out at the HIHR of the extended hadron hall, J-PARC.

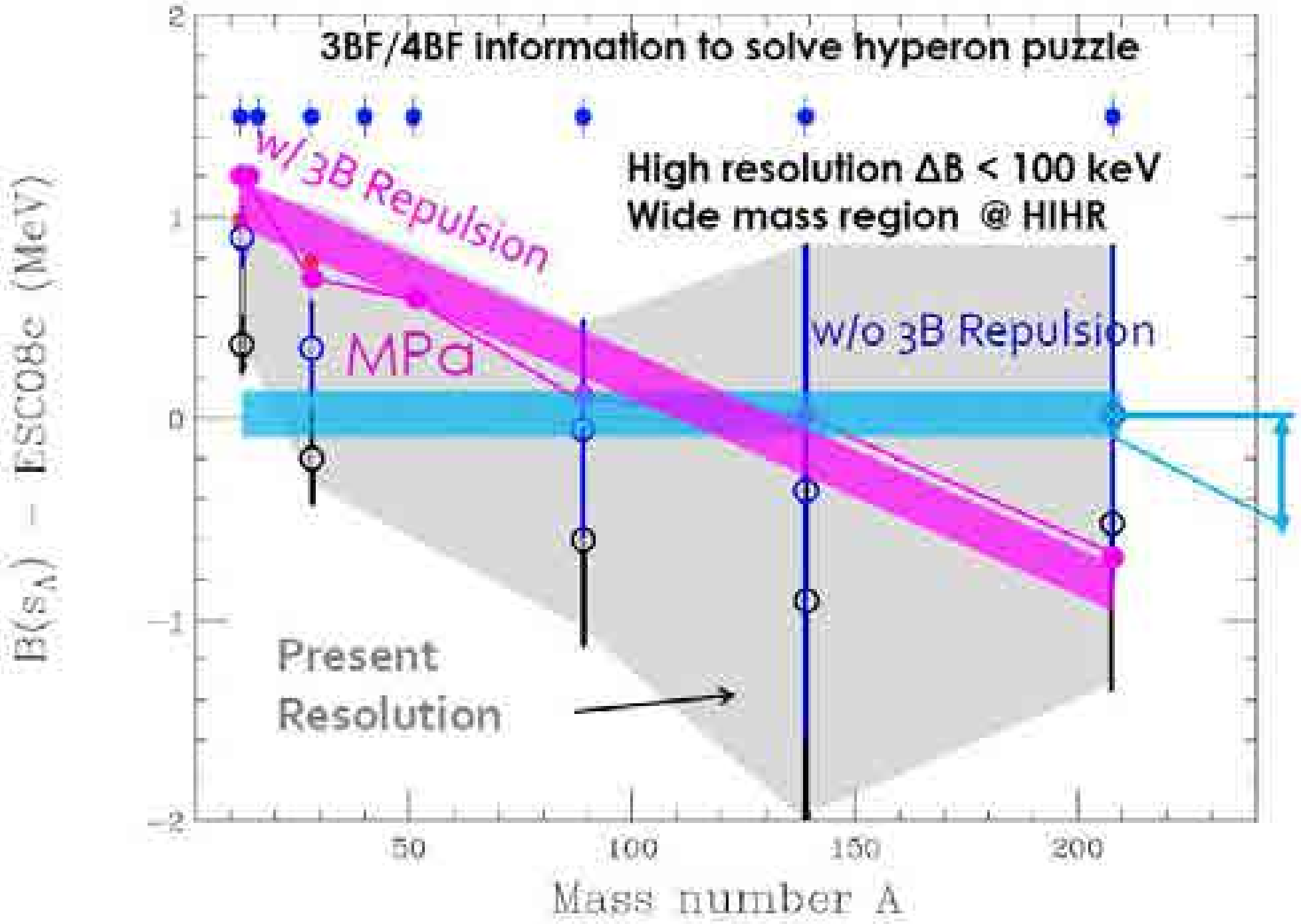

Figure 1: Expected effects of many-body repulsive force to various Λ hypernuclei with the Nijmegen potential [4]. The figure was originally prepared for KEK-PIP [10, 11].

## 2 Spectroscopy of Λ hypernuclei with the ($\pi^+$,$\mathbf{K}^+$) reaction at HIHR

A high intensity $\pi^+$ beam ($> 2.8 \times 10^8 \pi^+$s/pulse) will be available at HIHR. The $\pi$ beam is too strong to be momentum analyzed particle by particle, thus momentum dispersion matching technique is introduced [12].

Expected resolution at HIHR is compared with SKS at KEK-PS and HKS+HES at JLab (Table 1). Though further study is necessary, similar of better mass resolution can be expected ($<$ 0.5 MeV) at HIHR than JLab.

The expected yield of hypernuclei at HIHR is also compared with them at KEK-PS and JLab (Table 2). Based on the above estimation, 10 days of beamtime for one target will give 1500 counts of ground states of Λ hypernuclei. It should be mentioned major installation works are not necessary for such a study when HIHR is once constructed and it is quite different from the case for JLab experiments which need major installation works for a hypernuclear beamtime. Design of $\mathrm{K}^+$ spectrometer at HIHR is still under consideration and thus there is a room of optimization for hypernuclear yield at HIHR.

## 3 Summary

Planned HIHR at the extended hadron hall of J-PARC will enable us to carry out systematic study of binding energies of Λ hypernuclei with unprecedented precision in wide mass range. It will shed light on new types of hypernuclear studies such as tri-axial deformed hypernuclei with Λ probe and charge symmetry breaking by comparing with (e,e$'$K$^+$) data. Furthermore,

Table 1: Mass resolution comparison for $^{12}$C target

| | SKS @ KEK-PS | HKS+HES @ JLab | HIHR @ J-PARC Ex.HD |
|---|---|---|---|
| Beam | 1.1 GeV/$c$ $\pi^+$ | 1.5 GeV/$c$ $\gamma^*$ | 1.1 GeV/$c$ $\pi^+$ |
| Reaction | $^{12}$C($\pi^+$, K$^+$)$^{12}_{\Lambda}$C | $^{12}$C($\gamma^*$, K$^+$)$^{12}_{\Lambda}$B | $^{12}$C($\pi^+$, K$^+$)$^{12}_{\Lambda}$C |
| Beam Resolution (FHWM, $\Delta p/p$) | 1.1 MeV/$c$ ($1 \times 10^{-3}$) | 0.2 MeV/$c$ ($< 1 \times 10^{-4}$) | 0.2 MeV/$c$ ($2.3 \times 10^{-4}$) |
| Scat. Particles' Resolution (FHWM) | 1.1 MeV/$c$ ($1 \times 10^{-3}$) | 0.2 MeV/$c$ ($2 \times 10^{-4}$, K$^+$) | 0.2 MeV/$c$ ($2.3 \times 10^{-4}$) |
| Scat. Particles' Resolution (2$^{\rm nd}$) | NA | 0.36 MeV/$c$ ($4 \times 10^{-4}$, e$'$) | NA |
| Mass Resolution (Beam Optics) | 1.6 MeV | 0.45 MeV | 0.23 MeV (Mom. Dis. Match) |
| Straggling in Target | 1.0 MeV | 0.17 MeV (87.5 mg/cm$^2$) | 0.2 MeV (100 mg/cm$^2$) |
| Matrix offset due to z-displacement | | 0.09 MeV for C (0.68 MeV for 4 mm Li) | to be studied |
| Total Mass Resolution | 1.9 MeV | 0.5 MeV | 0.3 MeV + ? |

100 keV accuracy $B_\Lambda$ measurements to be realized for various targets at HIHR will clarify the nature of the 3-body $\Lambda NN$ repulsive force which is the key for solution of the hyperon puzzle.

In order to realize systematic study of $\Lambda$ hypernuclei at HIHR, further optimization of dispersion matching beamline and beamline optics studies including energy losses and matrix offset due to target thickness are necessary.

## References

[1] P.B. Demorest *et al.*, Nature, **467**, 1081 (2010).

[2] J. Antoniadis *et al.*, Science, **340**, 1233232 (2013).

[3] T. Furumoto *et al.*, Phys. Rev. C **79**, 001601(R) (2009).

[4] Y. Yamamoto, T. Furumoto, N. Yasutake and Th.A. Rijken, Phys. Rev. C **90**, 045805 (2014).

[5] D. Lonardoni, F. Pederiva, and S. Gandolfi, Phys. Rev. C **89**, 014314 (2014).

[6] D. Lonardoni, A. Lovato, S. Gandolfi, and F. Pederiva, Phys. Rev. Lett., **114**, 092301 (2015).

[7] M. Isaka, Y. Yamamoto, and Th.A. Rijken, Phys. Rev., C **95**, 044308 (2017).

[8] M. Isaka *et al.*, Phys. Rev. C **85**, 034303 (2012).

Table 2: Yields of hypernuclei for $^{12}$C target

| | SKS @ KEK-PS | HKS+HES @ JLab | HIHR @ J-PARC Ex.HD |
|---|---|---|---|
| Reaction | $^{12}C(\pi^+, K^+)^{12}_{\Lambda}C$ | $^{12}C(\gamma^*, K^+)^{12}_{\Lambda}B$ | $^{12}C(\pi^+, K^+)^{12}_{\Lambda}C$ |
| Beam (/sec) | $7.5 \times 10^5 \pi^+$ (3 M/spill) | $9.7 \times 10^9 \gamma^*$ (20 $\mu$A, HES) | $4.7 \times 10^7 \pi^+$ (280 M/spill, 50 kW) |
| Target Thickness (g/cm$^2$) | 0.9 | 0.088 ($\phi$ 12 mm) | 0.1 (Width $\pm$100 mm) |
| Solid Angle for K$^+$ (msr) | 100 (SKS) | 9 (SPL+HKS) | 10 (HIHR) |
| Efficiencies (incl. K loss) | 0.34 (5 m for SKS) | 0.27 (12 m for HKS) | 0.11 (12 m for QSQDD) |
| Straggling in Target | 1.0 MeV | 0.17 MeV (87.5 mg/cm$^2$) | 0.2 MeV (100 mg/cm$^2$) |
| Cross Section ($\mu$/sr) | 7.3 | 0.097 | 7.3 |
| Expected Yield(/h) | 30.4 | 26.8 | 6.6 |

[9] S.N. Nakamura it et al., JLab PAC44 proposal E12-15-008 (2016).

[10] S.N. Nakamura, Int'l WS on Ext. HD Facility of J-PARC, Tokai (2016).

[11] S. Sawada, Hadron Hall Extension Project, PIP meeting for KEK (2016).

[12] H. Noumi, HIHR, Int'l WS on Ext. HD. Facility of J-PARC (this WS), Tokai (2018).

# Systematic study of neutron-rich hypernuclei at HIHR beyond J-PARC E10

**Ryotaro Honda**[1]

[1]Dept. of Phys., Tohoku University

Neutron-rich hypernuclei, which are loosely bound systems, may reveal the specific aspect of the nature of nuclear physics, i.e., changing the property of the host environment by the impurity effect of Λ and the ΛN-ΣN mixing. So far, although several experiments were carried out to search for the neutron-rich hypernuclei using the double-charge exchange reaction, the clear peak structure has not bean observed yet. In order to overcome this situation, a new experimental idea for the neutron-rich hypernuclei spectroscopy using the $(\pi^-, K^+)$ reaction is introduced in this article. The experiment will be performed in the high intensity high resolution beam line in the extended hadron facility in J-PARC. As the result of the yield estimation, it was found that the quite long beam time about 230 days is necessary. Some reconsideration is essential.

## 1 Introduction

Neutron-rich nuclei, which show a different aspect of the nucleus such as the neutron halo and neutron skin, are an interesting subject in the nuclear physics. Implementation of Λ into neutron-rich nuclei introduces two more important properties. One is changing environment of host (core) nuclei owing to the glue-like role of Λ. The Λ particle can penetrate deeply inside the host nuclei as Λ is free from the Pauli blocking among nucleons. Surrounding nucleons are gathered by the attractive force of the ΛN interaction and environment of host nuclei such as a matter radius and a diffuseness may be changed. In general, as Λ and nucleons are weakly coupled, such the effect is small. However, it will be enhanced in a loosely bound system like a neutron-rich hypernuclei. Thus, Λ can be recognized as an impurity to control the diffuseness and to reveal a new aspect of nuclear system.

The other important property is the ΛN-ΣN mixing. Λ can be coupled with Σ inside nuclei, if the isospin of core nuclei is non-zero. It is the feature of the ΛN interaction that this mixing effect is large. This originates from the small mass difference between ΛN and ΣN comparing to that of between NN and ΔN. In the neutron-rich hypernuclei, it is expected that the three-body ΛNN interaction coming from the ΛN-ΣN mixing is important. Λ can be coupled with Σ without an excitation of core nuclei and the mixing effect is summed up coherently, if core nuclei have non-zero isospin. Akaishi *et al* suggested that the coherent Λ-Σ coupling is a key to understand the level energy of *s*-shell hypernuclei [1]. In addition, an extra attraction from the coherent coupling in the $^{6}_{\Lambda}$H hypernucleus was also suggested by Akaishi *et al* [2]. Therefore, the neutron-rich hypernuclei are suitable system to investigate the particle mixing of the baryon-baryon interaction.

The double-charge exchange reaction (DCX) such as the $(\pi^-, K^+)$ and $(K^-, \pi^+)$ reactions can be used to directly produce the neutron-rich hypernuclei and to measure its binding energy by the missing mass method. Two protons in the target are converted Λ and neutron by the DCX reaction, a neutron and proton ratio is drastically changed in the case of the light nuclei. So far, several experiments were performed using the $(\pi^-, K^+)$ and $(K^-_{\text{stopped}}, \pi^+)$ reaction in KEK-PS and DAΦNE [3, 4, 5]. Recently, an experiment to search for ${}^{6}_{\Lambda}$H using ${}^{6}$Li target was carried out in J-PARC (J-PARC E10) [6]. However, the peak structure of neutron-rich hypernuclei has not been observed yet. The reason why the peak has not been observed is the quite small production cross section. According to the past experimental result, the production cross section of neutron-rich hypernuclei via DCX is one thousandth smaller than those via the non-charge exchange reaction. In the KEK-PS E521 experiment, $1.05 \times 10^{12}$ $\pi^-$ was bombarded to the ${}^{10}$B target in total [5]. However, only 47 counts were observed below the binding threshold. The cross section per one event was quite small, about 0.23 nb/sr.

# 2 Experiment at HIHR

Author considers the high intensity high resolution (HIHR) beam line in the extended J-PARC hadron facility is an unique beam line in order to break through this situation. We don't need to measure the beam $\pi^-$ in spite of the secondary meson beam owing to the dispersion matching technique in HIHR. Then, it is possible to use a quite high intensity $\pi^-$ beam, that is $2.8 \times 10^8$ per spill (spill duration: 2s). Large number of beam injection is essential for the $(\pi^-, K^+)$ experiment by considering that $10^{12}$ beam accumulation was not enough in the past experiments. Thus, HIHR is the suitable beam line to search for the neutron-rich hypernuclei via the $(\pi^-, K^+)$ reaction. On the other hand, the $K^+$ spectrometer will be designed to achieve the high missing mass resolution of a few hundred keV. Then, we are able to search the peak structure with the better resolution comparing to that of the past experiment by choosing a reasonable target thickness. Furthermore, the missing mass spectroscopy using the $(\pi^-, K^+)$ reaction can achieve the background free condition [6]. Thus, the HIHR beam line the most suitable facility to investigate the neutron-rich hypernuclei. The better result is expected than that of J-PARC E10, which was performed using the superconducting kaon spectrometer (SKS) at the K1.8 beam line.

## 2.1 Physics impact

Figure 1 shows the $|N - Z|/(N + Z)$ ratio of the produced hypernuclei via the $(\pi^\pm, K^+)$ reactions, where $N$ and $Z$ represent the number of neutron and proton, respectively. The hypernuclei produced via the DCX reaction using light nuclus as a target are well neutron rich. Then, we study whether the neutron-rich hypernuclei is deeply bound owing to the extra attraction coming from the ΛNN interaction or not by the light neutron-rich hypernuclei. Deeply bound system due to the ΛNN interaction was theoretically suggested [2], but it is not confirmed experimentally yet. Measurement of the binding energy will give a conclusion for the existence of coherent coupling in the neutron-rich environment. In addition, as the mass number increases, the $|N - Z|/(N + Z)$ ratio decreases. Systematic study of the binding energy as a function of the $|N - Z|/(N + Z)$ ratio by picking up the several target nuclei may give a detailed information about the ΛN-ΣN mixing. The ΛN-ΣN mixing is an origin of the $k_F$ dependence of the ΛN interaction in the nuclear matter. This is essential knowledge to investigate the high dense matter such as the neutron star. However, the ΛN-ΣN mixing is still not well known. Systematic study of the neutron-rich hypernuclei may be a break through for this situation.

From the view point of the changing the host nucleus environment, the ${}^{12}$C target, which produces the ${}^{12}_{\Lambda}$Be hypernucleus, is an interesting subject. The core nucleus, ${}^{11}$Be, is well deformed and its ground state spin parity is $1/2^+$ in spite of the $p$-shell nucleus. However, by introducing $\Lambda$, as the overlap between $\Lambda$ and the core nucleus for $1/2^-$ state is larger because this negative parity state is less deformed, the $1/2^-$ state of $^{11}$Be gains larger binding energy than that of the $1/2^+$ state. Then, it is expected that the ground state parity of $^{12}_{\Lambda}$Be is negative [7]. This is called as the parity reversion. In this system, $\Lambda$ can be considered as a probe to measure the deformation parameter of the core nucleus. The new degree of freedom, deformation, will be given to the hypernucleus level energy by $^{12}_{\Lambda}$Be.

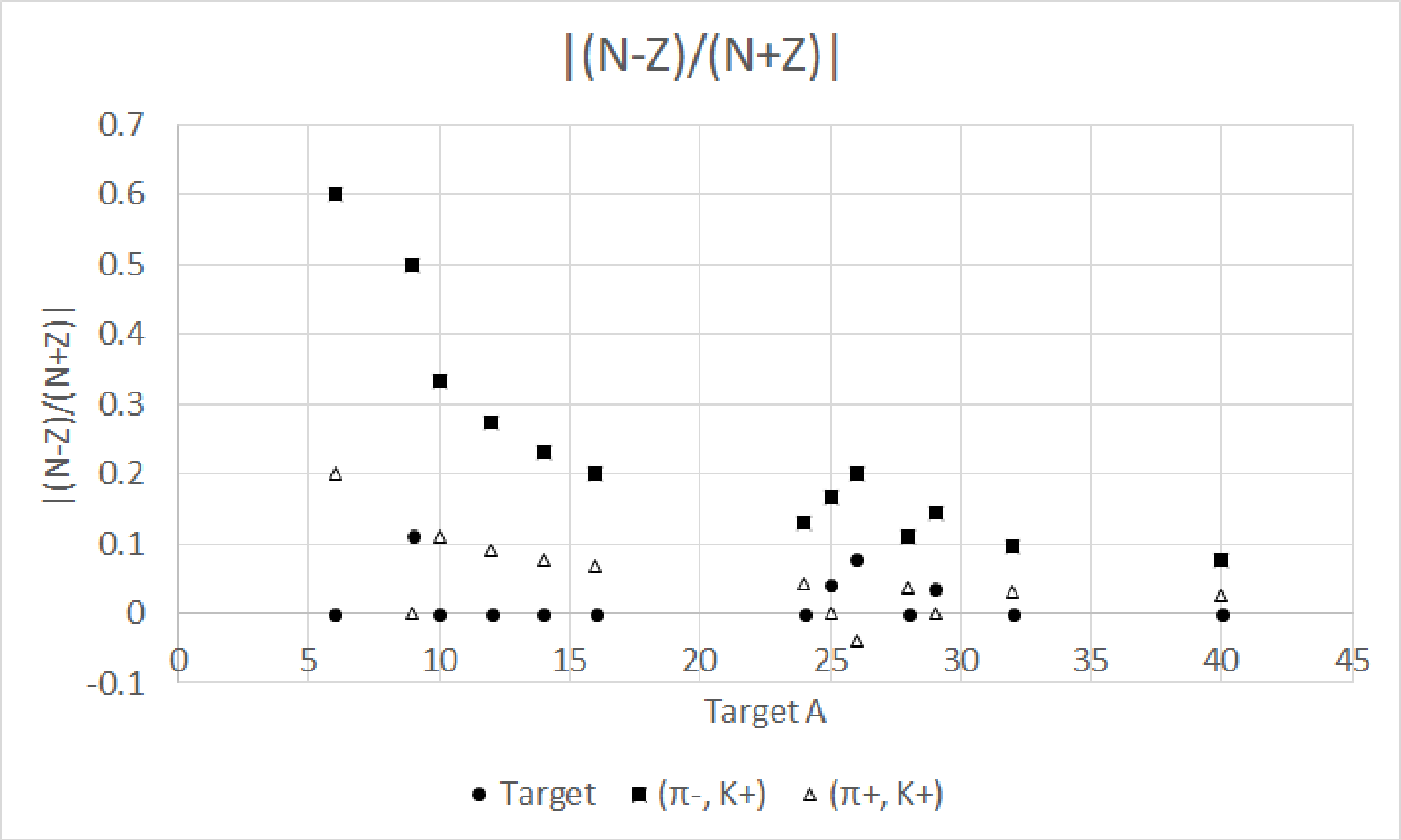

Figure 1: The $|N-Z|/(N+Z)$ ratio of hypernuclei produced via the $(\pi^{\pm}, K^+)$ reaction. Closed circle, closed square, and open triangle shows the ration of the target nuclei and hypernuclei produced via $(\pi^-, K^+)$ and $(\pi^+, K^+)$, respectively.

## 2.2 Experimental design

The HIHR beam line will equip the $K^+$ spectrometer, which completes dispersion matching. This experiment will be performed by using only the spectrometer system. No other system such as a surrounding detector is not considered at present. Author plans to use the $\pi^-$ beam with the momentum of 1.2 GeV/c. The beam intensity must be as high as possible. Although the beam intensity of $2.8 \times 10^8$ per spill is expected in the present plan, the higher intensity is desirable. Owing to the DCX reaction, the beam $\pi^-$ does not pass through the $K^+$ spectrometer. Then, it is expected to keep the low trigger rate if the beam intensity is increased.

Here, author estimates the yield. In the J-PARC E10 experiment, we assumed that the production cross section was 10 nb/sr according to the KEK-PS E521 result. However, this value was obtained by integrating all the event below the binding threshold. The cross section for each state may be smaller than 10 nb/sr. Then, author assumes the production cross section of 1 nb/sr for each hypernuclear state. In this estimation, author assumes the $^{12}$C target of 2.0 g/cm$^2$ in thickness. As the spectrometer systems are designed to achieve the missing mass resolution about a few hundred keV, a thinner target should be used originally. However, as the yield is not enough with a thin target, the thick target is considered. Other parameters used for the yield estimation are summarized in Table 1. As the result, it is found that $10^{15}$ $\pi^-$ is necessary to obtain 100 counts for the ground state. The required number of beam is three order of magnitude larger than the beam injection in KEK-PS E521 and J-PARC E10. The reason that such the massive beam accumulation is necessary originates from the $K^+$ spectrometer. The acceptance and the flight path length of the planed $K^+$ spectrometer in HIHR are ten times smaller and three times longer than those of the SKS complex used in the past experiments, respectively.

Unfortunately, the $(\pi^-, K^+)$ experiment at HIHR is unrealistic in the present plan if the production cross section is 1 nb/sr. As success of this experiment strongly depends on the cross section, a precise theoretical prediction is essential. From the view point of the experiment, the planed HIHR condition is not enough. We have to overcome this situation by improving the $K^+$ spectrometer or the beam intensity.

Table 1: Parameters used for the yield estimation.

| Cross section (nb/sr) | Beam intensity (per spill) | Target thickness (g/cm$^2$) |
|---|---|---|
| 1 | $2.8 \times 10^8$ | 2.0 |
| $K^+$ spectrometer acceptance (sr) | $K^+$ decay factor | Analysis and DAQ efficiency |
| 0.01 | 0.1 | 0.9 |

## 3 Summary

The neutron-rich hypernuclei are an important subject to investigate the changing the host nuclei environment and the ΛN-ΣN mixing. So far, several experiments were performed using the $(\pi^-, K^+)$ reaction and the $(K^-_{\text{stopped}}, \pi^+)$ reaction. However, the peak structure has not been observed yet in the missing mass spectrum. The HIHR beam line, which is planed in the extended hadron facility, is an unique experimental site to overcome this situation. Although author estimates the yield by assuming the production cross section of 1 nb/sr, it was found that the $10^{15}$ $\pi^-$ is necessary to obtain 100 counts for the ground state. It takes about 230 days to accumulate such the beam $\pi^-$. Some improvement for the beam intensity or the $K^+$ spectrometer are essential to shorten the beam time.

## References

[1] Y. Akaishi *et al.*, Phys. Rev. Lett., **84**, 3539 (2000).

[2] Y. Akaishi and T. Yamazaki, Frascati Phys. Ser., XVI, 59 (1999).

[3] M. Agnello *et al.*, Phys. Lett. B, **640**, 145 (2006).

[4] M. Agnello *et al.*, Phys. Rev. Lett, **108**, 042501 (2012).

[5] P. K. Saha *et al.*, Phys. Rev. Lett, **94**, 052502 (2005).

[6] R. Honda *et al.*, Phys. Rev. C, **96**, 014005 (2017).

[7] H. Homma *et al.*, Phys. Rev. C, **91**, 014314 (2015).

# Gamma-ray spectroscopy of hypernuclei produced by the $(\pi^-, K^0)$ reaction

**Takeshi Yamamoto**

KEK IPNS

The existence of the CSB effect was experimentally confirmed in $s$-shell mirror hypernuclear system. However, further precise data for $s$-shell as well as $p$-shell mirror hypernuclei is essential to study the CSB effect and underling $\Lambda N$ interaction. Precise data of gamma-ray measurements are available for proton-rich side mirror hypernuclei while no such data for neutron-rich side hypernuclei. Therefore, a gamma-ray spectroscopic experiment was proposed, in which the $(\pi^-, K^0)$ reaction will be used to produce neutron-rich side mirror hypernuclei. Reasonable yield can be realized with the SKS spectrometer, a Ge detector array (Hyperball-J) and a developing range counter system.

## 1 Introduction

The charge symmetry breaking (CSB) effect reported in the A=4 mirror hypernuclei is one of the hot topics in strangeness nuclear physics. Precise gamma-ray spectroscopy with an energy resolution of a few keV is a powerful tool to investigate such an effect. Recently, a gamma-ray spectroscopic experiment on ${}^4_\Lambda$He was performed at the J-PARC K1.8 beam line (J-PARC E13) for this purpose. ${}^4_\Lambda$He hypernuclei were produced by the non charge exchange $(K^-, \pi^-)$ reaction with a 1.5 GeV/$c$ kaon beam. The excitation energy of first excited state of ${}^4_\Lambda$He$(1^+)$ was successfully determined to be 1.406±0.004 MeV [1]. On the other hand, that of the mirror hypernucleus (${}^4_\Lambda$H) was reported as 1.09±0.02 MeV in average from old studies [2, 3, 4]. By comparing these results, we conformed the existence of CSB effect in the excitation energy. In theoretical studies based on widely accepted Nijimegen meson-exchange interaction model [5], such a large CSB effect could not be reproduced. It is pointed out that the $\Lambda N - \Sigma N$ mixing term is sensitive to the strength of the CSB effect. Study of the CSB effect in hypernuclear structure may become a new probe to investigate $\Lambda N$ interaction.

One of key points in the future experimental study is high precision measurement to compare with expected CSB effects of ~100 keV and also theoretical prediction via *ab-initio* calculation with an accuracy on a few keV[6]. In near future, precise measurement of the excitation energy of ${}^4_\Lambda$H$(1^+)$ (J-PARC E63) will be performed to experimentally establish the $s$-shell mirror hypernuclei system. Next step is systematic study of CSB effect in havier $p$-shell hypernuclear system, in which size of the CSB effect is expected to be 10-150 keV in theoretical case study [7]. Precise data of gamma-ray measurement are available for proton-rich side mirror hypernuclei while no such data for neutron-rich side hypernuclei. Therefore, new experimantal data for neutron-rich side hypernuclei are awaited. Charge exchange reactions, $(K^-, \pi^0)$ ,$(\pi^-, K^0)$ and $(e, e'K^+)$ reactions, are required to produce these hypernuclei. Taking account of experimental difficulty in using $(K^-, \pi^0)$ and $(e, e'K^+)$ reactions, we are planning the gamma-ray measurement via the $(\pi^-, K^0)$ reaction.

## 2 Proposed experiment

First measurement will be performed for $^{12}_{\Lambda}$B of which mirror hypernuclei, $^{12}_{\Lambda}$C, was well studied by precise gamma-ray spectroscopy (KEK-PS E566) [8]. $^{12}_{\Lambda}$C production via the non-charge exchange $(\pi^-, K^-)$ reaction was tagged by SKS spectrometer and gamma-rays were detected by Ge detector array (Hyperball-2). Most intense gamma-ray peak in the report, 161.5 ±0.3 ±0.3 keV with 172 ±25 counts, is originated from $^{12}_{\Lambda}$C$(2^- \rightarrow 1^-)$ transition, corresponding to the transition between the ground doublet states.

The proposed measurement follows the technique used in the $^{12}_{\Lambda}$C measurement with the different hypernuclear production reaction, the $(\pi^-, K^0)$ reaction. It will be the first time to apply the reaction for hypernuclear study while elementary $p(\pi^-, K^0)\Lambda$ production was already studied by previous experiments [9]. It is noted that, other reactions, $(K^-, \pi^0)$ and $(e, e'K^+)$ reactions, on $^{12}$C target also produce $^{12}_{\Lambda}$B, while there are technical difficulty to apply these reactions: (1) the $(K^-, \pi^0)$ reaction case, huge volume crystals will be necessary to detect high energy two gammas from the $\pi^0$ decay. Good energy and opening angle resolution will be required for the huge detector system to tagging hypernuclear production events. It is also difficult issue to avoid conflict of coverages of the high energy gamma detector system and Ge detectors. (2) the $(e, e'K^+)$ reaction case, good energy resolution will be obtained as reported in J-LAB experiments [10]. However, smaller production cross section, by a factor of $\sim 10^3$ than the $(\pi^-, K^0)$ reaction, requires $\mu$A order beam intensity in where Ge detectors can not be operated. The $(\pi^-, K^0)$ reaction seems to be rather easier to realize the gamma-ray spectroscopic experiment with reasonable production cross section. By applying the charge exchange $(\pi^-, K^0)$ reaction, a lot of neutron rich side hypernuclide can be produced as shown in Figure 1(a). Establishment of the tagging method with the $(\pi^-, K^0)$ reaction will extend gamma-ray data, which will be useful not only for study of CSB effect but also other issues related to neutron rich hypernuclei.

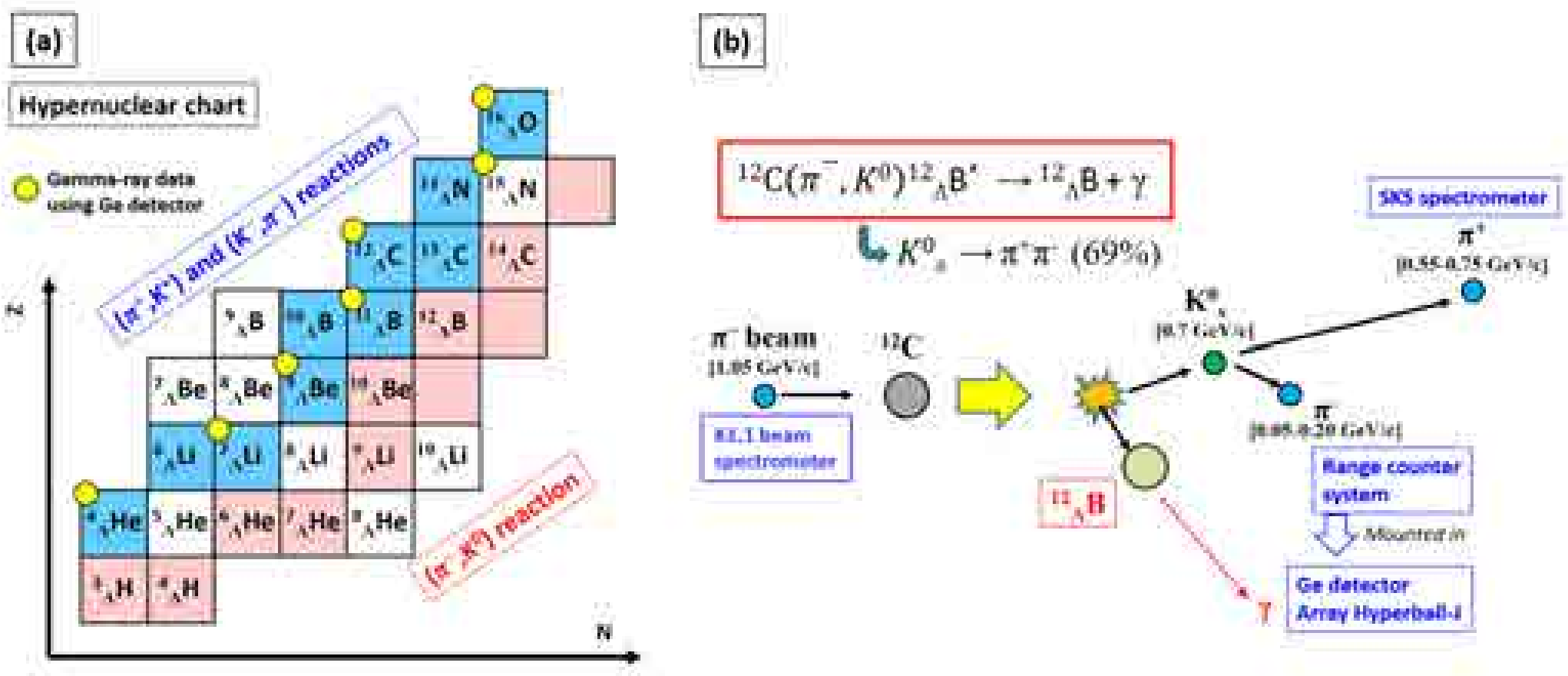

Figure 1: (a) A hypernuclear chart in which existing precise gamma-ray data are marked by yellow circles. Existing measurements employ the $(\pi^+, K^+)$ and $(K^-, \pi^-)$ reactions. The charge exchange $(\pi^-, K^0)$ reaction is necessary to produce mirror pair hypernuclei. (b) A schematic drawing of the $(\pi^-, K^0)$ reaction. High momentum $\pi^+$s and low momentum $\pi^-$s will be analyzed by the SKS spectrometer and the range counter system, respectively. Gamma rays will be detected by Hyperball-J.

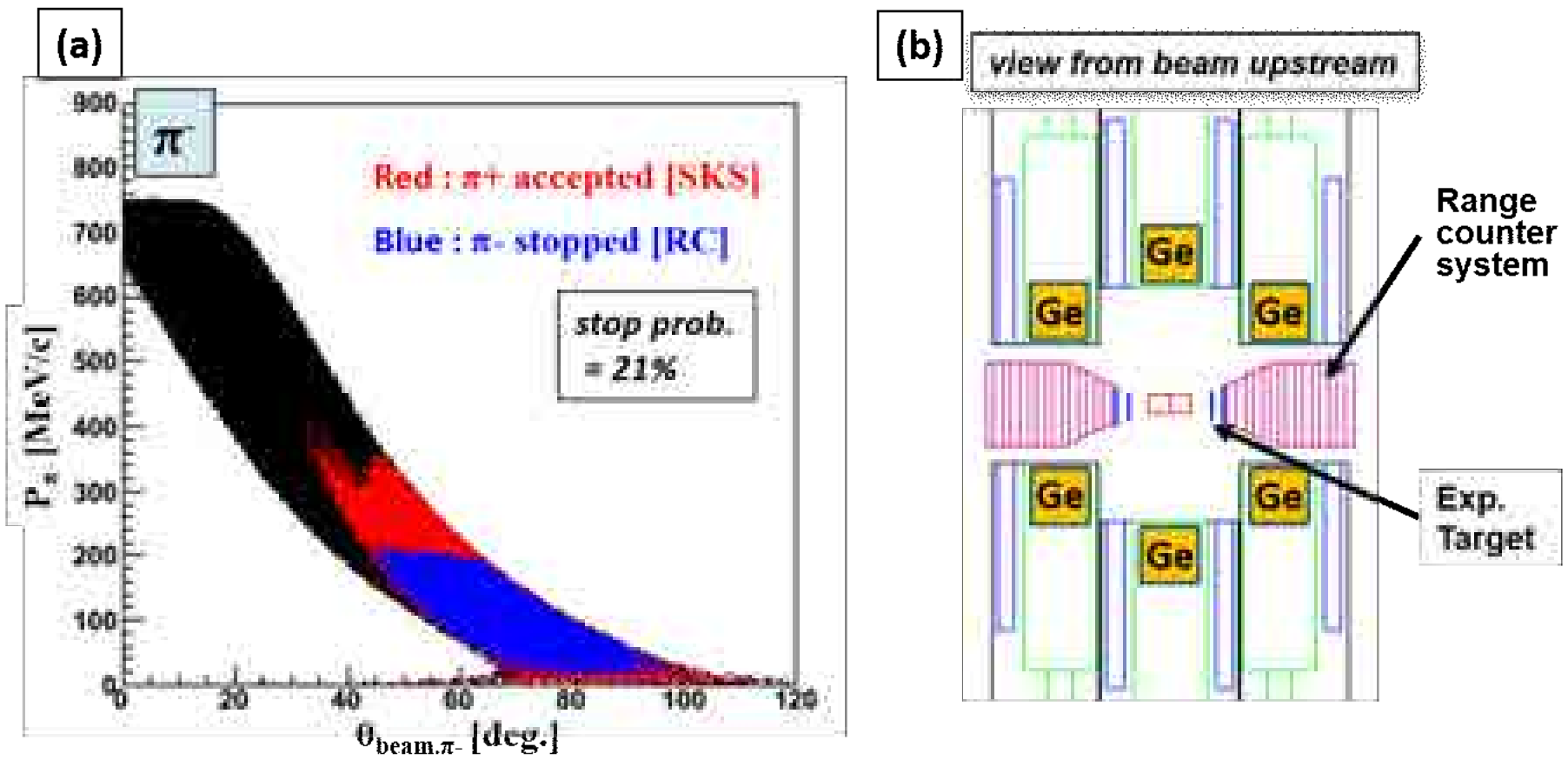

Figure 2: (a) correlation between momentum ($P_{\pi-}$) and emitting angle ($\theta_{beam.\pi-}$) of low momentum $\pi^-$. Red points shows events where $\pi^+$ is accepted by SKS spectrometer. Blue points shows $\pi^-$ stopped events. (b) cross-sectional view of the Hyperball-J with range counter mounted in. Range counters will be placed at side of the target to keep Ge detector coverage.

Figure 1(b) shows overview of the proposed measurement. This idea is based on weak decay experiments discussed in [11]. Beam pion momentum was chosen to be 1.05 GeV/$c$ for higher cross section of the elementary $(\pi^-, K^0)$ reaction. ${}^{12}_{\Lambda}$B will be produced via the reaction on the ${}^{12}$C target with emitting $K^0$ to forward angle as in the case of emitting $K^-$ in the ${}^{12}_{\Lambda}$C measurement. Two charged pions will be generated from $K^0_s$ decay (branching ratio = 69%) with wide angular distributions. We will concentrate on the case of emitting $K^0_s$ of which mean flight path is ∼4 cm, assuming not huge detector system for detecting both two pions. Reasonable coverage and energy resolution will be realized in the case of the $\pi^+$ has larger momentum, 0.6–0.7 GeV/$c$, than the other $\pi^-$, 0.1–0.2 GeV/$c$. Momentum analyzer will be the SKS spectrometer for $\pi^+$ and range counter for $\pi^-$, respectively. Required resolution is $10^{-3}$ for SKS spectrometer and ∼7% for range counter [12]. As shown in Figure 2, emitting direction of $\pi^-$ is 50–90 deg. from the beam axis when high momentum $\pi^+$ accepted with SKS spectrometer placed forward. Therefore, range counter should cover 50–90 deg. angle. It is noted that compact size range counter has a merit for realize reasonable coverage for both $\pi^+$ and gamma rays from ${}^{12}_{\Lambda}$B. The gamma-ray energy will be measured with recently developed Ge detector array, Hyperball-J. The array employs mechanical cooled Ge detector which can suppress effects of radiation damage by irradiating high intensity hadronic beam as in the case of proposed experiment.

Experimental setup is almost common with gamma-ray spectroscopic experiment on ${}^{4}_{\Lambda}$H (J-PARC E63) which will be performed at the K1.1 beam line. Additional compact size range counter, $20^W \times 10^H \times 30^D$, will be mounted inside Hyperball-J to maximize coverage for widely distributed $\pi^-$. The range counter will consists of multi layers of plastic scintillator with ∼5 mm thickness for each. $\pi^-$ stopped probability was estimated to be 21% of events where $\pi^+$ is accepted by SKS spectrometer. Coverage of the Hyperball-J will be reduced by 16% because part of Ge detectors will be unmounted to make space for the range counter. Yield estimation was done by comparing with the ${}^{12}_{\Lambda}$C measurement, in which the $(\pi^-, K^-)$ reaction, isospin symmetric pair of the $(\pi^-, K^0)$ reaction, was used. Major differences are follows: (1) branching ratio of $K^0_s \to \pi^+\pi^-$, (2) $\pi^+$ accept probability by the SKS spectrometer, (3) $\pi^-$ stopping probability in the range counter and (4) gamma/weak decay branching ratio of $2^-$ state. By assuming 2M Hz beam intensity with 5s accelerator cycle, which is limited by throughput of the Ge detector, yield rate of the proposed experiment is expected to be one order of magnitude lower than the $^{12}_{\Lambda}$C measurement. Expected duration of beam time is ∼6 months for 100 gamma-ray counts. Father upgrades are necessary to realize the measurement. There are possible upgrades as follows: (1) higher throughput of Ge detector by waveform readout method, (2) use of wider acceptance magnet for high momentum $\pi^+$, such as KURAMA type magnet and (3) optimization of detector configuration, range counters and Ge detectors. Especially, upgrade (1) is quite important and effective because it may allow us to use higher beam intensity with much better efficiency of Ge detector.

In present, the SKS spectrometer and Ge detector array (Hyperball-J) is ready to use for the measurement. Optimization of the detector system and development of the range counter system is ongoing. Further effort on readout system of Ge detector is necessary to realize the proposed experiment.

# 3 Summary

The CSB effect in *s*-shell hypernuclear system was experimentally studied by the precise gamma-ray spectroscopy. Further precise data for *s*-shell as well as *p*-shell mirror hypernuclei is essential to study the CSB effect and underling $\Lambda N$ interaction. Proposed measurement will provide gamma-ray data on neutron-rich side hypernuclei which can not be produced by previously used non-charge exchange reactions. The $(\pi^-, K^0)$ reaction will be suitable to be applied to the measurement with the new range counter system together with existing SKS spectrometer. Reasonable yield can be realized with the combination of Hyperball-J and the range counter. Development of the range counter system as well as upgrades of detector system is necessary to realize the proposed measurement.

### – Short summary of proposed experiment –

**Physics motivation:**

Precise, better than 10 keV accuracy, determination of *p*-shell mirror hypernuclear structure by $\gamma$-ray spectroscopic experiment via the $(\pi^-, K^0)$ reaction.

**Beam condition:**

$\pi^-$ beam, 1.05 GeV/$c$, 4–8M/spill, $\Delta p/p$=$10^{-2}$.

**Equipment:**

Beam line spectrometer, SKS spectrometer, Ge detector array(Hyperball-J), Range counter system (inside Hyperball-J).

**Expected duration of beam time:**

3∼6 months with possible upgrades.

# References

[1] T.O. Yamamoto *et al.*, Phys. Rev. Lett. **115**, 222501 (2015).

[2] M. Bedjidian *et al.*, Phys. Lett. **62B**, 476 (1976).

[3] M. Bedjidian *et al.*, Phys. Lett. **83B**, 252 (1979).

[4] A. Kawachi, Ph.D. thesis, University of Tokyo, (1997).

[5] A. Nogga *et al.*, Phys. Rev. Lett. **88**, 172501 (2002).

[6] D. Gazda, A. Gal, Nunc. Phys. **A954** 161 (2016).

[7] A. Gal, Phys. Lett. **B744**, 352 (2015).

[8] K. Hosomi *et al.*, PTEP, 081D01 (2015).

[9] M. Shrestha and D.M. Manley, Phys. Rev. **C86**, 045204 (2012).

[10] T. Gogami *et al.*, Phys. Rev. **C93** 034314 (2016).

[11] M. Agnello *et al.*, Nucl. Phys. **A954** 176 (2016).

[12] J.J. Szymanski *et al.*, Phys. Rev. **C43** 849 (1991).

# Hadronic atoms at J-PARC HEF

**Shinji Okada**[1]

[1]RIKEN

Hadronic atom is a unique tool to prove the strong interaction between a hadron and a nucleus at zero kinetic energy, which has a key role in studying the low-energy QCD system. Recently two different X-ray detectors are developed towards challenging kaonic-atom experiments: a large array of silicon drift detector and a novel superconducting microcalorimeter. In the K1.1BR beamline of the extended hadron experimental facility of J-PARC, hadronic-atom X-ray spectroscopies will be performed with the novel detector systems.

## 1 Introduction

Any negatively charged hadrons (e.g., $\pi^-$, $K^-$, $\bar{p}$, $\Sigma^-$, $\Xi^-$) can bind to an atomic nucleus via the Coulomb field. This system, commonly referred to as a "hadronic atom", is essentially a hydrogen-like atom in its electronic structure. A hadron e.g., $K^-$ is approximately 1000 times more massive than an electron. The Bohr radius of a kaonic atom is therefore three orders of magnitudes smaller compared to atomic hydrogen. Effects of the strong interaction between a hadron and an atomic nucleus appear in the most tightly bound energy level being most perturbed by the strong interaction as a shift from the purely electromagnetic value and a broadening due to absorption of the hadron by the nucleus. By measuring the energy and width of X-ray lines resulting from level transitions to those levels, one can probe the strong interaction between a hadron and a nucleus at zero kinetic energy.

QCD becomes non-perturbative at low energy. Instead, the effective field theories was used which rely on experimental inputs. The hadronic atom experiments have therefore a key role in studying the low-energy QCD system and have been conducted so far to collect data on a variety of targets [1].

## 2 Kaonic atom

The simplest kaonic atom i.e., kaonic hydrogen ($K^-p$) is of great interest, since the $1s$-atomic-state shift and width deduced from the X-ray spectroscopy of the $2p \rightarrow 1s$ transition is directly related to the real and imaginary parts of the complex $K^-p$ $S$-wave scattering length. Understanding of low-energy $\bar{K}N$ strong interaction has been substantially deepened by the most recent kaonic-hydrogen atom measurement [2] and its theoretical studies (e.g., [3]). This study revealed that the $\bar{K}N$ interaction is strongly attractive in the isospin I=0 channel, which creates extensive interest in studying "kaonic nuclear states": very recently, J-PARC E15 experiment shows a clear structure which could be interpreted as the $K^-$ nuclear bound state [4]. Thus the kaonic atom X-ray measurements become now again increasingly important since it provides unique information of the interaction at zero energy which is crucial for fundamental theories of $\bar{K}N$ and $\bar{K}$-nucleus systems, while the precision is still not enough to determine the iso-spin ($I$=0,1) dependent $\bar{K}N$ interaction and $K^-$-nucleus potential strength.

Towards improving the precision, two different experimental techniques are developed depending on the strong-interaction width of X-rays to be measured: 1) a large array of semiconducting detector for large width and 2) a novel cryogenic microcalorimeter for small width.

## 2.1 Silicon Drift Detector

It has been desired to measure the $1s$-atomic-state shift and width in the second simplest system, i.e., kaonic deuterium ($K^-d$) to extract the iso-spin ($I$=0,1) dependent $\bar{K}N$ scattering length. However it is experimentally difficult due to its low X-ray yield and the large natural width (500 $\sim$ 1000 eV for 8 keV).

Most recently, two experimental groups, the J-PARC E57 collaboration and SIDDHARTA-2 collaboration, are launching new scientific campaigns towards this challenging measurement at J-PARC hadron experimental facility (HEF) in Japan and DAΦNE in Italy. Both groups use large arrays of silicon drift detectors (SDD) together with secondary-charged particle tracking system or veto system to suppress background events. A SDD has good energy resolution ($<$ 200 eV FWHM at 6 keV) and time resolution (0.5 - 1.0 $\mu$s FWHM) with an effective area of 0.64 $cm^2$. A large solid angle, more than 200 $cm^2$ effective area, is covered with many SDD arrays.

## 2.2 Superconducting TES microcalorimeters

A recent theoretical calculation suggested a possibility of distinguishing two major theoretical potential models by determining the $2p$-atomic-state shift of kaonic helium-3 and -4 with a precision below 1 eV, where the strong-interaction width is predicted to be as small as 2 eV [5]. The J-PARC E62 experiment aimed at ultra-high resolution measurement of the $K^-$-$^{3,4}$He $3d$-$2p$ X-rays by introducing a cutting-edge cryogenic technology.

The X-ray spectrometer, superconducting transition-edge sensor (TES) microcalorimeter, is a highly sensitive thermal sensor [6, 7]. An energy deposition resulting in an X-ray hit is measured via the increase in the resistance of a superconducting thin film that is biased within the sharp phase transition edge between the normal and superconducting phases. This offers unprecedented high energy resolution being more than one order of magnitude better than that achieved in the past hadronic-atom experiments using conventional semiconductor detectors. Though the effective area is as small as 0.1 $mm^2$, the recent technological advances in multiplexed readout of multi-pixel TES arrays (more than 100 pixels) allow the performance of a precision kaonic atom measurement in a realistic data acquisition time.

Prior to the kaonic atom experiment, the HEATES collaboration demonstrated the feasibility of ultra-high resolution hadronic atom X-ray spectroscopy with TES spectrometer via pionic atom measurement at PSI [8]. The spectrometer has a 240-pixel TES array corresponding to about 23 $mm^2$ collecting area. Figure 1 (red line) shows an x-ray energy spectrum of pionic atom X-rays measured by the TES spectrometer. A sharp peak from the $\pi$-$^{12}$C $4f \rightarrow 3d$ transition together with a calibration peak of Fe $K_\alpha$ is observed. The achieved averaged energy resolution with TES is 6.8 eV FWHM at 6.4 keV under a high-rate pion beam intensity of 1.45 MHz. The resulting systematic uncertainty in the $\pi$-$^{12}$C $4f \rightarrow 3d$ transition x-ray energy was achieved to be less than 0.1 eV.

J-PARC E62 experiment was conducted in June 2018 and successfully observed distinct X-ray peaks from both $K^-$-$^{3,4}$He $3d$-$2p$ atoms with the same 240-pixel TES array. The achieved average energy resolution is 5 eV (FWHM) at 6 keV with the charged-particle beam off and 7 eV with the beam on. The time resolution is about 1 $\mu$s (FWHM) by merging data streams from beam-detection electronics into the TES's time-division-multiplexing DAQ system. This is an important milestone towards next-generation high-resolution hadronic-atom X-ray spectroscopies.

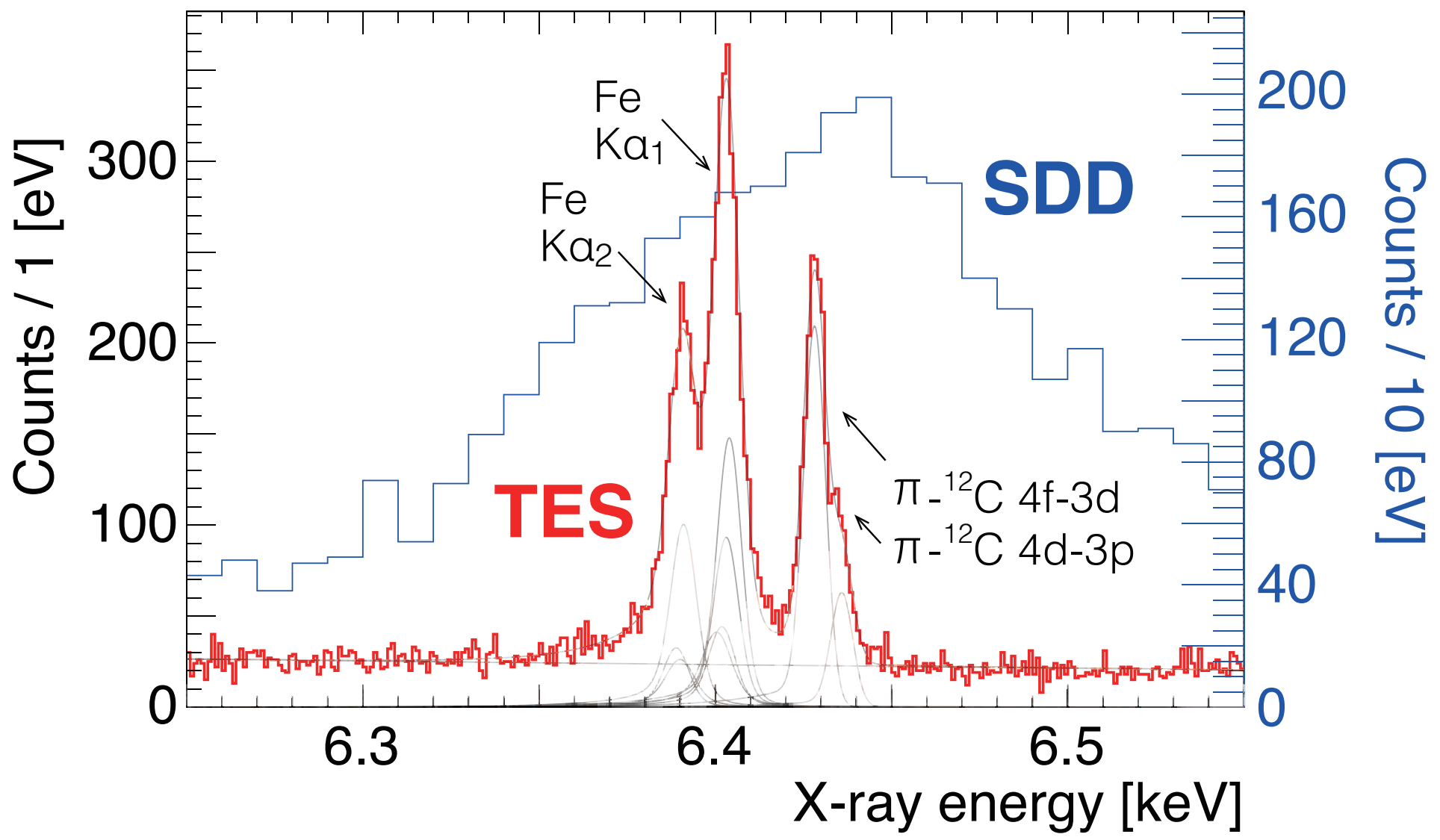

Figure 1: (Red line) Pionic atom X-ray spectrum together with a reference calibration peak, Fe $K_\alpha$, measured by the TES spectrometer. (Blue line) The same spectrum (but with almost no contamination of Fe $K_\alpha$) measured by the reference SDD. The FWHM energy resolution is $\sim$165 eV. [8]

## 3 Hadronic atom experiments at the extended J-PARC HEF

Hadronic atom experiments, especially for kaonic and sigma atoms, will be performed at the K1.1BR beamline of the extended HEF. This low-momentum beamline will be a short beamline, $\sim$ 20 m, which is essential for stopping the kaons to efficiently produce hadronic atoms.

Concerning the kaonic-atom experiment, the systematic study of heavier kaonic atoms is essential to determine $\bar{K}$-nucleus potential. It is suggested to improve the precision of the measurement on a small group of kaonic atom species, namely C, Si, Ni, Sn and Pb [9]. Moreover, the recent theoretical studies of strong-interaction effects in kaonic atoms suggest that analysing so-called 'lower' and 'upper' levels in the same atom could separate one-nucleon absorption from multinucleon processes [10]. About ten elements are identified as possible candidates from all of medium-weight and heavy nuclei [11]. Those measurements require the new TES spectrometer system for gamma-ray region ($<$ 400 keV). An average resolution of 53 eV at $\sim$100 keV was achieved for a 256 pixel spectrometer with a collecting area of 5 cm$^2$ [12]; and now the gamma-ray TES system is developing towards future exotic-atom experiments.

There are only a limited number of the data for sigma ($\Sigma^-$) atoms [1] despite being important for understanding the internal structure of neutron star. $\Sigma^-$ can be produced via stopped $K^-$ reaction, and then 170 MeV/c $\Sigma^-$ is stoped in the target. The precision X-ray spectroscopy of the $\Sigma^-$ atoms is therefore expected to be performed in the K1.1BR beamline. The $\Sigma^-$-p, $\Sigma^-$-He and more heavier sigma atoms X-ray spectroscopies will be conducted using semi-conducting detector arrays system covering a large solid angle and the multi-pixel TES X-ray spectrometers.

The low-momentum K- beam will be available more continually in the extended HEF of J-PARC requiring more production target. We aim to realize the high-precision hadronic atom X-ray spectroscopies with a combination of the world highest intensity kaon beam and the novel experimental technologies which allows a significant improvement on the accuracy of the X-ray spectroscopies. This will provide essential constraints on the theoretical description of the low-energy QCD system.

## References

[1] C.J. Batty, E. Friedman, and A. Gal, Phys. Rep. **287** (1997) 385-445.

[2] M. Bazzi, *et al.* (SIDDHARTA Collaboration), Phys. Lett. B **704** (2011) 113.

[3] Y. Ikeda, T. Hyodo, W. Weise, Phys. Lett. B **706** (2011) 63.;
Y. Ikeda, T. Hyodo, W. Weise, Nucl. Phys. A **881** (2012) 98.

[4] J-PARC E15 collaboration, Phys. Lett. B **789** (2019) 620-625.

[5] J. Yamagata-Sekihara, S. Hirenzaki, and E. Hiyama, private communication.

[6] K.D. Irwin and G.C. Hilton, "Transition-Edge Sensors", C. Enss (ed.), Cryogenic Particle Detection, Topics in Applied Physics, vol. **99**, Springer, 2005.

[7] W.D. Doriese et al., Review of Scientific Instruments **88** (2017) 053108.

[8] S. Okada et al., Prog. Theor. Exp. Phys. **2016**, 091D01.

[9] E. Friedman, International Journal of Modern Physics A **26** (2011) 468.

[10] E. Friedman, Hyperfine Interact **209** (2012) 127.

[11] E. Friedman and S. Okada, Nuclear Physics A **915** (2013) 170.

[12] D. A. Bennet et al., Rev. Sci. Instrum. **83** (2012) 093113

# Measurement of the ${}^{3}_{\Lambda}$H lifetime and of weak decay partial widths of mirror $p$-shell $\Lambda$-hypernuclei

**Alessandro Feliciello[1] and Elena Botta[1,2]**

[1]Istituto Nazionale di Fisica Nucleare (INFN), Sezione di Torino, Italy

[2]Dept. of Phys., Torino University, Italy

A proposal to accurately measure the hypertriton lifetime and to determine weak decay partial amplitudes for few selected neutron-rich $\Lambda$-hypernuclei belonging to the $p$-shell is outlined. The basic idea is to exploit for the first time the two-body ${}^{A}Z(\pi^{-},K^{0})^{A}_{\Lambda}(Z-1)$ reaction at the J-PARC K1.1 beam line and to add some specific detection capabilities to the existing SKS complex.

## 1 The physics cases

The accurate determination of the hypertriton (${}^{3}_{\Lambda}$H) lifetime ($\tau({}^{3}_{\Lambda}\mathrm{H})$) is today one of the key issues in strangeness nuclear physics [1]. Actually, this statement sounds surprising in itself. ${}^{3}_{\Lambda}$H is the lightest and, apparently, the simplest known $\Lambda$-hypernucleus. It is a bound nuclear system consisting of a proton, a neutron and a $\Lambda$ hyperon. However, the $\Lambda$ separation energy is $(0.13 \pm 0.05_{\mathrm{stat}} \pm 0.04_{\mathrm{syst}})$ MeV only [2] and then the ${}^{3}_{\Lambda}$H is the weakest bound $\Lambda$-hypernucleus as well.
On the basis of these considerations, it seemed plausible to assume that the behavior of the $\Lambda$ inside the ${}^{3}_{\Lambda}$H should not be very different from that in vacuum. In particular, the ${}^{3}_{\Lambda}$H lifetime value was expected to be very close to the one of the free $\Lambda$ particle ($263.2 \pm 2.0$) ps [3]. Such a naïve expectation was supported also on the theoretical ground [4, 5].

The first $\tau({}^{3}_{\Lambda}\mathrm{H})$ measurements were carried out in the decade 1963–1973 by exploiting two different visualizing techniques, namely photographic emulsions and He filled bubble chambers. Figure 1 shows the obtained results. The really poor quality of the data prevented to draw any firm conclusion about the effective $\tau({}^{3}_{\Lambda}\mathrm{H})$ value. Anyway, the weighted averages of the results turn out to be $203^{+40}_{-31}$ ps and $193^{+15}_{-13}$ ps in the case of emulsion and, respectively, of bubble chamber data [20]. Moreover, it is worth to note that these two mean values amount to $\approx 77\%$ and, respectively, to $\approx 73\%$ of $\tau(\Lambda_{\mathrm{free}})$. Since then, by taking into account also the measurements available for heavier $\Lambda$-hypernuclei, the assumption was that $\tau({}^{A}_{\Lambda}Z)$ was a smooth function of $A$ starting from a value close to the $\tau(\Lambda_{\mathrm{free}})$ and asymptotically approaching the 80% of such a value. Nevertheless, none of the theoretical approaches which were attempted was able to provide a satisfactory description of the experimental data trend over the overall $A$ range of the observed $\Lambda$-hypernuclei.

More recently, nearly 40 years after the last bubble chamber measurement, counter experiments originally designed to study heavy-ion collisions provided more precise values of $\tau({}^{3}_{\Lambda}\mathrm{H})$ (see again Fig. 1). These results were claimed to be unexpected and/or surprising, but their average value is actually $185^{+28}_{-33}$ ps (i.e. $\approx 70\%$ of $\tau(\Lambda_{\mathrm{free}})$), in total agreement with the old sets of measurements. However, these results had the merit of reopening the dormant debate about the ${}^{3}_{\Lambda}$H lifetime.
Moreover, the puzzle has been further fed by two recently achieved results.

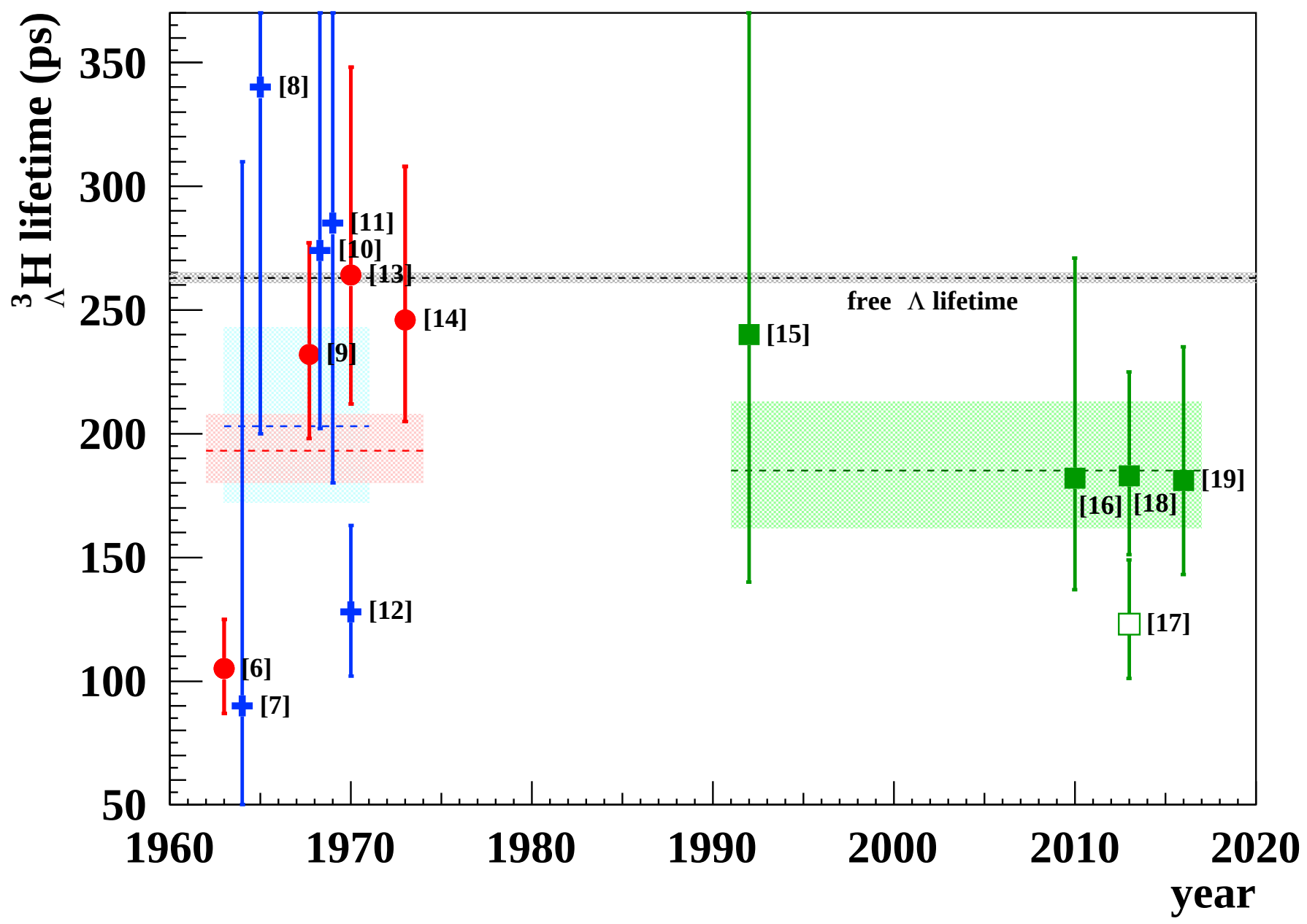

Figure 1: Chronological sequence of the published experimental data on $\tau(^{3}_{\Lambda}\mathrm{H})$. Red circles indicate results from He bubble chambers, blue crosses results from photographic emulsions and green squares results from counter experiments. For each point its Reference is given. Colored dashed lines and hatched areas represent the corresponding weighted averages with their errors. The open green square represents a result still classified as preliminary. (from Ref. [20])

The first one was announced by the STAR Collaboration, engaged in the analysis the Au-Au collisions at $\sqrt{s_{\mathrm{NN}}} = 200$ GeV collected at BNL/RHIC. By adding to the $^{3}_{\Lambda}\mathrm{H} \rightarrow {}^{3}\mathrm{H} + \pi^{-}$ data sample the $^{3}_{\Lambda}\mathrm{H} \rightarrow d + p + \pi^{-}$ events, they got $\tau(^{3}_{\Lambda}\mathrm{H}) = 142^{+24}_{-21} \pm 31_{\mathrm{syst}}$ ps [21]. Such a value, if confirmed, is $\approx 22\%$ lower compared to the previously published $\tau(^{3}_{\Lambda}\mathrm{H})$ value [16] and it amounts to $\approx 54\%$ only of $\tau(\Lambda_{\mathrm{free}})$.
Of completely opposite sign is the conclusion of the ALICE Collaboration, committed to the study of the Pb-Pb collisions observed at CERN/LHC at $\sqrt{s_{\mathrm{NN}}} = 5.02$ TeV. The analysis of the $^{3}_{\Lambda}\mathrm{H} \rightarrow {}^{3}\mathrm{H} + \pi^{-}$ channel led to $\tau(^{3}_{\Lambda}\mathrm{H}) = 237^{+33}_{-36} \pm 17_{\mathrm{syst}}$ ps. This result, announced during the HADRON 2017, Quark Matter 2018 and HYP 2018 Conferences but still unpublished, is higher than the previous ALICE determination [19] by $\approx 31\%$ and it is as high as the $\approx 90\%$ of $\tau(\Lambda_{\mathrm{free}})$ and fully compatible with it within the error.

The underlying doubt is then whether something could be wrong in the new measurements, in the sense that heavy-ion collisions where hypernuclei, especially $^{3}_{\Lambda}$H, are like “snowballs in Hell” [22] could not be the right context where to perform such a delicate measurement. Otherwise, the point could be that our present understanding of $^{3}_{\Lambda}$H structure is not correct. In other words, the $^{3}_{\Lambda}$H binding energy could not be as small as it is believed.

From the experimental point of view, a clear cut answer to such questions could be obtained by a new dedicated experiment only, possibly relaying on direct time measurement techniques.

As far as the second item of the possible future physics program is concerned, in the last ten years several experiments, that have successfully accomplished their scientific programs, demonstrated that the systematic study of the Λ-hypernuclei' decay modes is actually a powerful discovery tool [23, 24] and that it makes possible even an indirect spectroscopic study of the observed Λ-hypernucleus [23, 25, 26]. In particular, the FINUDA results put clearly in evidence the modulation effect that the nuclear structure has on the trend as a function of $A$ of the value of the partial decay width of both the mesonic ($\Gamma_{\pi^-}$) [27] and of the non-mesonic channels ($\Gamma_p$) [28]. Figures 2 and 3 show the current experimental situation about $\Gamma_{\pi^-}$ and, respectively, $\Gamma_p$ for $A \leq 16$ Λ-hypernuclei. There is an excellent agreement between the existing experimental points and the theoretical calculations presented in Ref. [29]. In addition, they predict a strong difference between the $\Gamma_{\pi^-}$ values in case of mirror Λ-hypernuclei pairs, like ${}^{12}_{\Lambda}$B and ${}^{12}_{\Lambda}$C (see again Figure 2).
The same effect should be observed for the $\Gamma_p$ values as well, even though in this case the amplitude of the variation is damped because of the larger momentum transfer involved in the non mesonic decay process with respect to the mesonic one (see again Figure 3).

The stimulating curiosity is that of verifying whether the predictions are correct by producing and by studying the ${}^{12}_{\Lambda}$B. Since the current experimental value of $\Gamma_p({}^{12}_{\Lambda}\mathrm{C})$ is affected by a quite large error, it could be also very useful to measure it again in the same experiment. The advantage is twofold: on the one hand the statistics will be significantly improved, on the other systematic errors will be very well kept under control. This way the comparison between the two measured quantities will be really meaningful.

Of course, the unknown $\tau({}^{12}_{\Lambda}\mathrm{B})$ will be measured as well.

## 2 The experimental setup

In order to produce ${}^{3}_{\Lambda}$H we plan to exploit, for the first time, the reaction

$$\pi^- + {}^3\mathrm{He} \to K^0 + {}^{3}_{\Lambda}\mathrm{H} \qquad (1)$$
$$\hookrightarrow K^0_S$$
$$\hookrightarrow \pi^+ + \pi^- \qquad (2)$$

on a liquid ${}^3$He target. As it is experimentally well known [30], the cross section of the charge-exchange reaction (1) is lower by at least a factor $\approx 10^3$ than the one of the ${}^3\mathrm{He}(K^-,\pi^0){}^{3}_{\Lambda}\mathrm{H}$ process. Nevertheless, we think that this drawback is overcompensated by the fact that when we take into account the decay chain (2), the final state is populated by charged particles only. The second advantage is that we don't need a large and generally very expensive calorimeter in order to detect high-energy $\gamma$ pairs following $\pi^0$ disintegration.

A strongly asymmetric topology of the decay process (2) can be selected. Then, the experimental apparatus should be able to detect $\pi^+$ in the forward direction and $\pi^-$ emitted in an angular range centered around $\theta = 90°$ with respect to the incoming $\pi^-$ beam axis. Finally, to directly measure $\tau({}^{3}_{\Lambda}\mathrm{H})$ it will be necessary to detect the decay products from the two- and/or the three-body channels:

$${}^{3}_{\Lambda}\mathrm{H} \to {}^3\mathrm{He} + \pi^-, \qquad (3)$$
$${}^{3}_{\Lambda}\mathrm{H} \to \mathrm{d} + p + \pi^-. \qquad (4)$$

Two facts make the K1.1 line of the J-PARC Experimental Hadron Facility (HEF) the ideal place where to pursue the experimental program outlined in the Sec. 1.

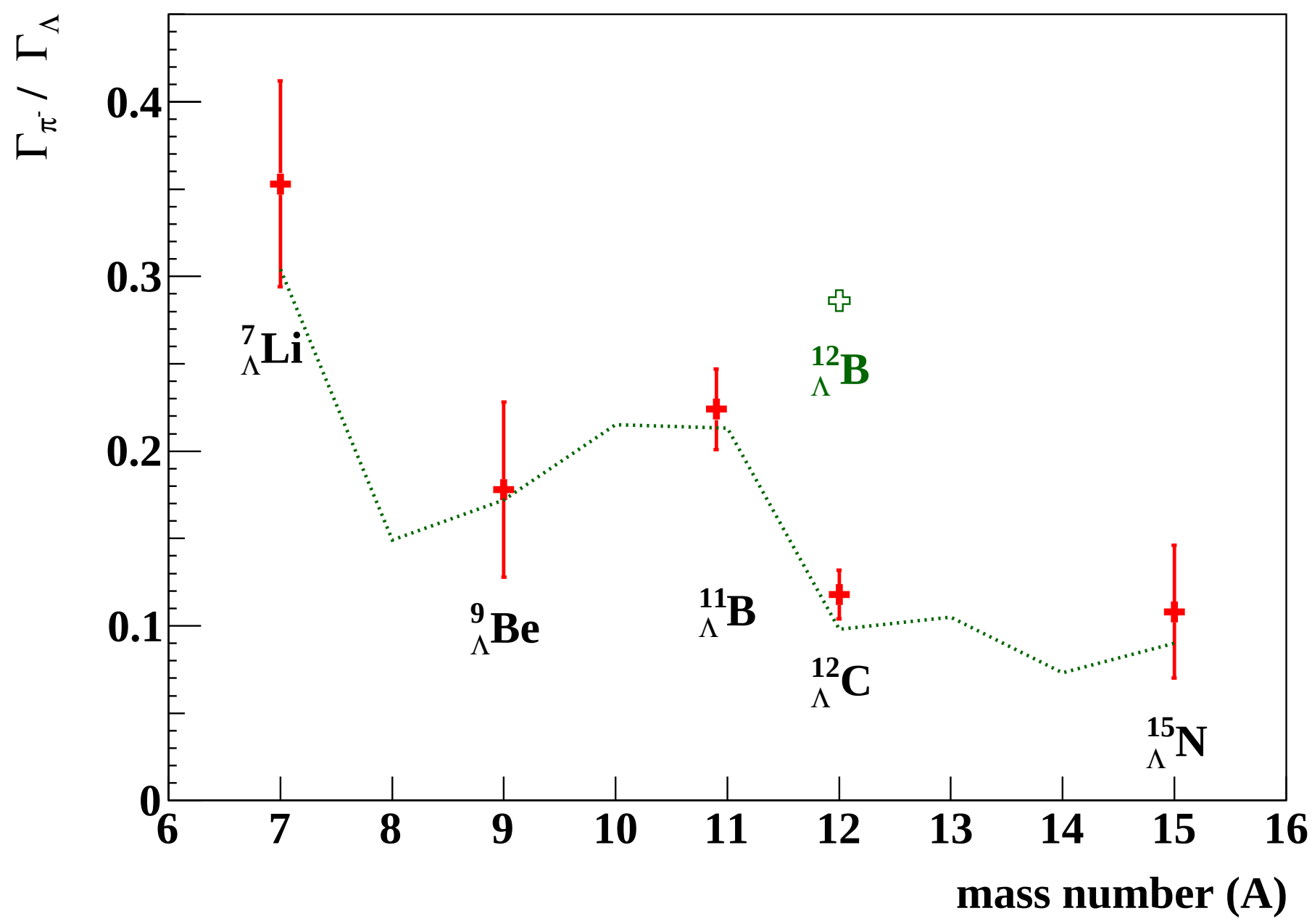

Figure 2: World average of the experimental $\Gamma_{\pi^-}/\Gamma_\Lambda$ values (red crosses) for different $p$-shell $\Lambda$-hypernuclei (adapted from Ref. [23]). The green dotted line and open cross represent the theoretical prediction of $\Gamma_{\pi^-}/\Gamma_\Lambda$ [29] for the already studied $\Lambda$-hypernuclei and, respectively, for the still unmeasured $^{12}_{\Lambda}$B.

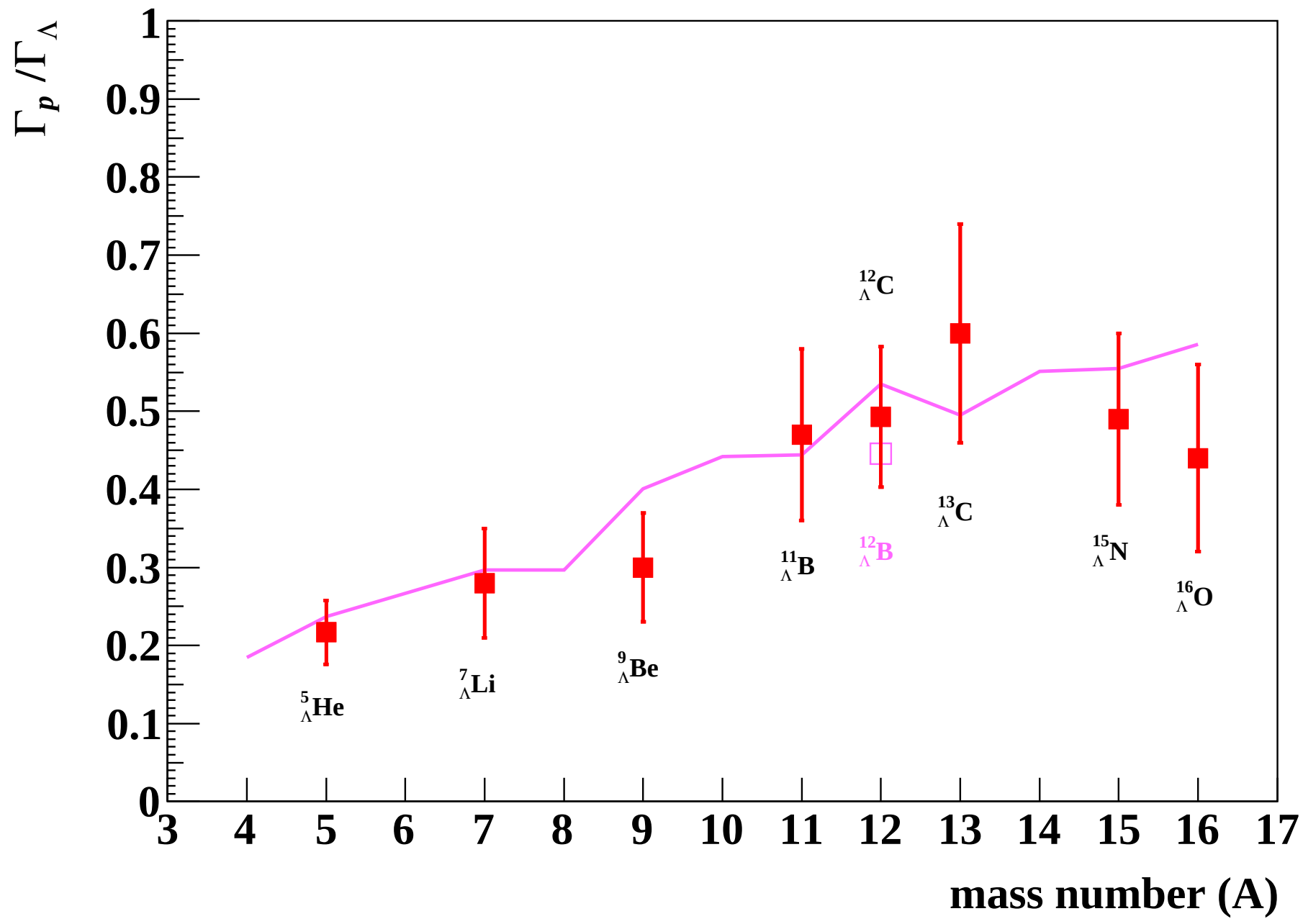

Figure 3: World average of the experimental $\Gamma_p/\Gamma_\Lambda$ values (red squares) for different $p$-shell $\Lambda$-hypernuclei (adapted from Ref. [23]). The magenta line and open square represent the theoretical prediction of $\Gamma_p/\Gamma_\Lambda$ [29] for the already studied $\Lambda$-hypernuclei and, respectively, for the still unmeasured $^{12}_{\Lambda}$B.

First, this new beam line will provide $\pi^-$ beam of excellent quality and intensity, suitable for the experimental needs.
Second, the Superconducting Kaon Spectrometer (SKS) complex was recently moved to the HEF K1.1 experimental area to carry on the already approved physics program on $\gamma$-spectroscopy of $\Lambda$-hypernuclei.

The first point is of course a mandatory requisite for the proposed measurement. The second circumstance permits, in principle, to design an experimental setup fulfilling the following important requirements:

- reinforcement of the synergic cooperation between non-Japanese researchers and J-PARC based Collaborations;
- mechanical integration of a new set of detectors with a preexisting hardware in order to reduce the cost and the completion time;
- modularity, in order to cope with the needs of different physics programs and to eventually permit a staged approach to the final detector configuration;
- optimization of the beam time allocation.

The liquid $^3$He target, placed in the K1.1 line final focus, will have a radius of 2–3 cm and a length of 7–8 cm, that is $\approx 1$ g/cm$^2$. It will be surrounded by a set of fast plastic scintillator slabs, arranged like the staves of a barrel. This detector, featuring a time resolution of less then 100 ps FWHM, will provide the stop time of the $^3_\Lambda$H decay products, while the start signal will be given by a beam scintillator or by a small hodoscope in case of high beam intensity. The trajectories of the charged particles following both the $K^0_S$ ((2)) and the $^3_\Lambda$H ((3) or (4)) decays will be precisely determined thanks to four pairs of low-mass drift chambers, installed immediately outside the scintillator barrel and placed in front of four modules of a fine layered range detector. The chambers will permit to measure the particle direction with a precision of the order of $\approx 100$ mrad, thanks to their design spatial resolution better than 300 $\mu$m FWHM. The range detector, possibly made of about one hundred 1 mm thick scintillator layers, will measure the energy of such particles and it will allow to identify them. The final goal is to achieve an energy resolution better than 2 MeV FWHM on the missing mass of the produced $\Lambda$-hypernucleus.

All this sub-detectors will cover a solid angle of $\approx 2\pi$ sr and will be placed around the target, upstream the SKS complex in its present configuration which will detect the forward emitted $\pi^+$ from $K^0_S$ decay (2). Figure 4 shows a schematic view of the outlined experimental setup.

The modular architecture will eventually permit a staged approach to the construction of the described apparatus. This way, it will be possible to cope with eventual mechanical clashes or with reduced budget problems by installing the detectors quadrant by quadrant (two at the minimum, either in the top-bottom or in the left-right configuration (see Table 1)). More details can be found in Ref. [20].

In order to carry on the second part of the proposed physics program, it will be sufficient to replace the liquid $^3$He target system only with a series of solid targets. $^{12}_\Lambda$B will be produced via the $^{12}$C($\pi^-$,$K^0$)$^{12}_\Lambda$B on graphite, machined in thin tiles of a typical thickness of $\approx 4$ mm, that is $\approx 1$ g/cm$^2$. It will be possible to install up to 4 of them along the $\pi^-$ beam axis and at an angle of 15–20 degrees with respect to it, as usually done in the past. Figure 5 shows how the above described experimental setup will be modified to perform measurement on solid targets. Also in this case, more details can be found in Ref. [20].

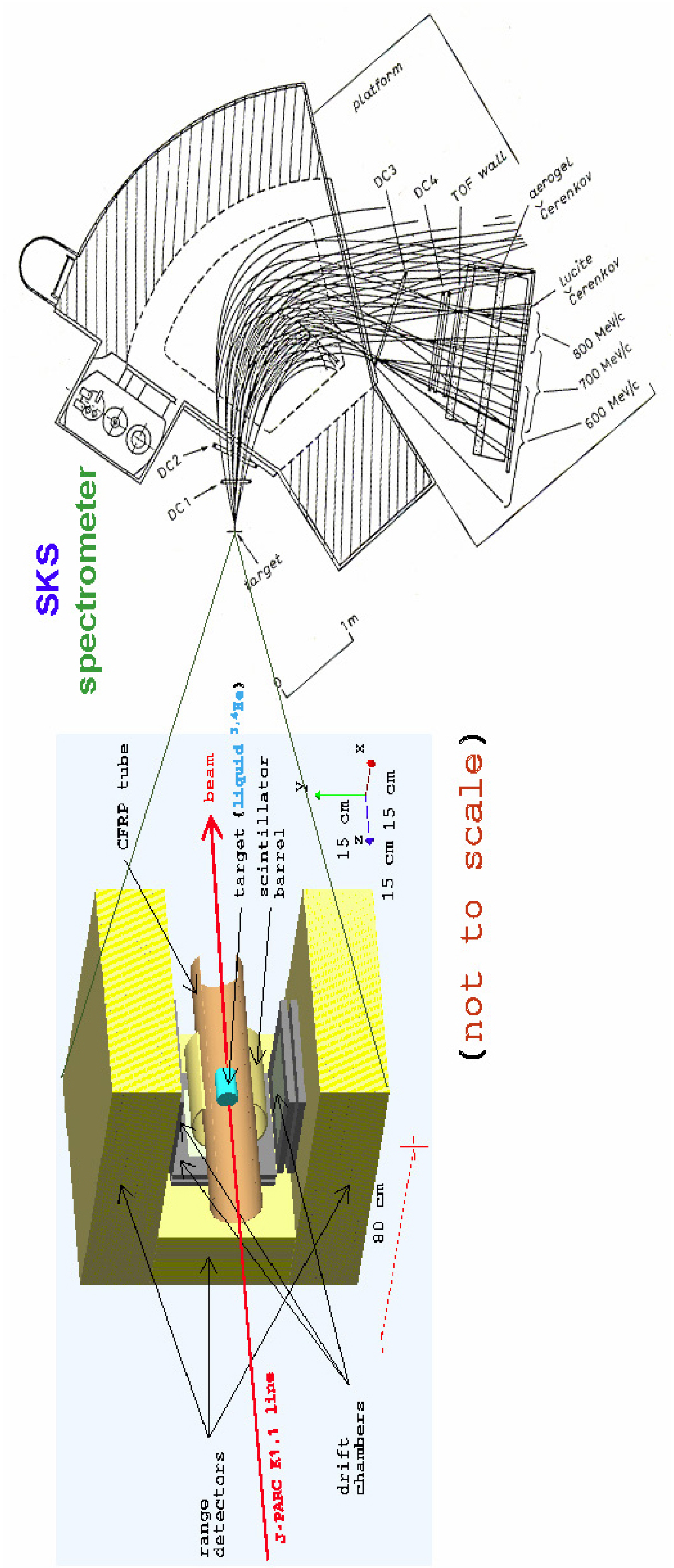

Figure 4: Schematic view of the proposed setup for the ${}^{3}_{\Lambda}$H lifetime measurement. One of the quadrant of the apparatus has been removed to permit to see the interior details.

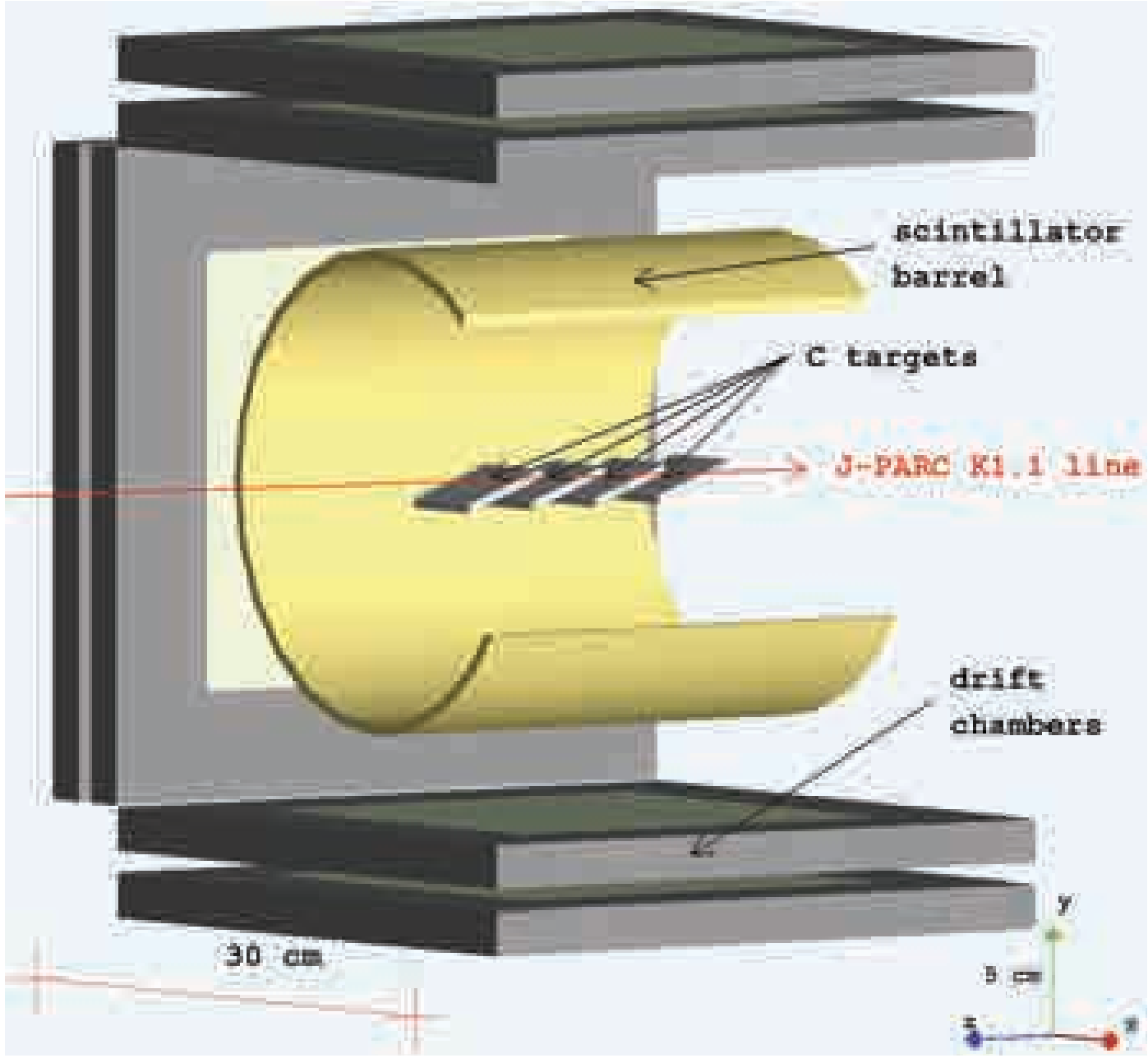

Figure 5: Magnified schematic view of the target layout for the decay measurement on $p$-shell $\Lambda$ with the $(\pi^-,K^0)$ reaction. One of the quadrant of the apparatus has been removed to permit to see the interior details.

# 3 Expected event rates and beam time requests

The main problem in evaluating the expected ${}^{3}_{\Lambda}$H yield is the fact that the cross section for process (1) is actually unknown. In order to get a plausible estimate of its value, one can remind that a prediction based on a DWIA calculation was done for the cross section of the ${}^{4}$He$(\pi^+,K^+){}^{4}_{\Lambda}$He reaction [31], which is the isospin symmetric of the ${}^{4}$He$(\pi^-,K^0){}^{4}_{\Lambda}$H process. It reaches the quite large value of $\approx 10$ $\mu$b/sr for incident $\pi^-$ momentum ranging between 1.0 and 1.1 GeV/c. Then a reasonable guess for the cross section of reaction (1) could be $\approx 5$ $\mu$b/sr.

Since the beginning of the J-PARC operation, the performance of the accelerator complex steadily increased. In particular the quality and, moreover, the intensity of $\pi$ secondary beams delivered to the different HEF experimental areas are approaching the design values. For this reason we prefer to evaluate the ${}^{3}_{\Lambda}$H and the ${}^{12}_{\Lambda}$B yields as a function of the number of available $\pi^-$. This way, the rate of the further beam improvement will determine the actual time necessary for a significant measurement.

The number of produced (detected) ${}^{3}_{\Lambda}$H in the apparatus acceptance is given by

$$\text{yield}({}^{3}_{\Lambda}\text{H}) = N_{\pi^-} \times \frac{T_{\text{tar}}}{A} \times N_A \times \frac{d\sigma}{d\Omega} \times \Omega_{\text{spe}} \times \varepsilon_{\text{spe}} \times \varepsilon_{\text{rec}} \approx 1 \times 10^4, \quad (5)$$

where

- $N_{\pi^-} = 5 \times 10^{13}$,

- $T_{\rm tar}$ ≡ liquid $^3$He target thickness = 1 g/cm$^2$,

- $A$ ≡ $^3$He atomic weight = 3,

- $N_A$ ≡ Avogadro constant,

- $\frac{d\sigma}{d\Omega}$ = 5 $\mu$b/sr,

- $\Omega_{\rm spe}$ ≡ spectrometer solid angle coverage (range detector + SKS) ≈ 0.05 sr,

- $\varepsilon_{\rm spe}$ ≡ BR($K^0 \to K^0_S$) × BR($K^0_S \to \pi^-\pi^+$) × ($\pi^-\pi^+$) pair detection probability ≈ 0.01,

- $\varepsilon_{\rm rec}$ ≡ reconstruction algorithm efficiency ≈ 0.5.

By taking into account the branching ratios values for $^3_\Lambda$H decay channels (3) and (4) it is possible to estimate the number of useful events for the $\tau(^3_\Lambda\mathrm{H})$ measurement. The decay products from two- and three-body will be detected by the range counters only, then the number of observed $^3_\Lambda$H is

$$\mathrm{yield(decaying}\ ^3_\Lambda\mathrm{H}) = \mathrm{yield}(^3_\Lambda\mathrm{H}) \times \mathrm{BR}(^3_\Lambda\mathrm{H} \to \text{2-b/3-b}) \times \Omega_{\pi^-} \times \varepsilon_{\pi^-} \times \varepsilon_{\rm rec} \approx 1 \times 10^3, \quad (6)$$

where

- BR($^3_\Lambda$H → 2-b/3-b) ≡ two- or three-body branching ratios ≈ 0.25 [5] or ≈ 0.40 [5],

- $\Omega_{\pi^-}$ ≡ range detector solid angle coverage for $\pi^-$ ≈ 0.5,

- $\varepsilon_{\rm rec}$ ≡ reconstruction algorithm efficiency ≈ 0.4.

Then, for both two- and three-body decay events we expect to have ≈ 1 × 10$^3$ entries in the corresponding time delay spectra, enough to get $\tau(^3_\Lambda\mathrm{H})$ with a statistical error of few percent.

In order to get the same number of detected $^{12}_\Lambda$B hypernuclei a lower number of incoming $\pi^-$ will be necessary. Actually, by using in (5) the measured cross section value of 15 $\mu$b/sr [32] for the reaction $^{12}\mathrm{C}(\pi^+,K^+)^{12}_\Lambda\mathrm{C}$ we got yield($^{12}_\Lambda$B) ≈ 1 × 10$^4$ with "only" 2 × 10$^{13}$ initial $\pi^-$ (see the third row of Table 1). Starting from this data sample and by resorting to the theoretical predictions available in literature for $\Gamma_{\pi^-}(^{12}_\Lambda\mathrm{B})$ (0.29 [33]) and for $\Gamma_p(^{12}_\Lambda\mathrm{B})$ (0.45 [29]) we estimated that we will collect ≈ 1.5 × 10$^3$ and ≈ 3.0 × 10$^3$ events which will permit to determine the mesonic and, respectively, the non-mesonic decay partial widths with a statistical error of few percent. The same statistical precision will be achieved on the measurement of $\tau(^{12}_\Lambda\mathrm{B})$ as well.

If on the one side the described physics program is very appealing, on the other it is very challenging. First of all it must be reminded that such an experimental approach has never attempted before. Then some aspects, first among all the cross section for reaction (1), are completely unknown and they could represent important factors of uncertainty. Moreover, the measurements are very demanding on both the human and the economic levels. Actually, they requires long data taking campaign and, just as an example, an expensive liquid $^3$He target.

Then, the most advisable strategy would be to start the experiment with the systematic study of the $^{12}_\Lambda$B decay process. This way we will have the opportunity to test the validity of the chosen experimental approach and we will gain important know-how to successfully carry on the remaining part of the physics program. In addition, exploiting the above described modularity of the design, we explored the possibility of performing the measurement with a reduced set of new detectors to be coupled to the SKS complex. We checked whether it still makes sense to perform the measurement without installing all the four quadrants of the range detector. Table 1 summarizes the outcome of this exercise, which aimed to demonstrate that it would be possible to carry out a sort of pilot run with still a good physics output. When we consider just one sector of the proposed detector assembly we got the number of detected ${}^{12}_{\Lambda}$B reported in the first row of Table 1. Clearly, it is insufficient in order to perform a significant measurement. As already anticipated at the end of Sec. 2, the minimum possible configuration consists of two modules. In this case, we will be able to measure $\Gamma_{\pi^-}({}^{12}_{\Lambda}\mathrm{B})$ and $\Gamma_p({}^{12}_{\Lambda}\mathrm{B})$ with a satisfactory statistical precision (see the second row of Table 1). Finally, the last row of Table 1 shows the minimum requirement of $\pi^+$ beam in order to re-measure the ${}^{12}\mathrm{C}(\pi^+,K^+){}^{12}_{\Lambda}\mathrm{C}$ reaction, as discussed in Sec. 2.

Table 1: Expected Λ-hypernuclei production rates for a given number of $\pi^-$ ($\pi^+$), for different targets and for different experimental configurations. The columns from 6 to 8 indicate the statistical significance of the measurement that can be achieved for the three main observables.

| beam request ($\times 10^{13}$ $\pi^-$) | target | thickness (g/cm$^2$) | exp. conf. | n. of detected | ${}^{A}_{\Lambda}Z$ | statistical significance $\tau({}^{A}_{\Lambda}Z)$ | $\Gamma_{\pi^-}({}^{A}_{\Lambda}Z)$ | $\Gamma_p({}^{A}_{\Lambda}Z)$ |
|---|---|---|---|---|---|---|---|---|
| 1 | ${}^{12}$C | $4 \times 1$ | 1/4 | $1.5 \times 10^3$ | ${}^{12}_{\Lambda}$B | poor | insufficient | poor |
| 1 | ${}^{12}$C | $4 \times 1$ | 1/2 | $3.0 \times 10^3$ | ${}^{12}_{\Lambda}$B | good | good | good |
| 2 | ${}^{12}$C | $4 \times 1$ | full | $1.0 \times 10^4$ | ${}^{12}_{\Lambda}$B | OK | OK | OK |
| 5 | L ${}^{4}$He | 1 | full | $1.5 \times 10^4$ | ${}^{4}_{\Lambda}$H | OK | OK | – |
| 5 | L ${}^{3}$He | 1 | full | $1.0 \times 10^4$ | ${}^{3}_{\Lambda}$H | OK | OK | – |
| ($\times 10^{11}$ $\pi^+$) | | | | | | | | |
| 1 | ${}^{12}$C | $4 \times 1$ | 1/2 | $3.5 \times 10^3$ | ${}^{12}_{\Lambda}$C | – | – | good |

The second step of the staged approach would be the measurement of $\tau({}^{4}_{\Lambda}\mathrm{H})$, which present less critical issues.

Finally, we will face the $\tau({}^{3}_{\Lambda}\mathrm{H})$ determination which, by the way, will hopefully rely on a more and more improved accelerator performance.
Actually, Table 2 shows how the duration of the data taking campaigns will crucially depend on the progress in the intensity of the beam delivered on the target.

# 4 Conclusions

We focused our attention on some selected arguments of particular interest in the field of hypernuclear physics. In our opinion, they deserve dedicated, well targeted experiments in order to provide a clear cut answer about the real value of the ${}^{3}_{\Lambda}$H lifetime and to definitely prove the strong effect of the nuclear structure on the size of the $\Gamma_{\pi^-}$ and of the $\Gamma_p$ decay partial widths of some selected neutron-rich Λ-hypernuclei belonging to the $p$-shell.

Table 2: Estimated beam time allocation as a function of the beam time intensity.

| delivered $\pi$ ($\times 10^{13}$) | $1.0 \times 10^7$ $\pi$/spill (present) | $1.5 \times 10^7$ $\pi$/spill | $1.0 \times 10^8$ $\pi$/spill | $1.0 \times 10^9$ $\pi$/spill (HIHR) |
|---|---|---|---|---|
| 1 | $6.9 \times 10$ d | $4.6 \times 10$ d | 7 d | $< 1$ d |
| 2 | $1.4 \times 10^2$ d | $9.3 \times 10$ d | $1.4 \times 10$ d | 1.4 d |
| 5 | $3.5 \times 10^2$ d | $2.3 \times 10^2$ d | $3.5 \times 10$ d | 3.5 d |

We are convinced that the ideal place to perform such an experiment is the K1.1 line of the HEF at J-PARC. To this purpose, we are proposing to build a relatively simple apparatus to be integrated with the existing SKS complex in order to perform such measurements.

# References

[1] T. Bressani, JPS Conf. Proc. **17**, 021002 (2017).

[2] D.H. Davis, Nucl. Phys. **A 754**, 3 (2005).

[3] C. Patrignani *et al.*, (Particle Data Group), Chin. Phys. **C 40**, 100001 (2017).

[4] M. Rayet, R.H. Dalitz, Il Nuovo Cim. **A 46**, 786 (1966).

[5] H. Kamada, J. Golak, K. Miyagawa, H. Witala, W. Glockle, Phys. Rev. **C 57**, 1595 (1998).

[6] M.M. Block *et al.*, Proc. of the International Conference on hyperfragments, St. Cergue 28-30 March 1963, p. 63.

[7] R.J. Prem and P.H. Steinberg, Phys. Rev. **136**, B1803 (1964).

[8] Y.V. Kang *et al.*, Phys. Rev. **139**, B401 (1965).

[9] G. Keyes *et al.*, Phys. Rev. Lett. **20**, 819 (1968).

[10] R.J. Phillips and J. Schneps, Phys. Rev. Lett. **20**, 1383 (1968).

[11] R.J. Phillips and J. Schneps, Phys. Rev. **180**, 1307 (1969).

[12] G. Bohm *et al.*, Nucl. Phys. **B 16**, 46 (1970).

[13] G. Keyes *et al.*, Phys. Rev. **D 1**, 66 (1970).

[14] G. Keyes *et al.*, Nucl. Phys. **B 67**, 269 (1973).

[15] S. Avramenko *et al.*, Nucl. Phys. **A 547**, 95c (1992).

[16] STAR Collaboration, Science **328**, 58 (2010).

[17] Y. Zhu, Nucl. Phys. **A 904**, 551c (2013).

[18] C. Rappold *et al.*, Nucl. Phys. **A 913**, 170 (2013).

[19] ALICE Collaboration, Phys. Lett. **B 754**, 360 (2016).

[20] M. Agnello, E. Botta, T. Bressani, S. Bufalino, A. Feliciello, Nucl. Phys. **A 954**, 176 (2016).

[21] STAR Collaboration, Phys. Rev. **C 97**, 054909 (2018).

[22] P. Braun-Munzinger, private communication (2015).

[23] E. Botta, T. Bressani, S. Bufalino, A. Feliciello, Rivista del Nuovo Cimento **38**, 387 (2015).

[24] A. Feliciello, JPS Conf. Proc. **17**, 021001 (2017).

[25] FINUDA Collaboration, Phys. Lett. **B 681**, 146 (2009).

[26] A1 Collaboration, Nucl. Phys. **A 954**, 149 (2016).

[27] FINUDA Collaboration, Nucl. Phys. **A 881**, 322 (2012).

[28] FINUDA Collaboration, Phys. Lett. **B 738**, 499 (2014).

[29] K. Itonaga, T. Motoba, Prog. Theor. Phys. Suppl. **185**, 252 (2010).

[30] O. Hashimoto, H. Tamura, Prog. Part. Nucl. Phys. **57**, 564 (2006).

[31] T. Harada, private communication (2006).

[32] H. Hotchi *et al.*, Phys. Rev. **C 64**, 044302 (2001).

[33] T. Motoba, K. Itonaga, Prog. Theor. Phys. Suppl. **117**, 477 (1994).

Measure Hypertriton lifetime with $^{3}He(K^{-}, \pi^{0})\,^{3}_{\Lambda}H$ reaction

**Yue Ma**[1]

[1]RIKEN, Japan

Three recent experiments based on heavy ion production method claimed a ~30% shorter lifetime for hypertriton than free hyperon, which is difficult to interpret. We plan to carry out a new measurement with a complementary experimental approach to pin down the ambiguities.

# 1 Introduction

Hyperon is a class of baryon that consists of strange quark. As an extension of nucleon, the study of hyperon-nucleon interaction (YN interaction) allows us to achieve a unified description for baryon-baryon interaction. Because of the short lifetime of hyperon, it is difficult to prepare hyperon beam. Our knowledge for YN interaction was mostly obtained through the study of hypernucleus: a hyperon bound by a nuclear core. The lightest hypernucleus is hypertriton ($^{3}_{\Lambda}H$) composed by one proton, one neutron and one hyperon. Similar to the role played by deuteron for nuclear physics, hypertriton provides the foundation of our understanding for YN interaction. It has been measured for a long time that the hyperon is very loosely bound by a deuteron core as $B_{\Lambda}$=130±50keV[1]. Consequently, one can expect that the wave function of $^{3}_{\Lambda}H$ processes a broad distribution in space and its lifetime should be very similar to that of hyperon in vacuum ($\tau$~263ps) because of less influence from the deuteron core. However, surprising results are reported by three recent experiments with heavy ion hypernucleus production as shown in Table 1.

Table 1: Summary for the recent measurements for $^{3}_{\Lambda}H$ lifetime

| Collaboration | Experimental Method | $^{3}_{\Lambda}H$ lifetime [ps] | Release date |
|---|---|---|---|
| STAR[2] | Au collider | 182±27 | 2010 |
| HypHI[3] | Fixed target | 183±37 | 2013 |
| ALICE[4] | Pb collider | 181±33 | 2015 |

# 2 Proposed experimental setup and beam condition

We will try to pin down the hypertriton lifetime puzzle by establishing an independent experimental approach to measure $^{3}_{\Lambda}H$ lifetime directly. Different from the heavy ion based experiments, we will employ strangeness exchange reaction, $^{3}He(K^{-}, \pi^{0})^{3}_{\Lambda}H$, as production method to generate hypertriton and directly measure the time distribution of $\pi^{-}$ emitted from $^{3}_{\Lambda}H \rightarrow {}^{3}He+\pi^{-}$ to obtain its mesonic weak decay lifetime.

We will carry out our experiment at J-PARC K1.8BR beam line. Fig.1 shows the schematic experimental setup. $K^{-}$ meson beam will be directed to the liquid $^{3}He$ target at 1GeV/c to populate hypertriton. A central tracking system (CDS) will be used to detect $\pi^{-}$ meson decayed from $^{3}_{\Lambda}H$. The CDS consists of a solenoid magnet, a central drift chamber (CDC) and a barrel of hodoscope (CDH). The details are given in [5]. Together with a charge veto counter, a forward Cherenkov based calorimeter ($PbF_2$ crystal) is used to tag the emission of high-energy gamma rays in order to select fast forward $\pi^0$ meson from strangeness exchange reaction. By requesting a cut condition to remove fast pions (300ps), we can suppress the contamination from hadron reactions and also guarantee the hypertritons decay at rest ($^3_\Lambda$H stopping time is ~100ps inside liquid 3He target). Because there is only one ground state for the mesonic weak decay of hypertriton, $\pi^-$ emitted from $^3_\Lambda$H has a unique magnitude in momentum space (p=114MeV/c), which can help us to identify the signal from other background channels involving minus pion emission such as quasi-free Λ hyperon mesonic decay. After selecting the $\pi^-$ from $^3_\Lambda$H mesonic weak decay, one can derive its lifetime by $\tau = t_{tof} - t_{beam} - t_\pi$ as illustrated in Fig.1. Based on production cross section calculated by Prof. Harada and our estimation, we could collect a few hundred $^3_\Lambda$H events and achieve a lifetime resolution of ~20ps with one month beam time**??**. However, due to the limited supply of $PbF_2$ crystal, we have to install this calorimeter close to the $^3$He target within CDS magnet. This will unavoidably increase the background events, whose effect is under investigation with GEANT4 simulation.

In summary, we propose to investigte Hypertriton lifetime with an independent approach. The detector development and detailed simulation are on going.

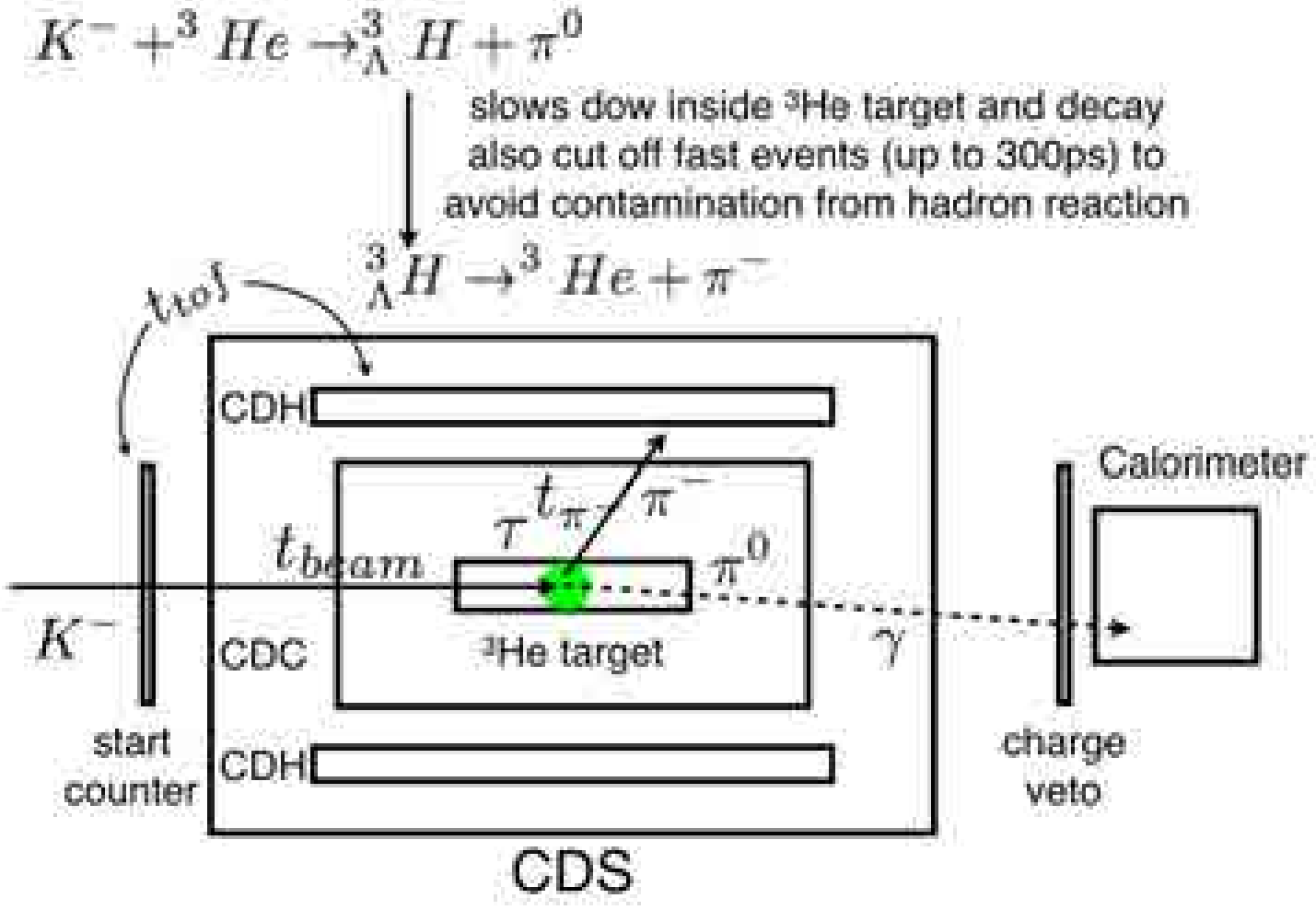

Figure 1: Schematic experimental setup

## References

[1] M. Juric *et al.*, Nucl. Phys. B, **52**, 1, (1973).

[2] B.I. Abelev *et al.*, Science, **328**, 58, (1973).

[3] C. Rappold *et al.*, Nucl. Phys. A, **913**, 170, (2013).

[4] ALICE collaboration, arXiv:1506.08453, 2015

[5] K. Agari *et al.*, Prog. Theo. Exp. Phys., **02B011**, (2012).

[6] Private communication with Prof. Harada, 2016

Study of $\Sigma N$ interaction by using the ultra high-resolution spectrometer

**Yudai Ichikawa**

ASRC, Japan, Atomic Energy Agency, Ibaraki 319-1195, Japan

We propose to measure a missing-mass spectrum around a $\Sigma N$ threshold of the $d(\pi^+, K^+)$ reaction by using the ultra high-resolution spectrometer. A clear enhancement was observed near the $\Sigma N$ threshold, so called "$\Sigma N$ *cusp*", for a long time ago. However, the dynamical origin of this enhancement remains unclear as yet. One of the key to make it clear is to improve the missing-mass resolution. We can achieve the missing-mass resolution of a few 100 keV by using a high-intensity and high resolution beam line (HIHR) at J-PARC. We can determine a scattering length of $\Sigma N$ system with isospin T = 1/2 from this measurement.

# 1 Physics motivation

The first experimental evidence of the "$\Sigma N$ *cusp*" was observed in $K^- d \to \pi^- \Lambda p$ reaction more than 50 years ago [1]. An enhancement near the $\Sigma N$ threshold (~2.13 GeV/$c^2$) was observed. This enhancement was confirmed by various experiments using $K^- d \to \pi^- \Lambda p$ and $\pi^+ d \to K^+ \Lambda p$ reaction. Recently, the "$\Sigma N$ *cusp*" has been intensively investigated in the $pp \to K^+ \Lambda p$ reaction at COSY. These results were summarized in Ref. [2]. At J-PARC, E27 collaboration reported the clear enhancement due to the "$\Sigma N$ *cusp*" in the inclusive missing-mass spectrum of the $d(\pi^+, K^+)$ reaction [3].

Generally, when a new threshold opens, it can make a cusp structure to conserve the flux and the associated unitary of $S$-matrix. However, such a cusp structure does not always appear at the threshold in the experimental cross sections. The similar cusp structure can be observed when there is a pole near the threshold [4]. In case of $\Sigma N$ system with isospin T = 1/2, it is theoretically predicted that a pole exists near the $\Sigma N$ threshold in a second or third quadrant of the complex plane of the $\Sigma N$ relative momentum. The pole in the second or third quadrant corresponds to be a deuteron-like unstable bound state or an inelastic virtual state, which was denominated in Ref. [4], of the $\Sigma N$ system, respectively [5]. Therefore, the structure of "$\Sigma N$ *cusp*" would not be a simple threshold effect but could be caused by the pole near the $\Sigma N$ threshold.

In Ref. [2], the spectra of the past experiments were fitted with a Breit-Wigner function after subtracting the continuum background. The obtained peak positions and widths were compared as shown in Figure. 1. Moreover, they also tried to fit with two Breit-Wigner functions because a shoulder at about 10 MeV higher mass could be seen in several data. The results of the two Breit-Wigner fit were summarized in Figure 2. Here, the obtained peak position and the width of J-PARC E27 experiment were 2130.5 ± 0.4 (stat.) ± 0.9 (syst.) MeV/$c^2$ and 5.3 $^{+1.4}_{-1.2}$ (stat.) $^{+0.6}_{-0.3}$ (syst.) MeV, respectively [3]. The shoulder could not be observed in the E27 spectrum, which might be due to the large quasi-free background.

It was discussed that the cusp structure could be expressed in terms of the scattering length of $\Sigma N$ (T = 1/2) system [13]. Note that the $\Sigma N$ (T = 1/2) scattering-length is complex reflecting the strong absorptive reaction of $\Sigma N \to \Lambda N$. Fig. 3 shows the calculated reaction rate as a function of invariant mass of $M_{\Lambda N} - M_{\Sigma} - M_N$ of the $K^- d \to \pi^- \Lambda p$ reaction. The calculated spectra with four choices of the scattering length $A_{\Sigma} = (a_{\Sigma}$ - $ib_{\Sigma})$ are shown. Thus, the "$\Sigma N$ *cusp*" may not necessarily distribute according to Breit-Wigner function. However, it seems that the obtained peak positions and width should be same for each data set, while the obtained values are not consistent each other as shown in Figure. 1 and 2.

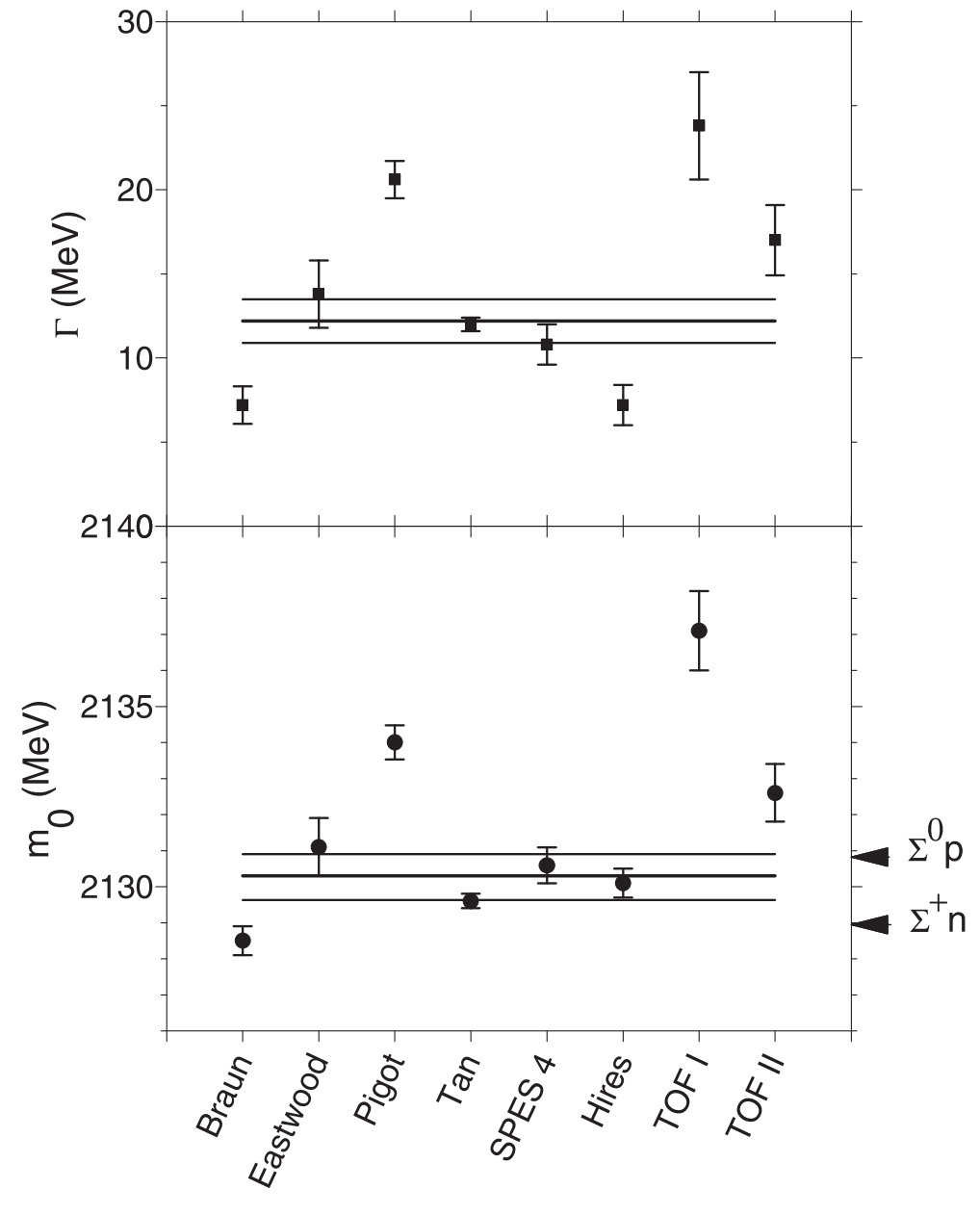

Figure 1: A summary of the peak positions ($m_0$) and widths ($\Gamma$) of one Breit-Wigner function taken from Ref. [2]. The fitting was done by the author of Ref. [2]. The values for the data set from Braun *et al.* [6], Eastwood *et al.* [7], Pigot *et al.* [8], Tan [9], SPES4 [10], HIRES [11], and TOF [12] are shown. The lines show the mean (thick line) and variance (thin line). The arrows show the thresholds of $\Sigma N$ system.

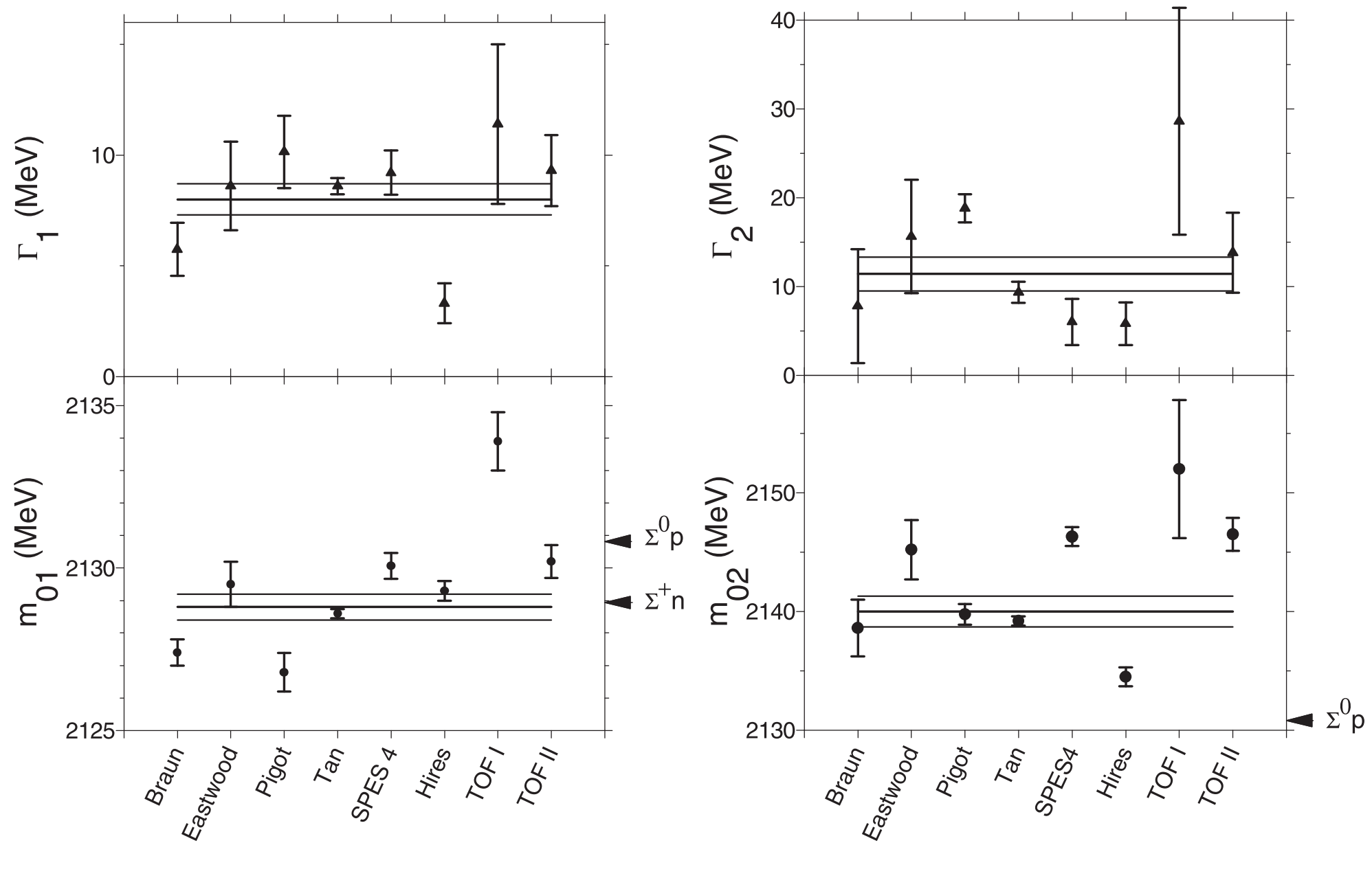

Figure 2: A summary of the peak positions ($m_{01}$, $m_{02}$) and widths ($\Gamma_1$, $\Gamma_2$) of two Breit-Wigner functions taken from Ref. [2]. The fitting was done by the author of Ref. [2]. The used data sets are same as Figure. 1. The lines show the mean (thick line) and variance (thin line). The arrows show the thresholds of $\Sigma N$ system.

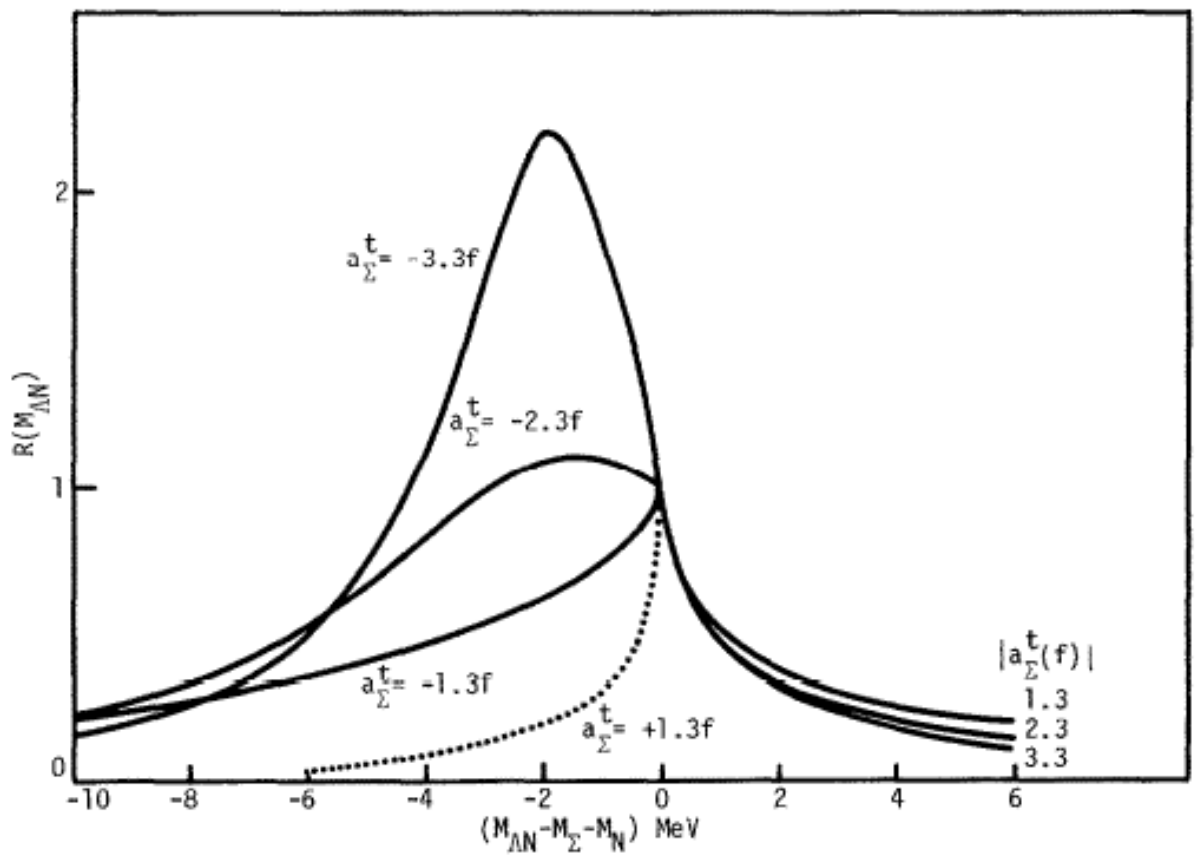

Figure 3: The calculated spectra for four choices of the scattering length taken from Ref. [13]. The real part $a_\Sigma$ was varied and the imaginary part $b_\Sigma$ was hold at 1.8 fm.

More realistic calculation was performed in Ref. [14]. In this calculation, the reaction amplitude of $K^-d \to \pi^-\Lambda p$ was expressed as,

$$T(K^-d \to \pi^-\Lambda p) = \int \{-\sqrt{\frac{1}{2}}N_1\phi^*_{\Lambda\Lambda}(r) + \text{ZERO }\phi^*_{\Lambda\Sigma^0}(r) + \text{PLUS }\phi^*_{\Lambda\Sigma^+}(r)\} \times \psi_d(r)\exp(iQ\cdot r)dr, \tag{1}$$

where $r$ denotes the separation between the two baryons, $Q = qm_N/m_{YN}$ and the coefficients are PLUS $= -\sqrt{\frac{1}{6}}M_0 + \frac{1}{2}M_1$, ZERO $= -\sqrt{\frac{1}{2}}M_1$. $M_T$ and $N_1$ are the amplitude of the elementary processes of $\bar{K}N \to (\Sigma\pi)_T$, where $T$ denotes the total isospin of $\Sigma\pi$, and $\bar{K}N \to \Lambda\pi$, respectively. $\phi_{ij}(r)$ are the components of the hyperon–nucleon ($YN$) wavefunctions for an outgoing plane wave in the channel $i$ with ingoing spherical wave in all channels $j$. For example, $\phi_{\Lambda\Sigma^0}$ corresponds to be the wavefunctions of outgoing wave of $\Lambda p$ with ingoing waves of $\Sigma^0 p$. $\psi_d$(r) denotes the wavefunction of the deuteron.

Fig. 4 (a) shows the calculated differential cross section $d\sigma/dm_{YN}$ around the $\Sigma N$ threshold in Ref. [14]. One boson exchange potential, so called NRS-F potential [15], was used and the $YN$ interaction parameters were not changed in this figure. Here, the elementary amplitude $N_1$ interferes with the terms ZERO and PLUS, and the phase of $N_1$ affects the final cross section. Thus, the phase $\phi$ of ($N_1$/PLUS) was varied and each mass distribution is shown in Figure 4 (a). In this calculation, they used a partial-wave amplitudes based on an energy dependent analysis taking into account all of the $\bar{K}N$ two-body data available at that time. Note that the magnitudes of $M_0$ and $M_1$, relative phase (in magnitude but not in sign), can be determined by the $\bar{K}N$ two body data. However, the $\bar{K}N$ data carries no information bearing on the phase of $N_1$ relative to those of $M_0$ and $M_1$. Namely, the phase information would be newly provided by the $K^-d$ reaction data.

As shown in Figure 4 (a), there are two prominent peak structures in all spectra due to the $\Sigma N$ thresholds (2128.9 MeV for $\Sigma^+ n$ and 2130.9 MeV for $\Sigma^0 p$). The intensities and sharpness of these peaks depend on the phase $\phi$, where these peak positions are almost on these thresholds. In the experimental data, these spectra should be smeared by the experimental resolution.

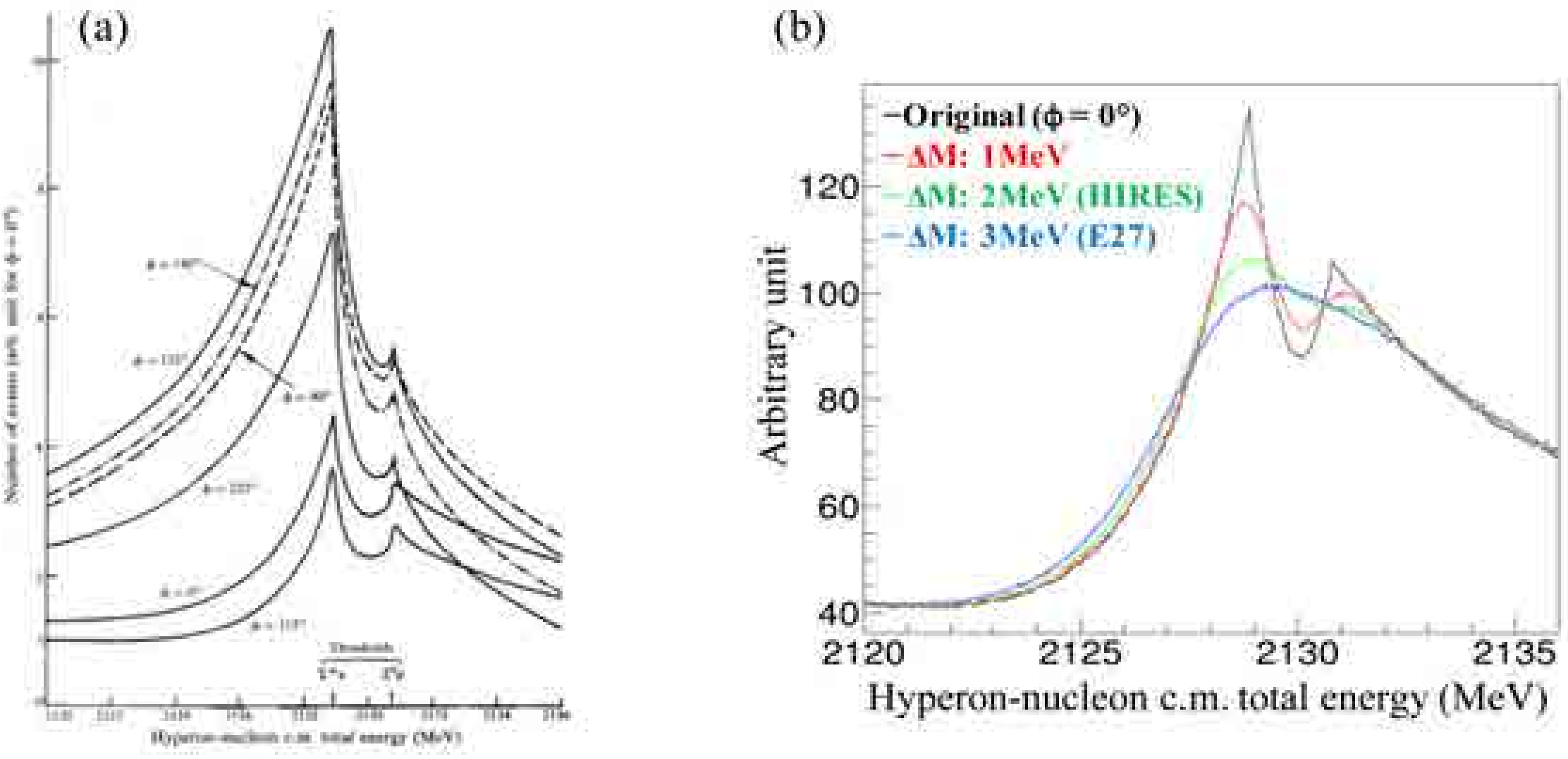

Figure 4: (a) The calculated $d\sigma/dm_{\Lambda p}$ distribution around the $\Sigma N$ threshold using NRS-F YN potential as a function of the phase $\phi$ between $N_1$ and PLUS taken from Ref. [14]. (b) The spectra smeared with the experimental resolution. A black line indicates the original spectrum in case of $\phi = 0°$ in Figure (a). Colored lines shows the smeared spectra with the experimental resolution of $\Delta M = 1$ (red), 2 (green), and 3 (blue) MeV in FWHM.

This effect is demonstrated in Figure 4 (b). The original spectrum in case of $\phi = 0°$ is shown as a black line. The spectra smeared with the mass resolution of $\Delta M = 1$, 2, and 3 MeV in FWHM are shown as red, green, and blue lines, respectively. The prominent two peak structure in original spectrum is disappeared in $\Delta M \geq 2$ MeV cases. The resolution of $\Delta M = 2$ MeV corresponds to be the best value of the past experiment achieved by HIRES at COSY [11]. Due to not sufficient resolution, it seems that the two peak structures have not been measured yet. Note that the shoulder corresponding to be $m_{02}$ and $\Gamma_2$ in Figure 2 should not be originated by this two peak structure. It is because the peak position $m_{02}$ is about 10 MeV higher than the $\Sigma N$ thresholds.

The phase $\phi$ should be different for each experimental data because the experimental conditions such as the beam momentum and selected scattering angle were different. Thus, the ratio of the intensities of $\Sigma^+ n$ and $\Sigma^0 p$ peaks should be different for each data condition, while the two peak structure was disappeared due to the experimental resolution. We interpret that it is the origin of the inconsistency of peak positions and widths of the past experiments. Namely, it is important to measure the spectrum with the very good resolution of $\Delta M \leq 1$ MeV in FWHM. We will carry out the spectrum fit as parameters of scattering length ($A_\Sigma$) and the phase $\phi$.

From the point of view of the $YN$ interaction, Harada *et al.* constructed complex $\Sigma N$ potentials assuming a simple two-range gaussian [16]. They constructed the $S$–matrix equivalent to the meson exchange potential (Nijmegen model-D). The $\Sigma N$ interaction has the strong isospin–spin dependence as shown in Figure 5. The attraction is stronger in the $(T, S) = (1/2, 1)$ and $(3/2, 0)$ states compared with the other two states. This characteristic feature is consistent with the comprehension of the quark-cluster model of baryon–baryon interaction developed by Oka *et al.* [17]. This isospin–spin dependence is a key point to understand $\Sigma$–hypernucleus with $A = 4$ system. Namely, the $\Sigma$–hypernucleus was observed by not $^4\text{He}(K^-, \pi^+)$ but $^4\text{He}(K^-, \pi^-)$ spectra [18] due to this strong isospin–spin dependence.

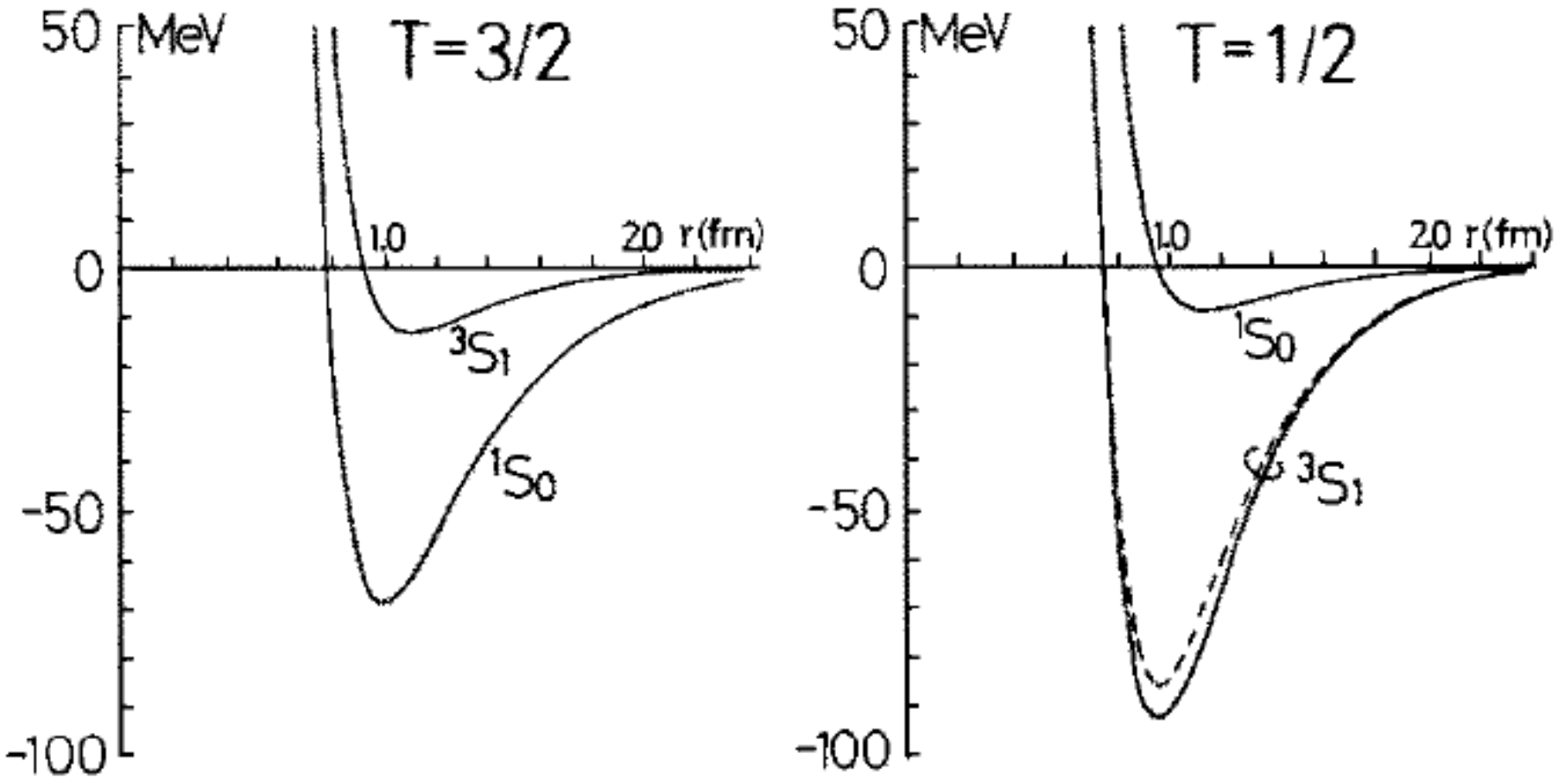

Figure 5: (a) The real parts of the $\Sigma N$ potential for two body $\Sigma N$ isospin($T$)–spin($S$) states $(T, S) = (3/2, 0), (3/2, 1), (1/2, 1)$, and $(1/2, 0)$ constructed by Harada *et al.*. This figure is taken from Ref. [16].

Here, the "$\Sigma N$ *cusp*" should be related the interaction of $\Sigma N$ with $T = 1/2$ because it was observed by selecting the final state as $\Lambda p$. Moreover, the "$\Sigma N$ *cusp*" was clearly observed by the $d(K^-, \pi^-)$ reaction selecting the forward scattering angle. Thus, this structure should be originated by the spin $S = 1$ component taking over the spin of target deuteron. As shown in Figure 5, the $\Sigma N$ potential with $(T, S) = (1/2, 1)$ has the strongest attractive pocket compared with the other states. It is a natural interpretation that the "$\Sigma N$ *cusp*" is originated by the strong attractive pocket of the $\Sigma N$ potential of $(T, S) = (1/2, 1)$ component. Therefore, the "$\Sigma N$ *cusp*" may be not the cusp but the deuteron like bound state of the $\Sigma N$ system. The precise experiment is necessary to reveal this question.

## 2 Purpose of the proposed experiment

The purpose of proposed experiment is to determine the scattering length of $\Sigma N$ system with $T = 1/2$ by fitting the obtained missing mass spectrum of $d(\pi^+, K^+)$ reaction. We will be able to confirm whether the deuteron like $\Sigma N$ system exists or not from the obtained scattering length. The key of this experiment is missing-mass resolution to obtain the two peak structure and estimate the phase $\phi$. The required resolution is $\Delta M \leq 1$ MeV in FWHM, which is almost two times better than the past experiment (HIRES at COSY [11]).

## 3 Experimental Method and Requested Beamtime

The proposed experiment will be performed at the HIHR. We can achieve the very good missing-mass resolution of a few 100 keV by using this spectrometer system. We will use $d(\pi^+, K^+)$ reaction at 1.69 GeV/$c$, which is same as J-PARC E27 experiment. The differential cross section of "$\Sigma N$ *cusp*" has already measured as $d\bar{\sigma}/d\Omega = 10.7 \pm 1.7 \mu$b/sr by E27. In this reaction, we interpret that "$\Sigma N$ *cusp*" is produced from two-step processes as, $\pi^+$"$N$"$\rightarrow \Sigma K^+, \Sigma$"$N$"$\rightarrow \Lambda p$, where “$N$” is the nucleon inside the deuteron. The main background is due to the quasi-free $\Lambda$ production as $\pi^+$“$n$”$\rightarrow \Lambda K^+$, where “$n$” is the neutron inside the deuteron. In this beam momentum, the cross section of the $\Lambda$ production is almost three times smaller than the $\Sigma$ production, which is a seed to produce the “$\Sigma N$ *cusp*”. The ratio of elementary cross sections ($\Sigma/\Lambda$) of this beam momentum is almost maximum. Therefore, we choose this beam momentum 1.69 GeV/$c$.

In this experiment, we aim to measure both inclusive and exclusive spectra of the $d(\pi^+, K^+)X$ and $d(\pi^+, K^+)\Lambda p$ reaction, respectively. We will install a multiplicity counter surrounding the target to select the final state as $\Lambda p$. In case of the background reaction such as the quasi-free $\Lambda$ production as $\pi^+(pn)_d \rightarrow K^+\Lambda p_s$, where $p_s$ is spectator proton, the multiplicity should be less than 2 ($m \leq 2$) as demonstrated in Figure 6(a). It is because the spectator proton does not give any signal to the multiplicity counter. The almost all of spectator proton can not escape the experimental target because its momentum is too small ($<$ 250 MeV/$c$). On the other hand, in case of the “$\Sigma N$ *cusp*” reaction, the two out going baryons share the momentum almost equally. The $\Lambda$ and proton of the final state are emitted back-to-back direction. Then, the $p\pi^-$ from the $\Lambda$ decay and emitted proton will be detected by the multiplicity counter. Thus, we can increase the signal noise ratio by selecting multiplicity $m = 3$ as shown in Figure 6(b).

The multiplicity counter system was adopted by the past experiment at CERN [8] and it worked successfully. They installed the multiplicity counter array surrounding the liquid deuterium target as shown in Figure 7(A). The obtained missing-mass spectrum for each multiplicity condition is shown in Figure 7(B). The spectrum with multiplicity $m_T \geq 1$ as shown in a) corresponds to be almost inclusive one. In this spectrum, there were two prominent structures around 2.07 and 2.16 GeV/$c^2$ due to the quasi-free $\Lambda$ and $\Sigma$ production, respectively. These quasi-free background components were suppressed by selecting higher multiplicity shown in b) or c). The narrow peak around 2.13 GeV/$c^2$ shown in dotted line was interpreted as a resonance, which is understood as the “$\Sigma N$ *cusp*” nowadays. A broad peak around 2.15 GeV/$c^2$ indicated in dashed-line was taken into account as the contribution of a conversion process of $\Sigma N \rightarrow \Lambda p$. This $\Sigma N \rightarrow \Lambda p$ conversion is a physics background, which is difficult to be suppressed. Anyway, they succeeded to improve the signal noise ratio by using multiplicity counter system. We will plan to install more segmented multiplicity counter than the past CERN experiment for the proposed experiment.

Note that J-PARC E27 could observe the “$\Sigma N$ *cusp*” in the inclusive spectrum due to the better missing-mass resolution (3.2 MeV/$c^2$ in FWHM), where the resolution of CERN experiment was 5.5 MeV/$c^2$ in $\sigma$. Thus, not only the exclusive spectrum but also the the inclusive one is important to derive the scattering length of $\Sigma N$ system.

We have two options for the experimental target. One is to prepare the liquid deuterium target same as E27 (option (A)). In case of E27, the liquid deuterium target of 1.99-g/cm$^2$ (0.166 g/cm$^3$ $\times$12 cm) was used. However, it may be difficult to use such a thick target to keep the good missing-mass resolution at HIHR because HIHR is not assumed to measure the beam trajectory. One option is to install the silicon strip detector (SSD) in front of the target and keep the low beam intensity about $10^7$/spill. Our acceptable missing-mass resolution is $\Delta M \leq 1$ MeV (FWHM) to observe the two peak structure as shown in Figure. 4 (b). Moreover, the energy-loss straggling in the target with 1.99-g/cm$^2$ is estimated to be 0.8 MeV in FWHM. In this estimation, not only the energy loss straggling but also the ambiguity of the path length inside the target due to the vertex resolution are included. Thus, the optimization of the target

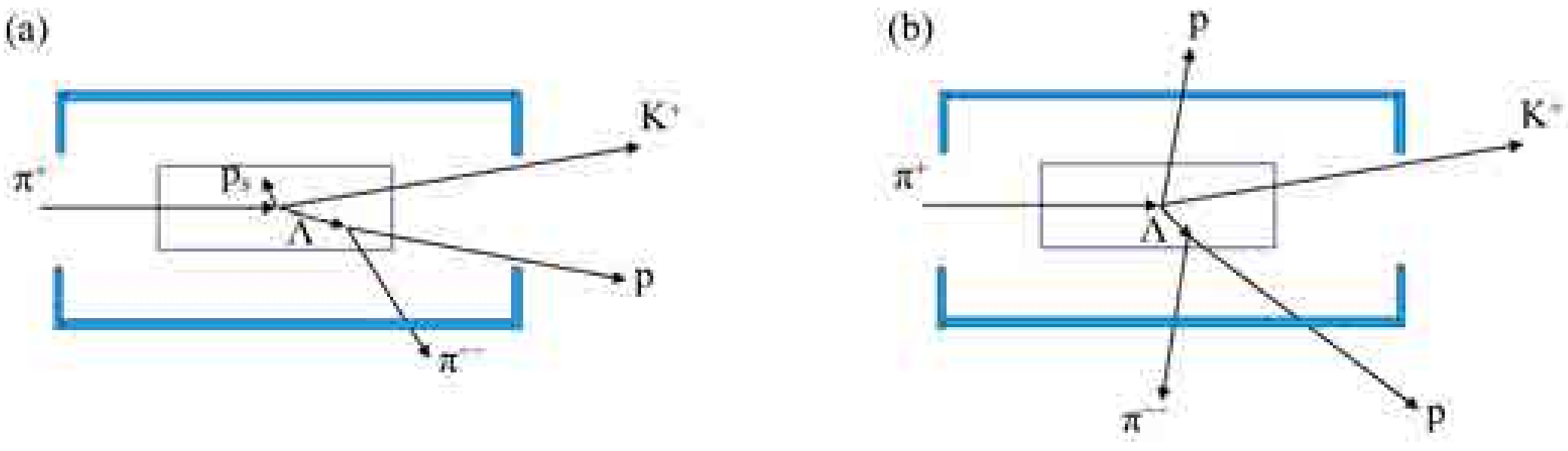

Figure 6: The schematic view of the background quasi-free (a) and "$\Sigma N$ *cusp*" reaction (b). The multiplicity counter and target are shown in light-blue and white boxes, respectively.

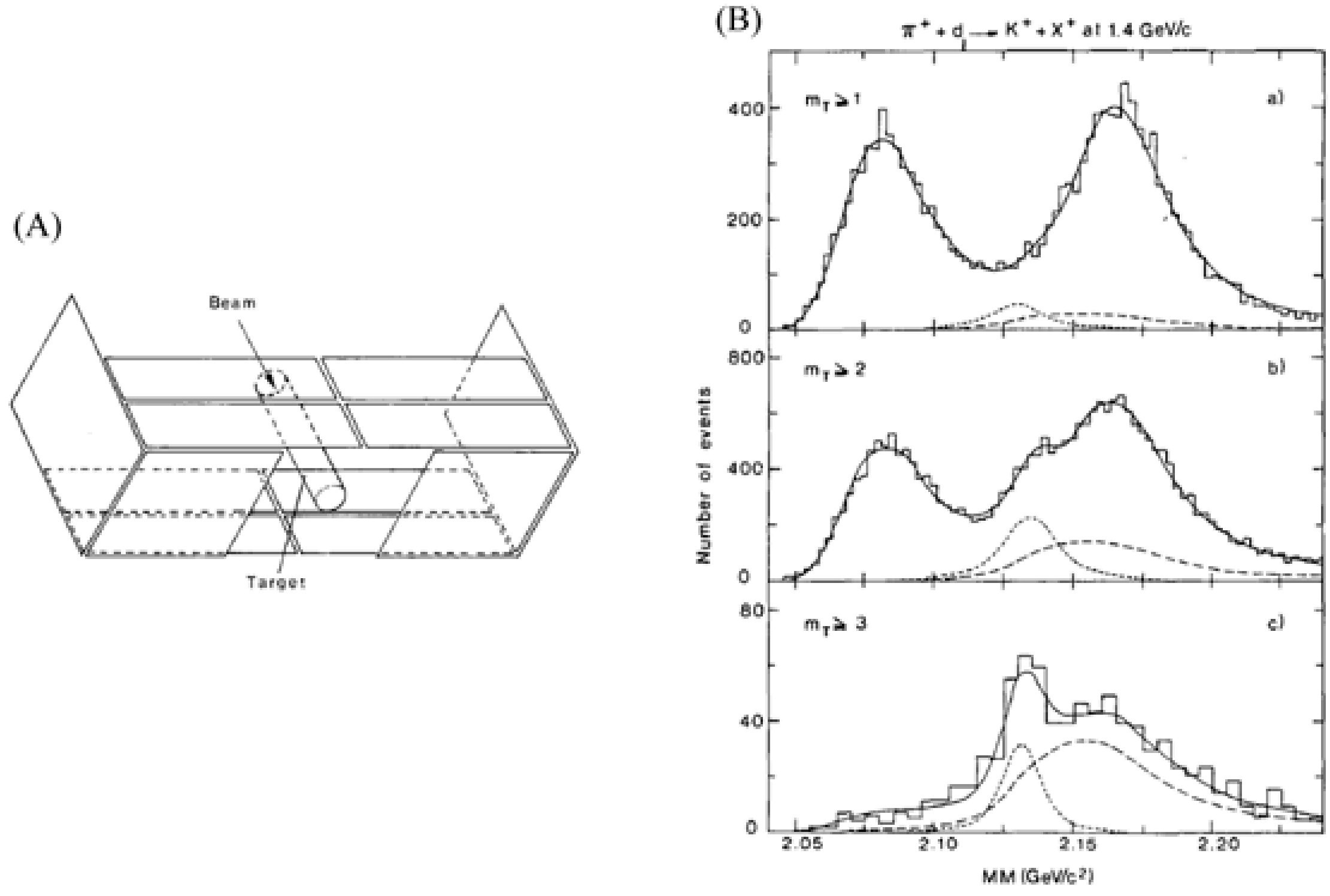

Figure 7: (A) The schematic view of multiplicity counter of past experiment at CERN [8]. (B) The missing-mass spectrum of the $d(\pi^+, K^+)$ reaction for each multiplicity condition of this past experiment taken from Ref. [8]. The doted line corresponds to be the signal of "$\Sigma N$ *cusp*" and the dashed line shows the component of conversion process of $\Sigma N \to \Lambda p$.

thickness considering the missing-mass resolution is necessary.

The other option is to use deuterated polyethylene ($CD_2$) target (option (B)). In this case, we will prepare very thin target and use high intensity pion beam about $1.8 \times 10^8$ /spill. An additional data with the carbon-graphite target will be necessary to subtract the carbon component of $CD_2$ spectrum. Moreover, it is not easy to operate the multiplicity counter under the high beam-rate condition ($> 10^8$ cps) because probability of a pile up event is expected to be high. The detail consideration and development including full simulation are necessary.

We would like to achieve almost same statistics as E27 in both options. Then, the beam time of 10 days (option (A)) or 30($CD_2$) + 10(C) days (option (B)) are requested. The detail comparison between E27 and propose experiment is summarized in Table. 1.

Table 1: The comparison between J-PARC E27 and proposed experiment. The requested beam time is estimated to achieve almost same statistics as E27.

| | J-PARC E27 | Option (A) | Option (B) |
|---|---|---|---|
| Target | $LD_2$ | $LD_2$ | $CD_2$ + C(Graphite) |
| Beam intensity [/spill] | $3.3 \times 10^6$<br>(Inclusive trig: prescale 1/3) | $\sim 10^7$ | $1.8 \times 10^8$ |
| Thickness [g/cm$^2$] | 1.99 | ($\leq$) 1.99 | 0.1 |
| Acceptance [msr] | 100 | 10 | 10 |
| Beam time [days] | 10 | 10 | 30($CD_2$) + 10 (C) |
| Mass resolution [MeV] | 3.2 | $\leq$1 | 0.3 |

Finally, we are also considering to carry out the experiment by using S–2S spectrometer, which is developed to search for a $\Xi$ hyper-nuclei [19], at K1.8 beam line. The momentum resolution of S–2S is estimated to be $\Delta p/p \sim 5 \times 10^{-4}$ in FWHM. When we assume the momentum resolution of the K18 beam line spectrometer of $\Delta p/p = 3.3 \times 10^{-4}$ in FWHM as the design value, the total missing-mass resolution is estimated to be $\Delta M \sim 1$ MeV. Here, the optimization of the target thickness of liquid deuterium target is necessary as same as option (A). In this S–2S option, we will take both $d(K^-, \pi^-)$ and $d(\pi^+, K^+)$ data because we can use $K^-$ beam at K1.8 beam line.

# 4 Summary

We will use the $d(\pi^+, K^+)$ reaction to determine the scattering length of $\Sigma N$ system and investigate whether the deuteron like $\Sigma N$ bound state exist or not. The proposed experiment will be carried out at HIHR or K1.8 beam line by using S–2S. For this measurement, the missing-mass resolution is the most important key. We can achieve the ultra high resolution (a few 100 keV) by using the HIHR spectrometer system. There is no past experiment which achieves the good mass resolution of $\Delta M < 2$ MeV in FWHM for the "$\Sigma N$ *cusp*" measurement. The ultra high-resolution spectrum with HIHR would play an important role to reveal the $YN$ interaction.

## References

[1] O.I. Dahl *et al.*, Phys. Rev. Lett., **6**, 3 (1961).

[2] H. Machner *et al.*, Nucl. Phys. A **901**, 65 (2013).

[3] Y. Ichikawa *et al.*, Prog. Theor. Exp. Phys. (2014) 101D03.

[4] A.M. Badalian, L.P. Kok, M.I. Polikarpov, and Yu.A. Simonov, Phys. Rep. **82**, 31 (1982).

[5] K. Miyagawa and H. Yamamura, Phys. Rev. C **60**, 024003 (1999).

[6] O. Braun *et al.*, Nucl. Phys. **B124**, 45 (1977).

[7] D. Eastwoood *et al.*, Phys. Rev. D **3**, 2603 (1971).

[8] C. Pigot *et al.*, Nucl. Phys. **B249**, 172 (1985).

[9] T.H. Tan, Phys. Rev. Lett. **23**, 395 (1969).

[10] R. Siebert *et al.*, Nucl. Phys. **A567**, 819 (1994).

[11] A. Budzanowski *et al.*, Phys. Lett. **B692**, 10 (2010).

[12] S. Abd El–Samad *et al.*, Eur. Phys. J. A **49**, 41 (2013).

[13] R.H. Dalitz, Nucl. Phys. **A354**, 101, (1981).

[14] R.H. Dalitz and A. Deloff, Aust. J. Phys. **36**, 617 (1983).

[15] M.M. Nagels, T.A. Rijken, and J.J. de Swart Phys. Rev. D **20**, 1633 (1979).

[16] T. Harada, S. Shinmura, Y. Akaishi and H. Tanaka, Nucl. Phys. **A507**, 715 (1990).

[17] M. Oka, K. Shimizu and K. Yazaki, Prog. Theor. Phys. Suppl. **137**, 1 (2000).

[18] T. Nagae *et al.*, Phys, Rev. Lett. **80**, 1605 (1998).

[19] T. Nagae *et al.*, http://j-parc.jp/researcher/Hadron/en/pac_1801/pdf/P70_2018-10.pdf

# Spectroscopy of light mesic nuclei such as $\eta$ and $\eta'$ mesons

**Kenta Itahashi**[1*] **and Hiroyuki Fujioka**[2]

[1]Cluster for Pioneering Research, RIKEN

[2]Department of Physics, Tokyo Institute of Technology

We propose measurement of $\eta$ and $\eta'$ meson-nucleus systems at HIHR, J-PARC. We consider pion-induced reactions for formation of the meson-nucleus systems and the spectroscopy. A resolution matching condition of the beamline serves unique opportunities with unprecedented resolution and statistical accuracy.

## 1 Introduction

It is known that in-medium properties of mesons provide information of low-energy region of the quantum chromodynamics (QCD). We propose to make spectroscopy of light mesons such as $\eta$ and $\eta'$ residing in nuclei which can be regarded as a high density medium. The mesons are expected to have attractive interaction with the nucleus and have larger widths than in the vacuum simply due to absorptive interaction with the nuclei. There are theoretical investigations of expected properties of the mesons in nuclei, some of which are expecting existence of the discrete bound states [1].

Spectroscopy of the mesons bound to nuclei yields unambiguous information on the interaction of the mesons with the nuclei. The spectral responses reflect the meson properties in the nuclear medium affected by the interaction. In case discrete peak structures are observed, the binding energies and widths are determined by the peak positions and widths. These information can lead to deduction of the underlying fundamental symmetry at a finite nuclear density. A good example is spectroscopy of pionic atoms where information on the chiral symmetry at finite density is deduced [2, 3, 4].

## 2 Experiment

In consideration of an experiment to make spectroscopy of mesic nuclei, we need to find suitable nuclear reactions to produce them. In case of $\eta$ and $\eta'$ mesic nuclei, proton, anti-proton and pion induced reactions should be examined. Here let us presume pion beam which has in principle advantages since the expected beam intensity is sufficient and the formation cross sections of the mesic nuclei are relatively high. We take account of $^{12}\mathrm{C}(\pi, N)$ reactions, where $N$ denotes a nucleon. There are two important factors for the formation of the mesic nuclei, the elementary meson formation cross sections and the momentum transfer ($q$) of the reactions. For production of $\eta$ in $(\pi, N)$ reactions, $q$ becomes zero at the incident $\pi$ momentum of about 0.95 GeV/$c$. For $\eta'$ in the $(\pi, N)$ reaction $q$ does not reach zero since the mass of the $\eta'$ is larger than the nucleon mass. We achieve $q < 200$ MeV/$c$ with the incident momentum $> 2.0$ GeV/$c$. There are theoretical spectra for $\eta$ and $\eta'$ production cross sections in the $(\pi, N)$ reactions at the incident energy of 0.95 GeV/$c$ and 1.8 GeV/$c$, respectively, in Ref. [1]. Anti-proton induced channels may have better kinematical conditions but there are no theoretical spectra so far due to lack of elementary reaction data.

*Email: itahashi@riken.jp

The expected experimental spectra are associated with background and reduction of the background is crucial. First experiment to search for $\eta$-nucleus bound states was conducted by using inclusive $(\pi^+, p)$ reaction with the beam momentum of 0.8 GeV/$c$ at a finite reaction angle of 15 degrees [5]. The observed spectrum does not show any distinct peak structures. The major part of the spectrum is due to multiple pion production. A possible method for the reduction of the background is by tagging the formation and decay of the mesic nuclei. $\eta$ meson in nuclear medium is known to strongly couples to $N^{-1}N^*(1535)$. Thus, a good tagging channel of the formation of the $\eta$ in nucleus can be decays of $N^*(1535)$ mainly into $N\pi$. An experimental spectroscopy at K1.8BR beamline has been considered using a $^{12}\mathrm{C}(\pi, n)$ reaction with tagging a back-to-back emission of $p\pi^-$ [6]. A similar experiment is feasible also at HIHR.

For the $\eta'$ in nuclei, one of the promising reactions is again $(\pi, N)$. A preceding experiment was performed at GSI using an inclusive measurement of $^{12}\mathrm{C}(p, d)$ reactions [7, 8, 9]. The measured data with extremely good statistical sensitivity showed a very smooth curve without any structures on it setting stringent constraints to the $\eta'$-nucleus interaction. Here we propose a $^{12}\mathrm{C}(\pi^+, p)$ reaction at the incident pion momentum of 1.8 GeV/$c$ tagging a channel of two nucleon absorption $\eta' NN \to NN$ which emits relatively high energy protons.

HIHR will serve better chances to perform the experiment with much better resolution and statistical sensitivities by the resolution matching ion optical conditions. We analyze pion beam momentum in the upstream section. The beam with the intensity of $\sim 10^8$/spill impinges on the carbon target. The forward emitted protons in the $(\pi^+, p)$ reactions are momentum-analyzed in the forward spectrometer. The beamline and the spectrometer need to be controlled to satisfy a resolution matching condition to eliminate contribution of the beam momentum spread to the Q-value resolution. We need a typical resolution of about 1-2 MeV (FWHM).

A detector system is installed surrounding the target to tag the decay of the meson-nucleus systems. A typical setup is a cylindrical detector system with $\sim 4\pi$ coverage. $p/\pi$ separation to $\sim 0.8$ GeV/$c$ is necessary with moderate momentum resolving power of $\Delta p/p$ of 5-10%. We roughly estimate necessary beam time to be about 90 shifts in total.

## References

[1] H. Nagahiro, Prog. Theor. Phys. Suppl., **186**, 316 (2010).

[2] K. Itahashi *et al.*, Phys. Rev. C **62**, 025202 (2000).

[3] K. Suzuki *et al.*, Phys. Rev. Lett. **92**, 072302 (2004).

[4] T. Nishi *et al.*, Phys. Rev. Lett. **120**, 152505 (2018).

[5] R.E. Chrien *et al.*, Phys. Rev. Lett. **60**, 2595 (1988).

[6] K. Itahashi *et al.*, "Spectroscopy of $\eta$ mesic nuclei by $(\pi^-, n)$ reaction at recoilless kinematics", `https://j-parc.jp/researcher/Hadron/en/pac_0707/pdf/LoI-itahashi.pdf` (2007).

[7] K. Itahashi *et al.*, Prog. Theor. Phys. **128**, 601 (2012).

[8] Y.K. Tanaka *et al.*, Phys. Rev. Lett. **117**, 202501 (2016).

[9] Y.K. Tanaka *et al.*, Phys. Rev. C **97**, 015202 (2018).

# $^{4}$n search with the $^{4}$He$(\pi^-, \pi^+)$ reaction at HIHR

**Hiroyuki Fujioka**[1]

[1]Department of Physics, Tokyo Institute of Technology

Very recently, candidate events of a tetraneutron resonant state, $^{4}$n, were observed at RIBF [1]. They used a heavy-ion double charge exchange reaction, ($^{8}$He,$^{8}$Be), on liquid $^{4}$He target. This finding has motivated us to investigate the tetraneutron state by another kind of double charge exchange reaction, namely the $(\pi^-, \pi^+)$ reaction. The detail is given in a Letter of Intent [2] and another one [3], in which a preparatory study with an $H_2{}^{18}O$ target at the existing K1.8 beamline is proposed. While old experiments at LAMPF and TRIUMF with low-energy $\pi$ beams ($\lesssim$ 200 MeV) did not observe any signature of tetraneutron, it is worth revisiting this reaction with a much higher-energy pion beam, because the mean free path of pions in nuclear medium gets longer and the effect of final state interaction might be less significant.

We plan to use the HIHR beamline, and the required kinetic energy (momentum) of the $\pi^-$ beam is 850 MeV (980 MeV/$c$), at which the cross section for the $^{18}$O$(\pi^+, \pi^-)^{18}$Ne$_{(\mathrm{g.s.})}$ reaction is expected to be locally maximal [6] and the same may hold for the $^{4}$He$(\pi^-, \pi^+)^{4}$n reaction. Since the formation cross section will be of the order of nb/sr or even less, a full-intensity ($\approx 1.6 \times 10^{8}$ /spill, which corresponds to 50 kW proton on 5 cm-thick Pt target) beam is mandatory. Under these conditions, the yield of a tetraneutron state will be 97 events/(2weeks)/(nb/sr), if we use a 2.0 g/cm$^{2}$-thick liquid $^{4}$He target.

## References

[1] K. Kisamori *et al.*, Phys. Rev. Lett. **116**, 052501 (2016).

[2] H. Fujioka *et al.*, Letter of Intent for J-PARC 50 GeV Synchrotron, "Search for tetraneutron by pion double charge exchange reaction on $^{4}$He", `http://j-parc.jp/researcher/Hadron/en/pac_1607/pdf/LoI_2016-18.pdf`

[3] H. Fujioka *et al.*, Letter of Intent for J-PARC 50 GeV Synchrotron, "Investigation of Pion Double Charge Exchange Reaction with S-2S Spectrometer", `http://j-parc.jp/researcher/Hadron/en/pac_1607/pdf/LoI_2016-19.pdf`

[4] J.E. Ungar *et al.*, Phys. Lett. B **144**, 333 (1984).

[5] T.P. Gorringe *et al.*, Phys. Rev. C **40**, 2390 (1989).

[6] E. Oset and D. Strottman, Phys. Rev. Lett. **70**, 146 (1993).

Exploring dense matter with heavy-ion collisions at J-PARC

**Hiroyuki Sako[1] for J-PARC-HI Collaboration**

[1]Advanced Science Research Center, Japan Atomic Energy Agency

Extremely dense matter as high as the neutron star core can be created in heavy-ion collisions at J-PARC energy. We are investigating a future project at J-PARC to accelerate heavy-ion beams, J-PARC Heavy-Ion Project (J-PARC-HI). We aim at exploring QCD phase structures such as the phase transition boundary and the critical point in the high density regime, and we can study equation-of state of dense matter to attack the super-heavy neutron star problem. Heavy-ion beams up to U at 10-20 GeV per nucleon at the world's highest beam rate of $10^{11}$ Hz would be achieved by introducing a new heavy-ion injector, and utilizing the existing RCS and MR synchrotrons. We study event-by-event fluctuations of conserved charges to search for the critical point, and collective flow to search for the phase transition. We also study dileptons to study chiral symmetry restoration. We can also study very rate mutli-strangeness system such as mutli-strangeness hypernuclei and strangelets. Heavy-ion spectrometers based on a toroidal and a dipole magnets are designed for these measurements and possible location is expected at the extended high momentum beamline in the extended Hadron Experimental Facility at J-PARC.

# 1 Introduction

We have been planning a future project to accelerate heavy-ion beams at at J-PARC, J-PARC Heavy-Ion Project (J-PARC-HI) [1]. In the current experimental programs at J-PARC, hadron and nuclear structures have been studied. Part of these studies are aimed for studying hadronic matter with the density higher than the normal nuclear density in order to study equation of state (EOS) for the problem of two solar mass neutron stars.

However, in order to reach the high density of $5-10$ times as high as the normal nuclear density, heavy-ion collisions are only the tool in the earth. With heavy-ion collisions at J-PARC energy, we will explore the phase structures in the QCD phase diagram in the high density regime as shown in Fig. 1.

We will also study properties of dense hadronic/partonic matter created by the collisions. In particular to clarify the equation of state is extremely important in order to solve the two-solar-mass neutron star problem. At J-PARC, due to the baryon stopping effect, the maximum baryon density reaches $5-10$ times as high as the normal nuclear density, according to particle transport models such as JAM [2].

The acceleration scheme of heavy-ion beams at J-PARC has been investigated as shown in Fig. 2. In this scheme, we introduce a new heavy-ion injector consisting of a heavy-ion linac and a booster ring, which accelerate a heavy-ion beam up to the same rigidity as a 400 MeV proton. Then, it is injected to the existing RCS and MR synchrotrons. Heavy-ion beams up to U at $10-20$ GeV per nucleon at the world's highest beam rate of $10^{11}$ Hz would be achieved by this scheme.

J-PARC-HI is compared to other heavy-ion facilities which plan to study dense matter physics as shown in Fig. 3. Fixed target experiments of J-PARC-HI and CBM experiment with SIS-100 synchrotron at FAIR will produce highest rate interactions, thus can perform extremely high statistics experiments, while J-PARC-HI will produce higher rates by a factor of 10. Their energy ranges overlap, while J-PARC-HI will have higher maximum energy up to 20 AGeV/c (corresponding to 6.2 GeV in nucleon-nucleon center-of-mass energy ($\sqrt{s_{NN}}$)).

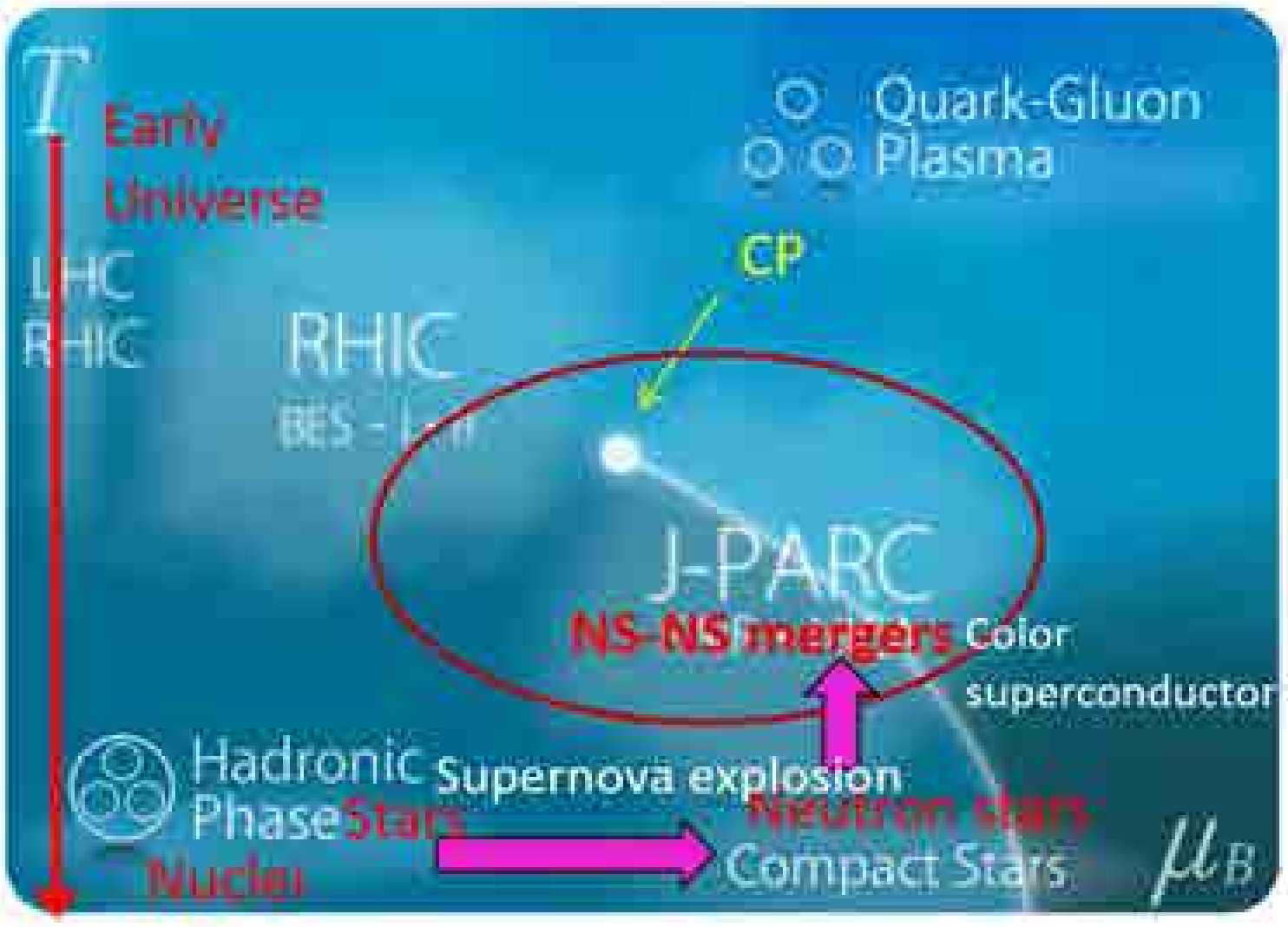

Figure 1: The QCD phase diagram.

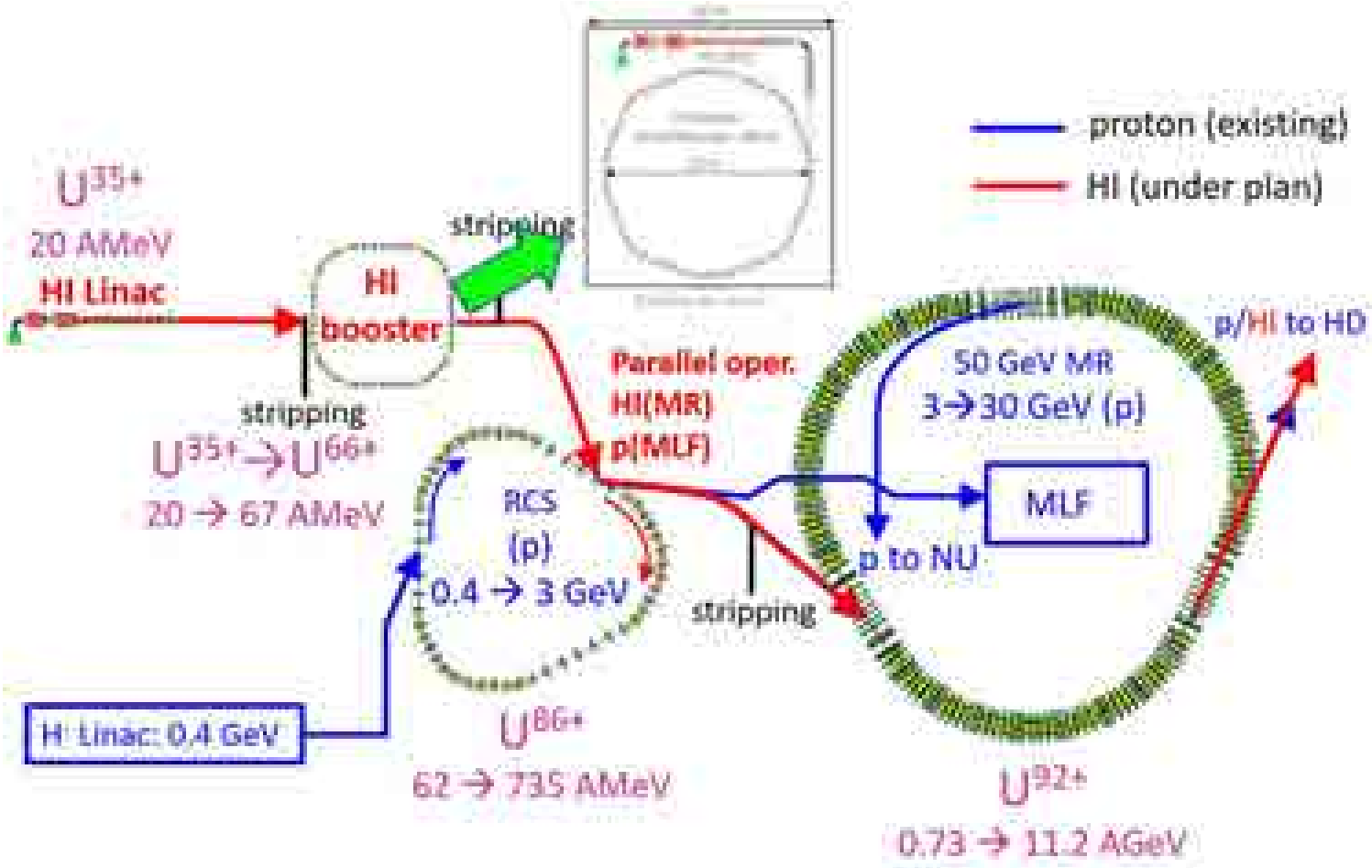

Figure 2: The heavy-ion acceleration scheme at J-PARC.

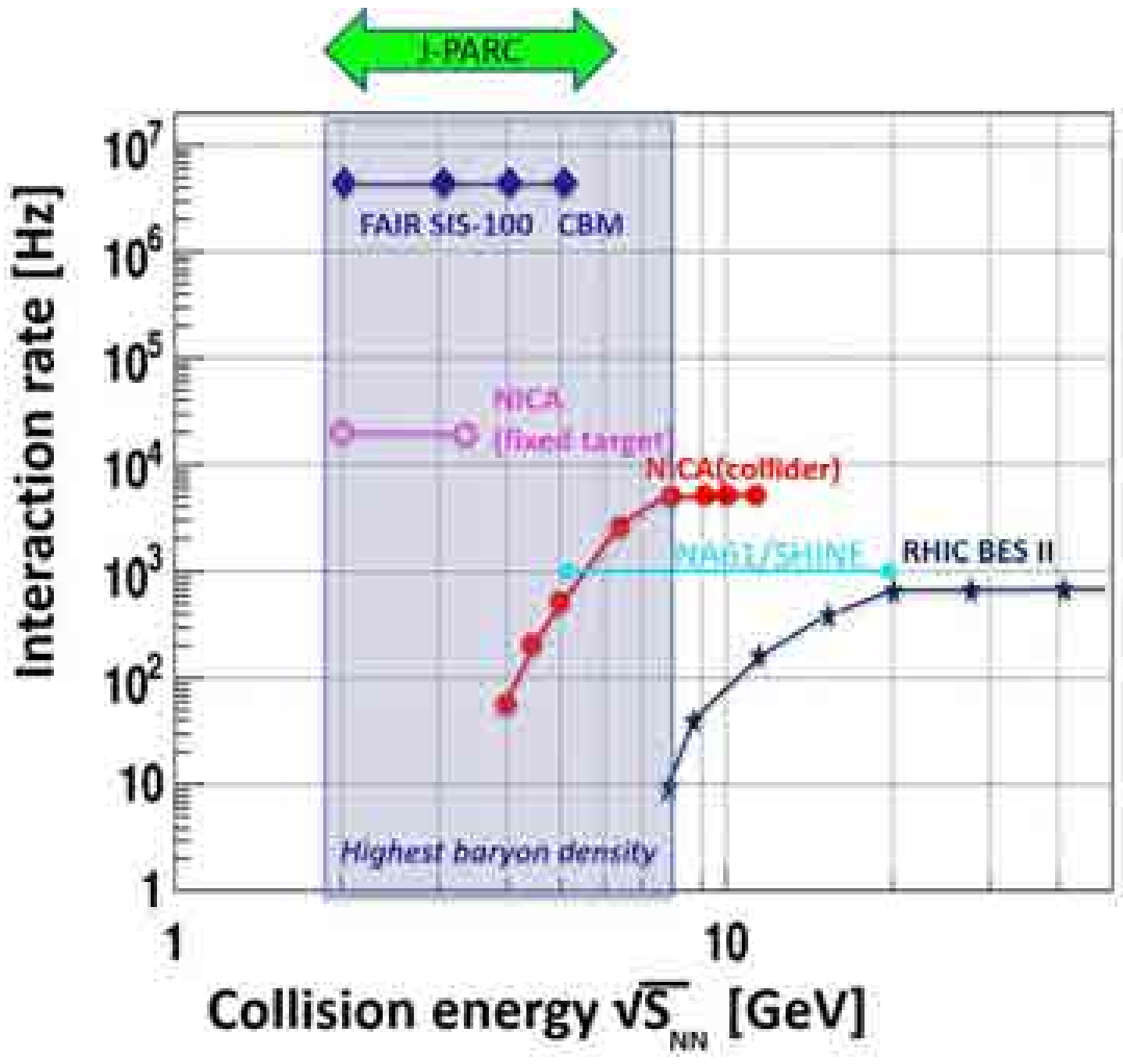

Figure 3: Beam rates and energy of world's heavy-ion facilities.

The particle multiplicity and the expected particle yields for 1-month beam time at $10^7$ Hz interaction rate are plotted in Fig. 4. In the heavy-ion experiments performed in the past at AGS around 10 AGeV/c, only particles above the red broken line were measured. However, many other observables shown below which are now believed to be the most important probes to study phase structures were left unmeasured. At J-PARC-HI, even for one month beam time, we could have good statistics for detailed studies for dileptons from vector mesons ($\rho$, $\omega$, and $\phi$), and single $\Lambda$ hypernuclei. We could also have in small statistics D mesons, $J/\psi$, or double $\Lambda$ hypernculei, and we can search for $|S| \geq 3$ hypernuclei and strangelets. These particles which can be measured for the first time in the J-PARC energies are very important to explore QCD phase structures, and to study the properties of dense matter.

## 2 Experimental Design

Event-by-event fluctuations of the conserved charge such as the net-baryon number and the net-charge number are considered to be one of the most important probes to search for the critical point. STAR Beam Energy Scan program (BES) observed an enhancement in net-proton 4-th order Cumulant (fluctuations) at the lowest energy ($\sqrt{s_{NN}} = 7.7$ GeV [3]. It is one of the most important goals for J-PARC-HI to measure fluctuations as a function of energy with very high statistics to discover the critical point. Also with higher-order fluctuations (6-th or 8-th), one could find the signature of the phase transition even if it is smooth cross-over transition. Note that one order higher fluctuations require 2-order higher statistics, and J-PARC-HI is suitable for high-statistics measurements.

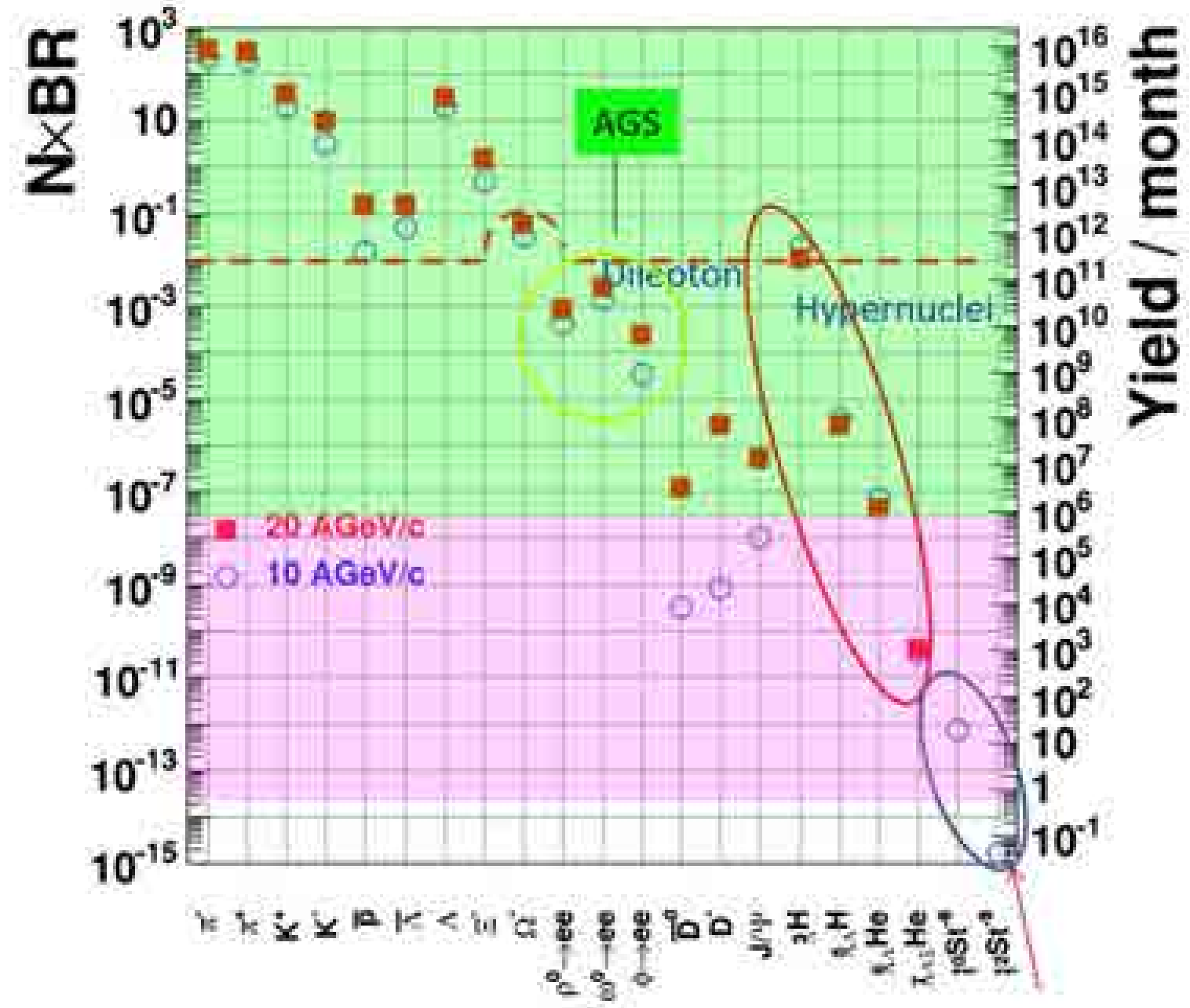

Figure 4: Particle multiplicity (left vertical axis) and expected particle yields per month at the $10^7$ Hz interaction rate (right vertical axis) expected at J-PARC-HI.

Dileptons are so called a “penetrating probe” which can carry directly the information of hot and dense matter. Dileptons from vector mesons are expected to signal chiral symmetry restoration expected in the QGP phase. By measuring dileptons in extremely high statistics at J-PARC-HI, we might be able to extract a measure for chiral symmetry, quark and gluon condensates, by comparing the spectrum shape with a model such as the QCD sum rules [4].

Collective flow reflects the initial pressure just after the heavy-ion collision, thus may include the information of EOS. Negative rapidity slopes of the first harmonics of the flow (directed flow, $v_1$) were observed between AGS to RHIC low energies, which may be an indication of softness of EOS, that could be related to the phase transition [5]. At J-PARC-HI, we will measure the energy dependence of flow precisely to constrain EOS. EOS studies are very important to solve the problem of two-solar mass neutron stars.

It is also very important to search for multi-strangeness systems, which may have similar properties to the matter in the neutron star core.

There is a strong relation of physics of J-PARC-HI to current hadron physics programs at J-PARC as shown in Fig. 5 In cooperation between the two experimental programs, we could study EOS from the normal nuclear density to dense matter complementary. For instance, $Y-N$ or $Y-Y$ interactions can be studied by hypernuclei in $K^-$ beams, and two particle momentum correlations in heavy-ion collisions.

A heavy-ion spectrometer at J-PARC-HI requires extremely high-rate capability with $10^{7-8}$ Hz interaction rates with the charged particle multiplicity of around 1000. We require continuous triggerless readout and online track reconstruction and data reduction. We also require large acceptance close to $4\pi$ in particular for flow and fluctuations measurements.

We design two kinds of spectrometers, based on a toroidal magnet (Fig. 6), and a dipole magnet (Fig. 7. The toroidal spectrometer (JHITS, J-PARC Heavy-Ion Spectrometer) consists of 4-layer Silicon Vertex Trackers, Ring Imaging Cherenkov Counter (RICH), GEM trackers, Time-of-Flight counter, Electro-Magnetic Calorimeter, Neutron Counter, and Muon Absorber and Trackers. The spectrometer has the $\phi$ symmetry, with almost $4\pi$ acceptance, and it has a magnetic-field free region before RICH, which is suitable for dielectron measurements. On the other hand, the toroidal magnetic field in the $\phi$ direction has large non-uniformity, which may be a disadvantage for fluctuation measurements.

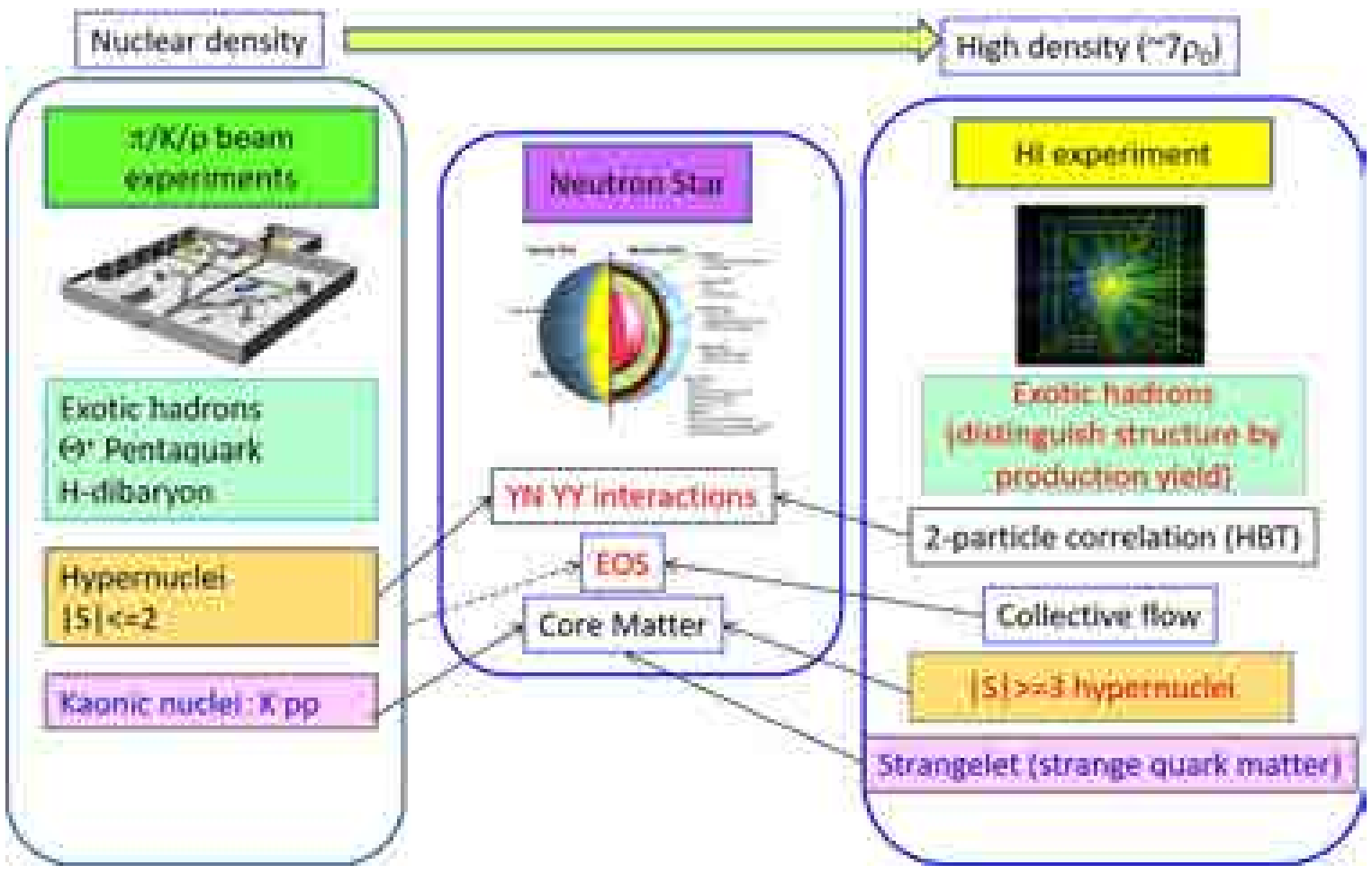

Figure 5: Physics extended by heavy-ion beams at J-PARC.

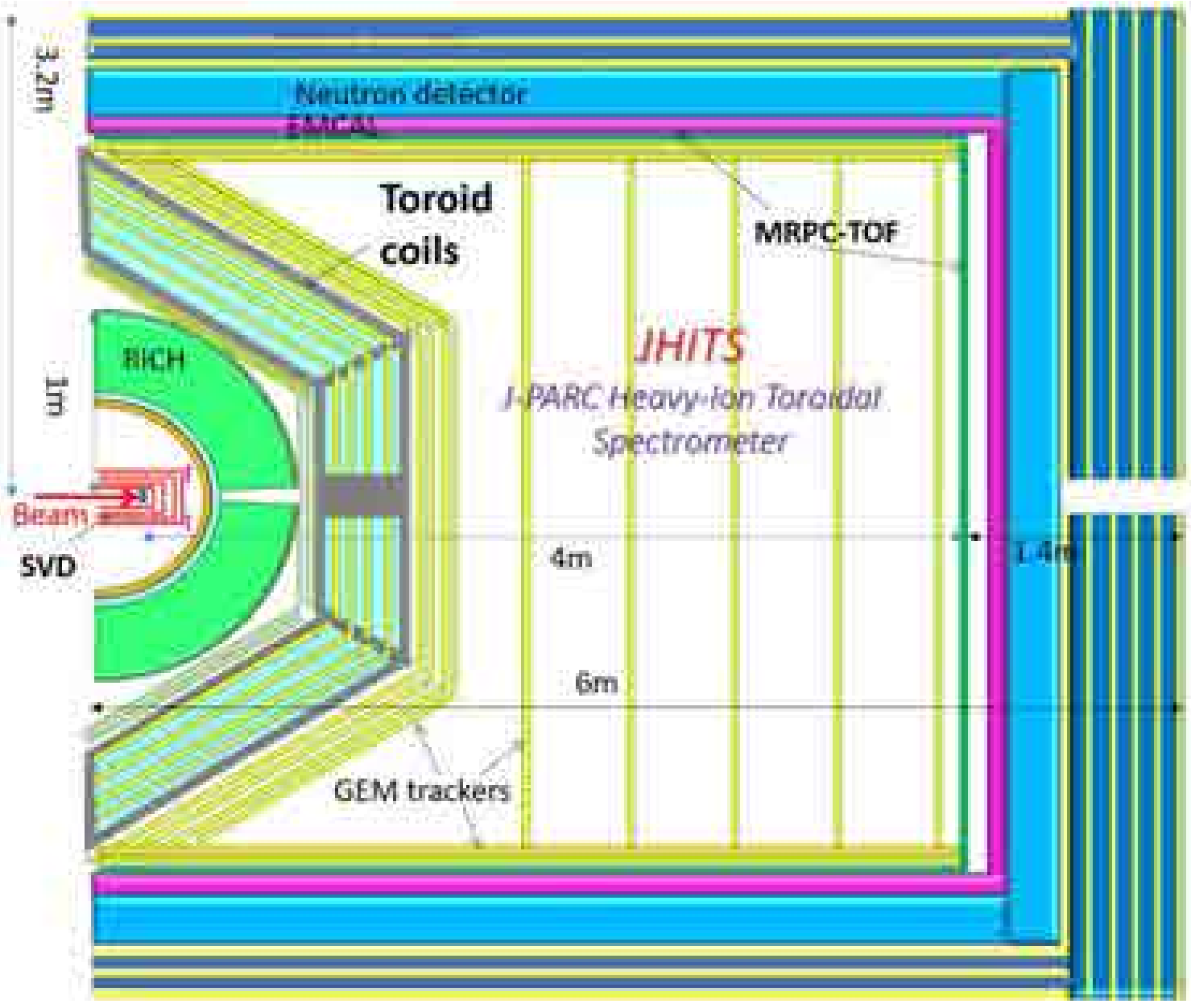

Figure 6: A heavy-ion spectrometer based on a toroidal magnet.

We are designing a new spectrometer based on a large dipole magnet with a Time Projection Chamber (TPC) as shown in Fig. 7. It is dedicated for charged hadron measurements. The silicon vertex trackers covers most forward angles at $\theta \leq 4°$, the TPC at $\theta > 4°$. MRPC (Multi-layer Resistive Plate Chamber) Time-of-Flight counters made of MRPC and their GEM trackers are positioned outside the TPC for charged particle identification. A neutron counter with a charge veto counter and a hadronic calorimeter is located in the downstream position aiming at the measurement of true baryon number fluctuations. The new dipole spectrometer has an advantage of uniform and large acceptance, which is suitable for fluctuations and flow measurements. The TPC limits the interaction rate up to around $10^6$ Hz, at which rates ALICE TPC (upgraded for Run3) and sPHENIX TPC will have similar track rates. For dimuon measurements, we will replace the TPC by a muon absorber and tracker system. Also, dielectron measurements may also be possible to install a RICH and an Electro-magnetic calorimeter downstream of the dipole magnet.

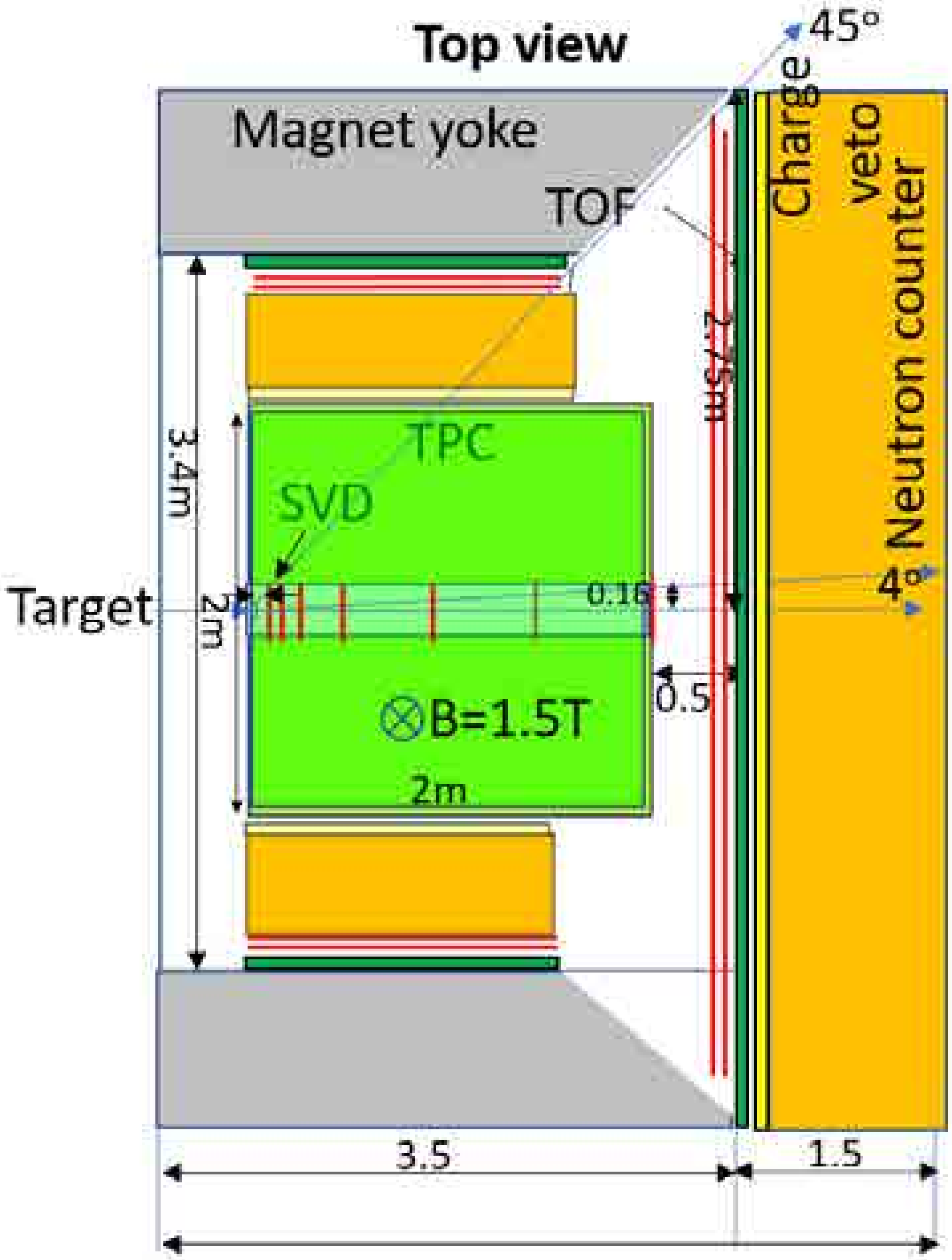

Figure 7: A heavy-ion spectrometer based on a dipole magnet.

Heavy-ion beams will be transported to Hadron Experimental Facility in the high-momentum beamline, which will be complete in 2019. The location for the spectrometers is considered to be in the downstream of the high-momentum beamline, in the extended facility, as shown in Fig. 8. Space of 7 m (horizontal) × 7 m (vertical) × 15 m (beam direction) is required for the spectrometers with the beam height of around 3 m is required, which should be considered in the extension project of Hadron Experimental Facility.

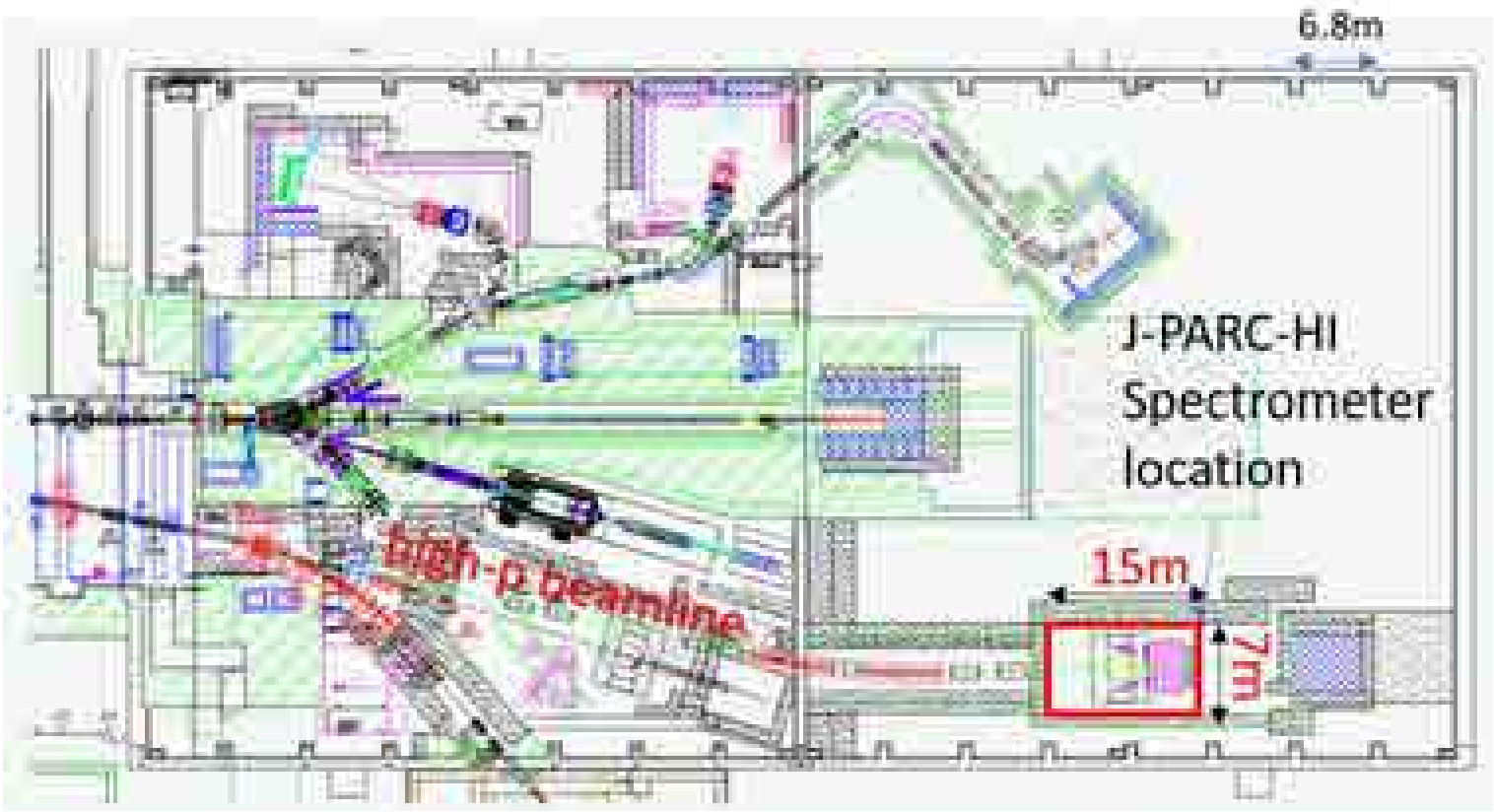

Figure 8: A planned location of the spectrometer of a heavy-ion collision experiment.

# 3 Conclusions

We have developed a plan for the future heavy-ion program at J-PARC (J-PARC-HI). The acceleration scheme of heavy-ion beams with a new heavy-ion injector consisting of a heavy-ion linac and a booster, RCS, and MR has been developed, which could produce the world's highest rate heavy-ion beams of $10^{11}$ Hz. Utilizing such high rate beams, J-PARC-HI has advantage of high-statistics measurements of higher-order fluctuations and dileptons to explore QCD phase structures, and measurements of multi-strangeness systems. We plan to propose the experimental program, and the accelerator design, and R&D works for them. We proceed with the proposal and the budget request aiming at the earliest possible start of the experiment in 2025.

# References

[1] Hiroyuki Sako *et al.*, Letter-Of-Intent of J-PARC-HI submitted to J-PARC PAC, (2016).

[2] Y. Nara *et al.*, Phys. Rev. **C61**, 024901 (1999).

[3] X. Luo, Nucl. Phys., **A956**, 75-82 (2016).

[4] R. S. Hayano and T. Hatsuda, Rev. Mod. Phys. **82**, 2949 (2010).

[5] Y. Nara *et al.*, Phys. Lett. **B769**, 543 (2017).

# High-momentum beam line and upgrade plan in the extended hadron hall

**Hiroyuki Noumi[1,2]**

[1]Reserach Center for Nuclear Physics, Osaka University

[2]Insitute of Particle and Nuclear Studies, High Energy Accelerator Research Organization (KEK)

Design performances of the high-momentum beam line for secondary beams and its extension in the extended hadron experimental facility are presented.

## 1 Introduction

High-momentum hadron beams greater than 2 GeV/c open new opportunities to study hadron nuclear physics, such as spectroscopic study of excited charmed baryons [1], at J-PARC. Here, we introduce design performance of the high-momentum beam line proposed at the hadron experimental facility. The high-momentum beam line can be extended in the extended hadron hall, where more flexible layout of the experimental setup will be possible and enhance physics cases.

## 2 Design Performance of High-momentum Beam Line

The high-momentum beam line is now under construction in the J-PARC Hadron experrimental facility. It is originally planned for an experimental studies on spectral modification of vector mesons in nuclei (E16[2]) to transport a small fraction of the primary proton beam at the branching point in the slow extraction beam line in the switch yard (Fig. 1). We designed the high-momentum beam line ion-optically so as to deliver secondary beams. The production target of 15 kW loss will be placed at the branching point to produce the secondary beams. The primary beam passing through the production target will be swung back to the primary beam line and transfered to the T1 primary target This is a so-called Beam Swinger Optics [3], for which two pairs of dipole magnets will be placed before and after the production target, as shown in Fig. 2. The magnetic field of the third dipole magnet placed just after the target is determined according to the secondary beam momentum. The fields of the other three dipole magnets will be adjusted so as to keep the primary beam on the right orbit going down to the T1 target. The orbit looks wavy. Owing to this, the production angle of negative particles can be set at zero degree at 20 GeV/c, where the production cross section can be maximum [7]. While, that of the positive particles is 3.9 degrees at 20 GeV/c. A schematic layout of the high-momentum beam line for the secondary beams is shown in Fig. 3. Produced secondary beam is collected by 2 quadrupole magnets and focused at the collimator. After the collimamtor, the beam is transported through 4 dipole, 5 quadrupole, and 3 sextapole magnets to the dispersive focal plane, where strong correlation between beam positions and momenta.are realized. Magnification and dispesion are $R_{11}$=0.708 and $R_{16}$=1.17 cm/%, respctively. Second order aberrations ($T_{122}$, $T_{126}$, and $T_{166}$) are controlled by 3 sextapole magnets. One can measure a beam particle momentum by a position sensitive particle detector placed at the dispersive focal plane. By using a ray tracing code TURTLE [6], we demonstrate a momentum resolution of as goog as ~0.1% with a spatial resolution of 1 mm, as shown in Fig. 4. The beam will be focused again at the experimental target. The beam envelope calcualted by TRANSPORT [5] to the second order is shown in Fig. 5. The beam length and acceptance are estimated to be 133.2 m and 1.5 msr%, respectively. The intensities of pion, kaon, and anti-proton are estimated as functions of momentum, as shown in Fig. 6. Here, we assume that a 20-mm thick gold target is irradiated by a 83.5-kW proton beam of 30 GeV. One expects more than $10^7$ negative pions per second at 20 GeV/c.

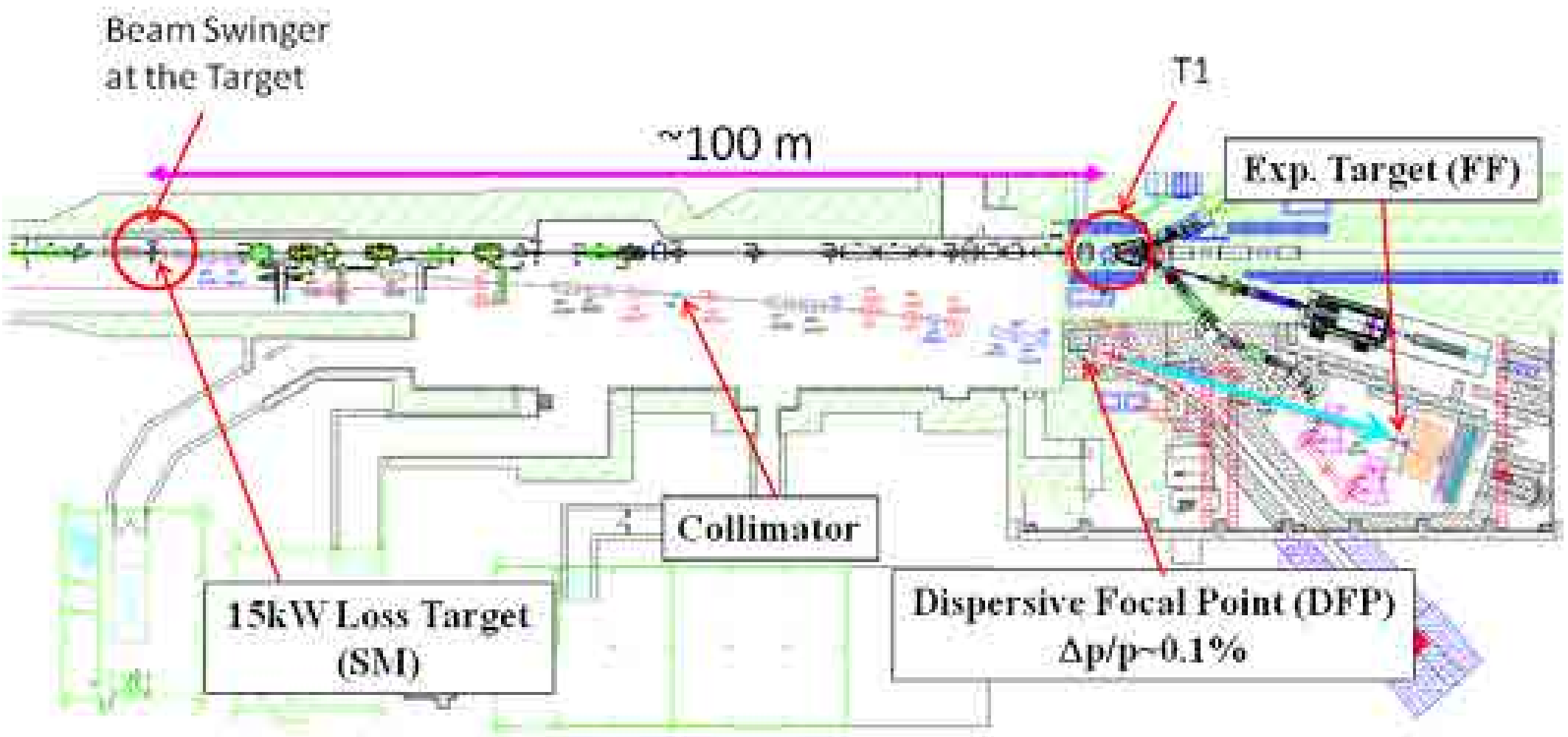

Figure 1: Plan view of the high-momentum beam line

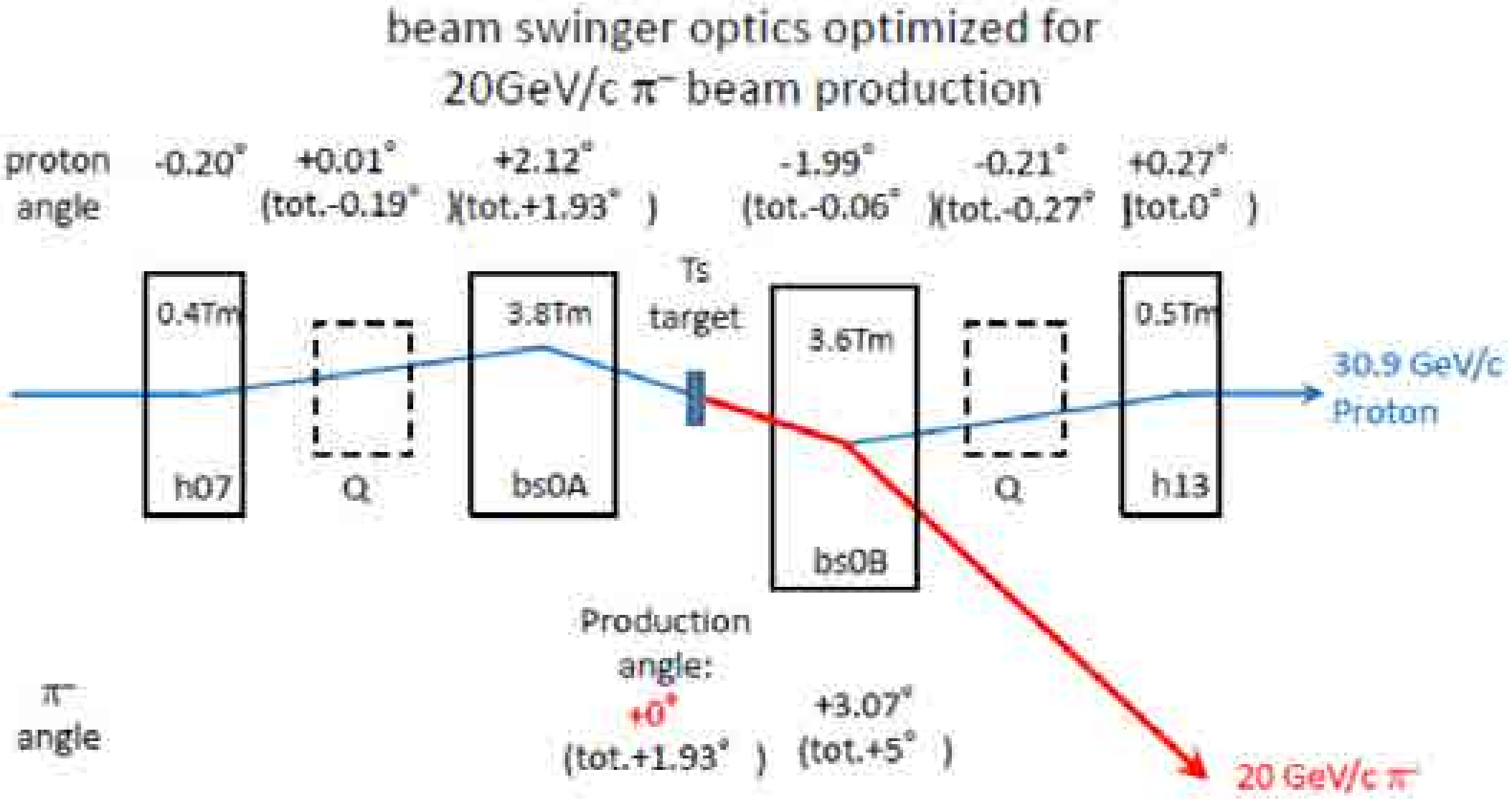

Figure 2: Magnet layout around the production target [8]

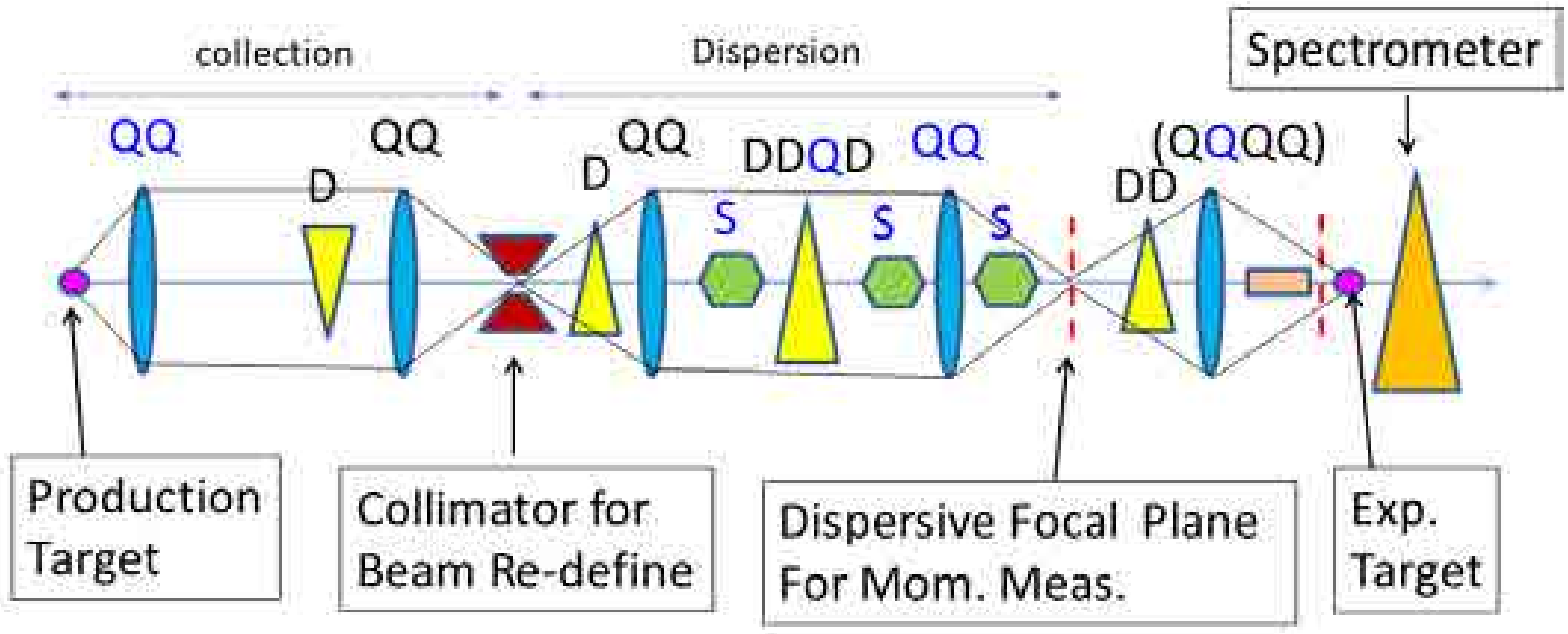

Figure 3: Schematic layout of the high-momentum beam line for the secondary beams

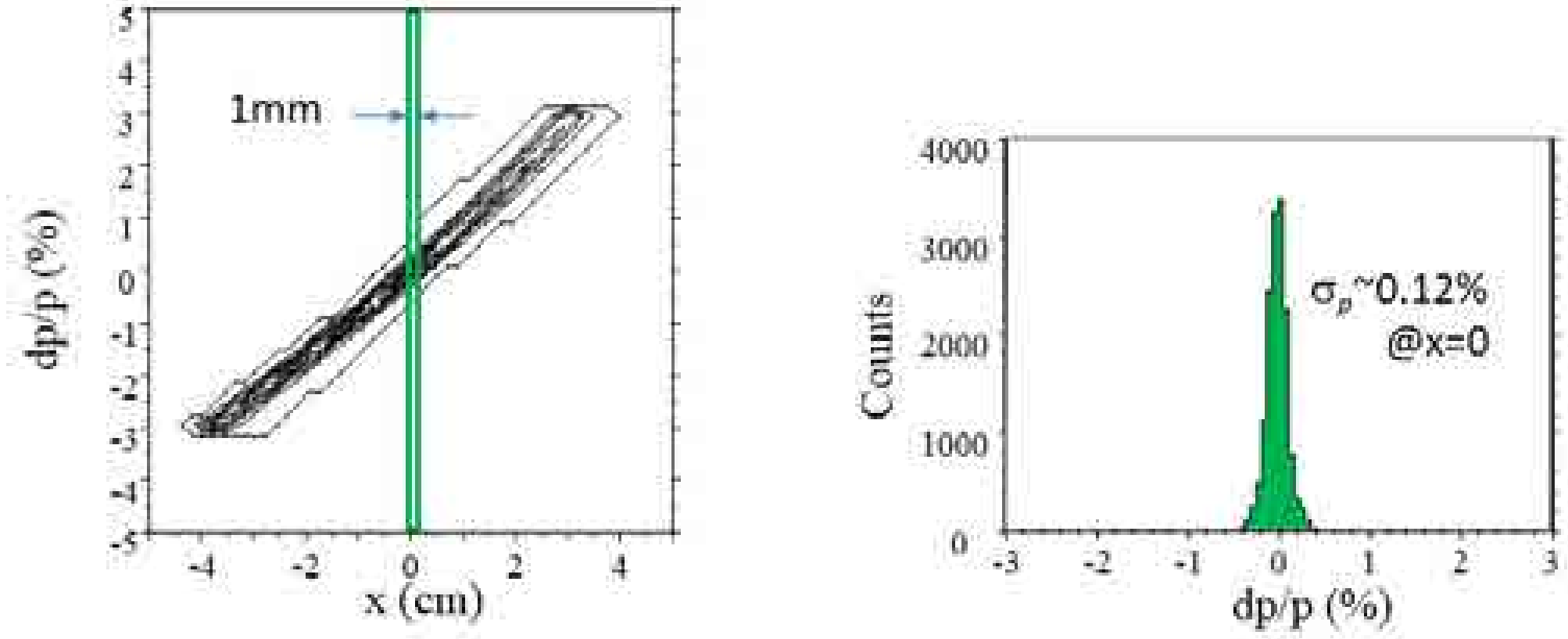

Figure 4: Expected momentum resolution

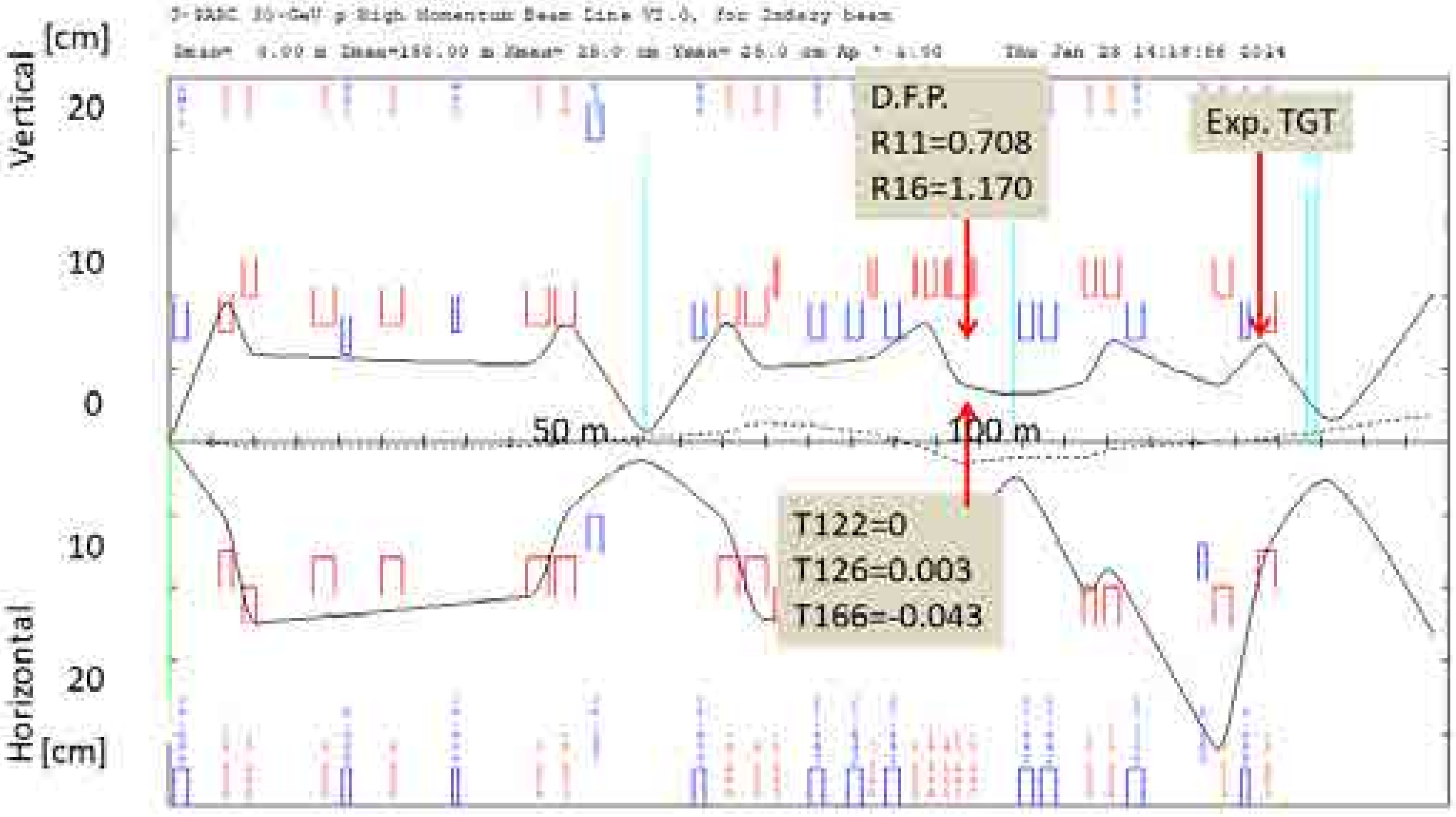

Figure 5: Beam envelope of the high-momentum beam line

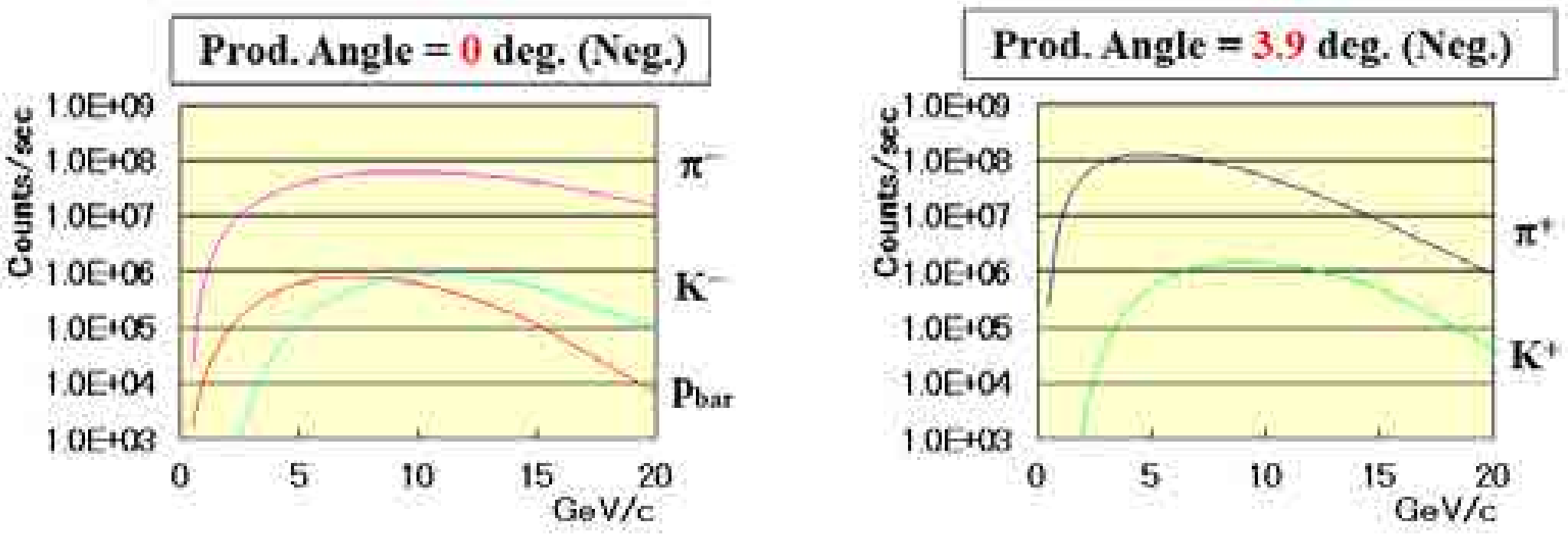

Figure 6: Expected beam intensities of pions, kaons, and anti-proton

# 3 Extension of High-momentum Beam Line in the Extended Hadron Hall

In the extended hadron hall, the high-momentum beam line can be extended. There are at lease two merits with a limited demerit, as summarized as follows.

*Merit 1*:
The experimental area will be enlarged in the extended experimental hall. It opens possible upgrade of the spectrometer system for various proposed experiments. The charmed baryon spectrometer will be 6∼7 m in total length in order to place tacking devices and large ring imaging cherenkov counter for momentum measurement and particle identification for high-momentum particles scattered at forward angles. Muon idenfication detector will be installed behind the spectrometer for application to an experimental study of the Drell-Yan process. A spectrometer system of 7m and 15 m in width and length can be accomodated in future.

*Merit 2*:
Achromatic beam (no mometum and angular disersions) can be realized at the experimetal target. No spatial correlation of the beam appear and the beam size will be minimized at the target.

*Demerit*:
The bea m line will be prolonged by approximately 40 m. Due to decay inflight, intensities of kaons decrease by factors of aprroximately 1/2 and 1/3 at 10 GeV/c and 5 GeV/c, respectively. Intensities of pions decrease too but the reduction factors will be as small as a few %.

A plan view of the extended high-momentum beam line shown in Fig. 7. The beam course will be slightly changed by switching off the bending magnet placed at the branching point to the COMET beam line. The dispersive focal plane will be set approximately 20 m downstream from the original position. The new focal plane will be located at the entrance of the experimental area in the current hadron hall, where one can access and maintain the focal plane detectors more easily. Magniication and dispersion parameters are $R_{11}$=1.441 and $R_{16}$=2.703, respectively. In the same manner as described above, we demonstrate a momentum resolution of ∼0.2% assuming a spatial resolution of 1 mm (Fig. 8). This would be improved by tuning the second order aberrations. After the dispersive focal plane, the beam will be bent to the extended hadron hall by 2 dipole magnets of 6.6 m each in length. Ion optical parameters are adjusted so as to eliminate both dispersion and angular disersion parameters at the exit of the second 6.6 m dipole magnet. The beam envelope calcualted by TRANSPORT [5] to the first order is shown in Fig. 9. The beam length and acceptance are estimated to be 174.6 m and 1.5 msr%, respectively. The intensities of pion, kaon, and anti-proton are estimated as functions of momentum, as shown in Figs. 10.

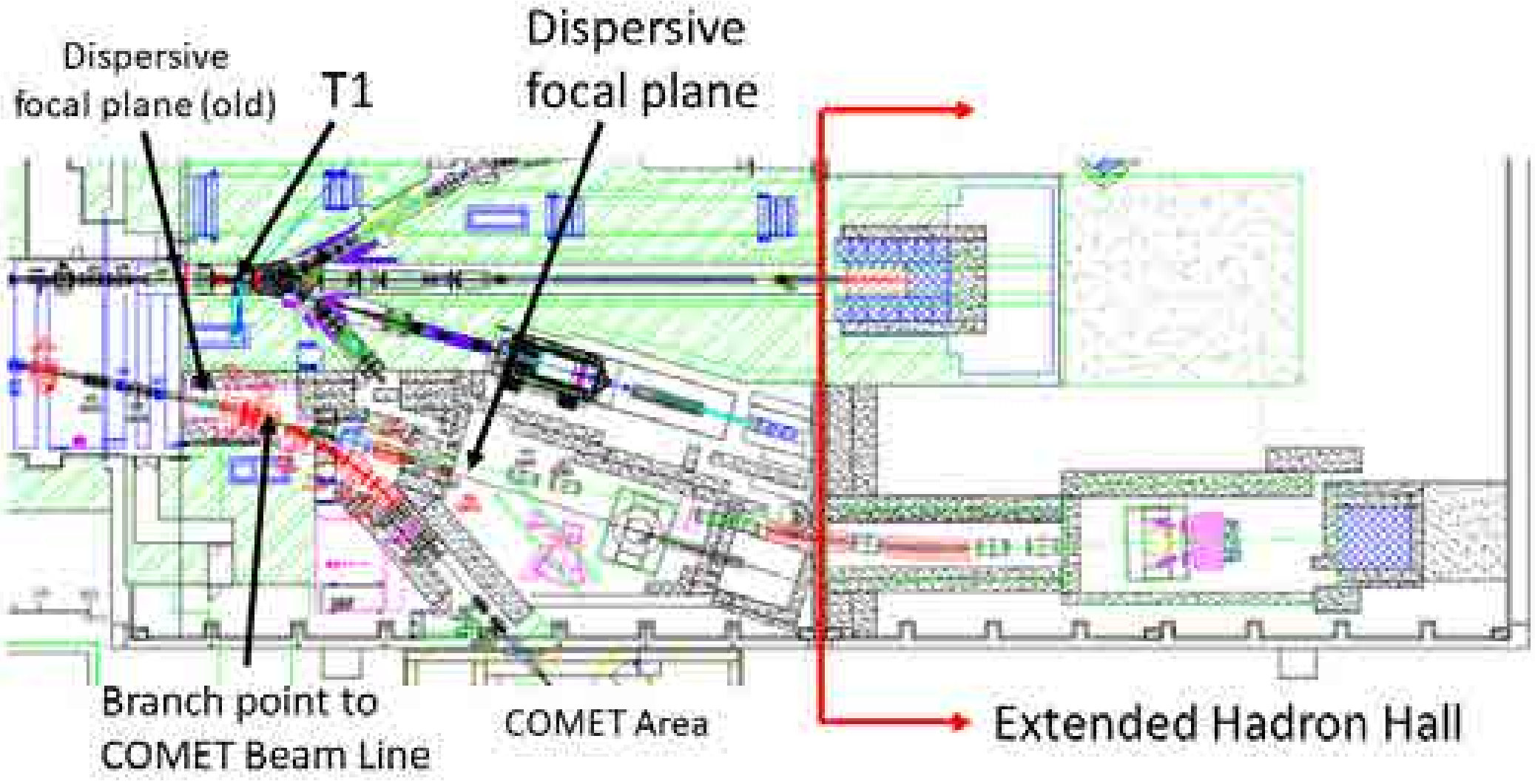

Figure 7: Plan view of the extended high-momentum beam line

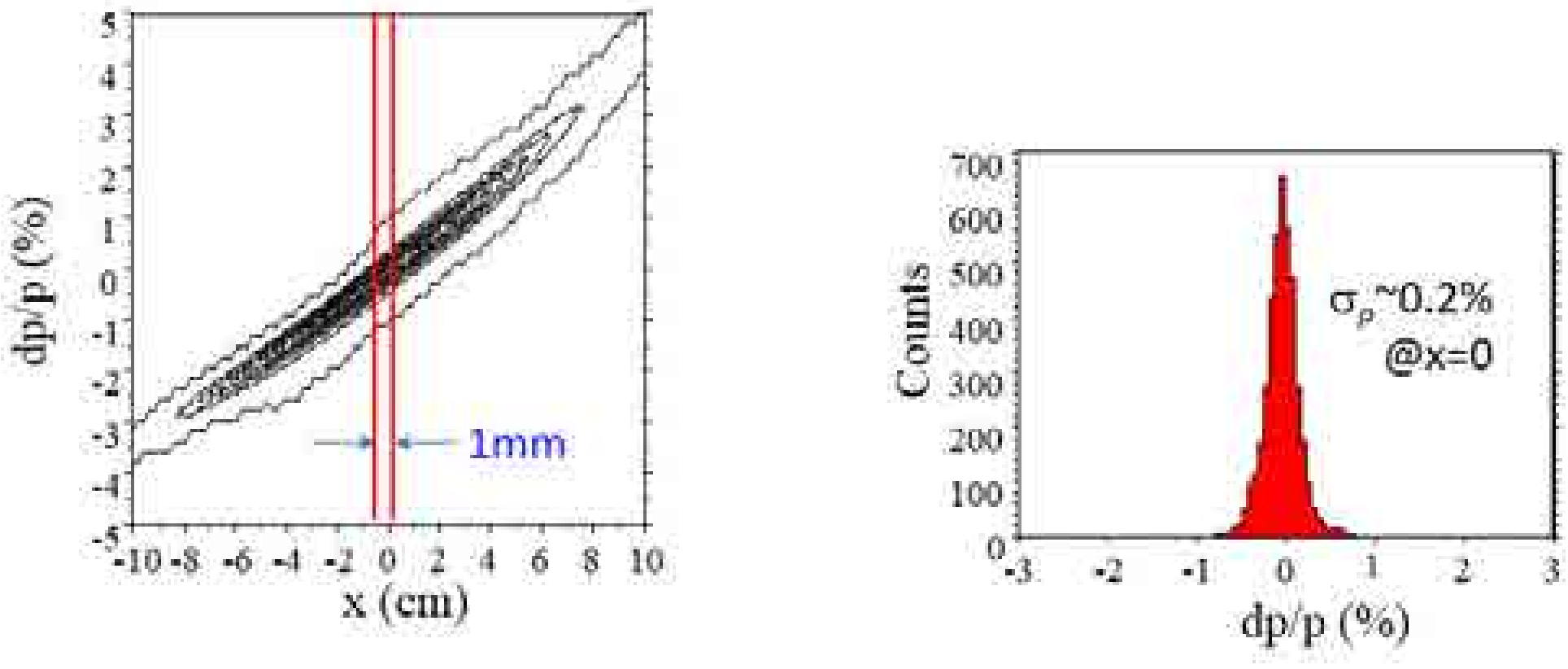

Figure 8: Expected momentum resolution

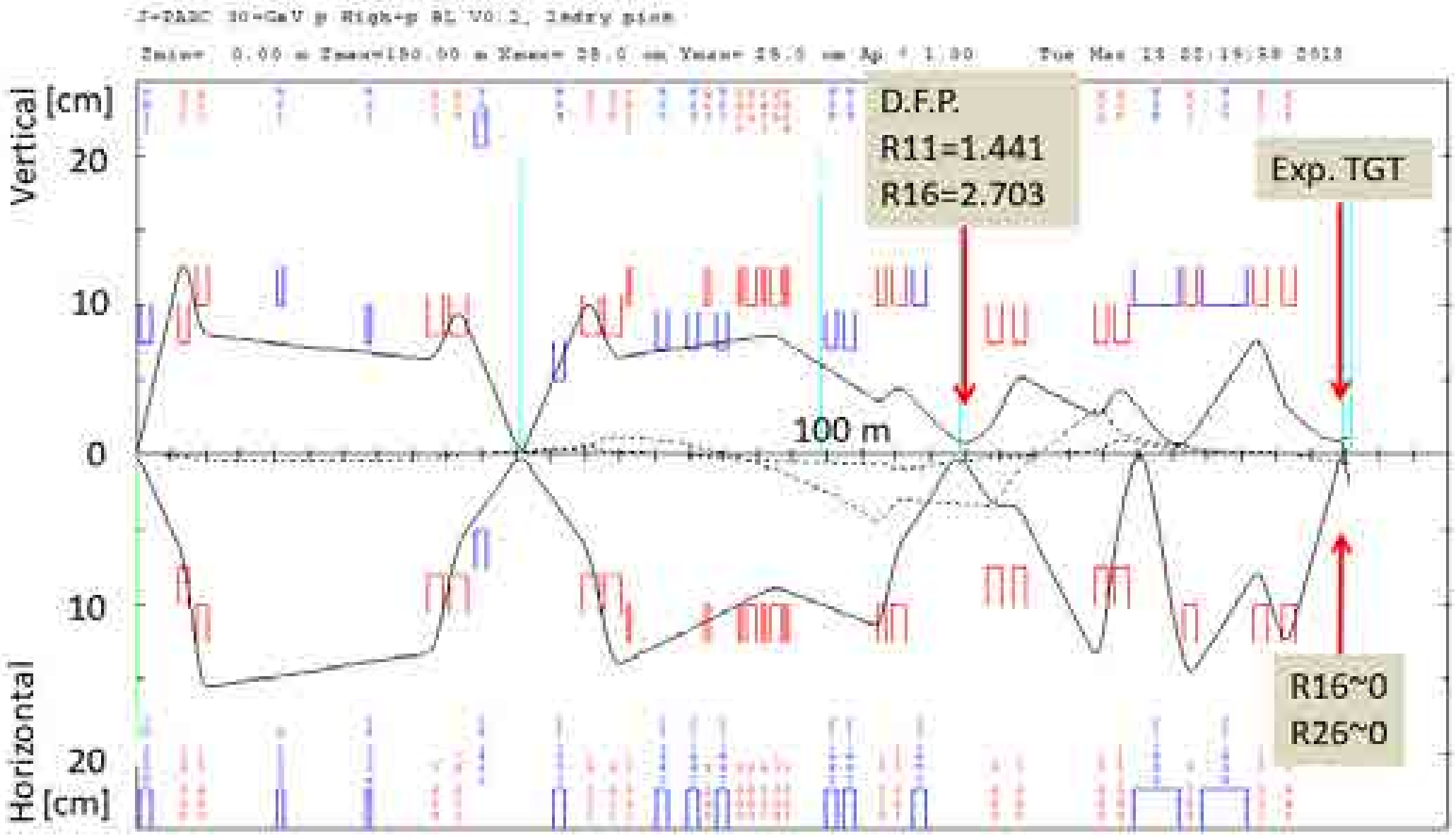

Figure 9: Beam envelope of the extended high-momentum beam line

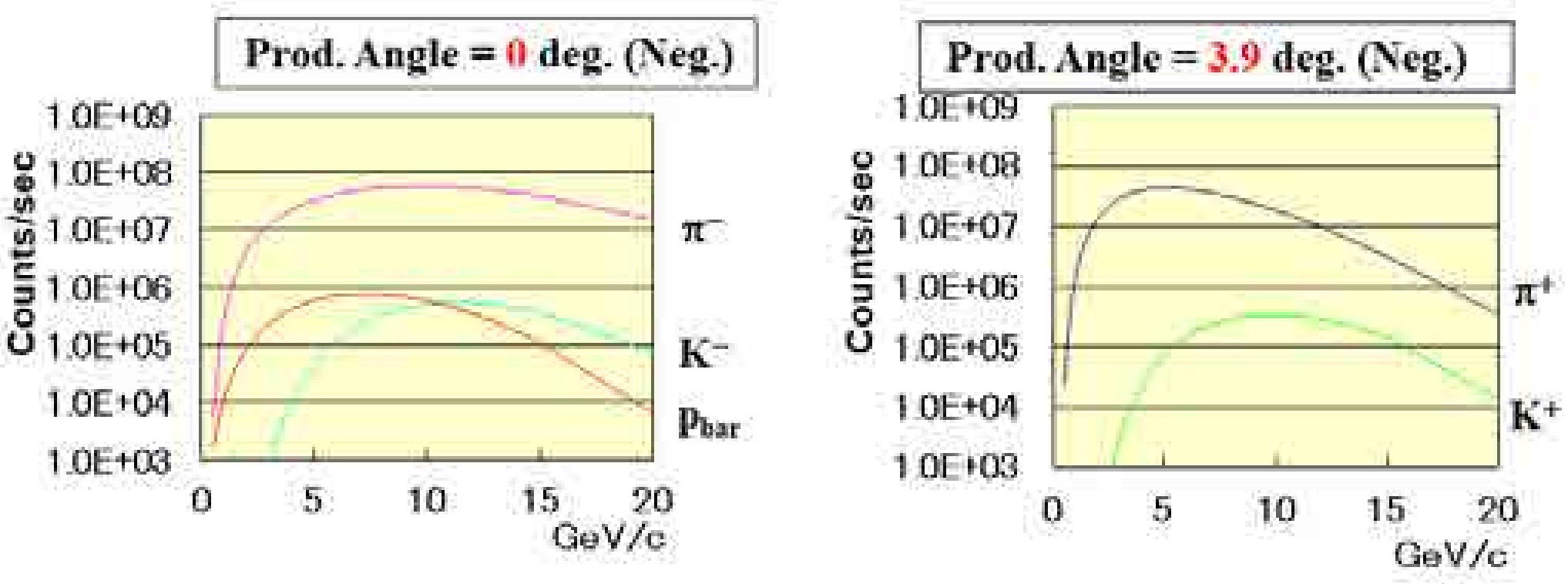

Figure 10: Expected beam intensities of pions, kaons, and anti-proton

# References

[1] J-PARC E50 Proposal, `http://www.j-parc.jp/researcher/Hadron/en/Proposal_e.html#1301`

[2] J-PARC E16 Proposal, `http://www.j-parc.jp/researcher/Hadron/en/Proposal_e.html#0606`

[3] K. H. Tanaka *et al.*, Nucl. Instr. Meth. A363, 114(1995).

[4] J. R. Sanford and C. L. Wang, BNL 11279 and BNL 11479, (1967).

[5] Urs Rohrer, Compendium of Transport Enhancements, http://aea.web.psi.ch/Urs_Rohrer/MyWeb/trancomp.htm; K. L. Brown, D. C. Carey, Ch. Iselin, and F. Rothacker, "Transport, a Computer Program for Designing Charged Particle Beam Transport Systems", CERN 73-16 (1973) and CERN 80-04 (1980).

[6] Urs Rohrer, Compendium of Transport Enhancements, http://aea.web.psi.ch/Urs_Rohrer/MyWeb/trancomp.htm;

[7] J. R. Sanford and C. L. Wang, BNL 11279 and BNL 11479, (1967); C. L. Wang, Phys. Rev. Lett. **25**,

[8] H. Takahashi, private communication, 2013.

# J-PARC E50 experiment

**Kotaro Shirotori[1], Jung-Kun Ahn[2], Shuhei Ajimura[1], Takaya Akaishi[3], Kazuya Aoki[4], Hidemitsu Asano[5], Wen-Chen Chang[6], Ryotaro Honda[7], Yudai Ichikawa[8], Takatsugu Ishikawa[9], Yusuke Komatsu[1], Yue Ma[5], Koji Miwa[7], Yoshiyuki Miyachi[10], Yuhei Morino[4], Takashi Nakano[1], Megumi Naruki[11], Hiroyuki Noumi[1], Kyoichiro Ozawa[4], Fuminori Sakuma[5], Takahiro Sawada[12], Yorihito Sugaya[1], Tomonori Takahashi[1], Kiyoshi Tanida[8], Natsuki Tomida[1], Takumi Yamaga[5] and Seongbae Yang[2]**

[1]Research Center for Nuclear Physics (RCNP), Osaka University

[2]Physics Department, Korea University

[3]Department of Physics, Osaka University

[4]Institute of Particle and Nuclear Studies (IPNS), High Energy Accelerator Research Organization (KEK)

[5]RIKEN

[6]Institute of Physics, Academia Sinica

[7]Physics Department, Tohoku University

[8]Advanced Science Research Center (ASRC), Japan Atomic Energy Agency (JAEA)

[9]Research Center for Electron Photon Science (ELPH), Tohoku University

[10]Physics Department, Yamagata University

[11]Department of Physics, Kyoto University

[12]University of Michigan

Charmed baryon spectroscopy is a key way to understand degree of freedoms to describe the hadron structure. The diquark correlation which is expected to be a degree of freedom of hadrons can be studied from the spectroscopy of charmed baryons. An experiment to investigate the charmed baryons was proposed at the J-PARC high-momentum beam line using an intense secondary hadron beam. The experiment via the $\pi^- \, p \to Y_c^{*+} \, D^{*-}$ reaction at 20 GeV/$c$ using the missing mass technique is performed for the systematic measurement of the excitation energy, the production rates and the decay products of charmed baryons. Mainly from the study of production of charmed baryons, the diquark correlation which is expectedly an essential degree of freedom to describe the hadron structure can be revealed.

## 1 Physics motivation

One of main subjects in the hadron physics is to understand how hadrons are originated by quarks. The constituent quark model well describes the properties of hadrons, such as hadron mass and magnetic moments of ground state hadrons. However, the constituent quark model doesn't well describe the excited states. There are many undiscovered states which are called as missing resonances. In addition, the exotic states which are recently discovered [1, 2] suggests that hadrons have rich internal structures which are beyond the naive quark model. Therefore, it is necessary to investigate effective degrees of freedoms of hadrons with extended constituents of internal structures of hadrons such as the diquark correlation as well as the hadron molecular picture.

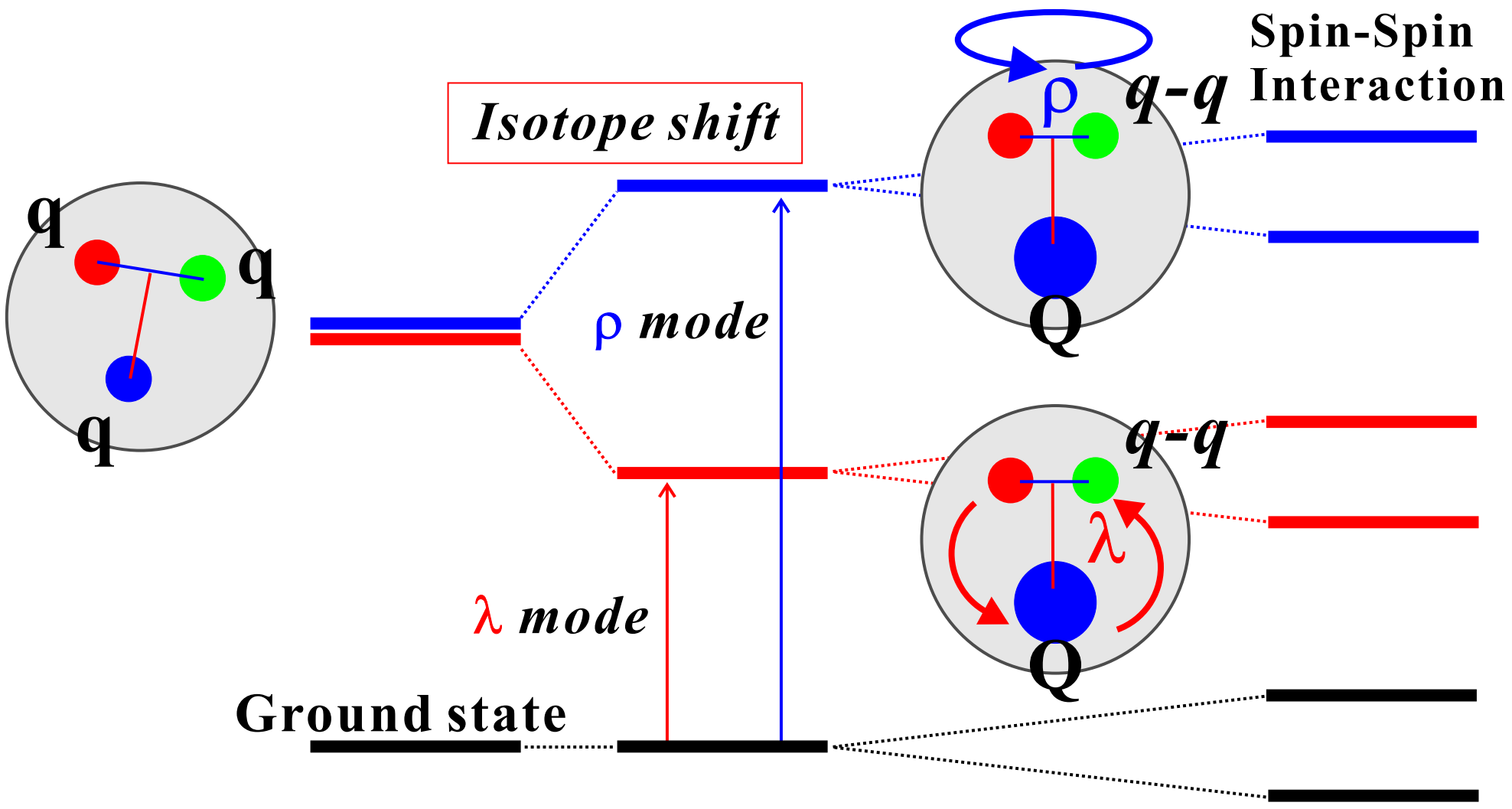

Figure 1: Brief images of the excitation spectrum of the charmed baryon. The level structure which is reflected by the $\rho$- and $\lambda$-mode excitation generated by the diquark correlation appear in the spectrum. It could not appear in light quark baryons because the $\rho$- and $\lambda$-mode excitations are degenerated due to the same dynamics of three diquark pairs.

In order to understand effective degrees of freedoms of hadrons, it is essential to study the interaction between quarks. The quark-quark correlation, namely diquark correlation, is expected to be a degree of freedom to describe hadrons. In the case of light baryons, three diquark pairs are correlated each other in an equal weight so that the extraction of the diquark correlation is not clear. When one quark in a baryon is replaced to a heavy quark, the diquark correlation is expected to be isolated and developed in the charmed baryon structure. The correlation of two light quarks is stronger than that of the other pair because the magnitude of the color-spin interaction between quarks is proportional to the inverse of the quark mass. As a result, two excitation modes, called isotope shift, emerge to the level structure as shown in Fig. 1. The $\rho$ and $\lambda$ modes are a rotation of the diquark and an orbital excitation between the diquark and the other heavy quark, respectively. Those excited states generated by the diquark correlation which could not be observed in light quark baryons due to the same dynamics of three diquark pairs appear in the excitation spectrum. From the study of the excitation spectrum of charmed baryons, we can study the $\rho$ and $\lambda$ excitation modes which are strongly related to the internal structure of the charmed baryons. Charmed baryons which have one heavy charm quark can provide unique opportunities to study diquark correlation.

The nature of the diquark correlation can appear not only in the level structure but also production rates and decay properties of charmed baryons. The production cross section is expected to be strongly related to the excitation modes. Charmed baryons are produced via the hadronic reaction, the $\pi^- \ p \rightarrow Y_c^{*+} \ D^{*-}$ reaction. For the production process, the $D^*$ exchange can be taken into account as shown in Fig. 2. The charm quark which is generated from the $q\bar{q}$ annihilation process is attached to a spectator diquark from the initial state in the proton wave function. Then, a charmed baryon forms from a diquark and a charm quark with an orbital excitation. In this reaction process, the production cross section can be described by the overlapping of wave functions between initial and final states. Therefore, the production rate has the relation to the spin/isospin configuration of charmed baryons and momentum transfer. Consequently, the production cross section has strong dependence on the excitation modes of charmed baryons [3]. It is a similar description for the hypernuclear production where the produced $\Lambda$ particles is attached to the core nucleus with orbital excitations. From the $D^*$ exchange reaction process as shown in Fig. 2, the forward production can produce the $\lambda$-mode excitation because only the orbital motion between the diquark and the charm quark can be excited by taken into account the single-step reaction. From the measurement of the $\pi^- \ p \to Y_c^{*+} \ D^{*-}$ reaction, we can observe the charmed baryon states with the different yield according to its internal structure. The production mechanism can gives us the unique information of the diquark correlation of charmed baryons.

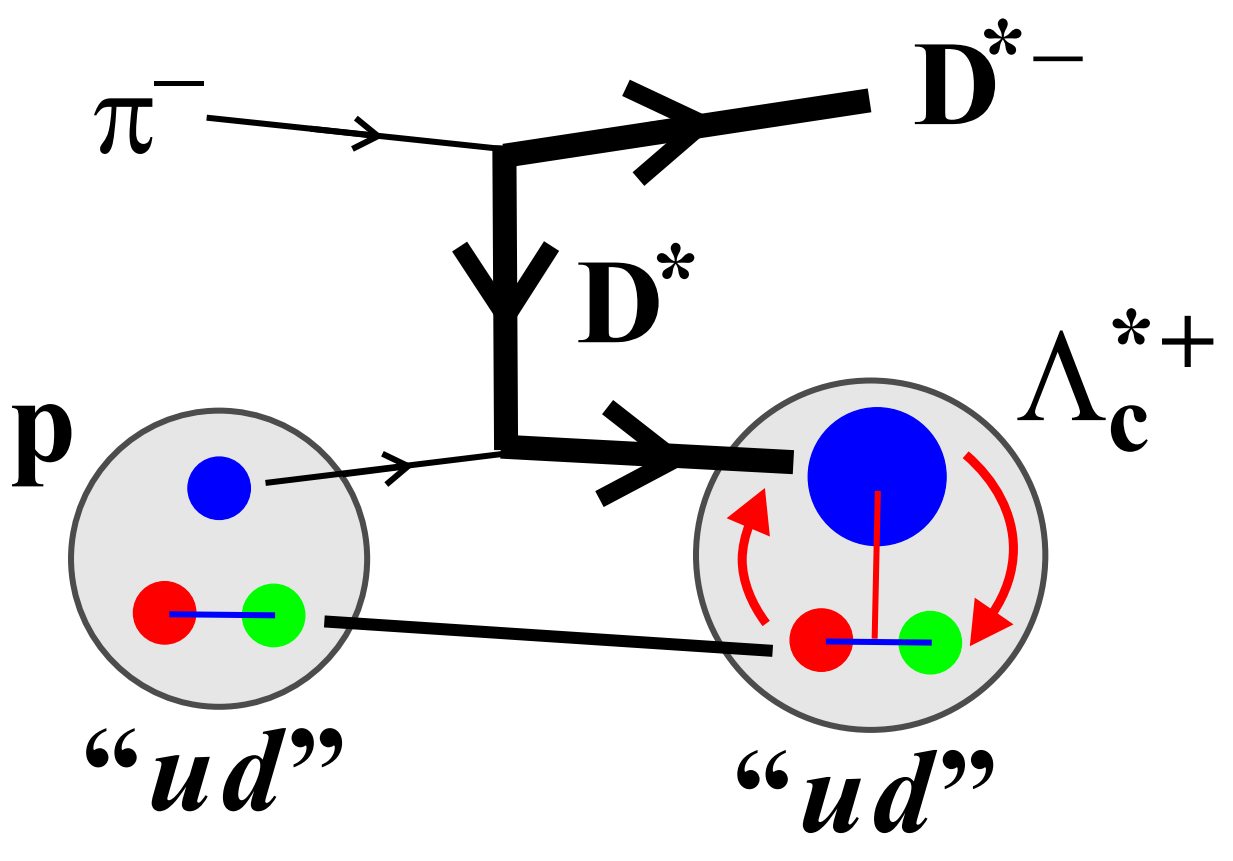

Figure 2: Diagram of the $D^*$ exchange process in the $\pi^- \ p \to Y_c^{*+} \ D^{*-}$ reaction. The generated charm quark from u quark is attached to a spectator diquark pair (”$ud$”) from the initial state in the proton wave function. The forward production which is mainly the single-step reaction can produce the $\lambda$-mode excitation because only the orbital motion between the diquark and the charm quark can be excited.

Figure 3 shows an expected missing mass spectrum simulated by the realistic experimental conditions. By assuming the ground state cross section of 1 nb, yields of charmed baryon states are expected to be ∼2000 counts per 100 days beam time. The main strangeness production background processes were simulated by the JAM code [4], while the PYTHIA code [5] was also used for the comparison. The most effective method of the background reduction is the $D^*$ tagging. Both the mass region of the $\bar{D}^0$ mass and the Q-value corresponded to the $\bar{D}^{*-}$ decay are selected. By using the $D^*$ tagging, the background reduction of $2 \times 10^6$ can be achieved from the JAM simulation result. The sensitivity of the production cross section was found to be 0.1−0.2 nb in the missing mass region of $2.2 - 3.4$ GeV/$c^2$.

For simulating the missing mass spectrum, the calculated [3] relative production rates which are summarized in Table 1 are input. In the expected spectrum, the strong dependence of the production rate can be observed. One of the subjects is to measure the production rate of the $\lambda$-mode doublet states which are enhanced in the missing mass spectrum, such as $\Lambda_c(2595)(\frac{1}{2}^-)$ and $\Lambda_c(2625)(\frac{3}{2}^-)$ and $\Lambda_c(2880)(\frac{5}{2}^+)$ and $\Lambda_c(2940)(\frac{3}{2}^+)$ for the $L = 1$ and $L = 2$ states, respectively. By taken into account the prediction from the $\lambda$-mode excitation, those $\lambda$-mode doublet states have production ratios with $R(\frac{3/2^-}{1/2^-}) \sim 2$ and $R(\frac{5/2^+}{3/2^+}) \sim \frac{3}{2}$, respectively. The expected result means that the production rates show the collective motion between diquark and charm quark in the charmed baryons. Therefore, the measurement of the production rates of charmed baryons will give us one of the strong evidences of the diquark correlation.

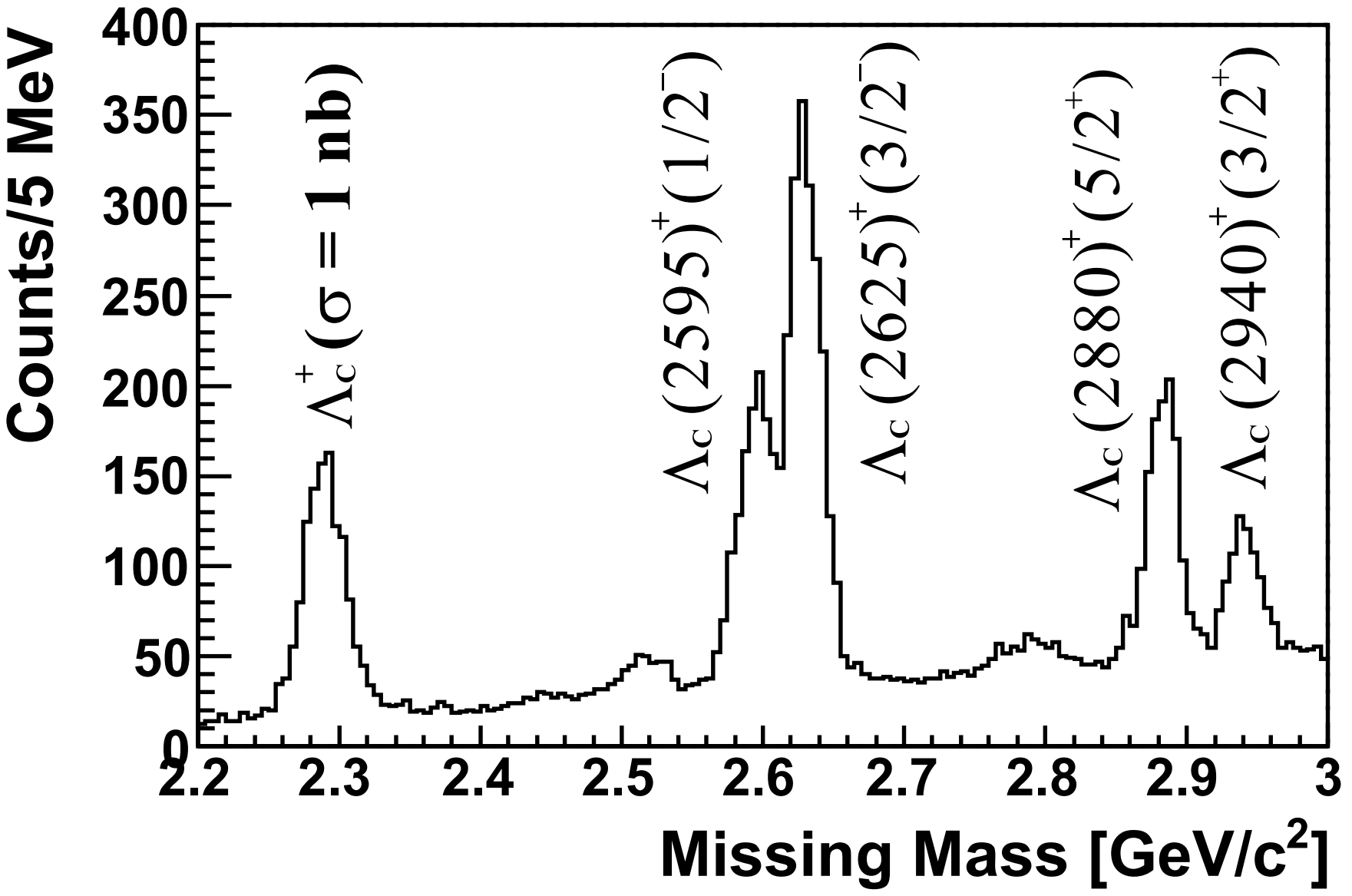

Figure 3: The expected missing mass spectra simulated with the known charmed baryon states. The production rate from the theoretical calculation [3] in the case of $\sigma(\Lambda_c^+)$ =1 nb are used for the simulation. The $\lambda$-mode doublet states, such as $\Lambda_c(2595)(\frac{1}{2}^-)$ and $\Lambda_c(2625)(\frac{3}{2}^-)$ and excited states with $\Lambda_c(2880)(\frac{5}{2}^+)$ and $\Lambda_c(2940)(\frac{3}{2}^+)$ are enhanced. The spin/parity of $\Lambda_c(2940)$ has not been determined yet. In this simulation, it is assumed to be $J_P = 3/2^+$.

Table 1: The relative production rates calculated by Ref. [3]. The production rates are normalized by the cross section of the ground state, $\Lambda_c^+(2286)$.

| $J^P$ | $\Lambda_c(\frac{1}{2}^+)$ | $\Sigma_c(\frac{1}{2}^+)$ | $\Sigma_c(\frac{3}{2}^+)$ | $\Lambda_c(\frac{1}{2}^-)$ | $\Lambda_c(\frac{3}{2}^-)$ | $\Sigma_c(\frac{3}{2}^-)$ | $\Lambda_c(\frac{5}{2}^+)$ | $\Lambda_c(\frac{3}{2}^+)$ |
|---|---|---|---|---|---|---|---|---|
| Mass [MeV/$c^2$] | 2286 | 2455 | 2520 | 2595 | 2625 | 2820 | 2880 | 2940 |
| Production rate | 1 | 0.03 | 0.17 | 0.93 | 1.75 | 0.21 | 0.86 | 0.49 |

The other approaches to reveal those excitation modes is to measure decay branching ratios as shown in Fig. 4. In the case of $\rho$ mode, the rotating diquark decays by emitting a pion. Then, the charmed baryon goes to a light-meson and a charmed baryon pair, such as the $\Lambda_c^{*+} \to \pi + \Sigma_c$ mode. On the other hand, in the $\lambda$ mode case, the charmed baryon decays to a charmed meson and a light nucleon pair, such as the $\Lambda_c^{*+} \to N + D$ mode. Difference of the decay branching ratios is expected for the decay property which provides us information of the excitation modes. The decay branching ratio of the $\rho$ mode is expected to be larger fraction ($\Gamma_{\pi\Sigma_c} > \Gamma_{ND}$) of the $\Lambda_c^{*+} \to \pi + \Sigma_c$ mode than that of $\Lambda_c^{*+} \to N + D$. The $\lambda$ mode decay is expected to be the opposite fraction ($\Gamma_{\pi\Sigma_c} < \Gamma_{ND}$) of the decay branching ratio compared with the $\rho$ mode. In particular, the absolute values of those decay modes can be measured by using the missing mass technique. The decay measurements give us the compliment information with the collider experiments such as Belle II and LHCb.

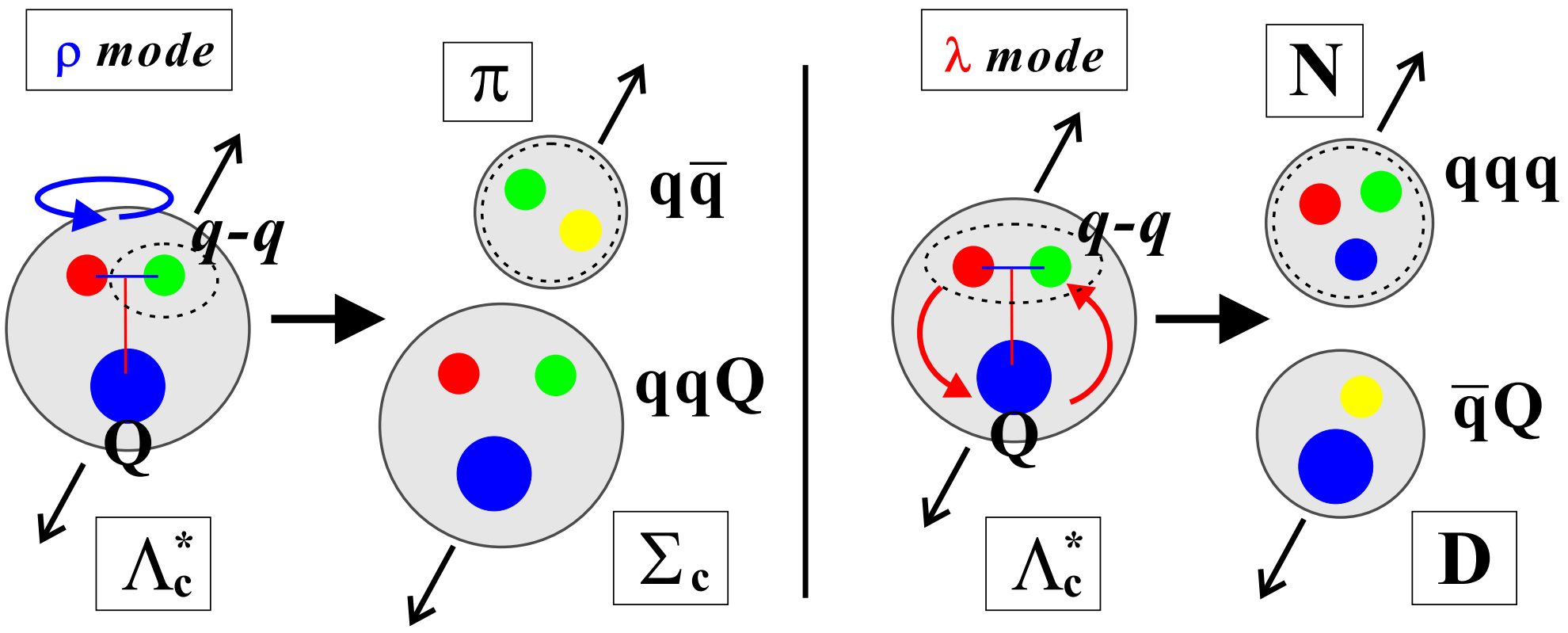

Figure 4: Brief image of the decay properties of the $\rho$- and $\lambda$-mode excitation modes. The decay branching ratio of the $\rho$ mode is expected to be larger fraction of the $\Lambda_c^{*+} \to \pi + \Sigma_c$ mode than that of the $\Lambda_c^{*+} \to N + D$ ($\Gamma_{\pi\Sigma_c} > \Gamma_{ND}$) mode. The $\lambda$ mode decay is expected to be the opposite fraction of the decay branching ratio compared with the $\rho$ mode ($\Gamma_{\pi\Sigma_c} < \Gamma_{ND}$).

## 2 Experiment at the J-PARC High-Momentum Beam Line

The purposes of experiment are to investigate the excited states of the charmed baryons [6]. The goal is to reveal the diquark correlation which is expected to be an essential degree of freedom to describe hadrons. Properties of charmed baryons are measured by the missing mass spectroscopy method by which all the excited states can be observed independent of the final state. The measurement of the production rate which strongly related to the spin/isospin configuration of charmed baryons can provide the unique information of the diquark correlation. In addition, the analysis of the decay particles and its decay chain which decays to the known states, the absolute decay branching ratio and the spin/parity of the excited states could also be determined. The systematic measurement of the excitation energy, the production cross section and the decay properties of the charmed baryon states can be performed.

The experiment will be performed at the high-momentum beam line which is being constructed until the end of 2019 in the J-PARC hadron experimental facility [7]. The high-momentum beam line will be upgraded to unseparated secondary beams with the high-intensity of more than $10^7$ /spill and the momentum of up to 20 GeV/$c$. The momentum resolution of the secondary beams is expected to be $\sim$0.1% by using the momentum dispersive optical method. The high-momentum secondary beams with both high-intensity and high-resolution can be delivered to the experimental area. For the charmed baryon production, the $\pi^- \, p \to Y_c^{*+} \, D^{*-}$ reaction is used with the beam momentum of 20 GeV/$c$. The decay chain of the $D^{*-}$ meson, $D^{*-} \to \bar{D}^0 \, \pi^-$ (branching ratio of 67.7%) and $\bar{D}^0 \to K^+ \, \pi^-$ (branching ratio of 3.88%), is detected for measuring missing masses of produced charmed baryons. $K^+$ and $\pi^-$ of 2–16 GeV/$c$ from the $\bar{D}^0$ decay and $\pi^-$ of 0.5–1.7 GeV/$c$ from the $D^{*-}$ decay are detected by a spectrometer. The decay measurement can also be performed by detecting decay products from the produced charmed baryon, $Y_c^{*+}$, such as the $Y_c^{*+} \to \Sigma_c^{*++,0}\ \pi^{-,+}$ and $Y_c^{*+} \to p\ D^0$ channels. The recoil momentum of $Y_c^{*+}$ is measured by the missing mass method so that the mass of the decay products ($\Sigma_c^{*++,0}$ or $D^0$) can be obtained as a missing mass by only detecting the emitted pion and proton with the momentum of 0.2−2.0 GeV/$c$.

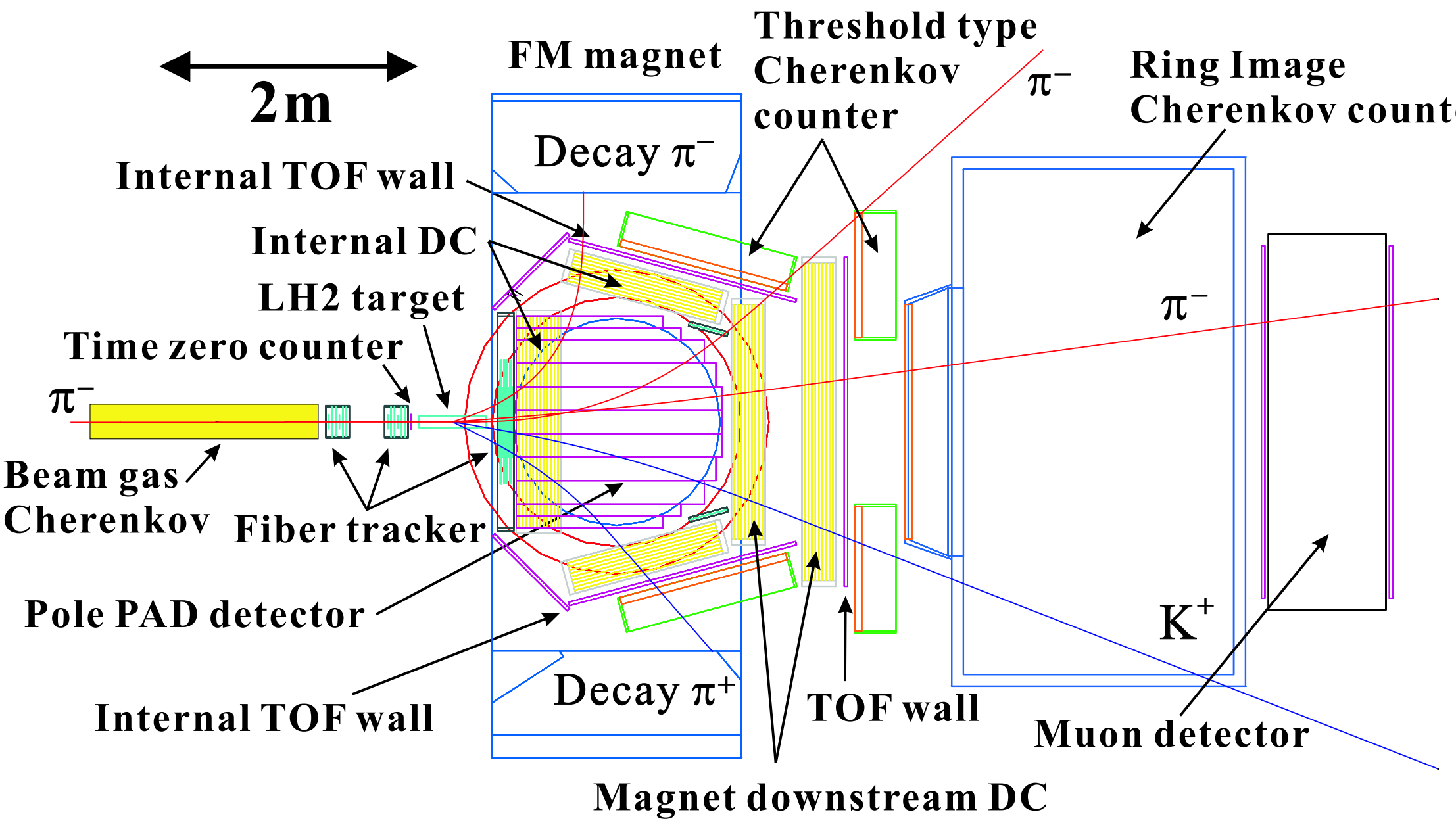

Figure 5: The schematic view of the charmed baryon spectrometer.

Figure 5 shows a conceptual design of the charmed baryon spectrometer (E50 spectrometer) [8]. In the case of the fixed target experiment with the high-momentum beam, all the generated particles, not only the scattered high-momentum particles from the $D^{*-}$ decay but also the decay products from the produced $Y_c^{*+}$, are scattered to the forward direction. Therefore, the dipole magnet system which commonly measures both the particles from the $D^{*-}$ decay for the missing mass method and the decay products from the produced $Y_c^{*+}$ for the decay measurement is used for the charmed baryon spectrometer.

Since the production cross section of charmed baryons is estimated to be $10^{-4}$ smaller than that of the strangeness production (10−100 $\mu$b of the $\pi^-\ p \to Y^*\ K^*$ reaction) [3], the charmed baryon production cross section of 1 nb was assumed for the experimental design. In order to obtain the production yield, the beam intensity of $6.0{\times}10^7$/spill is irradiated on the liquid hydrogen target. The high-rate detectors such as scintillating fiber trackers will be installed for the measurements of the beam momentum and profiles at focal plane and at the upstream of the experimental target, respectively. The timing of beam is measured by the time zero counter which consists of 1-mm fine-segmented acrylic Čerenkov radiator having a time resolution of 50 ps(rms). For the beam particle identification, the high-rate Ring-Image Čerenkov counter is planned to be used. The progress of development of the high-rate detectors shows better-than-expected performances so that the capability of using secondary beams by the beam detectors is expected to be $1.8{\times}10^8$ /spill.

The $D^{*-}$ meson is measured by the forward detector system, the scintillating fiber trackers at the downstream of the target, and the drift chambers and the Ring-Image Čerenkov counter for detecting the high-momentum $K^+$ and $\pi^-$ from $\bar{D}^0$ at the exit of the magnet. For measuring the slow $\pi^-$ from $D^{*-}$ at the forward region and the decay particles from $Y_c^*$ with the wide angular coverage, it is necessary to detect the particles inside of the magnet. The tracking detectors which have the large angular acceptance are installed at the downstream of the target. The horizontal and vertical directions can be covered by using the detectors installed around the magnet pole and the thin thickness detectors on the magnet pole face are installed, respectively. The acceptance for detecting the $D^{*-}$ decay was estimated to be 50−60% for the excitation states up to 1 GeV by assuming the angular distribution from the dominance of the t-channel production process. The momentum resolution of 0.2% was obtained at 5 GeV/$c$ so that the invariant mass resolution for reconstructing the $\bar{D}^0$ and $D^{*-}$ are estimated to be 5.5 MeV and 0.6 MeV, respectively. The missing mass resolution of excited states above 2.8 GeV/$c^2$ are estimated to be ∼10 MeV. For the decay measurement, both the polar and azimuthal angles are completely covered more than $\cos\theta_{CM} = -0.9 \sim -0.5$ for the $\Lambda_c(2940)^+ \to \Sigma_c(2455)^{++,0}\,\pi^{-,+}$ decay modes.

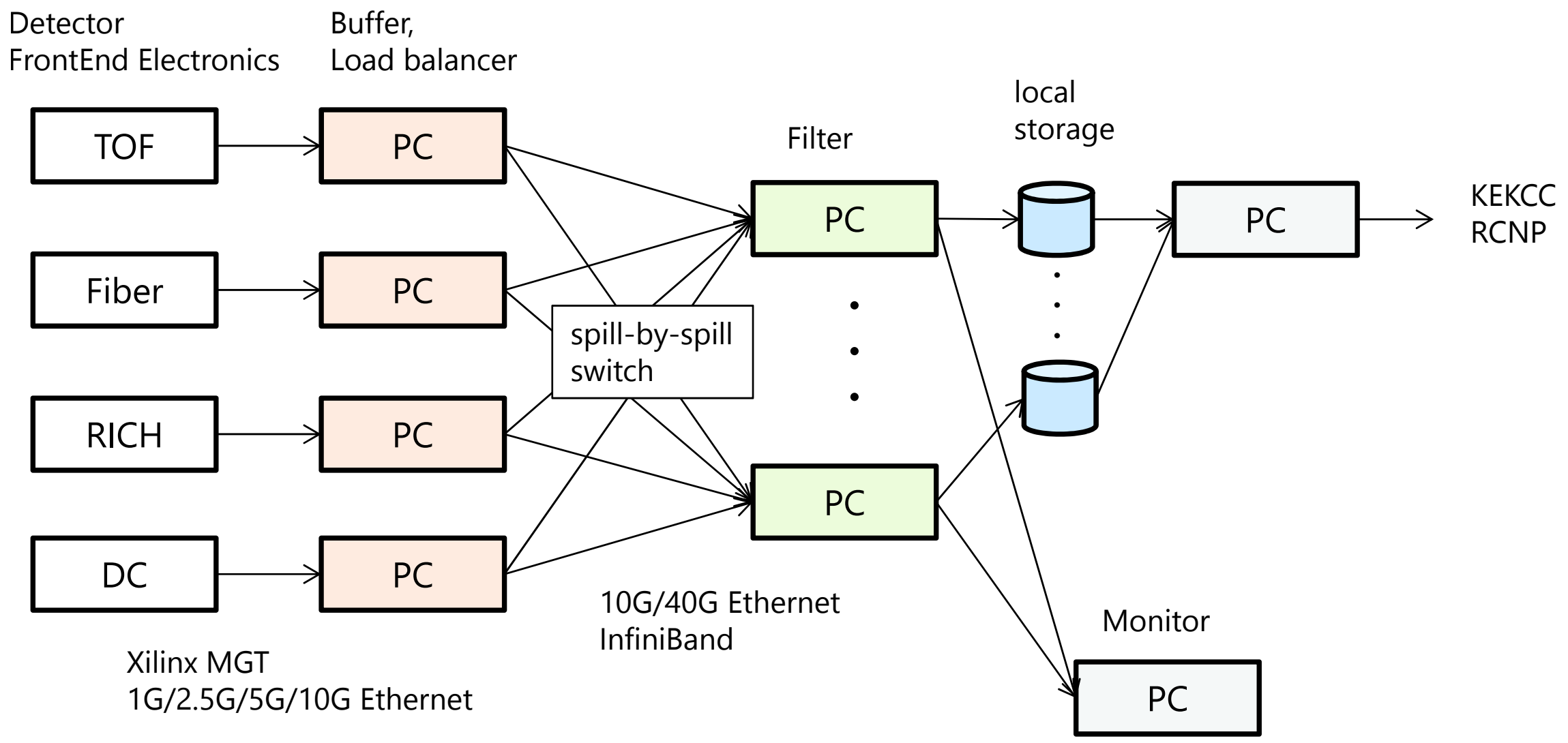

Figure 6: A conceptual view of DAQ system.

For the E50 spectrometer system, we adopt a streaming DAQ system whose main purpose is a trigger-less one. Because it is quite difficult to make a trigger by using information of the detected several particles with a momentum analysis from FPGAs, the trigger-less DAQ system is quite suitable and flexible to recorded events which include a lot of reaction channels. The DAQ system is one of the most essential components of the spectrometer system. By using the trigger-less DAQ system, we can take data of many reaction channels simultaneously. Therefore, it makes the E50 spectrometer to be the real multi-purpose purpose system.

Figure 6 illustrates a conceptual schematic of the DAQ system. Detector signals are digitized on front-end electronics modules placed near the detectors with a pipelined data transfer. A raw data size of ∼50 GB/spill is roughly expected for the beam rate of 60 M/spill. The data of front-end modules are transferred by a self-trigger (or a periodic trigger) with a time stump via FPGA's build-in multi-gigabit transceivers and/or 1G/2.5G/5G/10G Ethernet to buffer PCs. The PC nodes at the first stage plays as buffers and load balancers, which store the data and switch a destination of the data spill by spill. The PC nodes at the second stage plays as event filters. Each filter node performs the cellular automaton tracking, analyzes the RICH data (clustering and/or image reconstruction), scaler counting and writes the data into the local storage device. A part of data are visualized on monitoring PCs. Since the price of several tens to a hundred GB memories becomes not so expensive nowadays, each node can be equipped with an enough memory to store the whole data of one spill. The recorded data are forwarded to the KEKCC storage and the computing farm at RCNP. The filter nodes, local storages and monitoring PCs are placed at the counting room, whereas the buffer nodes are located at the experimental area. The experimental area and the counting room are connected with high speed optical links (10G/40G Ethernet, InfiniBand, etc.) in order to cope with ~50 GB/spill data rate, which corresponds to ~90 Gbps after the derandomization.

# 3 Physics with the E50 spectrometer

Owing to the large forward acceptance and the trigger-less DAQ system of the E50 spectrometer, the data of a lot reaction channels can be taken simultaneously. Since the data taking system having a flexibility, we can easily access to the physics channels which we are interested in. In addition, the secondary beam in the High-momentum beam line is unseparated so that we can use not only $\pi$ but also $K$ and anti-proton by using beam particle detector simultaneously. High-intensity beam and large forward acceptance also give us opportunities to produce a huge number of hyperon events. By measuring a missing mass of hyperon which is produced in the experimental target, we can use the hyperon production events as hyperon beams. The experiments using hyperon beams can also be performed by the E50 spectrometer. The following list shows examples of physics interests with the E50 spectrometer system.

- Hyperon spectroscopy experiment via the $\pi^- \ p \to Y^* \ K^*$ reaction to investigate the diquark correlation in the lighter quark system (diquark and strange quark system)
- Cascade baryon ($\Xi$) spectroscopy experiment [9] with $K^-$ beam for investigating the excited resonances which have not well studied yet and to study the diquark correlation in the system having one light quark and two heavier quarks
- Nucleon structure study from the Drell-Yan process with a high-intensity $\pi$ beam by detecting $\mu^{\pm}$ events by the muon detector installed in the most downstream of the spectrometer
- Omega baryon ($\Omega$) spectroscopy for pilot studies of the experiments which are planned at the K10 beam line
- Hyperon-Nucleon scattering experiment via the $\pi^- \ p \to \Lambda \ K^{*0}$ reaction for understanding properties of the YN interaction at the higher angular momentum regions (mainly for the p-wave region)
- Hyperon beam experiment for investigating the hadron structure from the production processes of ordinary and exotic states via the hyperon scattering method
- Search for pentaquark including $c\bar{c}$ component ($P_c$ state) from detecting its decay events such as $J/\psi + n$ and $Y_c + D^{(*)}$ channels
- Pilot studies of hadron in the nuclear medium via the $D\bar{D}$ production and $c\bar{c}$ mesons using the anti-proton beam

# 4 Summary

## 4.1 Physics motivation with its impact and expected result

The charmed baryon spectroscopy gives us opportunities to understand a proper degree of freedom of the hadron structure. For describing the charmed baryon structure, it is essential for the diquark correlation which is generated by the isolated two quarks from the charm quark. The excitation modes from the diquark correlation will be observed in the structure of the excited spectrum and the decay modes of excited states. The production cross section via the $\pi^- \; p \to Y_c^{*+} \; D^{*-}$ reaction also gives us information of the spin/isospin configuration of the charmed baryon structure. For investigating charmed baryons, we proposed a spectroscopy experiment to observe and investigate excited states of charmed baryons at the J-PARC high-momentum beam line. The systematic measurement of the excitation energy, the production rate and the decay products of charmed baryons will be performed by the spectroscopy experiment. From the experimental result of the production rates which show the the collective motion between diquark and charm quark in the charmed baryons, the diquark correlation which is expectedly an essential degree of freedom to describe the hadron structure can be established.

## 4.2 What kind of beam and equipments are necessary ?

The experiment is proposed to perform at the J-PARC high-momentum beam line by using secondary beams. The experiment can be performed at both the present high-momentum beam line by using secondary beams and the extended high-momentum beam line. For the measurements of charmed baryons, the charmed baryon spectrometer (E50 spectrometer) is used.

## 4.3 Particle, Energy(momentum), Intensity and Experimental setup

In the experiment, the $\pi^- \; p \to Y_c^{*+} \; D^{*-}$ reaction is used. The beam $\pi^-$ momentum and intensity are 20 GeV/$c$ and $6.0\times10^7$ /spill, respectively.

## 4.4 Expected duration of the beam time (yield estimation)

By assuming the cross section of 1 nb for the ground state ($\Lambda_c^+$), the yield of ~2000 counts is expected with beam time of 100 days ($8.64\times10^{13}$ $\pi^-$ on the target with 6-second duration). The yield calculation is as follows,

$$(1.0 \times 10^{-9} \times 10^{-24}) \times (4.0 \times 6.02 \times 10^{23}) \times (8.64 \times 10^{13}) \times (0.677 \times 0.0388) \times (0.7 \times 0.55) \sim 2000,$$

where thickness of the liquid hydrogen target is 4.0 g/cm$^2$, branching ratios of $D^{*-}$ and $D^0$ are 0.677 and 0.0388, respectively, spectrometer acceptance and detection efficiency are 0.70 and 0.55, respectively.

# References

[1] T. Nakano *et al.*, Phys. Rev. Lett. **91**, 012002 (2003).

[2] S. K. Choi *et al.*, Phys. Rev. Lett. **91** 262001 (2003).

[3] S. H. Kim, A. Hosaka, H. C. Kim, H. Noumi, K. Shirotori, Prog. Theor. Exp. Phys. 103D01 (2014).

[4] Y. Nara *et al.*, Phys. Rev. C **61** 024901 (2000).

[5] T. Sjöstrand, S. Mrenna and P. Skands, JHEP05 (2006) 026, Comput. Phys. Comm. **178** (2008) 852.

[6] H. Noumi *et al.*, J-PARC E50 Proposal (2012).

[7] K. H. Tanaka *et al.*, Nucl. Phys. A **835** 81 (2010).

[8] K. Shirotori *et al.*, PoS(Hadron 2013) 130 (2013).

[9] M. Naruki and K. Shirotori, LOI: $\Xi$ Baryon Spectroscopy with High-momentum Secondary Beam (2014).

Separated secondary beam lines

**Hitoshi Takahashi**
Institute of Particle and Nuclear Studies, KEK

Current design of the K10 beam line is presented. It aims at providing well separated secondary beams with the higher momentum than existing beam lines. There are two options for the method of the particle separation; One uses electrostatic separators and the other RF separators. In addition, a possibility to upgrade the high-p beam line to a RF-separated secondary beam line is discussed.

# 1 K10 beam line

The K10 beam line is designed to serve high-momentum separated secondary beams. Currently two options are planed for the particle separation; an electrostatic(ES)-separator option and a RF-separator option. The ES-separator option can be applied up to $4 \sim 6$ GeV/$c$, while the RF-separator option is suitable for the higher momentum. By making a common design at the front-end section of both options of the beam line, we can switch the separation method without accessing high-radiation area. The production angle of secondary beams is chosen to be 3 degrees, which is smaller than those of the K1.8 and K1.1 beam lines. This is because the production cross-sections of $K^-$ and $\bar{p}$ at 3 degrees are about 5 times larger than those at 6 degrees, according to the empirical formula by Sanford and Wang[3].

## 1.1 ES-separator option

The layout of the ES-separator option of the K10 beam line is presented in Fig.1, and the beam envelope calculated with the TRANSPORT code[1] is shown in Fig.2. The total length of the beam line is 82.8 m.

The beam line consists of three sections; the front-end section, the separation section, and the analyzing section. In the front-end section, secondary beams generated at the production target are extracted from the primary beam line, and focused vertically at the intermediate (IF) slit to reduce so-called "cloud $\pi$". The optics are also tuned to make beams almost achromatic at the IF slit. The separation of the secondary particles is performed by using three ES separators. Each of the ES separators have the effective length of 9 m, the electrode gap of 10 cm, and the electrostatic field of 75 kV/cm. In the separation section, parallel beams are made in both the horizontal and vertical directions to pass through such long ES separators, and then the beams are focused vertically at the mass slit (MS). They are analyzed with the beam spectrometer in the analyzing section, and finally focused both horizontally and vertically at an experimental target. In addition, there are two horizontal focal points in the beam line (HF1, HF2), where slits are located to determine the momentum bite and to increase the beam purity.

The beam intensity and purity of the beam line was estimated by using a lay-tracing code DecayTURTLE[2], assuming the beam loss of 25 kW at the production target and the spill repetition of 5.52 s. The result is summarized in Table 1.

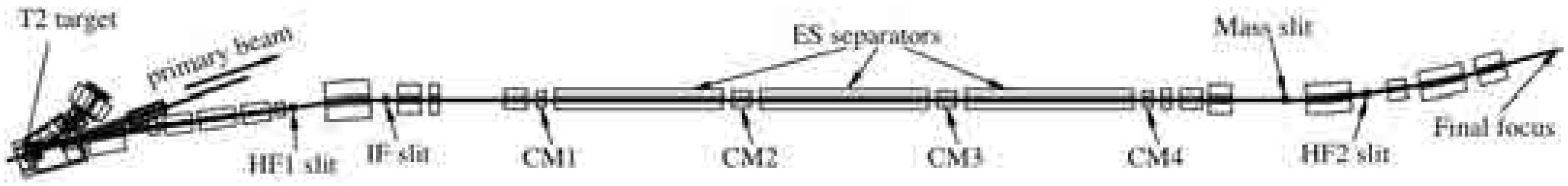

Figure 1: Layout of the ES option of the K10 beam line.

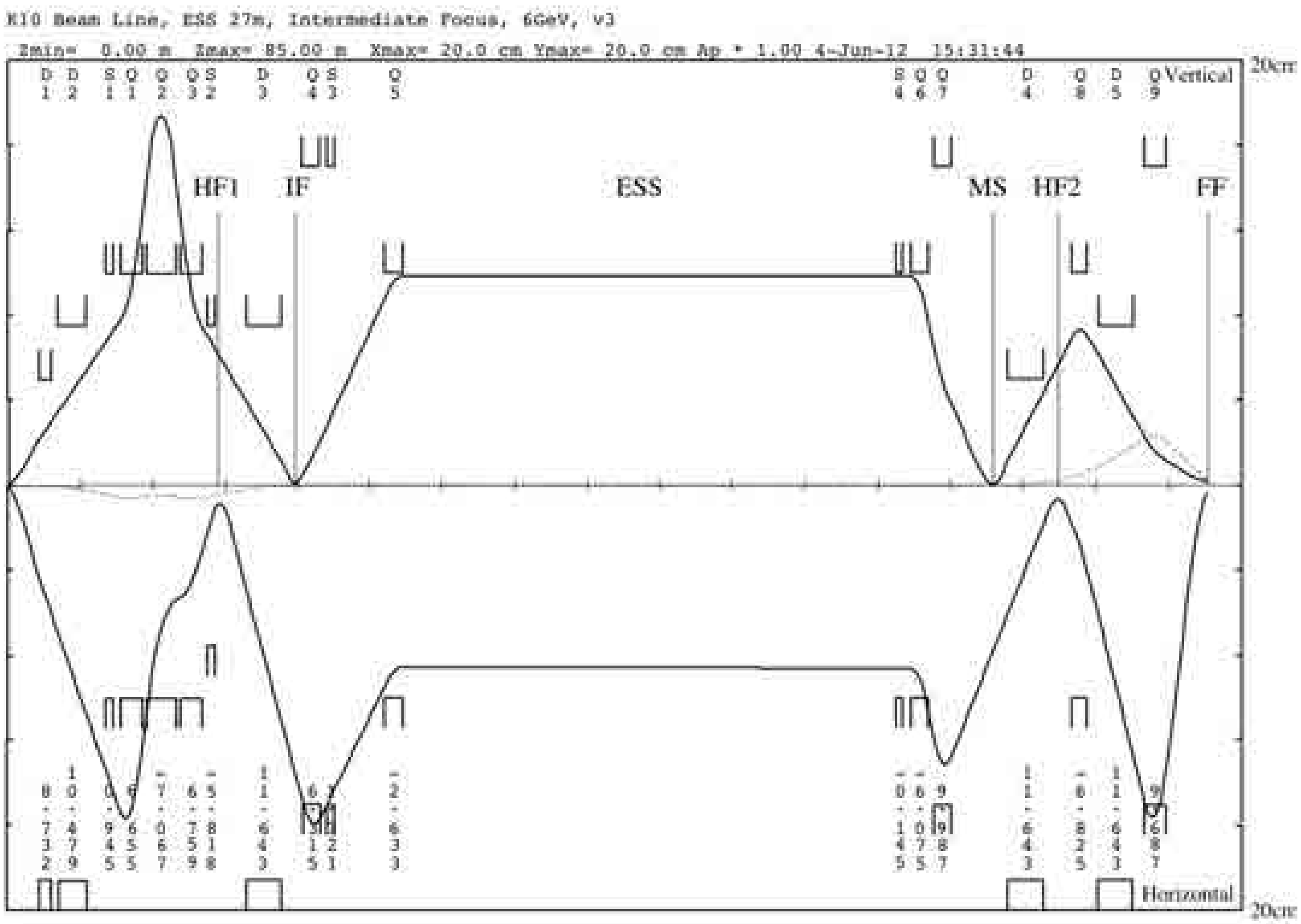

Figure 2: First-order beam envelope of the ES option of the K10 beam line. The upper and lower show the vertical and horizontal directions, respectively. The dash line indicate the horizontal dispersion.

Table 1: The expected intensity and purity of the ES option of the K10 beam line. The beam loss of 25 kW at the production target and the spill repetition of 5.52 s were assumed. The production cross-section was calculated by using Sanford and Wang formula. Decay muons and so-called "cloud-$\pi$" were not included.

| | 4 GeV/$c$ $K^-$ | 4 GeV/$c$ $\bar{p}$ | 6 GeV/$c$ $\bar{p}$ |
|---|---|---|---|
| acceptance [msr-%] | 0.33 | 1.2 | 0.55 |
| intensity [/spill] | $1.7 \times 10^6$ | $1.6 \times 10^7$ | $7.8 \times 10^6$ |
| purity ($K^- : \pi^-$ or $\bar{p} : \pi^-$) | 1.1:1 | 81:1 | 1:3.4 |

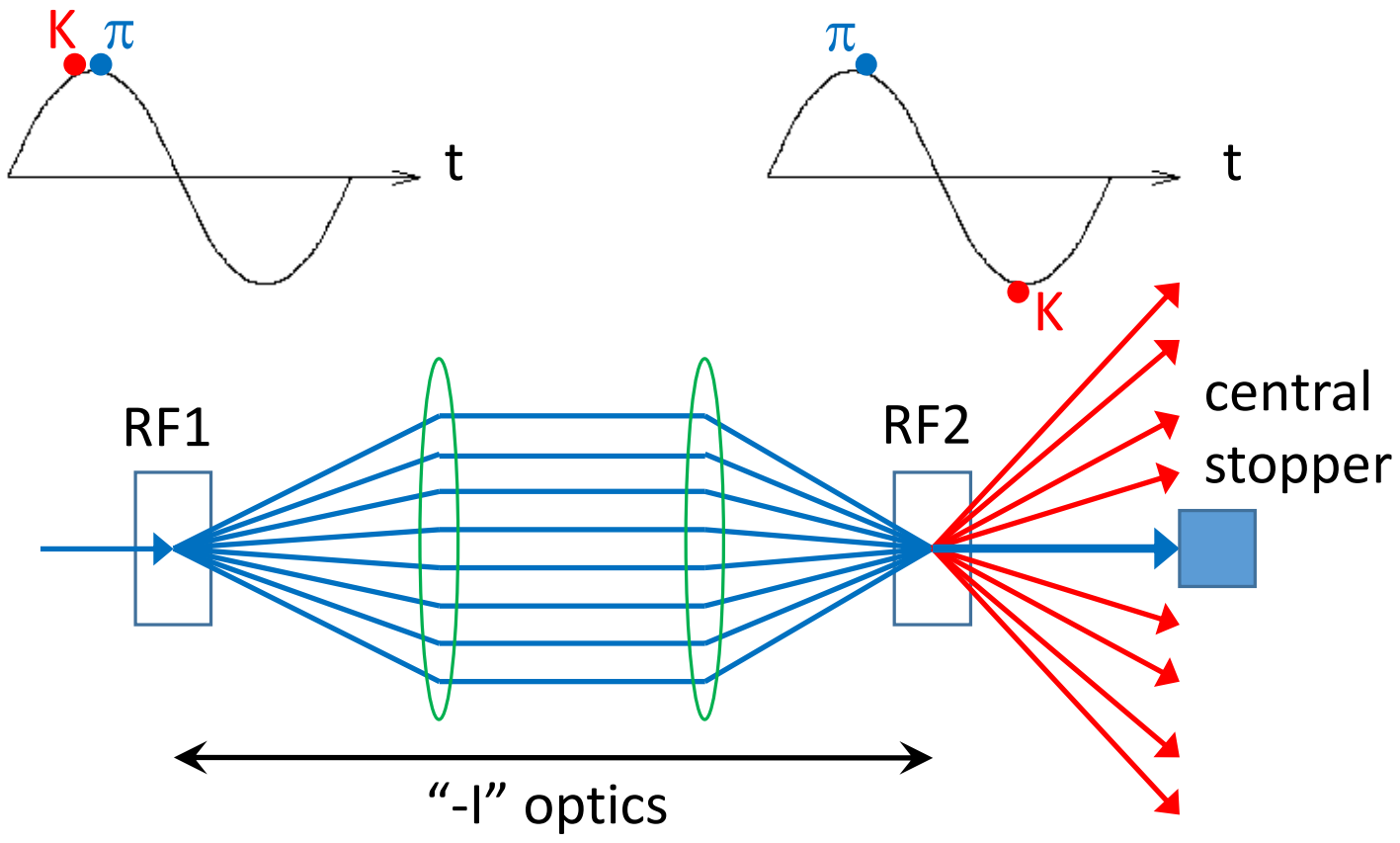

Figure 3: The principle of the particle separation using two RF cavities.

## 1.2 RF-separator option

Since the separation by an ES separator is proportional to $1/p^3$, the separation is much difficult for the momentum higher than 4 GeV/$c$ ($K^-$) or 6 GeV/$c$ ($\bar{p}$). Therefore, we need RF separators instead for such a momentum region.

The principle of the particle separation by using RF cavities is schematically shown in Fig.3. In this method, two RF cavities (RF1 and RF2) are located in the beam line, and the optics between the two cavities are set so that the transport matrix is equal to "$-I$". If the RF amplitude of the two cavities are the same, the sum of the beam deflection from the two cavities is

$$D = -A\sin(\omega t) + A\sin(\omega t + \Delta\phi) \quad (1)$$

$$= 2A\sin\frac{\Delta\phi}{2}\cos\left(\omega t + \frac{\Delta\phi}{2}\right), \quad (2)$$

where $\omega t$ is the phase at the first cavity, and $\Delta\phi$ the phase difference between the two cavities. The minus sign in the first line comes from the "$-I$" optics between the two cavities. The amplitude $A$ is given by

$$A = \frac{eEl}{pc\beta}, \quad (3)$$

where $e$, $p$, and $\beta$ are the charge, momentum, and velocity of the particle, respectively, and $E$ and $l$ denote the field gradient and the effective length of the cavity, respectively. When the phase at the second cavity is set to be same as that of the first cavity for an unwanted particle, namely $\pi$, the deflection of the particle is canceled in whichever phase it passes the first cavity, and it is absorbed with a central stopper downstream. On the other hand, the phase of the second cavity for particles with the other mass and velocity ($K^-$ or $\bar{p}$) differs by

$$\Delta\phi_w^u = \frac{2\pi fL}{c}\left(\frac{1}{\beta_w} - \frac{1}{\beta_u}\right) \quad (4)$$

$$\sim \frac{\pi fL}{c}\frac{m_w^2 - m_u^2}{p^2c^2}, \quad (5)$$

they are deflected by $2A\sin\frac{\Delta\phi_w^u}{2}$ in maximum depending on the phase at the first cavity, and pass outside of the central stopper. Here, $f$ is the RF frequency, $L$ the distance between the two cavities, $p$ the momentum, and $\beta_w/\beta_u$ and $m_w/m_u$ are the velocities and masses of the wanted/unwanted particles, respectively.

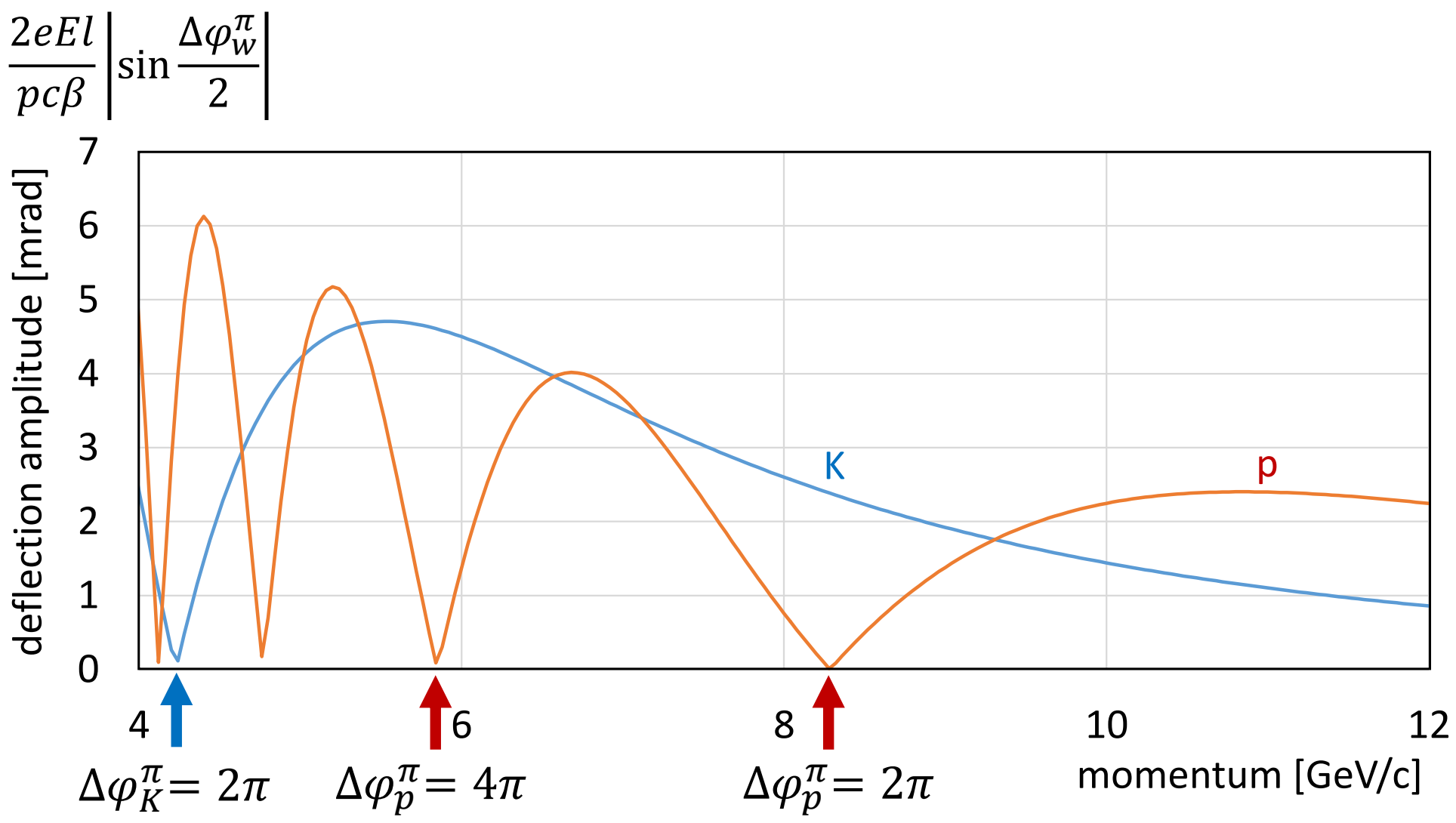

Figure 4: Momentum dependence of the deflection amplitude for $K^-$ (blue) and $\bar{p}$ (red) in the case that the RF phase of the two cavities are same for $\pi^-$. Since the deflection for $\pi^-$ is always canceled, usable is the momentum region in which the deflection for $K^-$ or $\bar{p}$ is large.

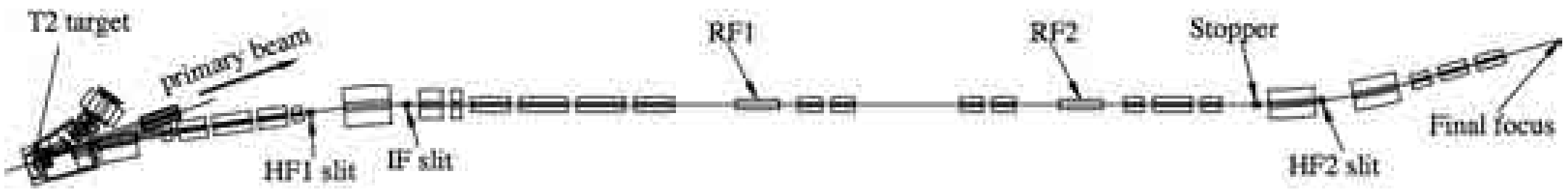

Figure 5: Layout of the RF option of the K10 beam line.

For example, assuming $\pi^-$ as unwanted particle, the momentum dependence of the deflection amplitude for $K^-$ and $\bar{p}$ in the case of $f = 2.857$ GHz and $L = 16.8$ m is plotted in Fig.4. The deflection for the wanted particle is also canceled in the momentum range corresponding to

$$\Delta\phi_w^u = 2n\pi \ \ (n = 1, 2, 3, ...), \tag{6}$$

whereas the deflection get maximum in the range satisfying

$$\Delta\phi_w^u = (2n-1)\pi \ \ (n = 1, 2, 3, ...). \tag{7}$$

The layout and beam envelope of the RF option of the K10 beam line are shown in Fig.5 and Fig.6, respectively. The total length of the beam line is 80.8 m, which is almost same as that of the ES option. The RF frequency and amplitude are set to 2.857 GHz and 6 MV/m, respectively. The effective length of the cavities is 2.25 m and the distance between the cavities is 16.8 m.

The front-end section from the production target to the IF is completely same as that in the ES option. In the separation section, the optics is tuned to obtain a parallel and narrow beam at the RF cavity, and the transport matrix of "$-I$" is realized between the cavities by using four quadrupoles with same field gradient. After analyzed with a beam spectrometer, the beam is focused to an experimental target both in horizontal and vertical directions.

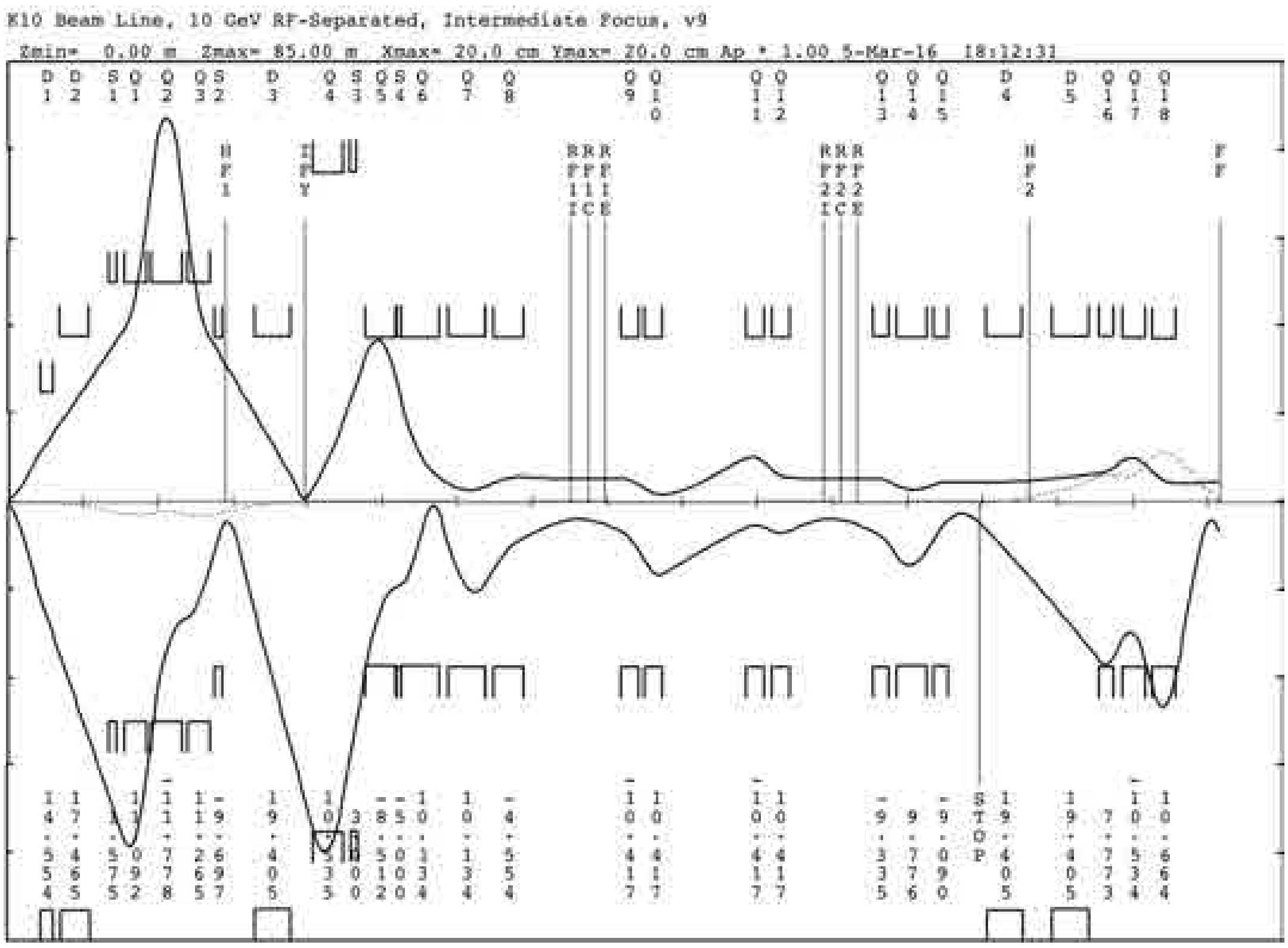

Figure 6: Same as Fig.2 but for the RF option.

Table 2: Same as Table 1 but for the RF option.

| | acceptance | intensity | putiry |
|---|---|---|---|
| | $(K^-/\bar{p})$ [msr-%] | $(K^-/\bar{p})$ [/spill] | $\pi^- : K^- : \bar{p}$ |
| 6 GeV/$c$ | 0.50 / 0.14 | $6.9 \times 10^6$ / $2.1 \times 10^5$ | 2.3:1:0.31 |
| 7 GeV/$c$ | 0.50 / 0.51 | $7.6 \times 10^6$ / $6.4 \times 10^6$ | 1.7:1:0.84 |
| 10 GeV/$c$ | 0.24 / 0.42 | $2.5 \times 10^6$ / $1.9 \times 10^6$ | 3.3:1:0.79 |

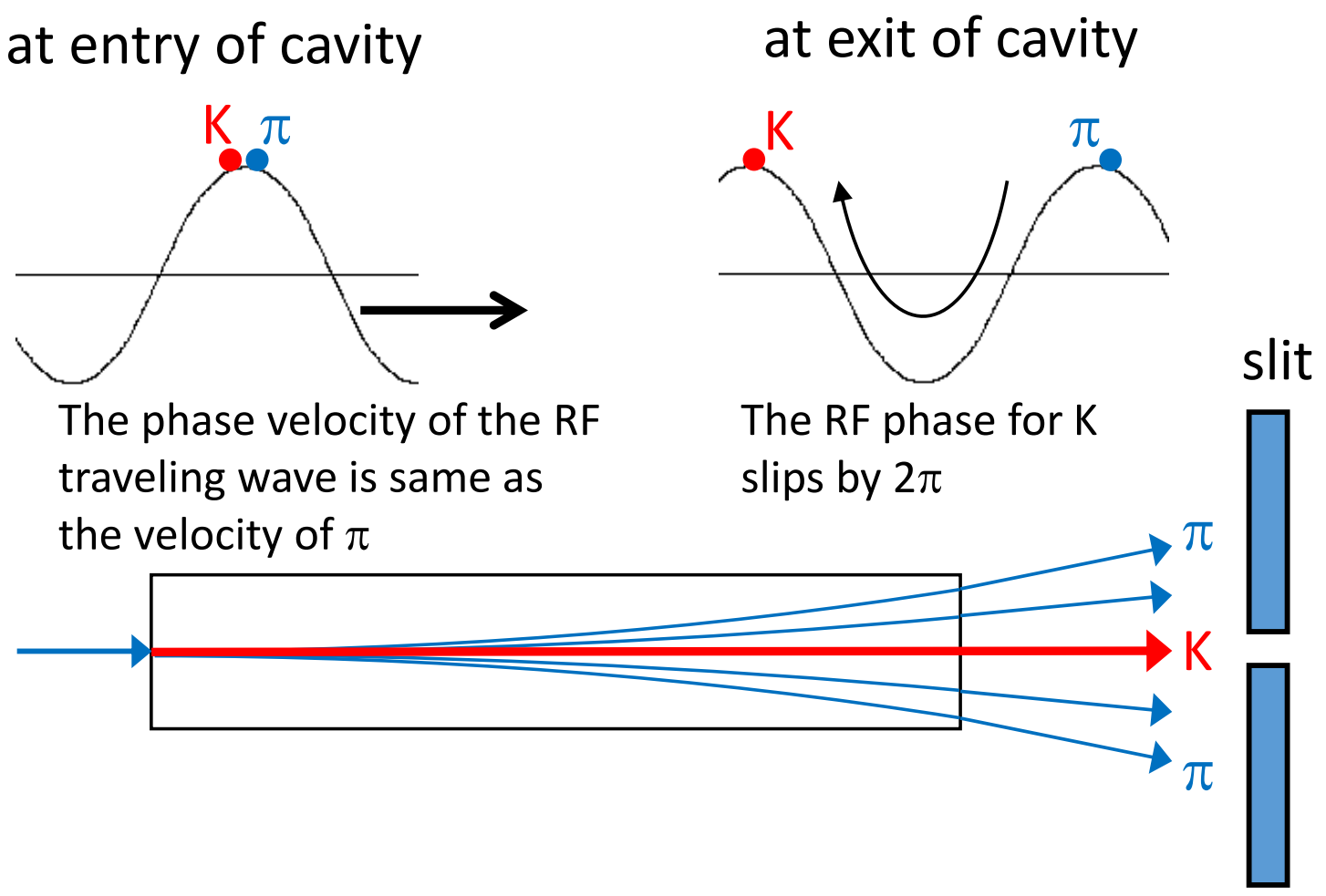

Figure 7: The principle of the particle separation using one RF cavity.

Table 2 lists the result of the beam simulation using DecayTURTLE. The assumption for the beam loss and spill repetition is same as that for the ES option.

## 2 RF-separated high-p-ex beam line

H. Noumi proposed the upgrade of the existing high-p beam line to an unseparated secondary beam line with the maximum momentum of 20 GeV/$c$[5]. Recently, he also proposed its extension to the extended area of the HD hall, which is referred to high-p-ex beam line[6]. The high-p-ex secondary beam line utilizes $\sim 10^8$ Hz of $\pi^-$ beam, while the intensities of $K^-$ and $\bar{p}$ are expected as 2 orders of magnitude lower than that of $\pi^-$. Because of the drop off of the $K^-$ yield at the momentum less than 4 GeV/$c$, it is desirable to separate beams at 5 GeV/$c$ or higher.

Since there is very limited space to locate RF cavities in the current configuration of high-p beam line, it is very difficult to insert two cavities without major rearrangement of the beam line. Therefore, we need the other method for the particle separation.

In the previous section, the phase slip between the beam particles and the RF traveling wave in a cavity was ignored, but this approximation is not satisfied when the beam momentum is low. The phase slip, however, depends on the velocity of the beam particle, that can be applied for particle separation using only one cavity[7].

Fig.7 presents the principle of the separation method using one RF cavity. When the phase velocity of the RF traveling wave is equal to the velocity of an unwanted particle, the particle feels the field gradient with the same phase as that at entry throughout the cavity, that gives the maximum deflection to the particle. On the other hand, for the wanted particle having a different velocity, the phase differs at entry and exit of a cavity. With the specific momentum in which the phase difference is $2\pi$, the deflection of the wanted particle by the cavity is always canceled independently of the phase at entry. Therefore, the wanted particle can be separated by using a slit downstream.

The equation of motion of a particle passing through a RF cavity is given by

$$\gamma m \frac{d^2 y}{dt^2} = eE \sin(\varphi + \Delta\varphi), \tag{8}$$

where $\varphi$ stands for the phase at entry. $\Delta\varphi$ is the phase difference between the particle and the RF wave, and can be written as

$$\Delta\varphi = kz - \omega t \tag{9}$$
$$= \frac{2\pi f}{c}\left(\frac{1}{\beta_\psi} - \frac{1}{\beta_u}\right) z. \tag{10}$$

Here, $k$ and $\omega = 2\pi f$ are the wave number and frequency of the RF, and $\beta_u$ and $\beta_\psi$ denote the velocity of the particle and the RF phase, respectively. By defining

$$\Delta k \equiv \frac{2\pi f}{c}\left(\frac{1}{\beta_\psi} - \frac{1}{\beta_u}\right), \tag{11}$$

Eq.(8) can be expressed as the differential equation with respect to $z$ of

$$\frac{d^2y}{dz^2} = \frac{eE}{\gamma m v^2}\sin(\varphi + \Delta k z). \tag{12}$$

The integration of Eq.(12) over the cavity length $l$ gives

$$\frac{dy}{dz} = \frac{eEl}{pc\beta_u}\sin\left(\varphi + \frac{\Delta k l}{2}\right)\frac{\sin\frac{\Delta k l}{2}}{\frac{\Delta k l}{2}}. \tag{13}$$

The deflection by a RF cavity, therefore, is reduced by a factor of $\sin\frac{\Delta\varphi_\psi^u}{2}/\frac{\Delta\varphi_\psi^u}{2}$, where

$$\Delta\varphi_\psi^u \equiv \Delta k_\psi^u l = \frac{2\pi f l}{c}\left(\frac{1}{\beta_\psi} - \frac{1}{\beta_u}\right). \tag{14}$$

Hence, the amplitude of the deflection is given by

$$\left|\frac{dy}{dz}\right| = \frac{eEl}{pc\beta_u}\left|\frac{\sin\frac{\Delta\varphi_\psi^u}{2}}{\frac{\Delta\varphi_\psi^u}{2}}\right|. \tag{15}$$

Fig.8 is the plot of the momentum dependence of the deflection amplitude for $K^-$ and $\bar{p}$ in the case of $\Delta\varphi_\psi^\pi = 0$, assuming $f = 8.857$ GHz, $E = 6$ MV/m, and $l = 3$ m. The positions of nodes of each curve correspond to the phase difference of $\Delta\varphi_\psi^{K/p} = 2\pi,\ 4\pi, \cdots$. Since the deflection for $\pi^-$ is maximum, the usable momentum range is only discrete and narrow region near these nodes. Therefore, the usability of the beam line is not appropriate.

The other option is setting $\Delta\varphi_\psi^\pi = 2\pi$. In this case, the deflection for $\pi^-$ is always canceled. Since the phase difference for $K^-$ or $\bar{p}$ is given by

$$\Delta\varphi_\psi^{K/p} = 2\pi - \Delta\varphi_\pi^{K/p}, \tag{16}$$

the deflection amplitude for $K^-$ or $\bar{p}$ depends on the momentum as shown in Fig.9.

One more possibility is the choice of $\Delta\varphi_\psi^{K/p} = 2\pi$, that means the deflection for $K^-$ or $\bar{p}$ is always canceled. The phase difference for $\pi^-$ is given by

$$\Delta\varphi_\psi^\pi = 2\pi + \Delta\varphi_\pi^{K/p}, \tag{17}$$

and its deflection amplitude is drawn in Fig.fig:RF1SeparationC. Compared with the previous example, the amplitude is smaller, because the denominator of the factor is larger; $2\pi + \Delta\varphi_\pi^{K/p}$

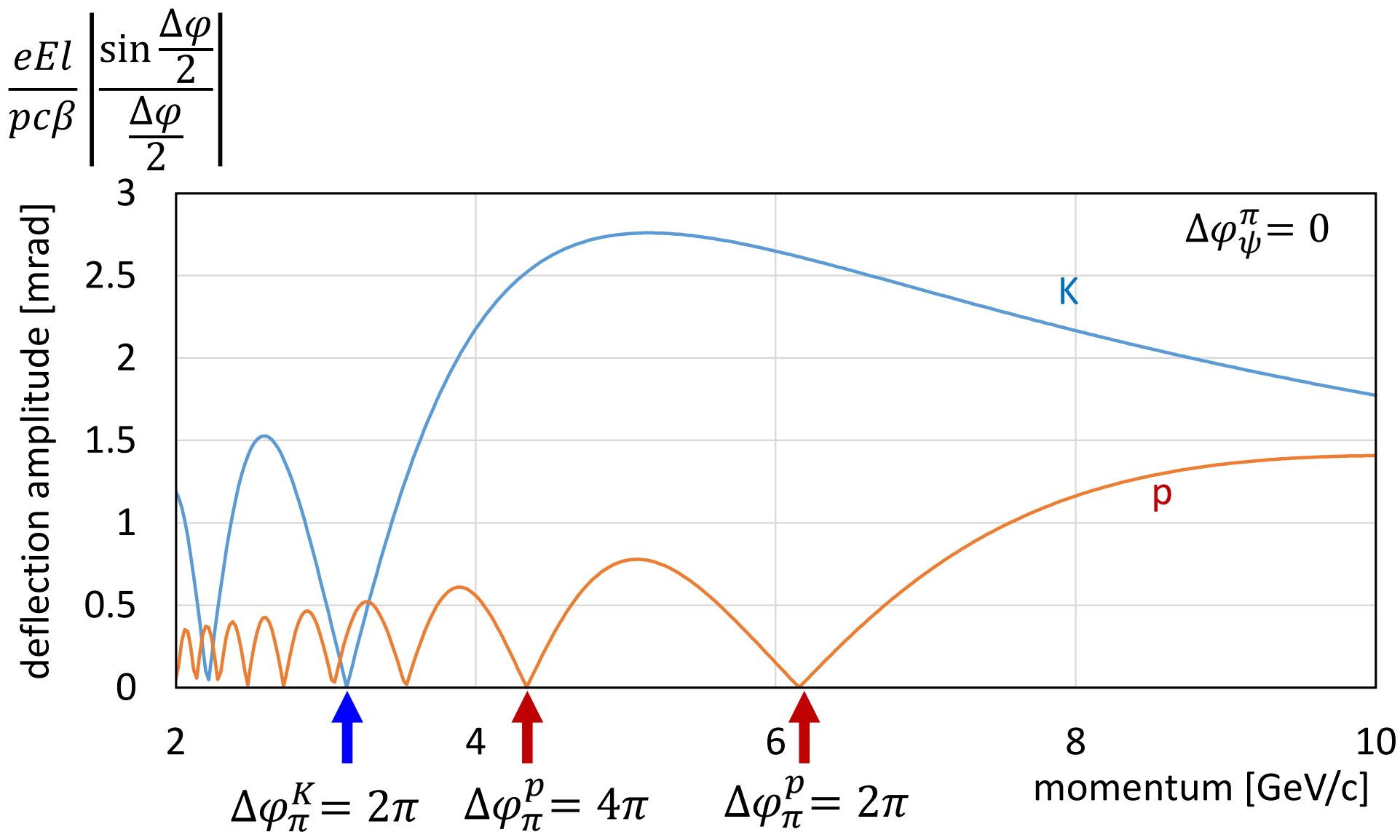

Figure 8: Momentum dependence of the deflection amplitude for $K^-$ (blue) and $\bar{p}$ (red) in the

on for

mall.

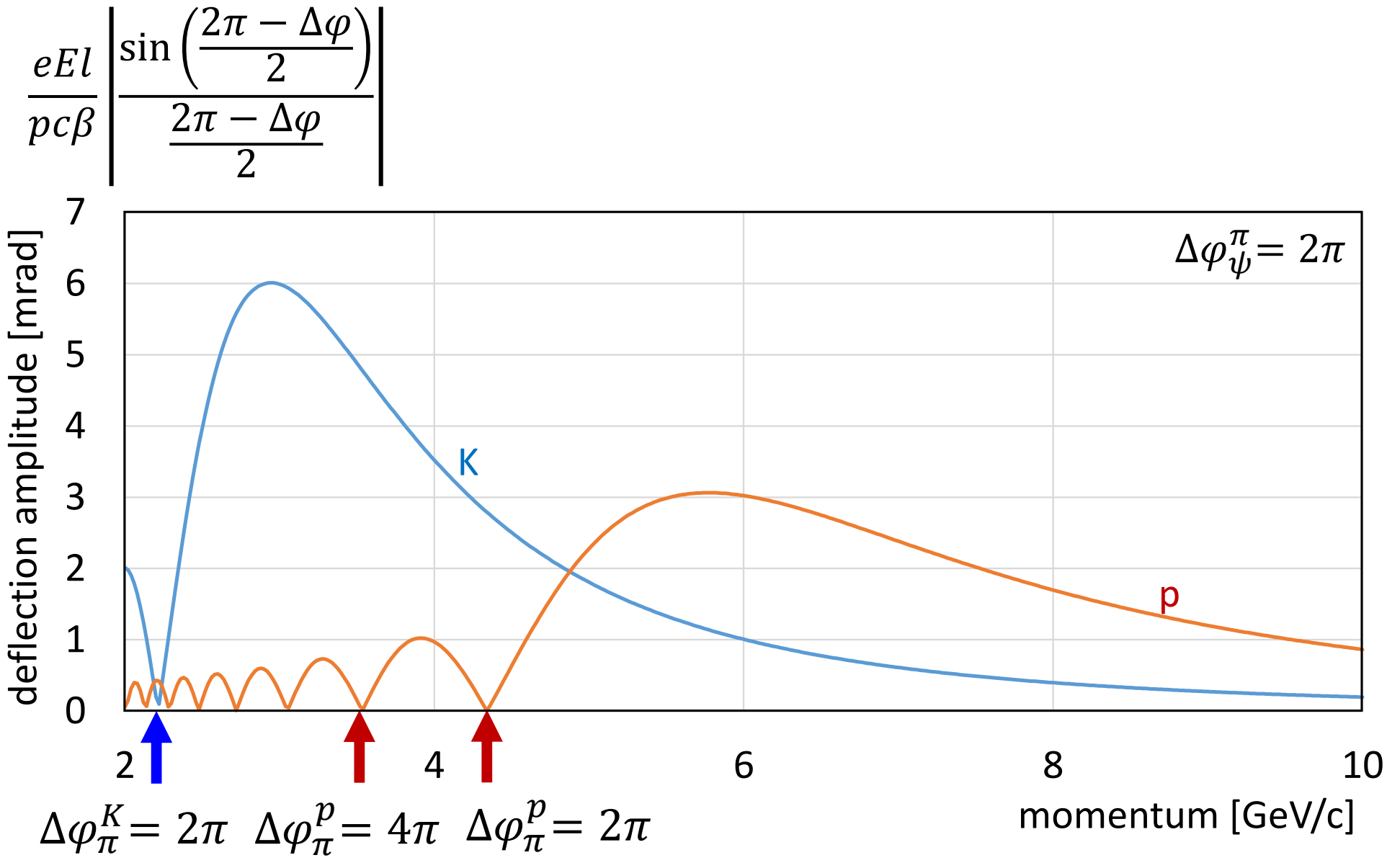

Figure 9: Momentum dependence of the deflection amplitude for $K^-$ (blue) and $\bar{p}$ (red) in the case of $\Delta\varphi_\psi^\pi = 2\pi$. Since the deflection for $\pi^-$ is always canceled, usable is the momentum region in which the deflection for $K^-$ or $\bar{p}$ is large.

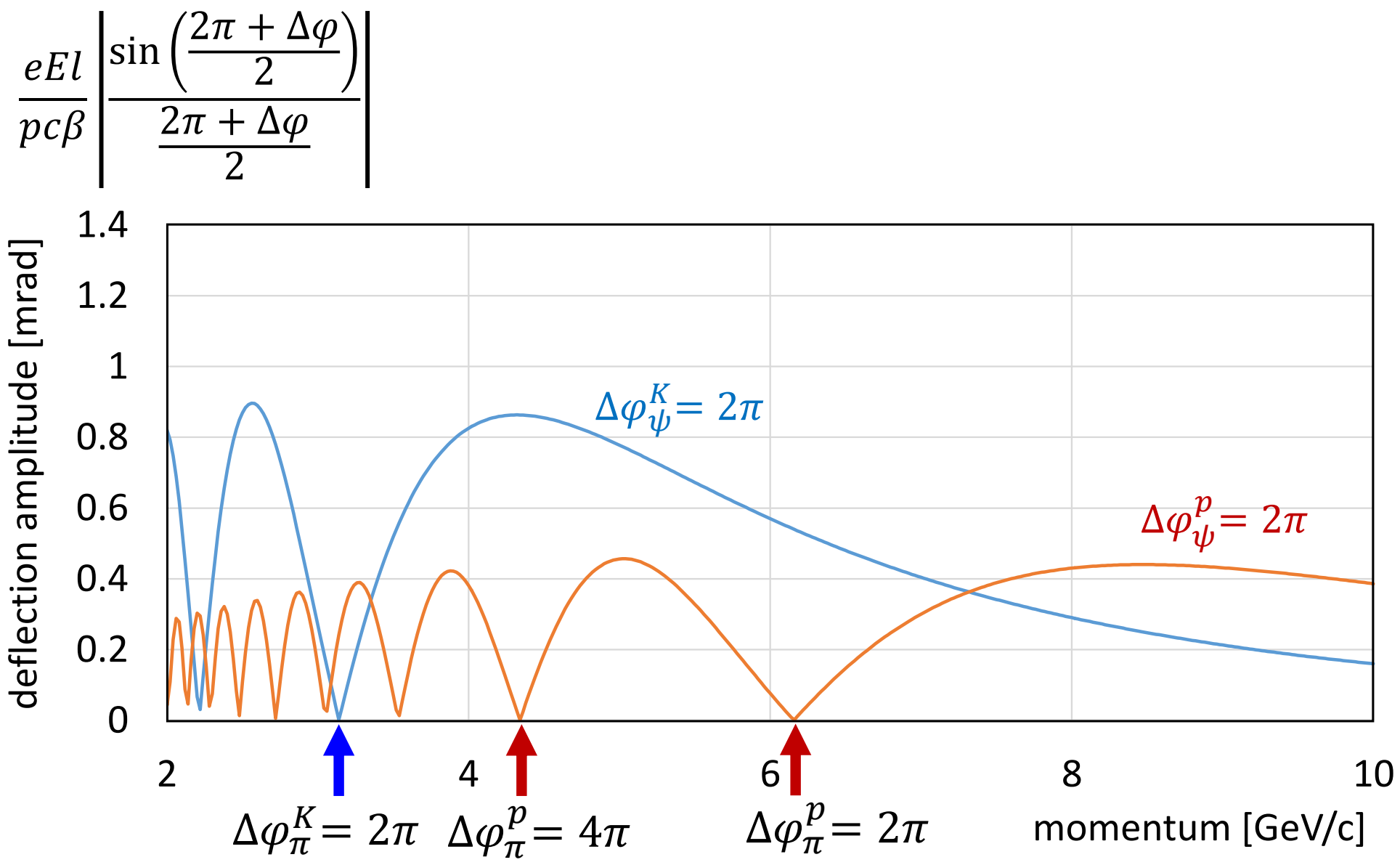

Figure 10: Momentum dependence of the deflection amplitude for $\pi^-$ in the case of $\Delta\varphi_{\psi}^{K} = 2\pi$ (blue) and $\Delta\varphi_{\psi}^{p} = 2\pi$ (red). Since the deflection for $K^-$ or $\bar{p}$ is always canceled, usable is the momentum region in which the deflection for $\pi^-$ is large.

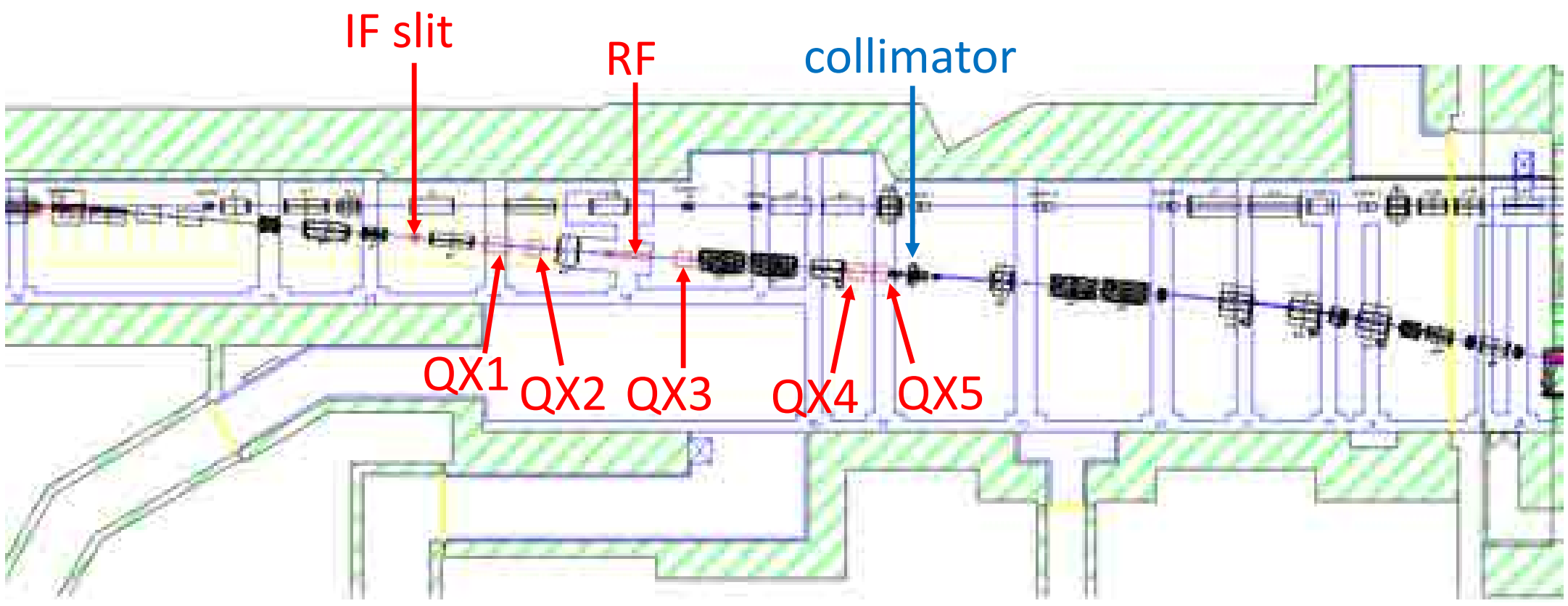

Figure 11: Layout of the midstream part of the RF-separated high-p-ex beam line. Five quadrupoles, a RF cavity, and a slit are added to the existing high-p beam line.

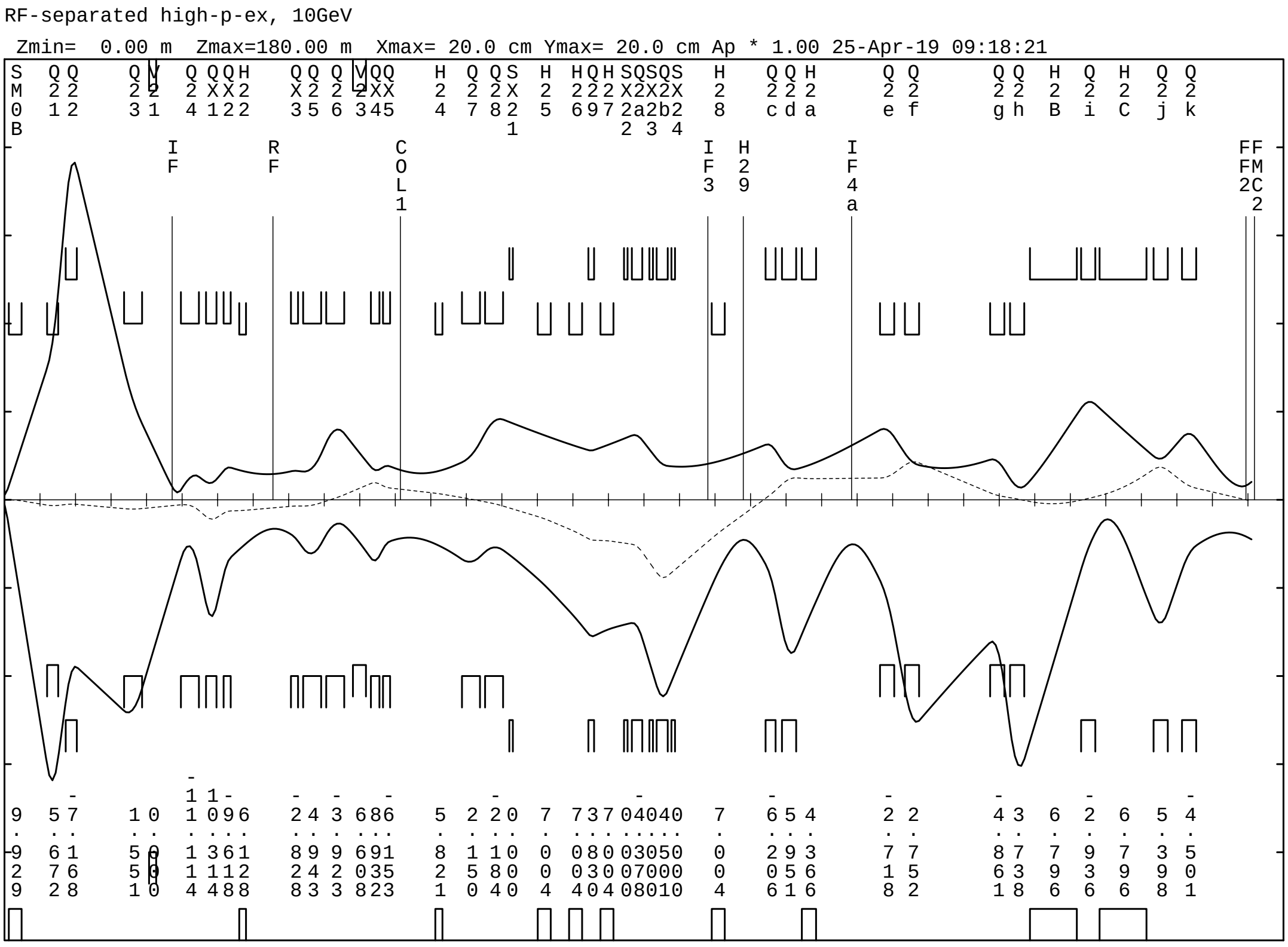

Figure 12: First-order beam envelope of the RF-separated high-p-ex beam line. The upper and lower show the vertical and horizontal directions, respectively. The dash line indicate the horizontal dispersion.

instead of $2\pi - \Delta\varphi_{\pi}^{K/p}$. However, it might be compensated by applying the higher field gradient, which can be obtained with recent developments of the RF technique.

Fig.11 shows a possible layout of the RF separation section of the separated high-p-ex beam line. The first-order beam envelope is presented in Fig.12. The RF separation is realized by adding five quadrupoles, a RF cavity, and a slit to the existing high-p beam line. The beam envelope is presented in Fig.12. Note that, since major rearrangement of existing components is not necessary, this new configuration does not affect current optics for the primary beam and the unseparated secondary beam.

# 3 Summary

The K10 beam line is designed to provide well-separated charged secondary beams with the maximum momentum of 10 GeV/$c$. By adopting electrostatic separators with the total length of 27 m, the beam line can separate $K^-$ mesons up to 4 GeV/$c$ and antiprotons up to 6 GeV/$c$. Another configuration is to employ two RF cavities and can serve higher-momentum beams.

A possible upgrade plan of the extended high-p beam line (high-p-ex) for separated beams is also discussed. By using the phase slip between the beam particles and the RF traveling wave, we can separate $K^-$ and $\bar{p}$ beams with the momentum higher than 5 GeV/$c$ with only one RF cavity.

Finally, we would like to point out that the one-cavity RF separation technique is also important for the K10 beam line. The shorter length of the beam line than that of the two-cavity option provides the more $K^-$ yield. Especially, it is quite important in the reduced plan of the hall extension, where the upstream part of the K10 beam line is shared with the HIHR beam line. As for the ES option, since the first stage of the ES separators can be shared with the HIHR, the line length is not so different from that of the original design of the K10. On the other hand, in the RF-separator option, RF cavities must be located after branched from the HIHR beam line, that makes the total length of the K10 longer. It might be compensated by applying the one-cavity method.

# References

[1] `http://aea.web.psi.ch/Urs_Rohrer/MyWeb/trans.htm`

[2] `http://aea.web.psi.ch/Urs_Rohrer/MyWeb/turtle.htm`

[3] J.R. Sanford and C.L. Wang, "Empirical formulas for particle production in p-Be collision between 10 and 35 BeV/c", BNL internal reports No.11299 and 11479 (1967).

[4] P. Bernard, P. Lazeyras, H. Lengeler and V. Vaghin, "Particle separation with two- and three-cavity RF separators at CERN", CERN Reports 68-29 (2968).

[5] H. Noumi, *et al.*, J-PARC E50 proposal.

[6] H. Noumi, in this proceedings.

[7] Ph. Bernard, H. Lengeler and J.Cl. Prelaz, "Some New Possibilities for RF-Separation at CERN", eConf C710920, 269 (1971).

# $\overline{\text{P}}$ANDA Experiment at FAIR

**Walter Ikegami Andersson**[1]
On behalf of the $\overline{\text{P}}$ANDA collaboration

[1]Department of Physics and Astronomy, Uppsala University

The $\overline{\text{P}}$ANDA experiment at FAIR is a future internal target detector aiming on studying the strong interaction. The new accelerators at FAIR will provide an antiproton beam with a momentum range of 1.5 – 15 GeV/c. This is similar to the proposed extension of the current hadron facility at JPARC. The $\overline{\text{P}}$ANDA experiment, as well as feasibility studies for measurements planned during the first phase of data taking, are discussed.

## 1 Introduction

Within the framework of the Standard Model, Quantum Chromodynamics (QCD) is the accepted theory of the strong interaction. QCD describes quarks and their interactions through the force mediator of the Standard Model, the gluon. At high energies, where the coupling constant $\alpha_s$ of the strong interaction is small, perturbation theory can be employed. However, as low energy scales, the coupling constant grows large. In this non-perturbative energy regime, alternative tools such as effective field theories and phenomenological models, must be used. This give rise to many interesting questions. One question concerns the relevant degrees of freedom; is it quarks and gluons or hadrons? Another open question is the comparisons of measured and predicted resonances. Furthermore, the Standard Model does not prohibit the existence of exotic bound states such as four- or five quark bound states. The future $\overline{\text{P}}$ANDA (antiProton ANnihilation at DArmstadt) experiment is designed to collect high quality data in the non-perturbative energy regime and test QCD.

## 2 The $\overline{\text{P}}$ANDA Experiment at FAIR

$\overline{\text{P}}$ANDA will be located in the Facility for Antiproton and Ion Research (FAIR), an accelerator facility currently under construction at GSI (Gesellschaft für Schwerionenforschung), Darmstadt, Germany. The High Energy Storage Ring (HESR), where $\overline{\text{P}}$ANDA will be situated, will provide an antiproton beam with a momentum range from 1.5 up to 15 GeV/c. Together with a proton target, center-of-mass energies from 2.0 up to 5.5 GeV/c$^2$ will be achieved. When FAIR is fully realised, HESR will have two modes of operation. High resolution mode, with a beam resolution of $\Delta p/p = 5 \times 10^{-5}$ and a luminosity of $\mathcal{L} \approx 2 \times 10^{31}$, and the high luminosity mode where $\mathcal{L} \approx 2 \times 10^{32}$ and the beam resolution is $\Delta p/p = 10^{-4}$. During the starting phase of PANDA, HESR will provide a luminosity of $\mathcal{L} \approx 10^{31}$ due to the lack of the accumulator preceding the HESR.

The $\overline{\text{P}}$ANDA detector, shown in Figure 1, is a general purpose spectrometer divided into two main parts. The forward spectrometer covers scattering angles of $\theta < 10°$. The Forward Tracking Station (FTS) will reconstruct charged tracks and a Shashlyk-type Calorimeter (FSC) will provide the necessary calorimetry. Particle identification will be provided by the Forward Time-of-Flight (FToF) detector and the aerogel Ring Imaging Cherenkov Counter (FRICH).

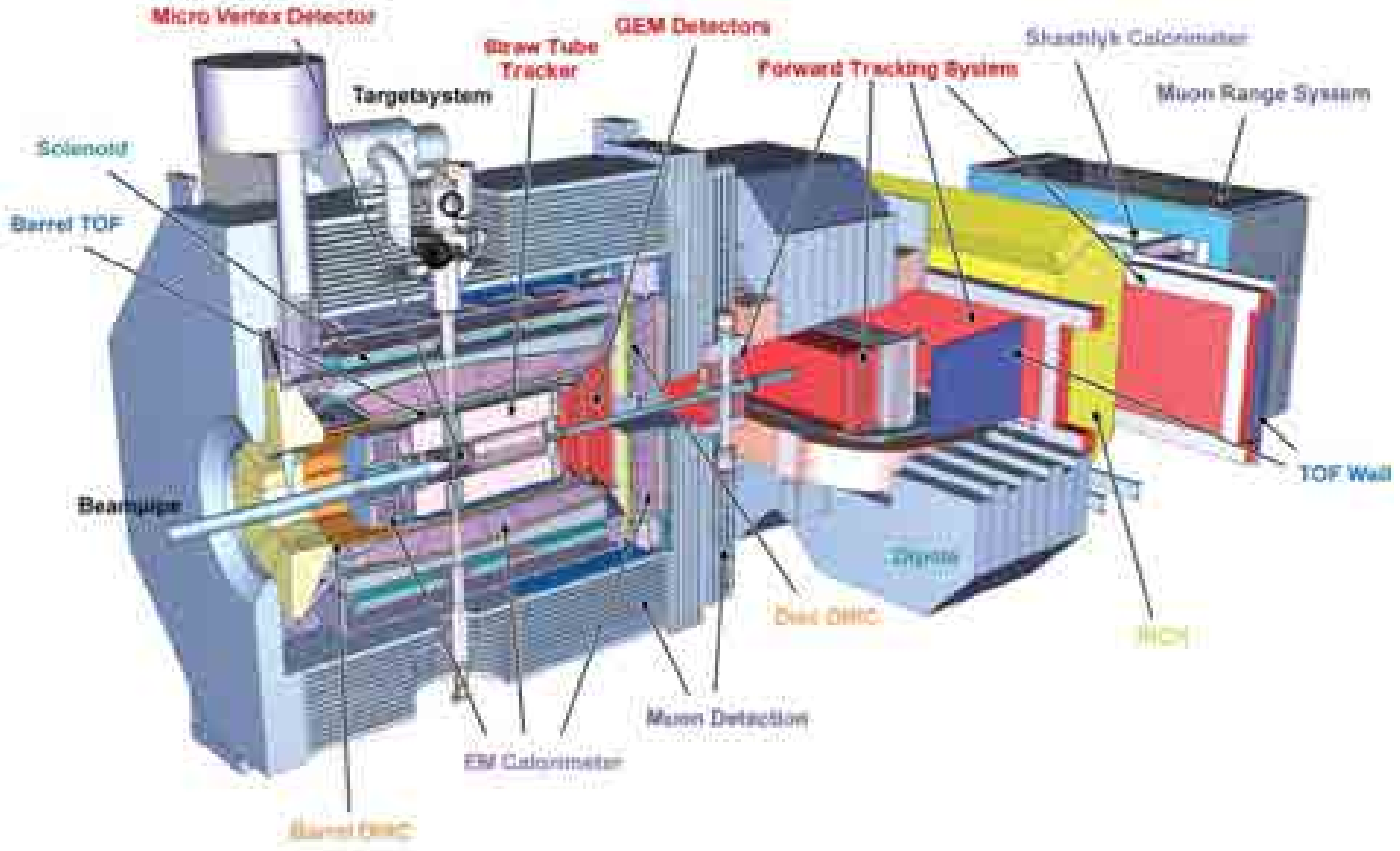

Figure 1: The $\overline{\text{P}}$ANDA detector setup.

The target spectrometer covers scattering angles $\theta > 10^\circ$ and is complemented by a solenoid magnet of 2T in the beam direction. It consists of a Micro Vertex Detector, which surrounds the interaction point. The Straw Tube Tracker surrounding the MVD provides the momentum reconstruction of charged particles. In the more forward region Gas Electron Multiplier (GEM) stations are placed to cover tracking in the $\theta < 22^\circ$ region where the STT coverage ends. The Electromagnetic Calorimeter (EMC), which consists of a barrel and two endcaps, will provide the calorimetry. A barrel Time-of-Flight (ToF) detector and the Internally Reflected Cherenkov light (DIRC) will provide particle identification. Muon chambers are placed outside the solenoid magnet. A detailed description of the $\overline{\text{P}}$ANDA detector is found in [1].

# 3 X(3872) Scan

Charmonium spectroscopy is a very active field where several new states have been discovered the past fifteen years. Observed charmonium states with masses below the $\overline{D}D$ threshold are in agreement with theoretical predictions. However, above the threshold, the spectrum becomes unclear. While many states have been measured by *e.g.* BaBar, Belle or BESIII, these states have not been mapped to the predicted charmonium spectrum.

The $X(3872)$ was first measured by the Belle collaboration in 2003 in the decay $X(3872) \to J/\Psi\pi^+\pi^-$ [2]. Since then, many experiments have confirmed its existence [3, 4, 5, 6, 7]. LHCb measured its quantum numbers to be $J^{PC} = 1^{++}$, making it an excited $\chi_{C1}(2P)$ state [8]. However, its mass being close to the $D^{*0}\overline{D}^0$ threshold makes it a molecule candidate. The line-shape of the resonance is the key to determine its nature. However, only an upper limit of the width $\Gamma < 1.2$ MeV is known, measured by Belle [9]. While other experiments are limited by the detector resolution, a $\overline{p}p$ machine could produce the $X(3872)$ in direct formation and perform a beam scan. A simulation study has determined the sensitivity and bias, defined in Equation 1 and 2, of a line-shape measurement assuming input Breit-Wigner widths of $\Gamma_0 \in [50, 70, 100, 130, 180, 250, 500]$. By performing $N_{MC} = 300$ toy experiment, it is concluded that $\overline{\text{P}}$ANDA will have a sensitivity better than 20% down to a width of 90 keV in the high luminosity mode and 120 keV for high resolution mode with a starting phase luminosity, as shown in Figure 2.

$$\frac{\Delta\Gamma_{\text{meas}}}{\overline{\Gamma}_{\text{meas}}} = \frac{\text{RMS}}{\text{Mean} + \Gamma_0} \pm \frac{\text{RMS}}{\text{Mean} + \Gamma_0}\sqrt{\frac{1}{2(N_{\text{MC}} - 1)} + \frac{\text{RMS}^2}{N_{\text{MC}} \cdot (\text{Mean} + \Gamma_0)^2}} \tag{1}$$

$$\frac{\overline{\Gamma}_{\text{meas}} - \Gamma_0}{\Gamma_0} = \frac{\text{RMS}}{\Gamma_0} \pm \frac{\text{RMS}}{\Gamma_0\sqrt{N_{\text{MC}}}}. \tag{2}$$

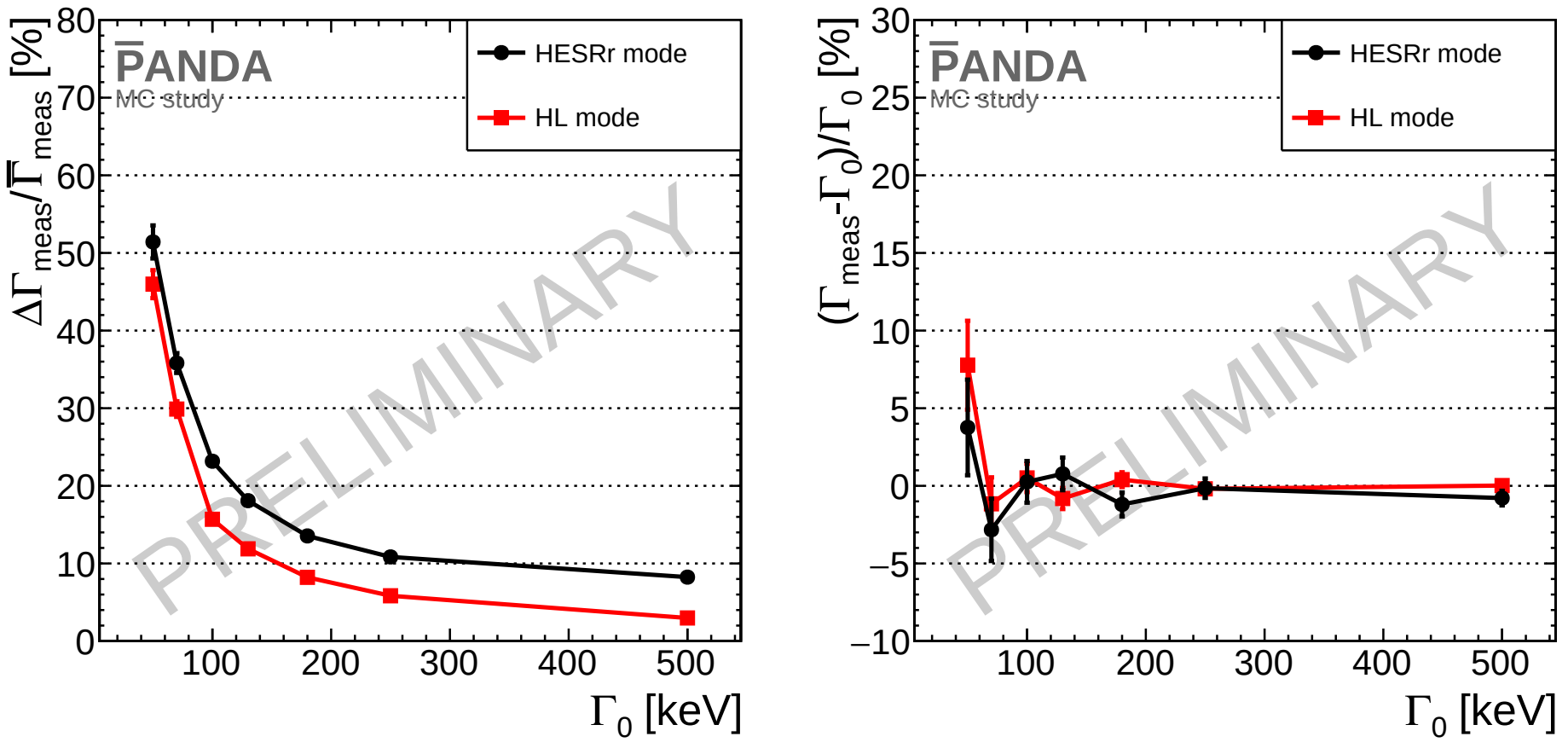

Figure 2: The expected sensitivity (left) and bias (right) for the high luminosity mode (red squares) and the high resolution mode (black circles) during the starting phase of HESR.

## 4 Hyperon Spectroscopy and Dynamics

Baryon spectroscopy explores spectra of bound quark states predicted by the quark model. Historically, baryon spectroscopy was crucial in the development of QCD. When the Eightfold Way was introduced in 1962, it predicted the existence of a yet unmeasured triple strange spin-3/2 particle with a mass of around 1680 MeV/c$^2$. Following the discovery of the $\Omega^-$, the quark model was formulated by Gell-Mann and Zweig [10][11]. Currently, several experiments are exploring baryon spectroscopy in both the light sector through $\pi N$ and $\gamma N$ reactions and in the heavy quark sector in *e.g.* B-factories and LHCb. The strangeness sector, however, is not nearly as active.

Various simulation studies have shown that $\overline{\text{P}}$ANDA can produce large amount of hyperons through $\bar{p}p \to \overline{Y}Y$ reactions. The expected reconstruction rates with the starting phase luminosity is shown in Table 1 [12, 13, 14]. One simulation study concerning the measurement of the $\Xi^-(1820)$ resonance in the $\bar{p}p \to \overline{\Xi}^+\Xi^-(1820)$ has shown that a reconstruction rate of 15 000 exclusive events per day can be achieved with the starting phase luminosity.

The availability of large data of $\bar{p}p \to \overline{Y}Y$ samples also opens up the possibility to probe QCD in the intermediate energy regime where the relevant degrees of freedom is not known. Attempts have been made to describe $\bar{p}p \to \overline{Y}Y$ reactions in both hadron pictures and quark and gluon pictures, *e.g.* [15, 16]. Using the Spin Density Matrix formalism, the angular distribution of the final state particles can be parametrised in terms of spin observables and decay asymmetry parameters. When considering *e.g.* $\Lambda \rightarrow p\pi^-$, the angular distribution is $I(\theta, \phi) = \frac{1}{4\pi}(1 + \alpha_\Lambda P \sin\theta \sin\phi)$, where the polarisation $P$ and the decay asymmetry $\alpha_\Lambda$ is accessible. When considering two hyperons simultaneously, the spin correlation becomes accessible and in *e.g.* $\Omega^- \rightarrow \Lambda K^-$, where $\Lambda \rightarrow p\pi^-$, the $\beta$ and $\gamma$ decay asymmetry parameters are also accessible [17]. measuring spin observables could shed light on the production mechanism and with high statistics, CP violation in the baryon sector could be tested through measurements of the asymmetry parameters.

Table 1: Expected production rates of hyperons at $\overline{\text{P}}$ANDA with luminosity $\mathcal{L} = 10^{31}\text{cm}^{-2}\text{s}^{-1}$. The cross sections $\sigma(\overline{p}p \rightarrow \overline{\Lambda}\Sigma^0)$ and $\sigma(\overline{p}p \rightarrow \overline{\Xi}^+\Xi^-)$ are extrapolated values. The cross sections $\sigma(\overline{p}p \rightarrow \overline{\Omega}^+\Omega^-)$ and $\sigma(\overline{p}p \rightarrow \overline{\Lambda}_c\Lambda_c)$ are based on theoretical predictions.

| $p_{\overline{p}}$ (GeV/c) | Reaction | $\sigma$ ($\mu$b) | Eff (%) | Decay | Rate |
|---|---|---|---|---|---|
| 1.64 | $\overline{p}p \rightarrow \overline{\Lambda}\Lambda$ | 64 | 10 | $\Lambda \rightarrow p\pi^-$ | 28 s$^{-1}$ |
| 4.0 | $\overline{p}p \rightarrow \overline{\Lambda}\Sigma^0$ | ∼40 | 30 | $\Sigma^0 \rightarrow \gamma$ | 50 s$^{-1}$ |
| 4.0 | $\overline{p}p \rightarrow \overline{\Xi}^+\Xi^-$ | ∼2 | 20 | $\Xi^- \rightarrow \Lambda\pi^-$ | 2 s$^{-1}$ |
| 12.0 | $\overline{p}p \rightarrow \overline{\Omega}^+\Omega^-$ | ∼0.002* | ∼30 | $\Omega^- \rightarrow \Lambda K^-$ | ∼4 h$^{-1}$ |
| 12.0 | $\overline{p}p \rightarrow \overline{\Lambda}_c\Lambda_c$ | ∼0.1* | ∼30 | $\Lambda_c \rightarrow \Lambda\pi^+$ | ∼2 d$^{-1}$ |
| 4.6 | $\overline{p}p \rightarrow \overline{\Xi}^+\Xi^-(1820)$ | 1 | 5 | $\Xi^-(1820) \rightarrow \Lambda K^-$ | 15000 d$^{-1}$ |

## 5 Summary

The $\overline{\text{P}}$ANDA experiment will open up many opportunities to study QCD. It will be possible to perform baryon spectroscopy in the strangeness sector, a field that is lacking experimental input. Simulation studies have shown that $\overline{p}p \rightarrow \overline{Y}Y$ reactions can be well reconstructed for single-, double-, triple- and even charmed hyperons. Furthermore, the production process can be studied with spin observables such as polarisation and spin correlation. It is also possible to study CP-violation in the baryon sector. A line-shape measurement of the $X(3872)$ state could shed light on the nature of the state. Since a beam scan measurement is not limited by the resolution of the detector, $\overline{\text{P}}$ANDA will be much more sensitive to the line-shape of resonances produced in direct formation. If JPARC realizes an antiproton beam with a momentum up to 10 GeV/c in the future, potential collaborations especially related to detector hardware and software R&D should be explored.

## References

[1] Kotulla M *et al.* PANDA Collaboration 2005 *PANDA Technical Progress Report*

[2] Choi S-K *et al.* Belle Collaboration 2003 *Phys. Rev. Lett.* **91**, 262001

[3] Auber B *et al.* Babar Collaboration 2005 *Phys. Rev. D* **71** 071103

[4] Acosta D *et al.* CDF II Collaboration 2004 *Phys. Rev. Lett.* **93** 072001

[5] Abazov V M *et al.* D0 Collaboration 2004 *Phys. Rev. Lett.* **93** 162002

[6] Aaij R *et al.* LHCb Collaboration 2012 *Eur. Phys. J. C* **72** 1972

[7] Chatrchyan S, Khachatryan V *et al.* CMS collaboration 2013 *J. High Energ. Phys.* **1304** 154

[8] Aaij R *et al.* LHCb Collaboration 2013 *Phys. Rev. Lett.* **110** 222001

[9] Choi S-K *et al.* Belle Collaboration 2011 *Phys. Rev. D* **84** 052004

[10] Gell-Mann M 1964 *A Schematic Model of Baryons and Mesons Phys. Lett.* **8** 214

[11] Zweig G 1964 *An SU(3) model for strong interaction symmetry and its breaking. Version 1* CERN-TH-401

[12] Pütz J 2016 *Proceedings of FAIRNESS 2016* Garmisch-Partenkirchen, Germany

[13] Thomé E 2012 *Multi-Strange and Charmed Antihyperon-Hyperon Physics for PANDA* (Ph. D. Thesis, Uppsala University)

[14] Grape S 2009 *Studies of PWO Crystals and Simulations of the* $\overline{p}p \to \overline{\Lambda}\Lambda, \overline{\Lambda}\Sigma^0$ *Reactions for the PANDA experiment* (Ph.D. Thesis, Uppsala University)

[15] Kohno M and Weise W 1986 *Phys. Lett. B* **179** 15

[16] Tabakin F and Eisenstein R A *Phys. Rev. C* **31**, 1857 (1985)

[17] E Perotti *Extraction of Polarization Parameters in the* $\overline{p}p \to \overline{\Omega}\Omega$ *Reaction* arXiv:1803.11482 (2018)

# Spectral changes of vector mesons in pion induced reactions as a signature for restoration of chiral symmetry

**Kazuya Aoki**

High Energy Accelerator Research Organization (KEK)

We propose experiment to measure the spectral change of vector mesons in nuclear medium using pion-induced reactions. These reactions will allow us exclusive measurement which has been never conducted and provide crucial information on chiral symmetry restoration.

## 1 Introduction

Spectral change of vector mesons in nuclear medium attracts interest as it is related to restoration of spontaneously broken chiral symmetry of QCD. Among various experimental efforts, dielectron measurement is one of the most promising way since the final state interaction is very small. J-PARC E16 experiment has been proposed to measure inclusive dielectron mass spectrum in $pA$ reactions at $E_{\rm kin} = 30$ GeV[1, 2]. We expect that the experiment provides us detailed information on spectral change in finite density environment in unprecedented quality. Here we propose a study on this matter to bring us one step forward, using pion induced reactions such as $\pi^- p \to \phi n$, $\pi^+ n \to \phi p$ and these reactions on nuclear targets. Detection of forward going neutron (proton) in $\pi^-(\pi^+)$ induced reactions will allow us to measure missing mass and the initial state can be identified. At the same time, the invariant mass of $\phi$ meson can be obtained. Such an exclusive measurement has not been performed and it provides qualitatively new information.

We propose three types of measurement using these reactions: inclusive measurement, missing mass measurement, and exclusive measurement. To perform the experiment, detailed knowledge of the reactions is important and we propose to do cross section measurements.

## 2 Proposed experiments

Low momentum pion beam allows us to utilize elementary reactions such as $\pi^- p \to \rho/\omega/\phi\ n$, and $\pi^+ n \to \rho/\omega/\phi\ p$ reactions. In this article we concentrate on $\phi$ meson production using $\pi^-$ beam. The cross section of $\phi$ meson of the $\pi^-$ induced reaction was measured near threshold [3] and found to be maximum around 1.6–2.0 GeV/$c$ and decrease as increasing momentum. On the other hand, the momentum transfer to the produced $\phi$ meson is lower as the beam momentum increases. Therefore, the beam momentum of around 2 GeV/$c$ is preferred.

### 2.1 Inclusive measurement

Beam: $\pi^- \sim 2$ GeV/$c$, $\sim 10^9$ /spill
Target: C (400$\mu$m), Cu (80$\mu$m), Pb (20$\mu$m)
Reaction: $\pi^- p \to \phi X$, $\phi \to e^+e^-$. ($e^+$ and $e^-$ are detected)
Detectors: E16 dielectron spectrometer

The total cross section of $\pi^- p \to \phi\ n$ was measured to be $21 \pm 7\ \mu$b with a beam momentum of 2.0 GeV/$c$. Although the total cross section is one order of magnitude smaller than that of $pp \to \phi X$ at $E_{\rm kin} = 30$ GeV, slowly moving $\phi$ meson ($\beta\gamma < 1.25$) production is dominant. We expect to collect 5.4 k $\phi$ which is an order of magnitude better statistics than that of the E325 experiment. We assume 100 days of physics running with a beam of $10^9 \pi^-$ / spill at $p_{\rm beam} = 1.8$ GeV/$c$ on Cu target.

### 2.2 Missing mass measurement

Beam: $\pi^- \sim 2$ GeV/$c$, $\sim 10^7$ /spill
Target: C (2000$\mu$m), Cu (800$\mu$m), Pb (200$\mu$m)
Reaction: $\pi^- A \to \phi n$, $\phi p \to K^+\Lambda$, $\Lambda \to p\pi^-$. (n, p, $K^+$ and $\pi^-$ are detected)
Detectors: Charged hadron spectrometer (such as E29) with E26 forward neutron counter

By detecting forward going neutron in $\pi^- A$ reactions, missing mass $M_{\pi^-,p,n}$ can be calculated. If there exists $\phi$ meson bound state, the state can be seen as a peak lower than the quasi free reactions. For the neutron counter, the one prepared for J-PARC E26 experiment [4] is assumed. It has a time resolution of better than 80 ps and is placed at 7 m downstream from the target. If $\phi$ is bound to the nucleus and the mass reduces, $K^+K^-$ channel may be closed. Therefore, as proposed in J-PARC E29 [5], $K^+\Lambda$ can be used to tag the reaction. For charged particle measurement, E29 charged hadron spectrometer is a candidate.

### 2.3 Combined measurement

Beam: $\pi^- \sim 2$ GeV/$c$, $\sim 10^9$ /spill
Target: C (400$\mu$m), Cu (80$\mu$m), Pb (20$\mu$m)
Reaction: $\pi^- p \to \phi n$, $\phi \to e^+e^-$. (n, $e^+$ and $e^-$ are detected)
Detectors: E16 dielectron spectrometer with forward neutron counter

Simultaneous measurement of initial state and final state has been never performed and will provide crucial information to understand the spectral change of vector mesons in nuclear medium. We expect exclusive measurement of $\sim 400$ $\phi$ mesons with a forward neutron counter in $\pm 60$ cm acceptance at 7 m downstream of the target and 180 days of physics running. For this measurement, $\pi^+$ induced reaction might be better considering the intensity of the beam.

### 2.4 Cross section measurement

Beam: $\pi^- \sim 2$ GeV/$c$, $\sim 10^7$ /spill
Target: LH, C (2000$\mu$m), Cu (800$\mu$m), Pb (200$\mu$m)
Reaction: $\pi^- p \to \phi n$, $\phi \to K^+K^-$. (n, $K^+$ and $K^-$ are detected)
Detectors: Charged kaon spectrometer (E29) with a forward neutron counter

The cross section of $\pi^- p \to \rho/\omega/\phi\ n$ has been measured to be $21 \pm 7$ $\mu$b ($25 \pm 8$ $\mu$b) with a beam momentum of 2.0 GeV/$c$ (1.8 GeV/$c$) and it has large uncertainty. The measurement was restricted to forward angle in CM system, whereas we are interested in the backward angle production. The measured angle distribution is consistent with isotropic, but again, the error bars are very large and we cannot draw definite conclusions based on the existing data. For detailed planning of the measurements, precise measurement of the cross section is essential. Therefore, we propose to measure the cross section of this reaction and its A dependence.

$K^+K^-$ channel will be used for this purpose since the channel has large branching ratio. J-PARC E29 spectrometer can be used for this measurement combined with a neutron counter as in the J-PARC E26 experiment.

We expect to collect 2k $\phi$ events with $10^7$ $\pi^-$/spill at beam momentum of 1.8 GeV/$c$ on Cu target for 10 days of physics running assuming that the acceptance is 0.15%.

### 2.5 Conclusions

We propose measurement of in-medium spectral change of vector meson using pion induced reactions. It requires high intensity beams and large scale detectors, therefore, hadron hall extension is crucial in realizing the experiment.

## References

[1] S. Yokkaich *et al.*, Proposal of J-PARC E16,
`https://j-parc.jp/researcher/Hadron/en/pac_0606/pdf/p16-Yokkaichi_2.pdf`

[2] S. Yokkaich *et al.*, Proposal of J-PARC E16 Run0,
`https://j-parc.jp/researcher/Hadron/en/pac_1707/pdf/E16_2017-10.pdf`

[3] H. Courant *et al.*, Phys. Rev. D, **16**, 1 (1977).

[4] K. Ozawa *et al.*, Proposal of J-PARC E26,
`https://j-parc.jp/researcher/Hadron/en/pac_1007/pdf/KEK_J-PARC-PAC2010-08.pdf`

[5] H. Ohnishi *et al.*, Proposal of J-PARC E29,
`https://j-parc.jp/researcher/Hadron/en/pac_1007/pdf/KEK_J-PARC-PAC2010-02.pdf`

# Intrinsic charm search at the J-PARC high momentum beam line

**Yuhei Morino**[1]

[1]KEK, High Energy Accelerator Research Organization, Tsukuba, Ibaraki 305-0801, Japan

## Short Summary

- Physics motivation with its impact and expected result
  intrinsic charm search, first measurement of backward J/$\psi$ in low energy p-A collisions.
- What kind of beam and equipments are necessary?
  30 GeV proton beam with $10^{10}$ per spill, the J-PARC E16 spectrometer
- Expected duration of the beam time (yield estimation)
  100 days(300 shifts) for the J-PARC E16 experiment and 20 days(60 shifts)
  for the intrinsic charm search

The existence of $|uudc\bar{c}\rangle$ Fock components in a proton, which is called *"intrinsic charm"*, was suggested in the early 1980's[1]. The intrinsic charm tends to have a large momentum fraction ($x$), unlikely *"extrinsic charm"* which is generated by gluon splitting perturbatively. In addition, parton distribution function (PDF) of the intrinsic charm can be different from the PDF of intrinsic anti-charm. These features of the intrinsic charm have been applied for possible solutions of various unexpected phenomena related to heavy quarks. However, the existence of the intrinsic charm remains still inconclusive, despite a number of experimental and theoretical studies to evaluate a probability of the intrinsic charm.

Additional experimental results are necessary to confirme the existence of the intrinsic charm. An identification of a clean and characteristic phenomenon of the intrinsic charm will be a smoking gun. The anomalous J/$\psi$ suppression of the yield per nucleon at large $x_F$ in high energy hadron-nucleus collisions is one of the most striking phenomena related with the intrinsic charm[2]. An introduction of *"soft"* production of J/$\psi$ due to the intrinsic charm can account for the anomalous J/$\psi$ suppression intutively[3]. On the other hand, the energy loss model, which assumes the color singlet model for J/$\psi$ production and the large energy loss for the color-octet $c\bar{c}$ pair in the nuclear matter, can also explain the anomalous suppression[4]. Since it is difficult to reject the energy loss model from the experimental results to date, the present J/$\psi$ suppression cannot be regarded as an evidence of the intrinsic charm.

The energy loss effect of the $c\bar{c}$ color-octet production becomes negligible in the case of backward production in low energy collisions, since the path length of the color-octet becomes significantly short. On the other hand, the intrinsic charm scenario predicts J/$\psi$ suppression at backward regions in a similar way to the forward J/$\psi$ suppression. Therefore, backward J/$\psi$ suppression in low energy collisions is the characteristic phenomenon of the intrinsic charm. The measurement of backward J/$\psi$ production in low energy hadron-nucleus collisions will provide a crucial information to judge the origin of the observed J/$\psi$ suppression, that is, the intrinsic charm or the energy loss.

The measurement of backward J/$\psi$ production in low energy hadron-nucleus collisions can be performed as the by-production of the J-PARC E16 experiment[5]. The E16 spectrometer can measure backward production of J/$\psi$. The energy of the proton beam at the high momentum beam line (30 GeV) is significantly lower than previous measurements of J/$\psi$ suppression, which are several hundred GeV. The cross section of J/$\psi$ via the hard processes gets considerably small in the case of low energy collisions. It leads that the fraction of the contribution from the intrinsic charm increases and backward J/$\psi$ production gets to be more sensitive to the intrinsic charm.

A model calculation which included nuclear PDF, perturbative J/$\psi$ production, soft J/$\psi$ production via the intrinsic charm and J/$\psi$ absorption in nuclei, was performed to evaluate the effect of the intrinsic charm on J/$\psi$ production. The effect of the energy loss of the $c\bar{c}$ color-octet was neglected since it was confirmed that the energy loss did not affect J/$\psi$ suppression pattern due to their short flight length by a toy calculation. A GEANT4-based Monte-Carlo simulation was also performed to evaluate a sensitivity of the measurement to the intrinsic charm in the case of full 26 modules installation. From the model calculation and the Monte-Carlo simulation, it was turned out that J/$\psi$ at $x_F \sim 0$ could not be reconstructed due to the geometrical acceptance, while the suppression degree at $x_F \sim 0$ gave the important baseline to extract the intrinsic charm effect.

The special run is proposed to cope with the inefficiency for J/$\psi$ at $x_F \sim 0$[6]. The targets are moved upstream by 23 cm near a vacuum film of a beam pipe. Data taking with 60 shifts (20 days) for the special run is necessary to collect enough statistics. The targets and the trigger condition will be optimized for the measurement of J/$\psi$ at $x_F \sim 0$ during the special run. The planned targets are 800 $\mu$m C and 400 $\mu$m Pb. The total radiation length of the special targets is about 5 times as thick as the total radiation length of the normal targets , leading to increasing of the trigger rate due to the backgrounds from the $\gamma$ conversion. When high energy deposit (0.7 GeV) in Lead-Glass calorimeter is required for the special run trigger, the trigger rate will be below 1 kHz. It satisfies the requirement of the E16 read-out system.

The statistics of the expected results were evaluated based on the cross section of J/$\psi$, the reconstruction efficiency, and the proposed run condition (300 shifts for the E16 experiment and 60 shifts for the special run). The sensitivity to the intrinsic charm was discussed with the expected uncertainty of the experiment. It was demonstrated that the measurement of backward J/$\psi$ production at the E16 experiment and the special run had the good sensitivity for the intrinsic charm with a probable value of $P_{IC}$.

## References

[1] S. J. Brodsky, P. Hoyer, C. Peterson, and N. Skai, Phys. Lett. **B 93** 451 (1980).

[2] M. J. Leitch *et al.*, Phys. Rev. Lett **84** 3256 (2000).

[3] R. Vogt, Phys. Rev. C **61** 035203 (2000).

[4] F. Arleo and S. Peigné, Phys. Rev. Lett **109** 122301 (2012).

[5] S. Yokkaichi, Lect. Notes. Phys.**C781** 161 (2009).

[6] Y. Morino *et al.*, J-PARC LOI, http://j-parc.jp/researcher/Hadron/en/pac_1801/pdf/LoI_2018_6.pdf

# $\Lambda$p scattering experiment via the $\pi^- p \to \Lambda K^*$ reaction using $D^*$ spectrometer at high-p

**Ryotaro Honda**[1]

[1]Dept. of Phys., Tohoku University

Today, the existence of 2 $M_\odot$ neutron star triggered the hyperon puzzle; the present nuclear physics cannot support such the massive neuron star. New investigation is attempted to solve this puzzle through the hypernuclear data. However, the present two-body $\Lambda$N interaction is not enough precise to proceed the further study due to the lack of the scattering data. In this article, an idea for the $\Lambda$p scattering experiment, which will be performed at the J-PARC high-p beam line in the future. In this experiment, we will use the $\pi^- p \to K^*(892)^0\Lambda$ for the $\Lambda$ production. We will measure the cross section of the $\Lambda$p scattering with a few percent accuracy within one month beam time by tagging the $\Lambda$ production using the $D^*$ spectrometer, which will be constructed for the J-PARC E50 experiment.

## 1 Introduction

The most effective way to experimentally investigate the two-body interaction is performing a scattering experiment. However, the $\Lambda$N interaction was historically studied through the hypernuclear experiments because the hyperon-nucleon (YN) scattering experiment was quite difficult. Hyperon decays before reaching detectors due to its short life time. Then, as the counter experiment was difficult, only several experiments were performed using a bubble chamber in 1970s. Recently, the lack of the scattering data impedes further progression of the strangeness nuclear physics.

The knowledge of the $\Lambda$N interaction is used to explain not only hypernuclei but also the neutron star, that is the most high dense system in our universe. The neutron star is a huge bound system consisting of neutron. Its Fermi energy is sufficiently large to produce a $\Lambda$ hyperon. Then, it is expected that $\Lambda$ spontaneously appears in the core of the neutron star. If the hyperons appear inside the neutron star, the particle density is increased because the hyperon is free from the Pauli blocking from nucleons. When the hyperons are mixed, the maximum mass of the neutron star is decreased as compared with the case of the pure neutron matter. Recently, the massive neutron start with 2 $M_\odot$ was found [1]. Such the massive neutron star cannot be supported by the present nuclear physics. This is called as the hyperon puzzle and one of the most important subject in the strangeness nuclear physics.

In order to support the massive neutron star, a lack of the repulsive force must be compensated. The many body force among $\Lambda$ and nucleons, that is $\Lambda$NN, $\Lambda$NNN, etc., is a candidate of the source of the extra repulsion. The existence of the extra repulsion coming from the many body force is about to be studied through the precise measurement of the hypernuclear binding energy. However, Isaka *et al.* suggested that the present two-body $\Lambda$N interaction is not enough precise to discuss the many body effect through the hypernuclear data [2]. The phase shift of the partial waves are not determined well. Especially, the uncertainty of the $p$ wave is the main reason to impede the discussion the many body effect. As this problem originates from the lack of the $\Lambda$p scattering data, the precise scattering data is desired in this situation. Here, the author describes the idea of an experiment for the $\Lambda$p scattering in J-PARC.

# 2 Λp scattering experiment

## 2.1 Experimental design

For the Λp scattering experiment, the first problem is how to produce Λ. As Λ is a neutral particle and there is no neutron target at rest, a neutral particle appears in the initial or the final state. For instead, $\gamma p \to K^+\Lambda$ or $\pi^- p \to K^0\Lambda$ are candidates to produce Λ. However, these two have problems. When we use the $\gamma$ beam, the production cross section becomes one hundredth smaller as compared with the case of the meson beam. In addition, the primary vertex position cannot be determined. On the other hand, when we adopt the $\pi^-$ beam, $K^0$ appears in the final states. As $K^0_S$ immediately decays to $\pi^-\pi^+$, no $s$ quark exists in the final state and it causes the large backgrounds. Therefore, the author introduces the reaction accompanying exited kaon, i.e., $\pi^- p \to K^*(892)^0\Lambda$. In this case, all the particle in the initial and the final states are charged particles. As $K^*(892)^0$ decays to $K^+\pi^-$, the tag of the $s$ quark production is easier. Furthermore, the cross section of this process is sufficiently large [3]. Thus, there are many advantages to use the reaction accompanying exited kaon.

Here, the author introduces the new Λp scattering experiment using the $D^*$ spectrometer, which will be constructed for the J-PARC E50 experiment, at the high-p beam line. The beam $\pi^-$ and the decay products from $K^*$ are measured by the high-p beam line spectrometer and the $D^*$ spectrometer, respectively. The production of Λ is tagged by the missing mass method. In addition, as shown in Fig. 1, the scattered proton and the decay particles from Λ are detected by the cylindrical detector, which surrounds the liquid hydrogen target. Their momenta (kinetic energies) are measured by this detector. As the momentum vector of Λ at the production is known by the spectrometer analysis, the momentum of the scattered proton can be calculated by assuming the Λp scattering kinematics. If the calculated momentum is consistent with the measured one, the event is recognized as the Λp scattering. For the Λp scattering experiment, it is also important to measure the decay products from Λ. In the case of the forward Λp scattering, the recoil energy of the scattered proton is small. Then, the proton will be stopped before passing through the detector. This fact limits the measurable range of $cos\theta_{\rm CM}$ for the Λp scattering. On the other hand, if we reconstruct the momentum vector of Λ in the final state by measuring the decay products, it is possible to confirm whether the Λp scattering is occurred or not by comparing the initial vector and the final vector. Therefore, we should prepare a detector system, which can measure the momentum of $\pi^-$.

The reason of choosing the high-p beam line as the experimental site is that the author is planing to use the beam momentum around 5-10 GeV/c. Fig 2 shows the relation between the momentum and the angle along with the beam axis for the produced Λ when the beam momentum is 10 GeV/c. If the beam momentum is enough high, the produced Λ has a wide recoil momentum range. This allows us to extend the Λ beam range to the high momentum side and gives an opportunity to investigate the higher partial waves. In the next, one can find that the production angle of Λ reaches 50 degree. As the beam momentum increases, the production angle becomes large. If the produced Λ has a large angle, the surrounding detector has an acceptance for the forward scattering without extension of the detector along with the beam axis. These are advantages to use the high momentum $\pi^-$ beam.

The expected beam rate is 60 M per spill, that is the design value of the $D^*$ spectrometer. If we take one month beam time, $10^7$ tagged Λ within the momentum range of interest will be obtained at least. This value is one order larger than the expected tagged Σ in the J-PARC

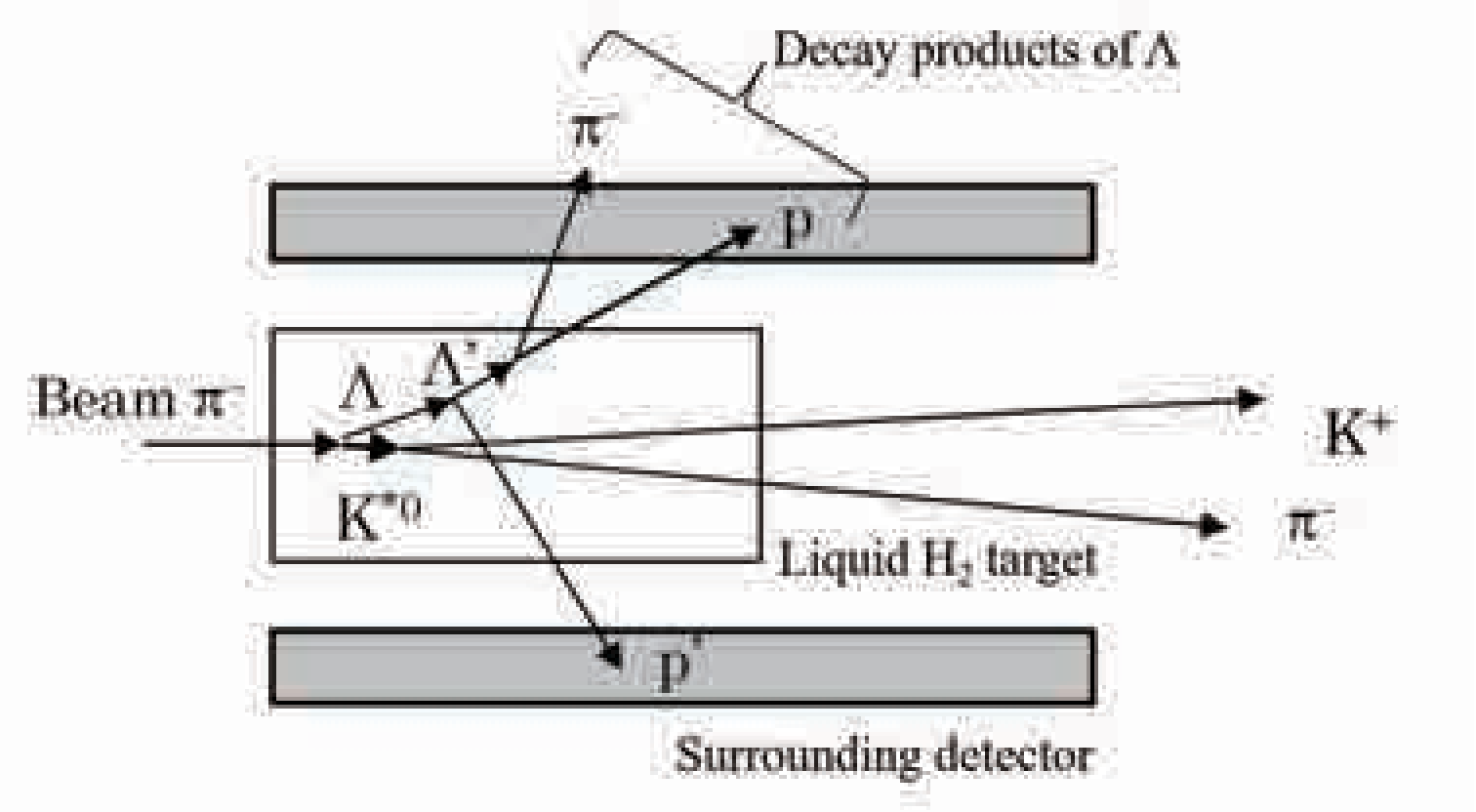

Figure 1: The schema of the $\pi^- p \to K^{*0}\Lambda$ reaction followed by the elastic Λp scattering in the liquid hydrogen target.

E40 experiment, which is the Σp scattering experiment at the K1.8 beam line.

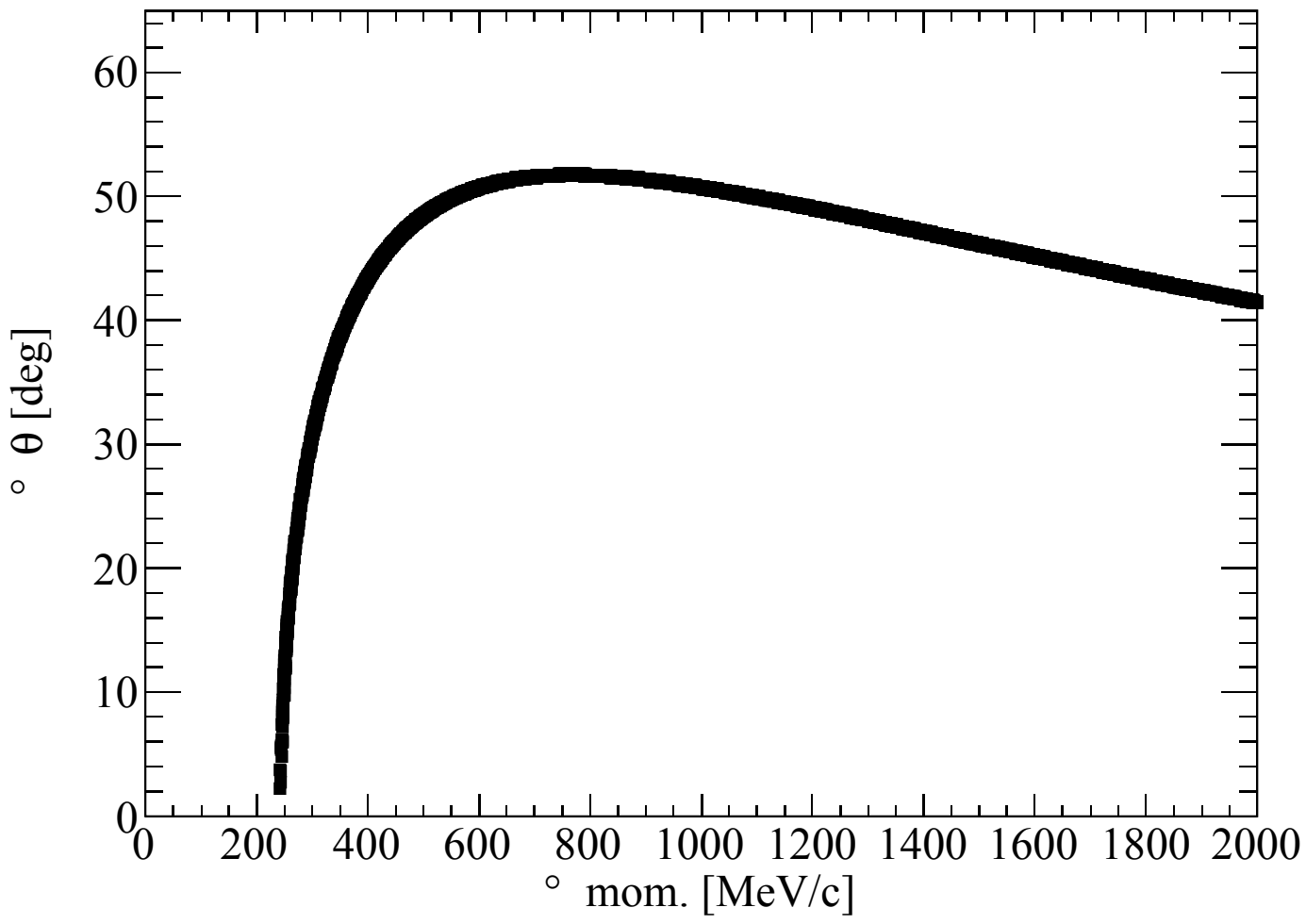

Figure 2: The relation of the production angle and the momentum of Λ via $\pi^- p \to K^*(892)^0\Lambda$ when the beam momentum is 10 GeV/c.

## 2.2 Expected result and impact

Fig 3 shows the total cross section of the Λp scattering measured by the past experiments and the theoretical calculation by the chiral effective field theory. If this experiment is successfully performed, a few percent statistical accuracy is expected. The data points shown in Fig. 3 will be drastically improved and give strong constraint to the theoretical models. In Fig. 3, one can find the ΣN cusp around $p_{lab} = 650$ MeV/c. The cusp shape strongly depends on the model. All the theoretical models predicted this shape, but it was not observed experimentally. Measuring the cusp shape precisely also gives a strong constraint to the models. In addition, Fijiwara *et al.* suggested that the amplitude of the cusp depends on the transition probability between ${}^3P_1$-${}^1P_1$. This is caused by the anti-symmetric LS force [5]. Precise measurement of the Λp scattering cross section may extract the amplitude of the anti-symmetric LS force.

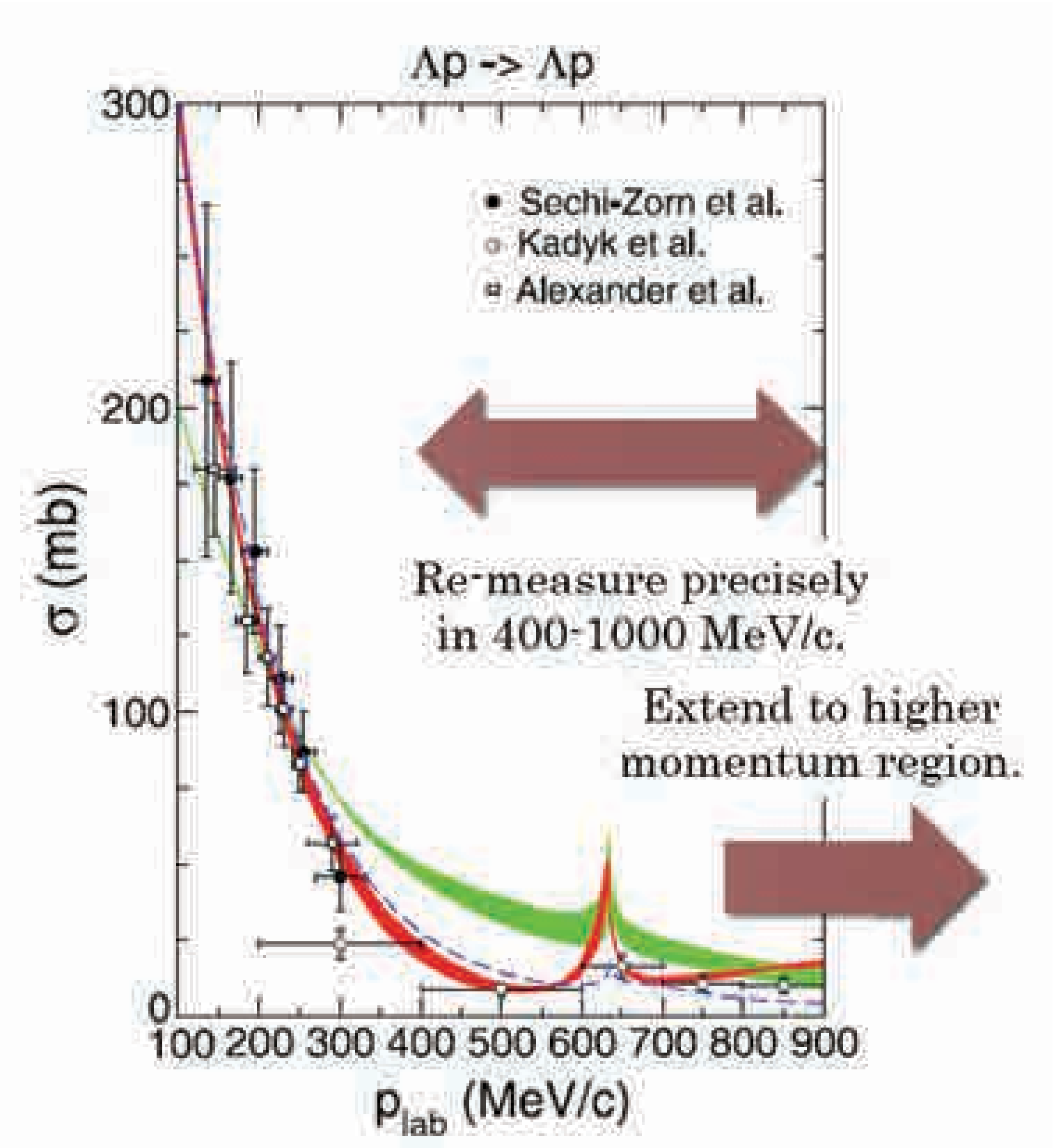

Figure 3: The total cross section of the elastic Λp scattering, which were calculated by the chiral effective field theory and measured by the past experiments, cited from Ref. [4]. The $x$ axis is the momentum of Λ. The red and green bands represent the cross section calculated by the chiral effective field theory. Markers show the measured cross sections.

## 3 Summary

The observation of the 2 $M_\odot$ neutron star triggered the hyperon puzzle; the present nuclear physics cannot support such the massive neuron star. The many body force between $\Lambda$ and nucleons is a candidate of the extra repulsion to support the massive neutron star and it will be investigated by the precise measurement of the hypernuclear level energy. However, the present two-body $\Lambda$N interaction is not enough precise to discuss the many body effect from the hypernuclei data due to the lack of the scattering data. Therefore, today, the $\Lambda$p scattering experiment is desired.

The author introduces the new experimental idea for the $\Lambda$p scattering at the J-PARC high-p beam line. The features of this idea are to use the reaction accompanying exited kaon instead of $K^0$ for the $\Lambda$ production and to use the high beam momentum of 5-10 GeV/c. We will obtain $10^7$ tagged $\Lambda$ with in one month beam time by using the $D^*$ spectrometer for the J-PARC E50 experiment. It is expected that the cross section of the $\Lambda$p scattering is determined with a few percent statistical accuracy.

## References

[1] P. B. Demorest *et al.*, Nature, **467**, 1081-1083 (2010).

[2] M. Isaka, Y. Yamamoto and Th. A. Rijken, Phys. Rev. C, **95**, 044308 (2017).

[3] M. Aguilar-Benitez *et al.*, Z. Physik C, Particles and Feilds **6**, 192-215 (1980).

[4] J. Haidenbauer *et al.,*, Nucl. Phys. A, **915**, 24-58 (2013).

[5] Y. Fujiwara *et al.*, Prog. in PNP, **58**, 439 (2007).

# Hyperon Beams for Hadron Spectroscopy

**Kotaro Shirotori**[1]

[1]Research Center for Nuclear Physics (RCNP), Osaka University

Ideas of experiments by using hyperon beams for investigating properties of hadrons from the production processes of ordinary and exotic states via the hyperon scattering method are proposed. The first idea is an investigation of the diquark correlation from the production rates of excited states by $\Lambda$ and $\Sigma$ beams. The second one is an the investigation of the hadron molecular state from the particular reaction channels of hyperon beams such as the $\Sigma\ \pi \to \pi\ \Sigma$ reaction by $\Sigma$ beams for studying the properties of $\Lambda(1405)$.

## 1 Physics motivation with its impact and expected result

In the J-PARC hadron experimental facility, various secondary beams such as pions, kaons and anti-protons are provided. Various experiments by using high-intensity secondary beams have been proposed. On the other hand, the experimental channels from hyperon beams have not been well used from the point of view to investigate properties of hadrons. The experiments by using hyperon beams have essential roles for studying properties of hadrons, because we can control properties of the projectile particles, degree of freedoms of spin and isospin components of constituent quarks inside of hyperon beams. In addition, the experiment using hyperon beams is expected to be one of the unique experiments in the world, because there may be no facilities except for J-PARC to use both intense hadron beams and large size experimental target having large diameter and thickness. Therefore, the information from experiments with hyperon beams can provide us quite unique knowledges of hadrons. In this documents, the experimental ideas by using hyperon beams are proposed.

### 1.1 Leading particle effect

For investigating properties of hadrons from hyperon beams, "Selectivity" is a key. In the high-momentum hadron reactions, a production process is contributed from many hadron resonances. There are no contributions from particular resonances so that we can assume the reaction processes from the point of view of quarks recombination between projectile and scattered hyperon and target nucleon. The properties of reaction processes are mainly taken into account from both initial and final states including degree of freedoms of the constituent quark components. For example, $\Lambda$ and $\Sigma$ have the initial states of good and bad diquark correlation, respectively. In the hyperon-nucleon scattering process with a high momentum, diquark are exchanged between projectile and scattered particles. The diquark correlation in the initial states can expected to affect the production of the final states which reflect fraction of wave functions of good and bad diquark correlation in the $\Lambda$ and $\Sigma$ particles. Therefore, the dependences from initial constituents of projectile particles can be expected in the production processes.

The selectivity to produce the hadron resonances is observed in the old CERN experiment (WA89) [1]. In the experiment, hadron resonances were produced by using $\Sigma^-$ beams with a momentum of 345 GeV/$c$. The fragmentation process is a main reaction in such high momentum region. The production enhancements of $\Sigma^*$ resonances were observed according to the dependence of spin (diquark) structure of projectile $\Sigma^-$. By comparing the production ratio between $\Sigma^{*+}$ and $\Sigma^{*-}$, $\Sigma^{*-}$ had larger production yield in the higher momentum fraction regions ($x_F = 0.3 - 0.8$) in the case of octet baryons ($\Sigma(1189)$ and $\Sigma(1660)$), while both $\Sigma^{*+}$ and $\Sigma^{*-}$ had the same yield in the case of decuplet baryons ($\Sigma(1385)$). The enhancement is called "Leading particle effect".

If the leading particle effect remains and appears in the region where hyperon beams have momentum of several GeV/$c$, we can use the effect to test the diquark correlation in the hyperon sector. In the case of the hyperon-nucleon scattering between ground states, $\Lambda/\Sigma$ ($\Lambda\ p \to \Lambda\ p \Leftrightarrow \Lambda\ p \to \Sigma\ p$, $\Sigma\ p \to \Sigma\ p \Leftrightarrow \Sigma\ p \to \Lambda\ p$,), the ratios of production rates are expected to be some particular values according to the diquark configuration of $\Lambda$ and $\Sigma$. By taken into account the SU(6) quark model, the fraction of wave functions of the diquark correlation in the ground state baryons are exactly determined. When we have scattering experiments between those ground states baryons, particular ratios of production rates are expected according to the fraction of wave functions of the good and bad diquark correlation in $\Lambda$ and $\Sigma$.

The test of the diquark correlation by using the leading particle effect can also be used for the investigation of excited baryons. By the study of the production process of the binary reactions ($\pi^-\ p \to Y_c^*\ D^*$ and $\pi^-\ p \to Y^*\ K^*$) from meson beams, we found the strong dependence of the production rate according to the diquark correlation of produced baryons [2]. From the analogy in the case the binary reaction, we can also expect the production rate depending on the internal structure of hyperon beams. By using different particles for example $\Lambda$ and $\Sigma$, it corresponds to control initial diquark states for producing hyperon resonances. From the naive expectation, for example, the $\lambda$-mode excited states having the good diquark can be produced with higher production rates by using $\Lambda$ beams, while the excited states having the bad diquark are produced with higher production rates by using $\Sigma$ beams, and vice versa.

It is expected for possibilities to control the production rate of hyperon resonances by changing beam hyperons. Therefore, the production rate may gives us information of internal structure of hyperon resonances. It means that one of the evidences of the diquark correlation in the hyperon sector if the production ratios are satisfied for the prediction assuming the diquark correlation. Hyperon beams can be used for the studies of excited hyperons by selecting the diquark correlation in the initial state in beam hyperons. Not only by using $\Lambda$ and $\Sigma$ but also $\Xi$ and $\Omega$, the similar investigation can be performed from hyperon beams with multi-strangeness.

## 1.2 Study of properties of molecular states

In order to produce excited baryons from hyperon beam, the inverse-reaction from $\Lambda$, $\Sigma$ and $\Xi$ can be one of the unique methods. For studying properties of excited states from the inverse-reaction processes, the particle-based picture is taken into account while the constituent quark picture is considered from the leading particle effect.

$\Lambda(1405)$ is known as an exotic resonance whose internal structure cannot be understood by the naive quark model. The Chiral unitary model tells us that $\Lambda(1405)$ is a hadron molecular state which has two resonance poles of $\bar{K}N$ and $\pi\Sigma$ from the calculation of the energy dependent $\bar{K}N$ potential. For understanding the unique properties of $\Lambda(1405)$, there are many approaches by using various experimental methods with $K$, $\gamma$ and proton beams. In those experiments, the virtual $\bar{K}N$ scattering is used for producing $\Lambda(1405)$ below the $\bar{K}N$ threshold. Then, the $\pi\Sigma$ decay modes are detected for measuring the line shapes of the invariant mass spectra of $\Lambda(1405)$. The $\bar{K}\ N \to \pi\ \Sigma$ channel has been well studies from many experiments. On the other hand, the $\pi\ \Sigma \to \pi\ \Sigma$ channel has not been well studied due to the difficulty to perform the $\Sigma\pi$ scattering experiment. By using $\Sigma$ beams on the nucleon target, we can perform the $\Sigma\pi$ scattering experiment from the reaction between $\Sigma$ and cloud pions around nucleons so that only the information of the $\pi\Sigma$ contribution can be extracted from the $\Sigma\ \pi \to \pi\ \Sigma$ reaction. The experiment by $\Sigma$ beams can give us unique information for understanding the internal structure of $\Lambda(1405)$. The $\Sigma$ beam can be the direct probe to study the $\pi\Sigma$ pole.

From the recent studies from both theory [5] and experiment [4], $\Xi(1620)$ and $\Xi(1690)$ are expected to be candidates of exotic hadrons having narrow widths. Mass of $\Xi(1620)$ and $\Xi(1690)$ are located near the threshold of mass of $\bar{K}+\Lambda$ and $\bar{K}+\Sigma$ so that those resonances are expected to be molecular states of $\bar{K}\Lambda$ and $\bar{K}\Sigma$, respectively. From the the $\Lambda\ p \to \Xi(1620)\ K^+\ p$ and $\Sigma^-\ p \to \Xi(1690)\ K^+\ n$ reaction, it is possible to access those resonance states by probing the $\bar{K}\Lambda$ and $\bar{K}\Sigma$ components. The reaction by $\Xi$ beams can also produce $\Xi(1620)$ and $\Xi(1690)$ via $\Xi\ p \to \Xi(1620)/\Xi(1690)\ p$ from the reaction between $\Xi$ and cloud pions around nucleons. By comparing reactions from both $\Lambda/\Sigma$ and $\Xi$ beams, the properties of $\Xi(1620)$ and $\Xi(1690)$ can be well investigated.

### 1.3 Experimental method

Owing to use a high-intensity secondary beams by the J-PARC facility, it is possible to produce huge number of hyperons by tagging its momentum from the missing mass technique. For example, by using the $\pi^-\ p \to \Lambda\ K^{*0}$ reaction on the liquid hydrogen target, by measuring the scattered $K^{*0}$ by a forward spectrometer system from the $K^{*0} \to K^+\ \pi^-$ decay channel, we can obtain the momentum vector of the produced $\Lambda$ so that $\Lambda$ is tagged as beam. Then, the hyperon-nucleon scattering events will be measured inside of the experimental target. By assuming the cross section of 50 $\mu$b for the the $\pi^-\ p \to Y\ K^*$ reaction [6] and the beam intensity of $1.8\times10^8$ /spill, the yield of hyperon beams is expected to be $\sim1.5\times10^8$/ day with a 6-second duration. It is enough to study various hadron channels having a cross section of order of 1 $\mu$b.

The E50 spectrometer [7] which has a large acceptance, a capability to use high-intensity beams and a flexible data taking system by a streaming DAQ scheme is suitable setup to obtain hyperon beams effectively. A thick liquid target with a diameter and a length of $phi$10 cm and 57 cm, respectively, is used for the experiment. The produced hyperons are scattered to be larger angle direction along the secondary beam direction due to a large momentum transfer by measuring the scattered mesons produced by high-momentum beams at the forward direction. Both thick experimental target and large scattering angles of hyperons effectively gives us a large number of hyperon-nucleon scattering events. For detecting hyperon-nucleon scattering events, the target surrounding detector system should be installed at the target region.

## 2 What kind of beam and equipments are necessary ?

The experiment is proposed to perform at the J-PARC high-momentum beam line by using secondary beams. The experiment can be performed at both the present high-momentum beam line by using secondary beams and the extended high-momentum beam line. In the extended high-momentum beam line, there is an advantage to use enough space of the experimental area for installing the target surrounding detector system. For the measurements of hyperons (hyperon beam tagging), the charmed baryon spectrometer (E50 spectrometer) is used.

The experiments with the $s = -1$ hyperons can be performed at the high-momentum beam line by using secondary beams. The pilot studies with $s = -2$ and $-3$ hyperons will be performed at the same beam line. For the dedicated studies which require higher yield of hyperon beams due to small cross sections, experiments are necessary to be performed at the K10 beam line with a multi-purpose spectrometer which is similar to the E50 spectrometer. By the high-intensity $K$ beam at the K10 beam line, hyperon beams can be produced with one order higher yield. In order to produce enough number of $\Omega$ beams, it is necessary to use the K10 beam line.

# 3 Particle, Energy(momentum), Intensity and Experimental setup

For obtaining hyperon beams having information of its momentum vector, it is available for any hyperon production channels which can be detected by the spectrometer. Beam particles of $\pi$, $K$ and anti-proton having proper momenta for experiments can be used for the production of hyperon beams. For example, in the case of using the $\pi^- \, p \to Y \, K^*$ reaction, the beam $\pi^-$ momentum and intensity of 2−20 GeV/$c$ and 1.0−2.0×$10^8$ /spill are used, respectively.

The detection of the hyperon scattering events is performed by using the target surrounding detector system which is installed around the target region of the E50 spectrometer. The target surrounding detector system has detector components as follows,

- Solenoid type magnet (the conflicting to the magnetic field of the E50 spectrometer magnet is acceptable due to no high precision momentum analysis experiment.)
- Surrounding tracking detector for measuring momentum of scattered particles
- High-resolution timing detector for identifying scattered particles with high efficiency
- Vertex detector for measuring positions of hyperon-nucleon scattering events.

# 4 Expected duration of the beam time (yield estimation)

By assuming the cross section of 100 $\mu$b for the the $\pi^- \, p \to Y \, K^*$ reaction, the yield of hyperon beams is expected to be 3.1×$10^8$ /day with a 6-second duration as follows,

$$(100.0 \times 10^{-6} \times 10^{-24}) \times (4.0 \times 6.02 \times 10^{23}) \times (2.6 \times 10^{12}) \times (0.5) \sim 3.1 \times 10^8,$$

where thickness of the liquid hydrogen target is 4.0 g/cm$^2$, the beam intensity is 1.8×$10^8$ /spill, and spectrometer acceptance including detection efficiency is 0.50.

The yield of production of the excited state is expected to be 5.7×$10^3$ events by assuming the production cross section of 1 $\mu$b in the 30-days beam time as follows,

$$(1.0 \times 10^{-6} \times 10^{-24}) \times (4.0 \times 6.02 \times 10^{23}) \times (0.5 \times 3.1 \times 10^8 \times 30.0) \times (0.5) \sim 5.7 \times 10^3,$$

where thickness of the liquid hydrogen target is 4.0 g/cm$^2$, hyperon beam survival efficiency is 0.50, and target surrounding detector acceptance including detection efficiency is 0.50.

It is enough to investigate exotic hadrons having a cross section of order of 1 $\mu$b. If we use the $K^- \, p \to Y \, \pi$ reaction at the K10 beam line, it is expected for one order higher yield.

## References

[1] M. I. Adamovich *et al.*, Eur. Phys. J. C **22** 255-267 (2001).

[2] S. H. Kim, A. Hosaka, H. C. Kim, H. Noumi, K. Shirotori, Prog. Theor. Exp. Phys. 103D01 (2014).

[3] D. Jido *et al.*, NPA **725** 181 (2003)., T. Hyodo and D. Jido, Prog. Part. Nucl. Phys. **67**1 55-98 (2012).

[4] Belle collaboration, private communication.

[5] T. Sekihara, Prog. Theor. Exp. Phys. 091D01 (2015).

[6] D. J. Crennell *et al.*, Phys. Rev. D **6** 1220 (1972).

[7] K. Shirotori *et al.*, PoS(Hadron 2013) 130 (2013).

# Creation of $K^-K^-pp$ in view of Kaonic Proton Matter at J-PARC

**T. Yamazaki[1,2], Y. Akaishi[2,3], F. Sakuma[2], T. Yamaga[2], K. Tanaka[3], Y. Ichikawa[4], K. Suzuki[5], , M. Hassanvand[2,6]**

[1]Physics Department, University of Tokyo

[2] Nishina Center, RIKEN

[3]Institute for Particle and Nuclear Physics, KEK

[4] Advanced Science Research Center (ASRC), Japan Atomic Energy Agency (JAEA)

[5] Stefan Meyer Institute, Austrian Academy of Sciences

[6] Department of Physics, Isfahan University of Technology, Isfahan, Iran

We propose to produce a new neutral baryonic composite $K^-K^-pp$ as a precursor of kaonic proton matter (KPM): $(K^-p)_m^{I=0}$ multiplet that has recently been predicted [1, 2]. Following the DISTO example of a dense $K^-pp$ production in a p-p reaction with a door-way of $\Lambda^* = K^-p$ in high-energy $pp$ collisions at $T_p = 2.85$ GeV: $p + p \to K^+ + \Lambda^* + p \to K^+ + [K^-pp] \to K^+ + \Lambda + p$, we extend this super-coalescent reaction to double door-ways in pp collisions at $T_p \approx 8$ GeV:

$$p + p \to (K^+\Lambda^*) + (K^+\Lambda^*) \to K^+ + K^+ + [[\Lambda^*] - [\Lambda^*]] \to K^+ + K^+ + \Lambda + \Lambda.$$

This is shown to be feasible at J-PARC as well as at FAIR by accommodating a high-momentum ($T_p \approx 8$ GeV) proton beam line.

## 1 Introduction

Very recently a high-density neutral baryonic matter, *Kaonic Proton Matter (KPM)*,

$$(KPM)_m = [(K^-p)^{I=0}]_m \tag{1}$$

has been proposed [1, 2]. The KPM is a collective name for a neutral baryonic matter, which is composed of pairs of only protons and negative kaons ($\Lambda^* \equiv (K^-p)^{I=0}$) with a multiplicity $m$. Its simplest form with a multiplicity $m = 2$ is a double-kaonic di-proton: $K^-K^-pp = [(K^-p)^{I=0}]_{m=2}$. The present note is aimed at a precursor research plan toward *macroscopic astrophysical study of Kaonic Proton Matter (KPM)* based on *microscopic accelerator experiments* at J-PARC.

## 2 Toward KPM

The KPM, $(KPM)_m = [(K^-p)_m^{I=0}]$, is most likely a QCD condensed matter (quark-gluon bound states, QGB, in contrast to as QGP, quark gluon plasma) which is not stable at normal conditions on the earth. Such "unstable" matter can, however, be produced at an accelerator laboratory as metastable kaonic proton composites, such as $K^-p$ and $K^-pp$, but they decay and vanish in a short time, and thus, multiple cumulative production of larger-$m$ multiplets appears to be impossible. Nevertheless, it is pointed out from series of theoretical and experimental studies of kaonic nuclear interactions [1, 2, 3, 4, 5, 6, 7] that the basic attractive $\bar{K}N(I = 0)$ interaction increases with the surrounding nucleon density (due to the chiral symmetry restoration [8, 9, 10, 11] as well as the Heitler-London type multi-body covalence bonding [12, 13, 14]) so that the KPM mass per baryon decreases with $m$, approaching a long-lived or even a stable matter, as shown in Fig. 3 of [1].

In this way the hot quark-gluon plasma phases quickly passed away, and reach stable baryonic and quark-gluon bound states. The calculation given in [1, 2] indicates that the KPM becomes a perfectly stable baryonic ground state with $m > 8$. However, how to reach such stable KPM from any particle sources with conceivable reactions? This sounds to be a very difficult and virtually impossible problem, but we challenged to it in [1, 2]. We argued that the formation of such stable KPM matter might be possible in the early universe where primordial quarks and gluons were born at ultra high temperatures and densities, and eventually (but rapidly enough) reach lower temperatures by collisions and expansions till the temperatures become around an energy gap (say, 0.1 MeV, $\epsilon < \epsilon_{gap} \approx 10^5$ eV) below which nuclear interactions become cease so that quiet universes are realised. By then, most of anti-quarks are going to annihilate and the population of anti-matter is rapidly disappearing and baryogenesis of the whole universe is supposed to be nearly achieved. However, if some exceptionally remaining anti-quarks get trapped by deep holes of QCD bound states of $\bar{q}q$, they may survive as long-lived "impurities" that produce trapped "$\bar{q}q$ QCD atoms". The components of KPM (mainly, $K^- + p$, and hadrons) with strong interactions, and so, they seem to be different from so called Dark Matter. Before reaching KPM they pass through very violent interactions with many residues, but most of the residues have eventually gone. In the mean time the internal energy and momentum of KPM fell down below some threshold ones ($\approx$ 0.1 MeV) and KPM looks like a matter of very small energy gap so that they cannot shine by photon irradiation any more. This is a very distinct phenomena, because the total and partial charges are all zero. So, they might show very exotic properties similar to Dark Matter.

In [1, 2] we also investigated the stability and stiffness of KPM against external and internal collisions by calculating possible weak decays from $K^-$ and $p$ (for instance), which are embedded in very high density medium, and showed that the simultaneous weak decays from such dense composites are extremely retarded. Multi-body weak decays are virtually forbidden. It may act as a catalysis for the formation of a large scale metastable or even a stable KPM.

## 3 Increased $\Lambda^* - p$ interaction and predicted $\Lambda^*$ strangelets

The basic interaction, $\bar{K}N$, depends on the medium. The most important effects are as follows.

A) The CSR effect of the $I = 0$ $\bar{K}N$ interaction by the chiral symmetry restoration in dense nuclear matter [8, 9, 10, 11].

B) The Heitler-London type covalence effect of the migration of $K^-$*bosons* among protons [12, 13, 14, 15].

The multiplicative results of the above two kinds of effects cause substantial increase of the binding energy and nuclear shrinkage. Calculation of the ground state of a strangelet for variable multiplicity was carried out by the variational method "ATMS" [16]. As shown in Fig. 3 of [1], the binding energy per $m$ increases and the level energy decreases. For $m > 8$, the binding energy of $\Lambda^*$ becomes smaller than the neutron energy so that the neutron decay channel is closed and the nucleus becomes gentle and stabilised.

The most important unique properties of KPM lie in the formation of a high-density multi-body system with strong attractions among them, which result from the Heitler-London type exchange covalence force caused by migrating $K^-$ bosons among protons. They are too strong to allow the conventional relativistic mean-field (RMF) approach [27]. Four $\Lambda^*$'s formed in mean-field regime can possess only 4 covalent bonds, whereas those without mean-field constraint are allowed to form 12 covalent bonds (see Fig. 3 of [1]). So, the KPM system in principle cannot be constructed by the RMF model, that is strongly claimed in [28]. The authors of [28], however, argued that strong mutual attractions of $\Lambda^*$ would cause *unrealistic* saturations of the binding energies and densities of $\Lambda^*$'s, criticising the predicted formation of KPM. Contrarily though, we have clarified in the realistic calculations that the ever-increasing mutual attractions among $\Lambda^*$'s, which we considered in details in [1], are really the reason for creating the exotic situation that is not touched by [28]. Furthermore, the CSR effect additionally enhances the $I = 0$ $\bar{K}N$ attractive interaction.

## 4 Experimental information on the simplest composite $K^-pp$

We found theoretically [1], [22] that the simplest kaonic di-baryon state ($K^-pp$, a single precursor of the $m = 1$ KPM) can be highly populated in head-on collision of two high-energy protons, as

$$p + p \rightarrow K^+ + [[p - \Lambda^*]], \tag{2}$$

the intermediate compound being immediately followed by

$$[[p - \Lambda^*]] \rightarrow K^-pp \rightarrow p + \Lambda, \tag{3}$$

where $\Lambda^* = K^-p$ is the known $\Lambda$ (1405) resonance state that plays an important role as a doorway in this hard collision. In this reaction the $\Lambda^* - p$ pair is produced at a short distance and time, and thus, a dense $K^-pp$ state is easily accommodated by the $\Lambda^* - p$ pair. We can name this reaction " **super coalescent** " reaction. This situation is shown theoretically to take place when the distance of the $\Lambda^* - p$ pair is equivrated to a minimum value (namely, to a larger density). This relation (cross section versus the incident energy) can be used to " measure " the size of the produced $K^-pp$. In fact, the analysis of DISTO experimental data at $T_p = 2.85$ GeV [17, 18, 19] demonstrated the presence of a large bump in $M(K^+\Lambda)$ vs $M(p\Lambda)$ Dalitz plots of three dynamical variables, $p$, $\Lambda$ and $K^+$. It revealed a large population of $p + \Lambda$, proving not only its large population but the density of the formed $K^-pp$ to be so high as to cause a " **super coalescence** ".

Recently, the J-PARC E27 group performed another type of experiment on $d(\pi^+, K^+)$, to find the same $K^-pp$ by a totally different "inverse" reaction, which had been studied theoretically in comparison with the direct "$p + p$" reaction [13, 14, 25]. In an inclusive reaction, three dominant production channels of $\Lambda, \Sigma$ and their resonance states were clearly seen, whereas no expected signal for $K^-pp$ was observed in the inclusive spectra. On the other hand, when tagging by proton emission was applied, a distinct peak showed up nearly at the same mass as the $K^-pp$ peak of the DISTO experiment, but this peak intensity is only a few % of the total continuum. This is consistent with the expectation that the cross section for the $K^-pp$ signal in the inverse reaction as in the E27 reaction is calculated to be roughly 2% compared with the case of the DISTO experiment. This is understood, since there is no such super coalescence effect in the inverse E27 experimental arrangement. Thus, the two different experiments, DISTO and J-PARC E27, are jointly interpreted to show nearly the same peak intensities (after the efficiency correction by tagging) that reflects the two different efficiencies of dynamical origins.

Table 1: Observed energy levels of $K$-$pp$ by different reaction processes. $E$: energy (MeV), $M$: Mass (MeV/$c^2$), and $\Gamma$: width (MeV).

| Name | $E$ | $M$ | $\Gamma$ | Exp Group Ref. | Reaction | State suggested |
|---|---|---|---|---|---|---|
| X(2267) | -103 | 2267 | | DISTO (2010) [17] | $p(p, K^+)M(p\Lambda)$ | 1s |
| X(2273) | -95 | 2273 | | J-PARC E27 (2015) [20] | $d(\pi^+, K^+)M(p\Lambda)$ | 1s |
| X(2324) | -47 | 2324 | | J-PARC E15 (2018) [21] | $^3\text{He}(K^-, n)M(p\Lambda)$ | 2p |

Table 1 lists the mass, width and binding energy of each observed state. The two experimental results on the $K^-pp$ mass are nearly the same ($M(K^-pp)$ around 2.26 GeV/$c^2$), which indicate a binding energy for $K^-pp$ to be about 100 MeV. If we assume the observed states are the ground state of $K^-pp$, the binding energy for $K^-pp$ is approximately twice as large as the original prediction ($\approx$ 50 MeV)[4, **?**, 32]. The origin of these discrepancies might come from the assumed theoretical $\bar{K}N$ interactions that are taken from the Particle Data Group values. The observed experimental results can be understood if the $\bar{K}N$ interactions that are responsible for the ground state of $K^-pp$ are 17 % larger compared with the assumed PDG ones.

More recently, the J-PARC E15 experiment studied $^3\text{He}(K^-, n)\Lambda p$ reactions [21]. The experiment revealed another kind of peak in the invariant mass spectrum of $M(p\Lambda) \approx -40$ GeV/$c^2$. In this reaction there are still more interesting aspects to be solved in E27 and E15. Very recently, an additional E15 experiment of J-PARC reported a $K^-pp$ state mass as observed to be $M(K^-pp, E15) = 2.324 \pm 0.003$ $(stat)$ GeV/$c^2$ [21]. This is clearly distinct from the other two peaks that have already been mentioned, and suggests an excited state to be populated, if the two states with $M \approx 2.26$ GeV/$c^2$ are for the ground state. Further experiments with more statistics are waited for. Table 1 summarises the presently obtained energy levels of $K^-pp$ that are statistically robust experimental values [17, 20, 21].

We note a caution between the level assignment and the observed transition energy that the new E15 is now reporting. Only one robust assignment of the three observations listed in Table would be to assign an excited state at $M = 2324$ MeV to which the observed 2324-MeV transition is assigned to an excited state.

## 5 Toward the experimental production of $K^-K^-pp$ by the super coalescent direct reactions

It was shown theoretically [23, 24] that the important composite $K^-K^-pp$ ($m = 2$) can be produced by increasing the projectile proton energy from 3 GeV to 7 GeV in the $p + p$ experiment used for the DISTO experiment. Since we know that the p + p collision at 3 GeV produces one pair of $\Lambda^* + K^+$ with close proximity to a participating proton [13, 14] we predict that the 7 GeV protons will produce two pairs of $K^+ + \Lambda^*$ at one hard (short-distance) collision (see Fig. 1). Thus, we expect an efficient production of two $\Lambda^*$ doorways that become direct sources for $K^-K^-pp$, as shown below.

$$p + p \to [K^+ + \Lambda^*] + [K^+ + \Lambda^*] \to 2 \times K^+ + [[\Lambda^* - \Lambda^*]], \quad (4)$$

$$[[\Lambda^* - \Lambda^*]] \to K^-K^-pp \to \Lambda + \Lambda. \quad (5)$$

Calculation of the invariant-mass $M(\Lambda\Lambda)$ spectra for the parent composite nucleus $K^-K^-pp$ is shown in Fig. 2 and 3. Strikingly, it shows a distinct peak of $\Lambda + \Lambda$ that grows when the produced composite state ($K^-K^-pp$) besomes deeper bound. This is the same mechanism as predicted in connection with the DISTO experiment. Furthermore, as demonstrated in the DISTO mechanism, the produced $K^-pp$ state as observed in J-PARC E15 experiment decays largely to two-body final states $X \to \Lambda + p$ with substantial reduction of three-particle decay channels: $X \to \Sigma + \pi + p$. This feature, "enhanced $\Lambda - \Lambda$ emission with suppressed three-body ($\Sigma + \pi + p$) continuum" looks very encouraging in planning and commissioning future experiments to search for direct production of $\Lambda - \Lambda$ not only in $p - p$ but also in heavy-ion collisions [25, 26].

Figure 2 shows partial-wave contributions with spin weights ($L = 0, 1, 2$ and Total) to the differential cross sections in the standard case: $T_p = 7.0$ GeV, obtained from [24]. The cross section is dominated by $L = 0$, which is peaked at around 2630 MeV. The $L = 1$ contribution is not large, but is still visible at $M(K^-K^-pp) = 2720$ MeV/$c^2$.

Differential cross sections for various ground-state energies, $E$, of the $K^-K^-pp$ system for $T_p = 7.0$ GeV, $\Gamma = 150$ MeV are shown in Fig.3. It is a striking property of the $p - p$ direct collisions, namely, the dominant s-wave peak grows up, as the bound state energy increases.

The $K^-K^-pp$ (ground-state) cross section versus the incident proton energy in lab. is shown in Fig. 4. This diagram will guide the experimentalists to choose the incident energy.

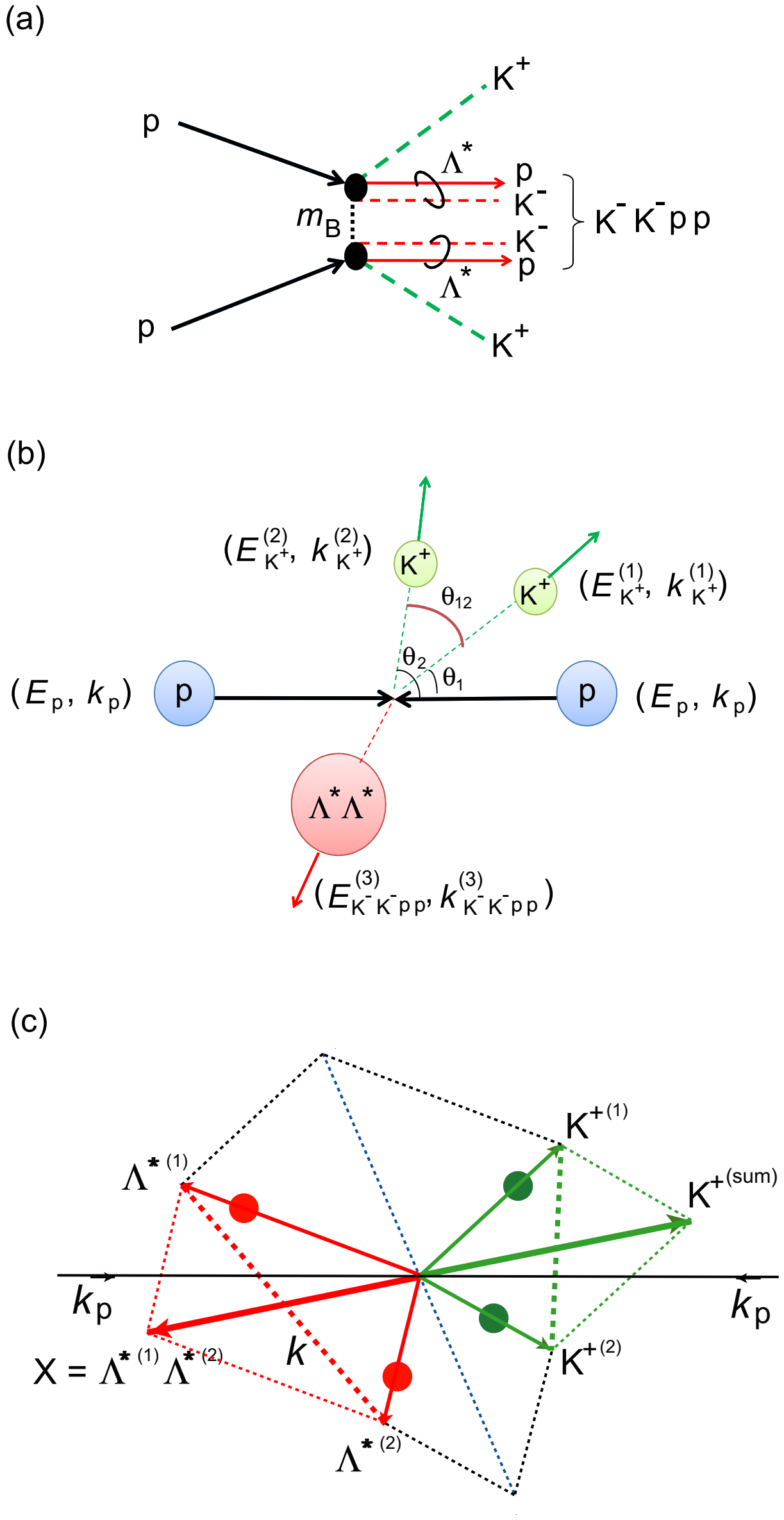

Figure 1: Expected Projected $IM(\Lambda\Lambda)$ and $IM(\Sigma\pi)$ spectra. a) Impact of p+p collisions in the super-coalescent reactions. b) Kinematics. From [24].

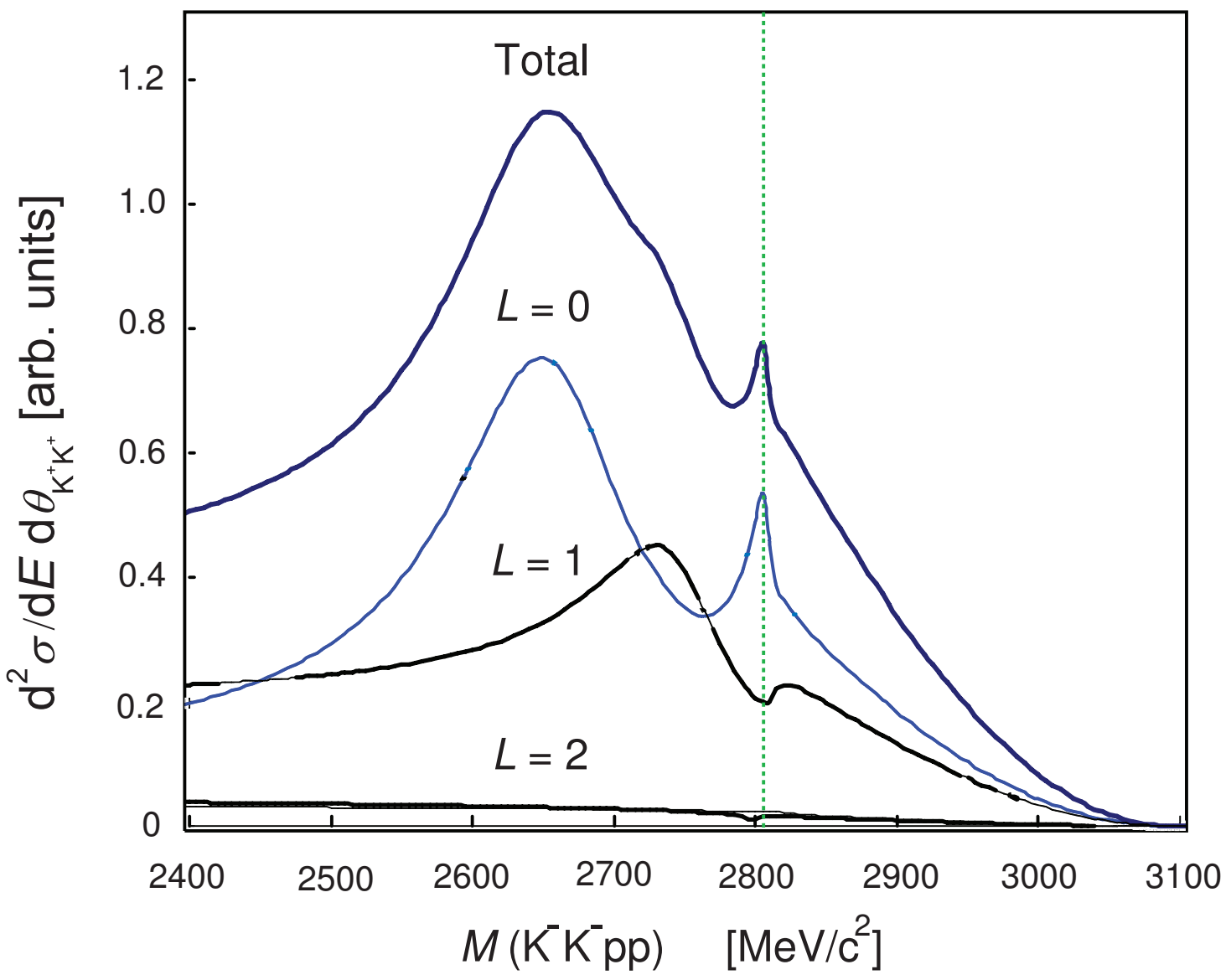

Figure 2: (Colour online) Partial-wave contributions with spin weights ($L = 0, 1, 2$ and Total) to the differential cross sections in the standard case: $T_p = 7.0$ GeV, $b = 0.3$ fm and $\theta_{12} = 180$.
From [24].

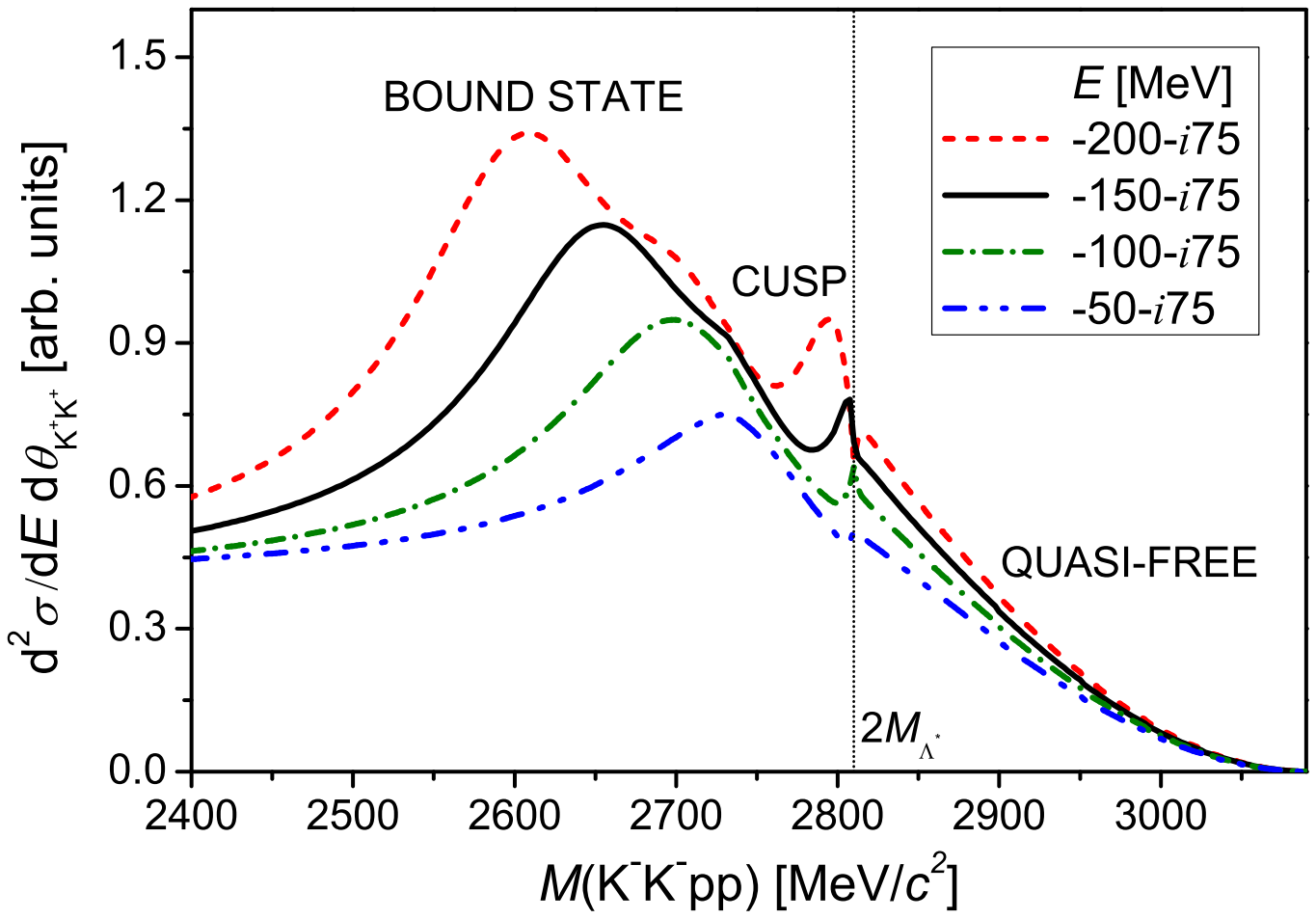

Figure 3: (Color online) Differential cross sections for various ground-state energies, $E$, of the $K^-K^-pp$ system for $T_p = 7.0$ GeV, $\Gamma = 150$ MeV, $b = 0.3$ fm and $\theta_{12} = 180$. For the cases of the $BS/QF$ ratio = 0.43, 1.09 and 1.21 for $E = -100, -150$, and $-200$ MeV, respectively. From [24].

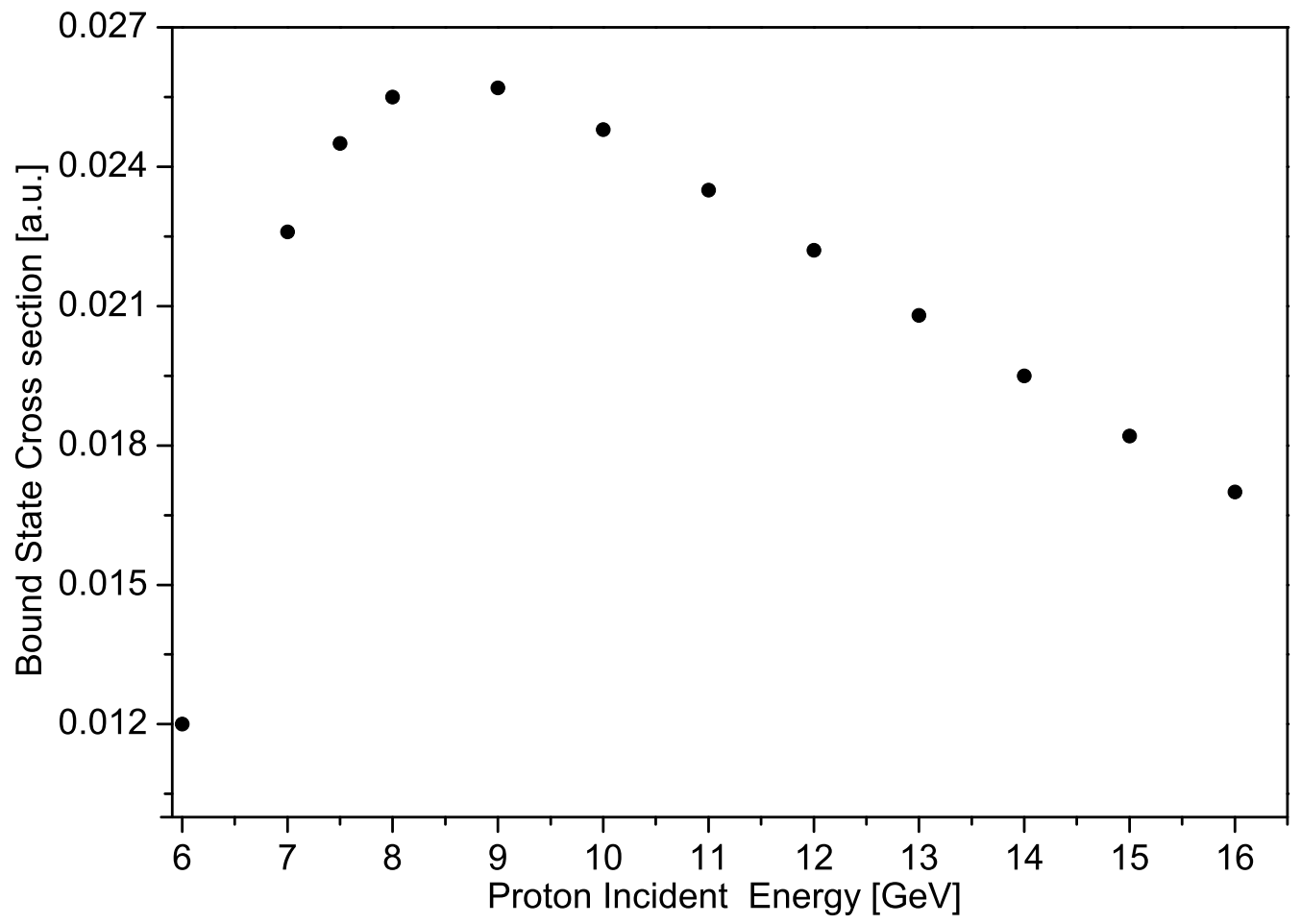

Figure 4: The $K^-K^-pp$ (ground-state) cross section versus the incident proton energy in lab.

# 6 Future scopes

Besides the above mentioned special Super-Coalescent reaction mechanisms, there are ordinary reactions used to populate $K^-$ induced reactions. For instance, using a $K^-$ incident beam one can produce $K^-K^-pp$ together with $K^0$ as a spectator:

$$d + K^- \to K^0 + \Lambda^*\Lambda^* \to K^0 + K^-K^-pp \to K^0 + \Lambda + \Lambda \tag{6}$$

In such arrangements, also with other light projectiles and targets various phenomena will be studied. In the near future, it will become possible to study multi-particle emission from HI-HI collisions. After passing the phase-1 of $K^-K^-pp$ experiments, we proceed to survey experiments of $m = 3, 4, 5$, by changing the incident beam energy. It would be very exciting to observe the cross-over to the emission of metastable many-body $\Lambda^*$ related emissions. Our first ultimate goal would be to detect emission of nearly stable $\Lambda^*$ clusters beyond a high-energy limit.

Through various international collaborations with the CERN LHC-ALICE experiments, the CBM (Condensed Baryonic Matter) of FAIR, number of new phenomena in relation with the KPM search will be explored. When an impurity object is produced in a nuclear compound medium, the object may find a comfortable space and time for very short lived intermediate lifetimes.

A preliminary layout of the new beam lines at J-PARC Hadron Hall is under discussion. Many of the new experimental projects at J-PARC may have common goals and interests with some of the FAIR [33, 34]. International collaborations are to be stimulated and pursued.

### 6.1 Particle, energy, intensity, and experimental setup

We need slowly extracted incident proton beam from an acceleration ring to several GeV/c protons. Typical beam momenta to be used for experiments are varied.

Incident beam species: at first, 1) proton, 2) eventually, deuteron, 3) alpha particles, etc.

Standard devices for hadronic reaction/decay experiments. Can be commonly used with other experiments.

For presenters of proposing experiment:
- Items we can collaborate in view of physics and R&D
- Items we can collaborate in view of physics and R&D

### 6.2 Preliminary and exploratory stages

The present team consists of the following subgroups.

* SubGroup A: Theoretical studies of the reaction processes.

* SubGroup B: Design and construction of experimental devices. Simulation studies to establish experimental procedures.

* SubGroup C: Design and construction of a slow extraction ring and beam lines.

The authors acknowledge the support of Japanese Monbu-Kagakusho for the support. They are grateful fo Prof. Norbert Hellmann of GSI-FAIR for encouraging discussion.

## References

[1] Y. Akaishi and T. Yamazaki, Phys. Lett. B 774 (2017) 522-526.

[2] Y. Akaishi and T. Yamazaki, arXiv:1903.10687v2[nucl-th].

[3] Y. Akaishi and T. Yamazaki, Phys. Rev. C **65** (2002) 044005.

[4] T. Yamazaki and Y. Akaishi, Phys. Lett. B **535** (2002) 70.

[5] A. Doté, H. Horiuchi, Y. Akaishi and T. Yamazaki, Phys. Lett. B **590** (2004) 51; Phys. Rev. C **70** (2004) 044313.

[6] T. Yamazaki, A. Doté, Y. Akaishi, Phys. Lett. B **587** (2004) 167.

[7] Y. Akaishi and T. Yamazaki, Int. J. Mod. Phys. A **24** (2009) 2118.

[8] Y. Nambu and G. Jona-Lasinio, Phys. Rev. **122** (1961) 345; **124** (1961) 246.

[9] T. Hatsuda and T. Kunihiro, Prog. Theor. Phys. **74** (1985) 765.

[10] U. Vogel and W. Weise, Prog. Part. Nucl. Phys. **27** (1991) 195.

[11] G.E. Brown, K. Kubodera, and M. Rho, Phys. Lett. B **192** (1987) 273.

[12] W. Heitler and F. London, Z. Phys. **44** (1927) 455.

[13] T. Yamazaki and Y. Akaishi, Proc. Jpn. Acad. B **83** (2007) 144.

[14] T. Yamazaki and Y. Akaishi, Phys. Rev. C **76** (2007) 045201.

[15] Y. Akaishi, T. Yamazaki and M. Hassanvand, in preparation.

[16] Y. Akaishi, M. Sakai, J. Hiura and H. Tanaka, Prog. Theor. Phys. Suppl. No.56 (1974) 6.

[17] T. Yamazaki *et al.*, Phys. Rev. Lett. **104** (2010) 132502.

[18] P. Kienle *et al.*, Eur. Phys. J. A **48** (2012) 183.

[19] K. Suzuki, T. Yamazaki, M. Maggiora, and P. Kienle, Prog. Exp. Theor. Phys., JPS Conf. Proc. **17** (2017) 082003.

[20] Y. Ichikawa *et al.*, Prog. Exp. Theor. Phys. 021D01 (2015).

[21] S. Ajimura *et al.*, arXiv:1805.12275v2 [nucl-ex].

[22] S. Maeda, Y. Akaishi and T. Yamazaki, Proc. Jpn. Acad. B **89** (2013) 418-437.

[23] T. Yamazaki, Y. Akaishi and M. Hassanvand, Proc. Jpn. Acad. B **87** (2011) 362 .

[24] M. Hassanvand, Y. Akaishi and T. Yamazaki, Phys. Rev. C **84** (2011) 015207.

[25] Y. Akaishi, private communication, 2018.

[26] F. Sakuma, to be published, 2018

[27] D. Gazda, E. Friedman, A. Gal and J. Mareš, Phys. Rev. **C 77** (2008) 045206.

[28] J. Hrtankova *et al.*, arXiv:1805.11368v1 [nucl-th]

[29] L.C. Gomes, J.D. Walecka and V.F. Weisskopf, Ann Phys. **3** (1958) 241.

[30] ALICE Collaboration, CERN-PH-EP-2015-025 (02 Feb 2015); CERN-PH-EP-2015-069 (13 Mar 2015).

[31] P. Braun-Munzinger and J. Stachel, arXiv:1101.3167v1 [nucl-th] (17 Jan 2011).

[32] S. Maeda, Y. Akaishi and T. Yamazaki, JPS Conf. Proc. **17** 082007 (2017).

[33] H. Tamura and K.H. Tanaka, Reopening of Research Activities on Strangeness Nuclear Physics at J-PARC, Nucl. Phys. News, 27 (2017) 21.

[34] N. Herrmann, private communication (2018).

# Measurement of the cross sections of $\Omega^- p$ scatterings

**Hitoshi Takahashi**

Institute of Particle and Nuclear Studies, KEK

We would like to perform an $\Omega^- p$ scatterings experiment to investigate the octet-decouplet baryons interaction. By utilizing a high momentum kaon beam at the K10 beam line and the J-PARC E50 spectrometer with additional detectors surrounding a liquid hydrogen target, we can obtain more than 1000 events of $\Omega^- p$ elastic scatterings in 1 month of beam time.

## 1 Physics motivation

By introducing the strangeness, the nucleon can be treated as a member of the octet baryons under flavor SU(3) symmetry. The octet-octet baryons interaction has been investigated very actively in past three decades. On the other hand, the $\Delta$ baryon also plays a very important role in ordinary nuclear physics, for example, three-body nuclear force and the equation of state of neutron stars. The $\Delta$ is a member of the decouplet baryons. Only the $\Omega^-$ is a long lived particle among the decouplet baryons and its interaction with a nucleon can be measured directly. In this meaning, the experimental determination of $\Omega N$ interaction together with the construction of theoretical framework of the decouplet-octet baryons interaction will give us a new knowledge on the baryon-baryon interaction. Under flavor SU(3) symmetry, the decouplet-octet baryons interaction can be classified as $\mathbf{10} \otimes \mathbf{8} = \mathbf{35} \oplus \mathbf{27} \oplus \mathbf{10} \oplus \mathbf{8}$ (Fig. 1).

The experimental measurements of elastic and inelastic scattering cross sections give us direct information on interactions of the particles. Compared to abundant data of $NN$ scatterings, however, current hyperon scatterings data are extremely limited. For $S = -1$ hyperons, there are limited data of $\Sigma N$ and $\Lambda N$ scatterings measured by old bubble-chamber experiments and those using scintillating-fiber active targets at KEK PS [1]. The J-PARC E40 experiment is now being prepared to measure the cross sections of $\Sigma p$ scatterings with 100 times higher statistics than past experiments [2]. As for $\Xi N$ scatterings, more limited number (several tens in total in the world) of events were reported [3]. Currently no data for $\Omega N$ scatterings.

In theoretical studies, there are several calculations on the $\Omega N$ interaction. However, the results varied in a wide range as summarized in Table. 1.

We, therefore, would like to measure the cross sections of $\Omega^- p$ scatterings for the first time at the extended hadron experimental facility.

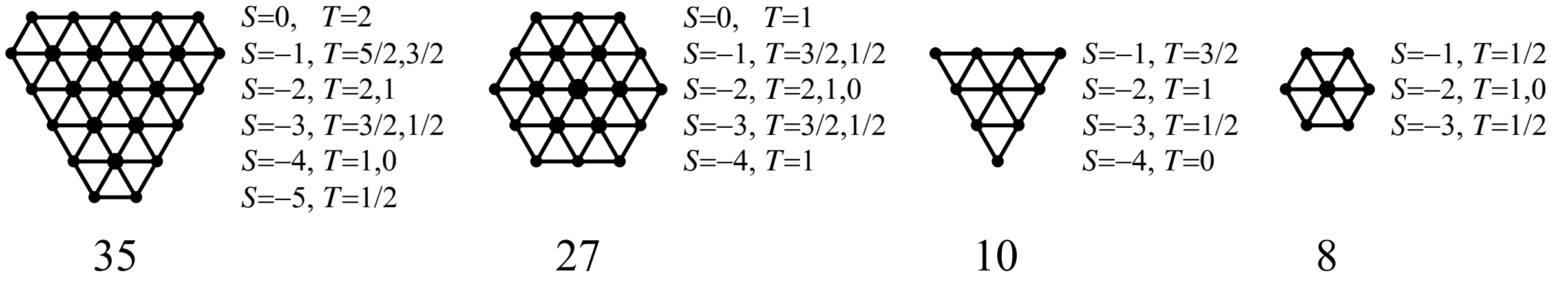

Figure 1: Irreducible representations for decouplet(**10**)-octet(**8**) baryons system based on flavor SU(3) symmetry.

## 2 Experimental method

We plan to utilize 6 GeV/$c$ $K^-$ beam at the K10 beam line to produce $\Omega^-$ hyperons. RF separator(s) would be required to obtain a moderate purity for such a high momentum. A liquid hydrogen target is used both for the production and for scattering of $\Omega^-$ hyperons.

The experimental technique to detect $\Omega^- p$ scatterings is quite similar to that used in the J-PARC E40 experiment. The $\Omega^-$ production via $K^- p \to \Omega^- K^+ K^{*0}$ reaction is identified with J-PARC E50 spectrometer, which has a large acceptance, a multiple tracking capability, and good particle identification in a wide momentum range [9]. In order to detect the recoil protons and the decay daughters of $\Omega^-$ hyperons ($\Omega^- \to \Lambda K^-$ and $\Lambda \to p\pi^-$) after $\Omega^- p$ scatterings, we will add some trackers and counters surrounding the target.

## 3 Yield estimation

According to the past experiments [10], the inclusive cross section of $K^- p \to \Omega^- X$ is about 1.5 $\mu$b at the $K^-$ momentum of 6 GeV/$c$. We assume 0.5 $\mu$b for the cross section of the $K^- p \to \Omega^- K^+ K^{*0}$ reaction, although the exclusive cross section of the reaction is unknown. The cross section of $\Omega^- p$ elastic scattering is assumed not to be so different from that of $\Xi^- p$ elastic scattering [3], and the value of 5 mb is adopted.

The intensity of a $K^-$ beam is expected as $7 \times 10^6$ /spill with a 25 kW beam loss at the production target. The thickness of the liquid hydrogen target is 10 g/cm$^2$, so the effective thickness of the target for $\Omega^- p$ scattering is 5 g/cm$^2$. The efficiency to identify the $\Omega^-$ production is assumed to be same as that in the planed $\Xi^*$ spectroscopy experiment [11], where the acceptance of E50 spectrometer, the live time of data acquisition, the tracking efficiency, and the efficiencies for pions and kaons identifications are considered. As for the detection of $\Omega^- p$ scatterings, we assume the overall efficiency to be 10%. The branching ratios of $\Omega^- \to \Lambda K^-$, $\Lambda \to p\pi^-$, and $K^{*0} \to K^+\pi^-$ decays are also taken into account.

With these assumption, the estimated yield of $\Omega^- p$ elastic scatterings is about 40 events/day. With the 30 days beam time, we can obtain more than 1000 scatterings. It should be noticed that $\Xi^- p$ scatterings can also be measured simultaneously in this experiment. Since the production cross section of $\Xi^-$ is 2 orders of magnitude larger than that of $\Omega^-$, much richer data of $\Xi^- p$ scatterings can be obtained.

Table 1: Theoretical calculations on $\Omega N$ systems.

| model | result |
|---|---|
| relativistic quark model [4] | deeply bound ($B = 140$ MeV) |
| chiral SU(3) quark model [5] | weakly or deeply bound ($B = 4 \sim 50$ MeV) |
| quark delocalization color screening model [6] | deeply bound ($B = 62$ MeV) |
| (extended) chiral SU(3) quark model [7] | weakly bound or unbound |
| lattice QCD [8] | weakly bound ($B = 18.9(5.0)(^{+12.1}_{-1.8})$ MeV) |

## 4 Summary

The $\Omega^-$-nucleon interaction is a new interaction which is not included in those between flavor SU(3) octet baryons investigated experimentally and theoretically so far. The K10 beam line at the extended hadron experimental facility will provide the high momentum $K^-$ beam with a high intensity and purity, and is suitable for carrying out the $\Omega^- p$ scattering experiment. We expect more than 1000 events of $\Omega^- p$ elastic scatterings can be detected in a month with the J-PARC E50 spectrometer with additional detectors around the liquid hydrogen target.

## References

[1] R. Engelmann, H. Filthuth, V. Hepp, E. Kluge, Phys. Lett. **21**, 587 (1966).
F. Eisele, H. Filthuth, W. Föhlisch, V. Hepp, G. Zech, Phys. Lett. **37B**, 204 (1971).
J.K. Ahn *et al.*, Nuclear Phys. **A648**, 263 (1999).
J.K. Ahn *et al.*, Nuclear Phys. **A761**, 41 (2005).
Y. Kondo *et al.*, Nuclear Phys. **A676**, 371 (2000).
T. Kadowaki *et al.*, Eur. Phys. J. **A15**, 295 (2002).
J. Asai *et al.*, Jpn. J. Appl. Phys. **43**, 1586 (2004).

[2] K. Miwa *et al.*, J-PARC E40 proposal,
`http://j-parc.jp/researcher/Hadron/en/pac_1101/pdf/KEK_J-PARC-PAC2010-12.pdf`

[3] G.R. Charlton, Phys. Lett. **B32**, 720 (1970).
R.D.A. Dalmeijer *et al.*, Nuovo Cimento Lett. **4**, 373 (1970).
R.A. Muller, Phys. Lett. **B38**, 123 (1972).
J.M. Hauptman *et al.*, Nucl. Phys. **B125**, 29 (1977).
J.K. Ahn *et al.*, Phys. Lett. **B633**, 214 (2006).

[4] T. Goldman *et al.*, Phys. Rev. Lett., **59**, 627 (1987).

[5] Q.B. Li, P.N. Shen, Z.Y.Zhang, and Y.W. Yu, Nucl. Phys. **A683**, 487 (2001).

[6] H. Pang *et al.*, Phys. Rev. C **69**, 065207 (2004).

[7] L.R. Dai, D. Zhang, C.R. Li, and L. Tong, Chin. Phys. Lett., **24**, 389 (2007).

[8] F. Etminan *et al.*, Nucl. Phys. **A928**, 89 (2014).

[9] H. Noumi *et al.*, J-PARC E50 proposal,
`http://j-parc.jp/researcher/Hadron/en/pac_1301/pdf/P50_2012-19.pdf`

[10] R.J. Hemingway *at al.*, Nucl. Phys. **B142**, 205 (1978).
D. Scotter *et al*, Phys. Lett. **B26**, 474 (1968).
J.K. Hassall *et al*, Nucl. Phys. **B189**, 397 (1981).

[11] M. Naruki and K. Shirotori, Letter of Intent for J-PARC,
`http://j-parc.jp/researcher/Hadron/en/pac_1405/pdf/LoI_2014-4.pdf`

# The $\Omega^*$ spectroscopy experiment

**Hitoshi Takahashi**
Institute of Particle and Nuclear Studies, KEK

We would like to perform the spectroscopy of $\Omega^*$ resonances, which is complementary to the $\Xi^*$ spectroscopy for understanding the substructure of hadrons. With the secondary beams from the high-p beam line, we can measure the production cross sections of $\Omega^*$ baryons within 1 month, while the K10 beam line will provide sufficiently higher yields for the measurements of the angular distributions.

## 1 Physics motivation

Recently, the effective degrees of freedom in hadrons have been deeply discussed. Baryons which contain one (Qqq) or two (QQq) heavy quarks are expected to have two distinct excitation modes coming from the spatial parametrization concerning a diquark (qq or QQ) contribution. The $\Xi^*$ spectroscopy experiment is being planed at the J-PARC hadron experimental facility aiming at observing $\Xi^*$ resonances in messing mass and/or invariant mass spectra [1]. Since the $\Xi$ baryons are composed of two heavy ($s$) quarks and a light ($u$ or $d$) quark, their mass spectrum is expected to reflect the diquark structure in hadrons. On the other hand, the $\Omega$ hyperons are constituted with three same heavy quarks and the diquark contribution is not expected. In contrast to $N^*$ and $\Delta^*$ consisting of three light quarks, currently observed $\Omega^*$ resonances have relatively narrow widths [2, 3, 4], and can be detected by the missing mass and/or invariant mass spectroscopy. In this sense, the $\Omega^*$ spectroscopy could complement the $\Xi^*$ spectroscopy.

Although many $\Omega^*$ resonant states are predicted in various theoretical models, only three states were reported experimentally and two of them are omitted from the Summary Table by the Particle Data Group [5]. Figure 1 summarizes the states predicted theoretically as well as experimental data. We, therefore, would like to carry out the $\Omega^*$ spectroscopy to confirm the known three states and to search for missing states.

## 2 Experimental Method

The planed $\Xi^*$ spectroscopy experiment will utilize 5 GeV/$c$ $K^-$ and $\pi^-$ beams [1]. Our $\Omega^*$ measurement can be performed simultaneously only by increasing the beam momentum to 7 $\sim$ 8 GeV/$c$. The $\pi^-$ beam mode of the $\Xi^*$ experiment uses the $\pi^- p \to \Xi^{*-} K^{*0} K^+$ reaction, whereas we use $K^- p \to \Omega^{*-} K^{*0} K^+$ reaction.

The detectors and the target, therefore, are completely same as those for the $\Xi^*$ spectroscopy; J-PARC E50 spectrometer system [14] and a liquid hydrogen target. The E50 spectrometer has a large acceptance, a multiple tracking capability, and good particle identification in a wide momentum range.

## 3 Yield Estimation

The production cross sections of $\Omega(2250)$ and $\Omega(2470)$ are 0.63 $\mu$b [3] and 0.29 $\mu$b [4], respectively, for the $K^-$ beam momentum of 11 GeV/$c$. We assume 0.1 $\mu$b for the cross section of the $K^-p \to \Omega^{*-}K^+K^{*0}$ reaction. The detection efficiency of $\Omega^{*-}$ resonances is expected to be same as that in the planed $\Xi^*$ spectroscopy experiment [1], where the acceptance of E50 spectrometer, the live time of data acquisition, the tracking efficiency, and the efficiencies for pions and kaons identifications are considered. The thickness of the liquid hydrogen target is also same as the $\Xi^*$ experiment, e.g. 4 g/cm$^2$. The intensity of 7 GeV/$c$ $K^-$ beam is expected as $1 \times 10^5$ /spill at the high-p ($\pi$20) beam line, and $7 \times 10^6$ /spill at the K10 beam line.

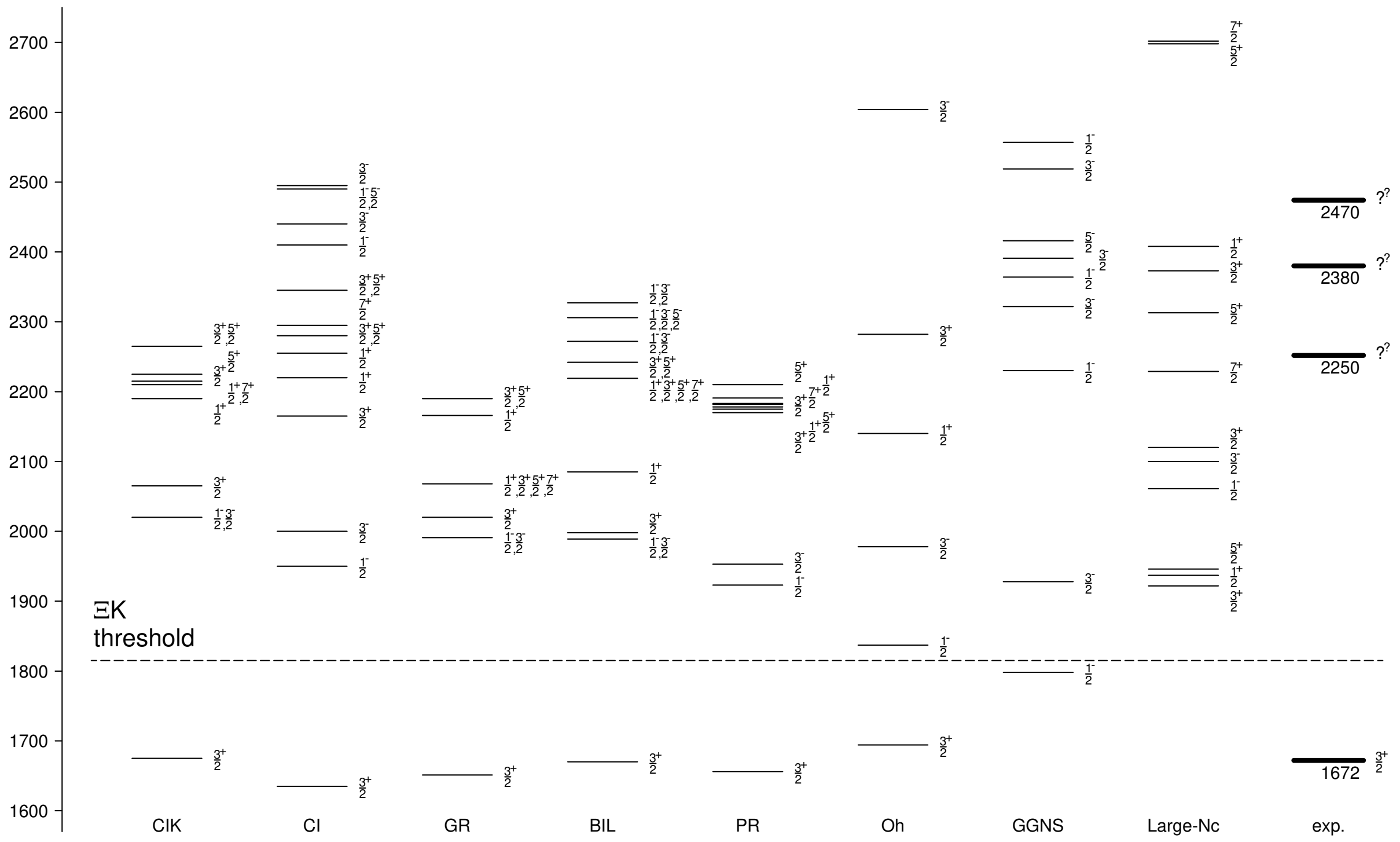

Figure 1: Low-lying Ω baryon spectrum predicted by the non-relativistic quark model (CIK) [6], the relativized quark model (CI) [7], the Glozman.Riska model (GR) [8], the algebraic model (BIL) [9], the recent nonrelativistic quark model (PR) [10], the Skyrme model (Oh) [11], the SU(6) spin-flavor symmetric model (GGNS) [12], and large Nc analysis [13]. The experimental data were from the particle listings by the Particle Data Group [5].

Then, the estimated yield of $\Omega^{*-}$ resonances is about 70 /day at the high-p beam line, and about 4900 /day at the K10 beam line. The 1 month beam time at high-p beam line would be sufficient to measure the production cross sections of $\Omega^*$ states. The K10 beam line would be suitable for further studies such as the measurements of the angular distributions.

## 4 Summary

The $\Omega^*$ spectroscopy experiment is complementary to the $\Xi^*$ spectroscopy for the investigation of the effective degrees of freedom in hadrons. The $\Omega^*$ experiment can be divided to two stages. In the first stage, we will utilize unseparated (or possibly RF-separated) secondary beams from the high-p ($\pi$20) beam line to measure the production cross sections of $\Omega^*$ states. In the second stage, the K10 beam line in the extended hadron facility will provide about 70 times more statistics of the $\Omega^*$ production, which will be sufficient to measure the angular distributions.

## References

[1] M. Naruki, K. Shirotori, Letter of Intent for J-PARC,
`http://j-parc.jp/researcher/Hadron/en/pac_1405/pdf/LoI_2014-4.pdf`

[2] S.F. Biagi *et al.*, Z. Phys. C, **31**, 33 (1986).

[3] D. Aston *et al.*, Phys. Lett. **B194**, 579 (1987).

[4] D. Aston *et al.*, Phys. Lett. **B215**, 799 (1988).

[5] C. Patrignani *et al.* (Particle Data Group), Chin. Phys. C **40** 100001 (2016).

[6] K.-T. Chao, N. Isgur, G. Karl, Phys. Rev. D **23**, 155 (1981).

[7] S. Capstick, N. Isgur, Phys. Rev. D **34**, 2809 (1986).

[8] L.Ya. Glozman, D.O. Riska, Phys. Rep. **268**, 263 (1996).

[9] R. Bijker, F. Iachello, A. Leviatan, Ann. Phys. **284**, 89 (2000).

[10] M. Pervin, W. Roberts, Phys. Rev. C **77**, 025202 (2008).

[11] Y. Oh, Phys. Rev. D **75**, 074002 (2007).

[12] D. Gamermann, C. García-Recio, J. Nieves, L.L. Salcedo, Phys. Rev. D **84**, 056017 (2011).

[13] C.E. Carlson, C.D. Carone, Phys. Lett. **B484**, 260 (2000).
C.L. Schat, J.L. Goity, N.N. Scoccola, Phys. Rev. Lett. **88**, 102002 (2002).
J.L. Goity, C. Schat, N.N. Scoccola, Phys. Lett. **B564**, 83 (2003).
N. Matagne, Fl. Stancu, Phys. Rev. D **71**, 014010 (2005).
N. Matagne, Fl. Stancu, Phys. Rev. D **74**, 034014 (2006).

[14] H. Noumi *et al.*, J-PARC E50 proposal,
`http://j-parc.jp/researcher/Hadron/en/pac_1301/pdf/P50_2012-19.pdf`

# Possible measurement of CP violation in hyperon decay at the K10 beamline

**Yuhei Morino[1]**

[1]KEK, High Energy Accelerator Research Organization, Tsukuba, Ibaraki 305-0801, Japan

## Short Summary

- Physics motivation with its impact and expected result
  CP vaiolation of hyperon weak decay .
- What kind of beam and equipments are necessary?
  several GeV $K^-$ and $K^+$ separated beam with $< 10^7$ per spill,
  the J-PARC E50 spectrometer
- Expected duration of the beam time (yield estimation)
  100 days(300 shifts) at least

The origin of charge-parity (CP) violation is still one of the most important puzzles in particle physics. The mounts of CP violation allowed within the standard model is too small to account for the matter dominance in our universe. CP violation has been measured in a number of processes, such as K and B mesons decay. Although CP violation is expected to be sensitive to beyond the standard model, the observed CP violation is still largely consistent with the standard model. Measurements of CP violation in various systems are important to clear its origin, since they have each different sensitivity to possible new physics. Hyperon nonleptonic weak decays also provide an important information for the origin of CP violation.

CP violation in hyperon nonleptonic weak decay can be measured via asymmetry of angular distributions of decay daughters between hyperon and anti-hyperon. While the predicted asymmetry is very small within the standard model, it could be one or two orders of magnitude larger if new physics contributions are appreciable[1]. The largest asymmetry is predicted for $\Omega$ among of them[2]. Several experiments have measured CP violation in hyperon weak decay. The most precise measurement has been performed for $\Xi \to \Lambda\pi \to p\pi\pi$ decay by Hyper CP experiment. They collected large statistics of unpolarized $\Xi$ and observed large CP violation which is far from the standard model prediction with $2\sigma$ significance, although the result is still preliminarly[3]. All other measurements of CP asymmetry in hyperon weak decay have reported still zero-consistent results.

A experiment with a kaon beam has a great advantage in measurements of hyperon decays due to their large cross section in K$p$ collisions. Table 1 shows a summary of hyperon cross sections. When the energy of the kaon beam is above several GeV, the flight length of the produced hyperons becomes enough long so that the hyperons will decay outside the target material even if we use a rather thick target. It leads that the angular distributions of the decay particles are not affected by the material budget of the targets and the hyperons can be identified clearly by the vertex cut. Therefore, the K10 beamline is a suitable place to measure CP violation in hyperon weak decay.

Table 1: Summary of hyperon cross sections.

| reaction | beam momentum | cross section |
|---|---|---|
| $K^- \, p \rightarrow \Omega^- X$ | 7 GeV/c | $\sim$2 $\mu b$[4] |
| $K^- \, p \rightarrow \Xi^- X$ | 7 GeV/c | $\sim$160$\mu b$[4] |
| $K^+ \, p \rightarrow \Xi^+ X$ | 12.7 GeV/c | $\sim$ 10 $\mu b$[5] |

When we assume a beam intensity of $7.6\times10^6$ per spill for 7 GeV/c kaons[6] and 1 cm W target, more than one billion $\Xi^-$ and $\Xi^+$ will be produced with 100 days data taking, which is larger than the full statistics of the Hyper CP experiment. More than hundred millions $\Omega^-$ and $\Omega^+$ also will be produced with the same run conditions, which is several tens times larger than the Hyper CP experiment. Therefore, the K10 beamline can provide the environment for the confirmation of the large CP asymmetry in $\Xi$ decay and the significant improvement of the sensitivity for $\Omega$ decay. In addition, when the parent hyperons are polarized, there is an observable, called as a parameter "B", which is much more sensitive to CP violation than the conventional observable[7]. Since the produced hyperons in K$p$ collisions are polarized, the extraction of the parameter B is very interesting. While the further consideration is necessary to discuss the feasibility of the measurement for CP asymmetry, measurements of hyperon decays will provide rich physics opportunities at the K10 beamline due to its capability to produce enormous hyperons.

# References

[1] X.-G. He, H. Steger, G. Valencia, Phys. Lett. **B 272**, 411 (1991).

[2] J. Tandean, Phys. Rev. **D 70**, 076005 (2004).

[3] C. Materniak, Nucl. Phys. Proc. Suppl. **187** 208 (2009).

[4] J. K. Hassall, *et al.*, Nucl. Phys. **B 189** 397 (1981).

[5] S. L. Stone, *et al.*, Phys. Lett. **B 32** 515 (1970).

[6] http://www.rcnp.osaka-u.ac.jp/ jparchua/share/WhitePaperJ160827.pdf

[7] J. F. Donoghue, X.-G. He, S. Pakvasa, Phys. Rev. . **D 34**, 833 (1986).

# Charmonium in nucleus at J-PARC

**Hiroaki Ohnishi**[1]

[1] Research Center for Electron Photon Science(ELPH), Tohoku University

## 1 Introduction

The strong interaction between elementary particles has been described very well by the quantum chromodynamics (QCD). The missing piece of the element on QCD, i.e., Higgs Boson, has been discovered at CERN/LHC in 2013; therefore, the theory of strong interaction is now completed. The strong interaction created many various types of matter, such as hadron, nuclei, and very high-density nuclear matter such as neutron stars. Those systems should also be described based on the theory of strong interaction, i.e., QCD, however, it is not possible because of the nature of the QCD at low energy where perturbative QCD cannot be applied. Thus, it is complicated to solve problems and to understand the connection between elementary particles, i.e., quarks and gluons to hadron or hadron to extreme high-density matter.

Even the first step, how the hadron emerged from elementary particles, is not clearly understood yet. Hadrons are known to be an excitation of the vacuum governed by QCD; thus, the meaning of understanding of the hadron is indeed the understanding of the QCD vacuum itself. Therefore, not only more experimental efforts to understand hadron phenomena but also strong theoretical supports for the hadron/nuclear physics are still mandatory.

One of the questions need to understand and explored by the experiment is the mechanism of how a hadron acquires its mass from the QCD vacuum. Nowadays, the mechanism itself is known to be the spontaneous breaking of chiral symmetry. The theory tells that due to the chiral symmetry breaking the vacuum will be filled with quark and anti-quark pairs, i.e., $\bar{q}q$ condensation. On such condition, the expectation value of $\bar{q}q$ condensation, $< \bar{q}q >$, will be one of the order parameters which characterized QCD vacuum. Moreover, not only chiral symmetry breaking but also gluon condensation in QCD vacuum, $< \alpha_s G^2 >$, fulfill a significant role in the generation of hadron mass. For example, the mass splitting of $S^3_1 - S^1_0$ state charmonium will be able to explain by the value of gluon condensation[1]. Besides, in the reference[2], the contribution of gluon condensation on proton mass is evaluated to be 24%. Thus, indeed, gluon condensation is also significant components on the mass of a hadron. However because the chiral symmetry is already broken spontaneously on the world, and gluons are already condensed in QCD vacuum, direct experimental evidence of the mechanism, i.e., how the mass of the hadron is generated, is not presented yet.

Here, we will concentrate on the gluon condensation, i.e., $< \alpha_s G^2 >$, in QCD vacuum. The value of the $< \alpha_s G^2 >$ is expected to be reduced when we change the condition of the vacuum, If $< \alpha_s G^2 >$ is an essential source of the hadron mass, the mass of the hadron will be reduced or changing in high temperature and/or high-density matter. Hereafter, we will be focusing on the high-density environment, inside in nucleus.

What is the best probe to investigate the change of gluon condensation in a nucleus? The spectral function of vector meson in a nucleus will be one of the best probes to identify such effect, i.e., modification of the spectral function of vector meson in a nucleus. There are two reasons. First one is decay width of vector meson, such as $\phi, \omega$ and charmonium($J/\Psi, \Psi(2S)$, etc.) is quite narrow. Thus it will be sensitive to the changing of spectral function in the nuclear matter. Secondly, vector meson will be able to decay to di-lepton channel($e^+e^-$ or $\mu^+\mu^-$), therefore we will be able to reconstruct spectral function via di-lepton invariant mass, where final state interaction of leptons is expected to be very small compared with meson decay with hadrons in the final state. Thus we will be able to measure clean spectral function via di-lepton channel rather than a hadron decay channel. However, light vector mesons, i.e., $\phi$ and $\omega$, have a problem to probe gluon condensation. In light quark sector, i.e., hadron which contained $u$, $d$ and $s$ quarks as valence quark, chiral symmetry breaking is a dominant contribution to generating hadron mass. Therefore light vector mesons in a nucleus will be the best probe to investigate chiral symmetry breaking, in other words quark condensation, but not for gluon condensation.

Thus most promising probe to investigate the change of gluon condensation in a nucleus will be charmonium which is made from only heavy quark and anti-quarks, namely charm quark and its antiparticle. Because the mass of charm quark is much higher than light quarks, i.e., $u$, $d$, and $s$ quarks, the dynamics, i.e., spontaneously breaking of chiral symmetry, which is very important for the world in SU(3) is not relevant. Moreover, the dynamics of charmonium will be sensitive directly with gluon exchange, which is known as QCD van der Waals interaction.

In this paper, we will discuss the possibility to investigate the changing of gluon condensation in nuclear matter via measuring the spectral function of charmonium reconstructed by di-lepton in a nucleus. Hereafter, we will be focusing on di-muon instead of di-electron. Because the production cross section on charmonium is quite low(nanobarn), we need thick nucleus target to achieve a high luminosity, i.e., to correct enough number of the signal from charmonium decay. Due to the photon conversion, multiple scattering, pair production of electrons in such thick nucleus target, di-electron measurement with a thick target is not feasible. Thus we choose muon as a probe of the measurement.

## 2 Charmonium production with anti-proton beam

The key to the measurement to detect modification of charmonium spectra function is to maximize the probability of the charmonium decay probability in nuclear media. One assumed that the particle which has lifetime $\tau$ traveling with $\beta$, the effective decay length can be expressed by $L_{eff} = \beta\gamma c\tau$, where $\gamma$ is Lorentz factor. Thus, elementary processes which will be able to produce charmonium with small $\beta\gamma$ value is indeed optimal for this measurement. Among many physics processes which will be able to produce charmoniums, such as p-p collisions or photo production, we conclude that the best process will be anti-proton($\bar{p}$)-p annihilation processes, because the charmonium production threshold is much lower than the others. For example, the threshold momentum of $\bar{p}$ beam to produce J/$\Psi$ is $\sim$ 4.08 GeV/$c$, where a $\beta\gamma$ value of the produced J/$\Psi$ is only 1.3. In case of $\Psi(2s)$ state, the threshold momentum is $\sim$ 6 GeV/$c$, and $\beta\gamma$ value of produced $\Psi(2s)$ state is found to be $\sim$1.2. Therefore the measurement of the charmonium spectral function in nuclear media will be feasible at K10 beamline currently planned to construct one of the four new beamlines as a J-PARC hadron hall extension project.

There are theoretical predictions about the charmonium spectral function in a nuclear matter. In the reference [3], detail theoretical prediction was made. The result shows that the only small modification, in order of a few MeV in $\Delta m$, is expected in J/$\Psi$. However, as large as ~100 MeV and ~140 MeV shift of pole position are predicted for $\Psi(2s)$ and $\Psi(3770)$ state respectively. Moreover, the measured background, such as the Drell-Yan process and decayed muon from $\pi$ and $K$ cannot be reached to such high invariant mass region with this incident $\bar{p}$ momentum. Therefore "background free" experiment will be possible. This fact will be a significant advantage for this measurement.

The design of the di-muon spectrometer is now under discussion. A very preliminary version of the spectrometer design and GEANT4 based detector model with a J/$\Psi \rightarrow \mu^+\mu^-$ event are shown in Figure 1. To avoid muons from pion and Kaon decay, we are planning to use small solenoid magnetic spectrometer ( 1.4 m diameter) surrounding by iron- scintillator based muon identification detector. The optimization of thickness and contents, such as a fraction of iron and scintillator, is under the progress.

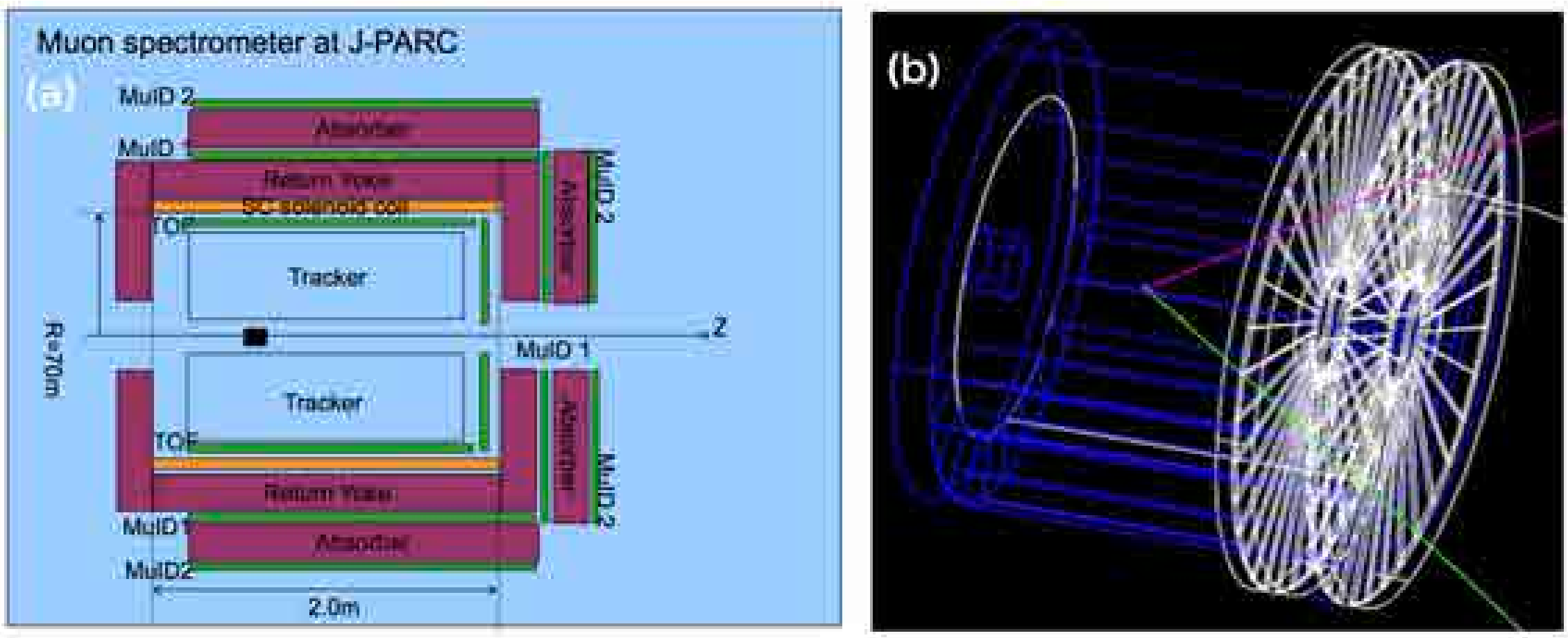

Figure 1: (a) Conceptual design for the Spectrometer and (b) GEANT4 based spectrometer design with J/$\Psi \rightarrow \mu^+\mu^-$ a event

The expected event rate is evaluated by the following assumptions.

- expected $\bar{p}$ beam intensity is $\sim 10^7 \bar{p}$/spill, spill = 3/s.
- 2 g/cm$^2$ carbon target
- detector acceptance and reconstruction efficiency are evaluated by the GEANT4 based simulation. overall efficiency is assumed to be 30%.
- Experiment running time is assumed to be 30 days.

The obtained spectra together with the expected background from pion decay are shown in Figure 2. Accurate J/$\Psi$ signal measurement can be possible only one month of the beam time, however, to accumulate enough statistics for $\Psi(2S)$ or higher mass charmonium, more longer beam time, at least three months or even more will be required. Detail confirmation of those number is under discussion.

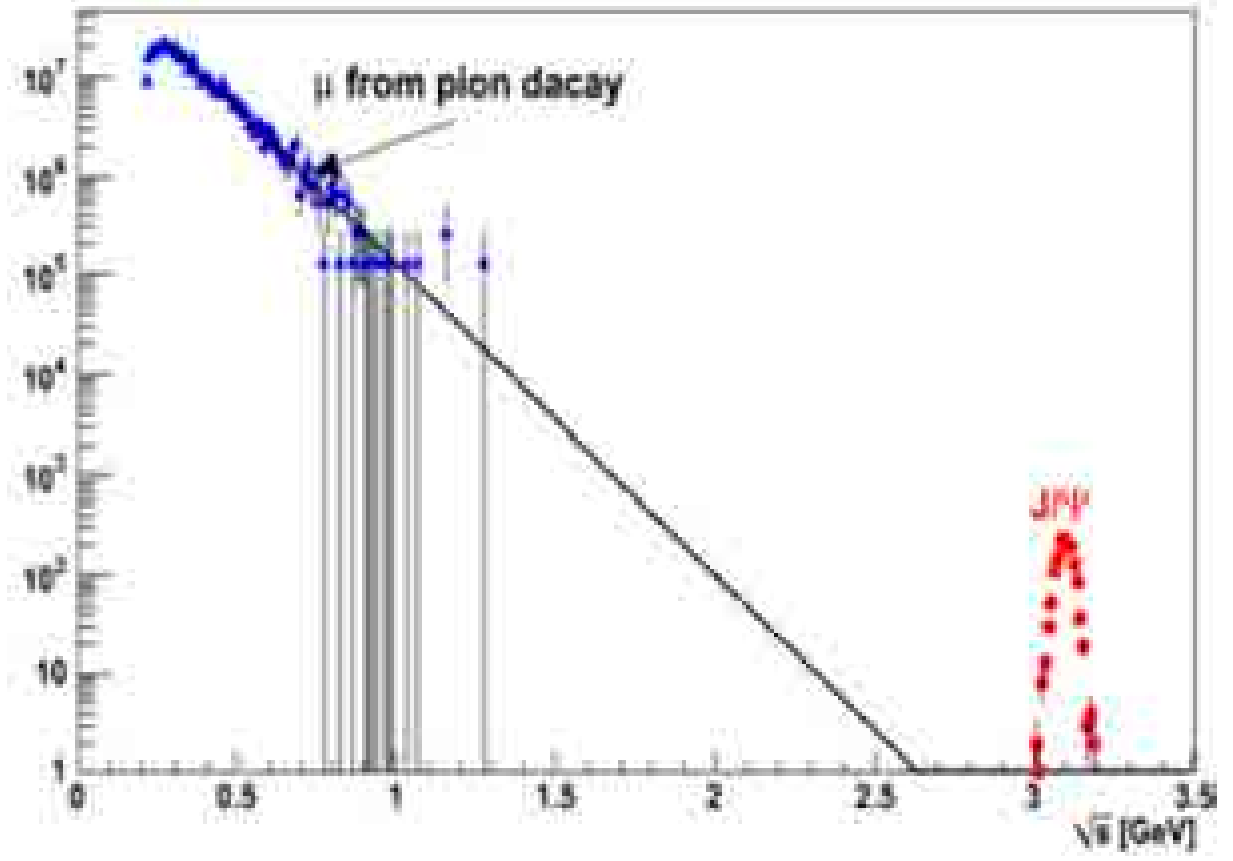

Figure 2: Reconstructed di-muon signal together with expected background from pion decay

## 3 Summary of beam particle, momentum and intensity required for the measurement

High momentum and high-intensity antiproton beam are needed for the measurement. The beam momentum needs for this measurement is more than 4 GeV/$c$ and up to 6-10 GeV/$c$, i.e., above the production threshold for $\Psi(3770)$. As we discussed above, at least $\sim 10^7$ $\bar{p}$/spill and one month of beam time are mandatory to reach reasonable measurement for J/$\Psi$. However, the production cross section for $\Psi(2S)$ and $\Psi(3770)$ is small compared with J/$\Psi$; we will need more than one-month running time. The very preliminary estimation tells that we need at least three months to access $\Psi(2S)$ information. However, this value needs to be confirmed.

## Acknowledgement

This work was supported in partly by Grants-in-Aid for Scientific Research (Nos. 26287057 and 24105711) from the Minister of Education, Culture, Sports, Science and Technology, Japan.

## References

[1] Stephan Narison, Phys.Lett.**B387**162-172(1996).

[2] Xiangdong Ji,Phys.Rev.Lett.**74**1071(1995).

[3] G. Walf *et al.*,Phys.Lett.**B780**25,(2018).

[4] https://panda.gsi.de

# New KL beam line

**J-PARC KOTO Collaboration**[1]

[1]Collaboration between Arizona State, Chicago, Chonbuk, KEK, Jeju, JINR, Korea, Kyoto, Michigan, NDA, Okayama, Osaka, Saga, NTU, and Yamagata.

A precise measurement of the branching fraction of a rare decay of the long-lived neutral kaon, $K_L \to \pi^0 \nu\overline{\nu}$, will be made at the new KL beam line from a new production target located in the extended hall.

## 1 physics motivation

The $K_L \to \pi^0 \nu\overline{\nu}$ decay, which directly breaks the CP symmetry, occurs once in tens of billions of $K_L$ decays due to the complex phase in the diagrams from the strange quark in $K_L$ to the down quark (Fig. 1). In the Standard Model of particle physics the origin of the phase is the imaginary part $\eta$ in the Kobayashi-Maskawa matrix element $V_{td}$, and the predicted branching fraction is $(3.0 \pm 0.3) \times 10^{-11}$ [1]. The 10% uncertainty comes from the uncertainties in the parameters of the Kobayashi-Maskawa matrix ($|V_{ub}|$, $|V_{cb}|$, $\gamma$) that should be determined by the experiments of B meson decays; its pure theoretical uncertainty, 2.5%, is small [2]. In other meson decays, the effects of higher-order electroweak interactions or hadronic interactions cannot be controlled so well as in $K_L \to \pi^0 \nu\overline{\nu}$.

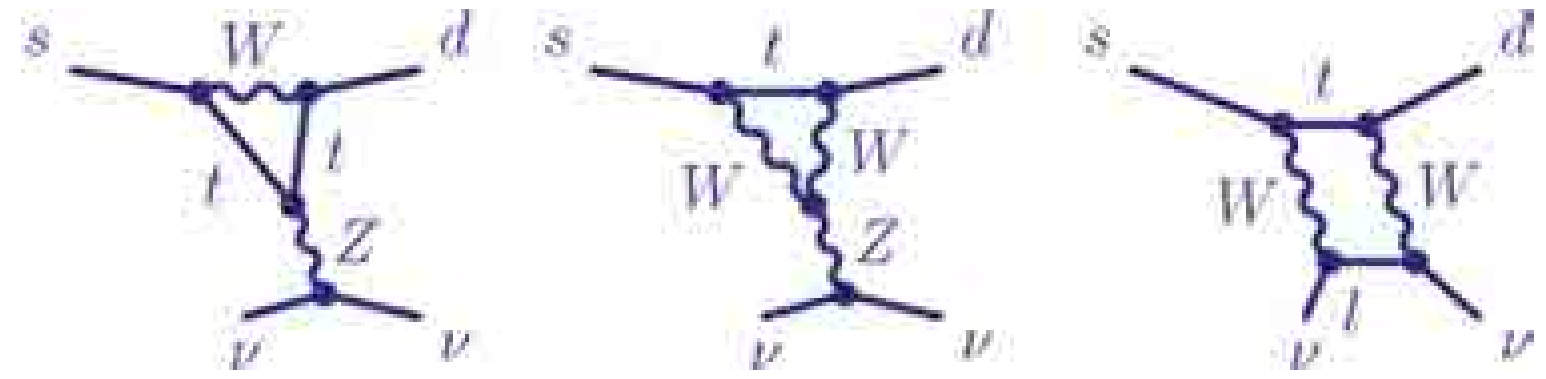

Figure 1: Diagrams of the $K_L \to \pi^0 \nu\overline{\nu}$ decay in the Standard Model.

The $K_L \to \pi^0 \nu\overline{\nu}$ decay is sensitive to the contribution of new particles such as those in Super Symmetric theories. Since direct production of new particles would not be discovered so easily in LHC, new physics in much higher energy-scale, e.g. 50 TeV corresponding the distance of $4 \times 10^{-21}$ m, will be investigated indirectly through rare decays [3]. Figure 2 shows how the branching fractions of the $K_L \to \pi^0 \nu\overline{\nu}$ decay and the corresponding rare charged-kaon decay $K^+ \to \pi^+ \nu\overline{\nu}$ would be shifted, due to new physics effects, from the Standard Model predictions.

A $K_L \to \pi^0 \nu\overline{\nu}$ experiment can also search for the two-body decay $K_L \to \pi^0 X^0$, where $X^0$ is a new invisible neutral boson. It was recently pointed out [6] that the limits on $X^0$ in the mass region close to the nominal $\pi^0$ mass ($116 < M_{X^0} < 152$ MeV/$c^2$) had not been imposed from the charged-kaon experiments in the past.

## 2 current status

The $K_L \to \pi^0 \nu\overline{\nu}$ decay has, so far, never been detected experimentally; the upper limit on the branching fraction obtained by the E391a experiment [7] performed at KEKPS from 2004 to 2005 was $2.6 \times 10^{-8}$ at the 90% confidence level (C.L.). In the Hadron Experimental Facility (HEF) of J-PARC the KOTO experiment [5, 8] is being conducted; the collaboration consists of 69 physicists from Japan, Korea, Russia, Taiwan and USA and aims to observe $K_L \to \pi^0 \nu\overline{\nu}$ for the first time. No other $K_L \to \pi^0 \nu\overline{\nu}$ experiment is present as of 2019.

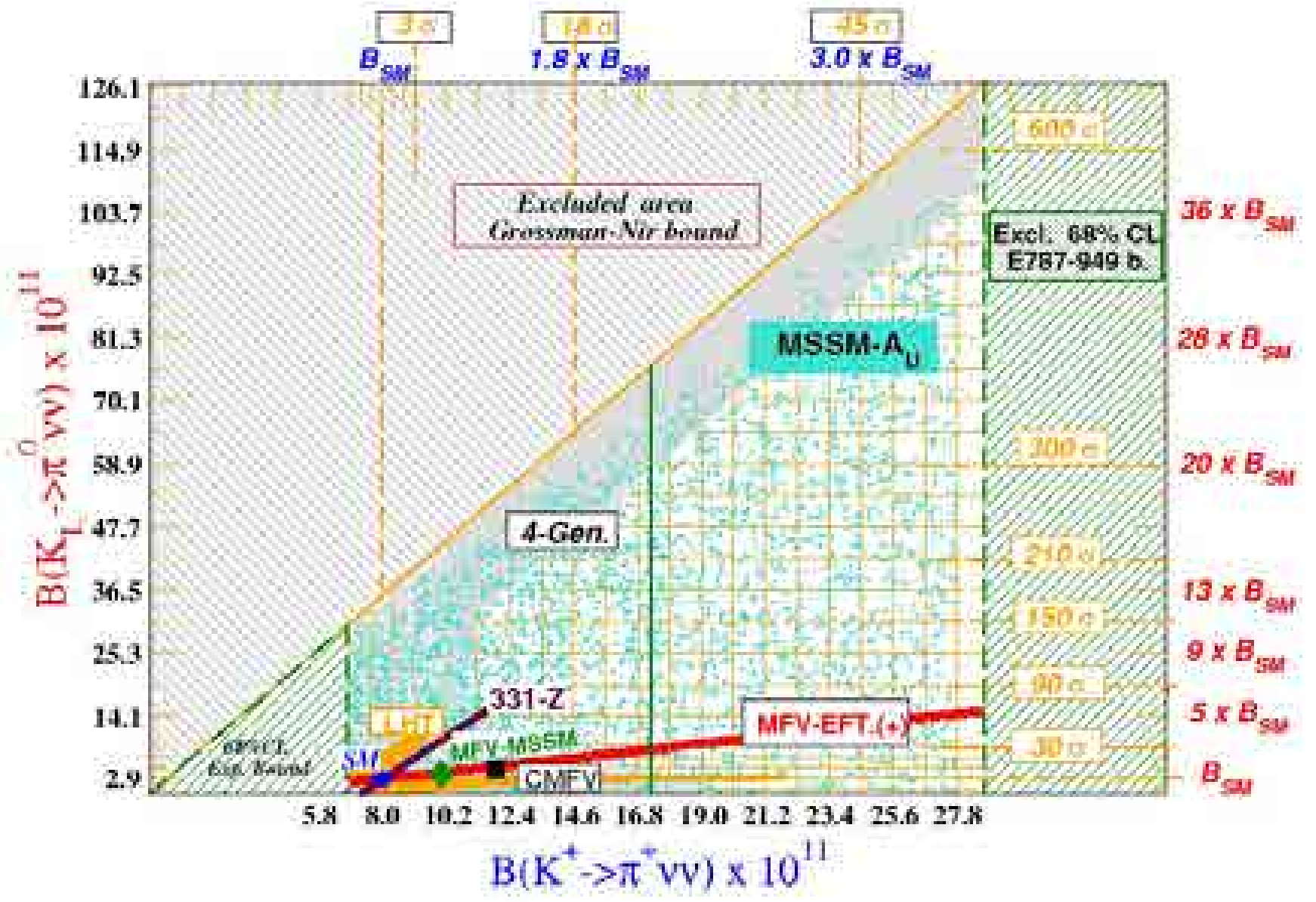

Figure 2: Plot to present how the branching fractions of rare kaon decays would be shifted from the Standard Model predictions ($SM$) due to the effects of new physics models ($MFV$、$CMFV$、$MSSM$、$LHT$、331-$Z'$、4-$Gen.$) [4, 5].

In 2009 the neutral beam line for the KOTO experiment was constructed (Fig. 3) in the current HEF hall. Neutral particles produced in the T1 target are extracted to the 16-degree direction, and the length of the beam line is 21 m. From 2010 to 2012, before and after the earthquake, the KOTO detector (Fig. 4) was built at the KL experimental area inside the hall. In the detector, two photons from $\pi^0$ in the final state of the $K_L \to \pi^0 \nu\overline{\nu}$ decay are measured in the CsI calorimeter located at the downstream side. Furthermore, the $K_L$ decay region is evacuated and is hermetically surrounded by photon counters and charged-particle counters to ensure that no other particle is in the final state [1]. From January 2013 the commissioning started. In May the first physics run [9] was conducted; with only four-days worth of data, it achieved the single-event sensitivity comparable to the E391a experiment. On April 24 in 2015 the user operation of HEF restarted, and KOTO resumed the physics data taking. The results from the 2015 data, published in January 2019 [10], improved the E391a sensitivity by an order of magnitude and set so far the most stringent upper limit on the $K_L \to \pi^0 \nu\overline{\nu}$ branching fraction: $3.0 \times 10^{-9}$ at the 90% C.L. KOTO upgraded the Veto counters in 2016 and collected data from 2016 to 2018, which corresponds to 1.4 times larger than the data in 2015. The analysis is currently undertaken intensively. In the autumn of 2018, KOTO upgraded the calorimeter to significantly suppress the bakground due to neutrons. A data taking with the new detector system started in February 2019.

[1] Thus, the counters are called "Veto" counters.

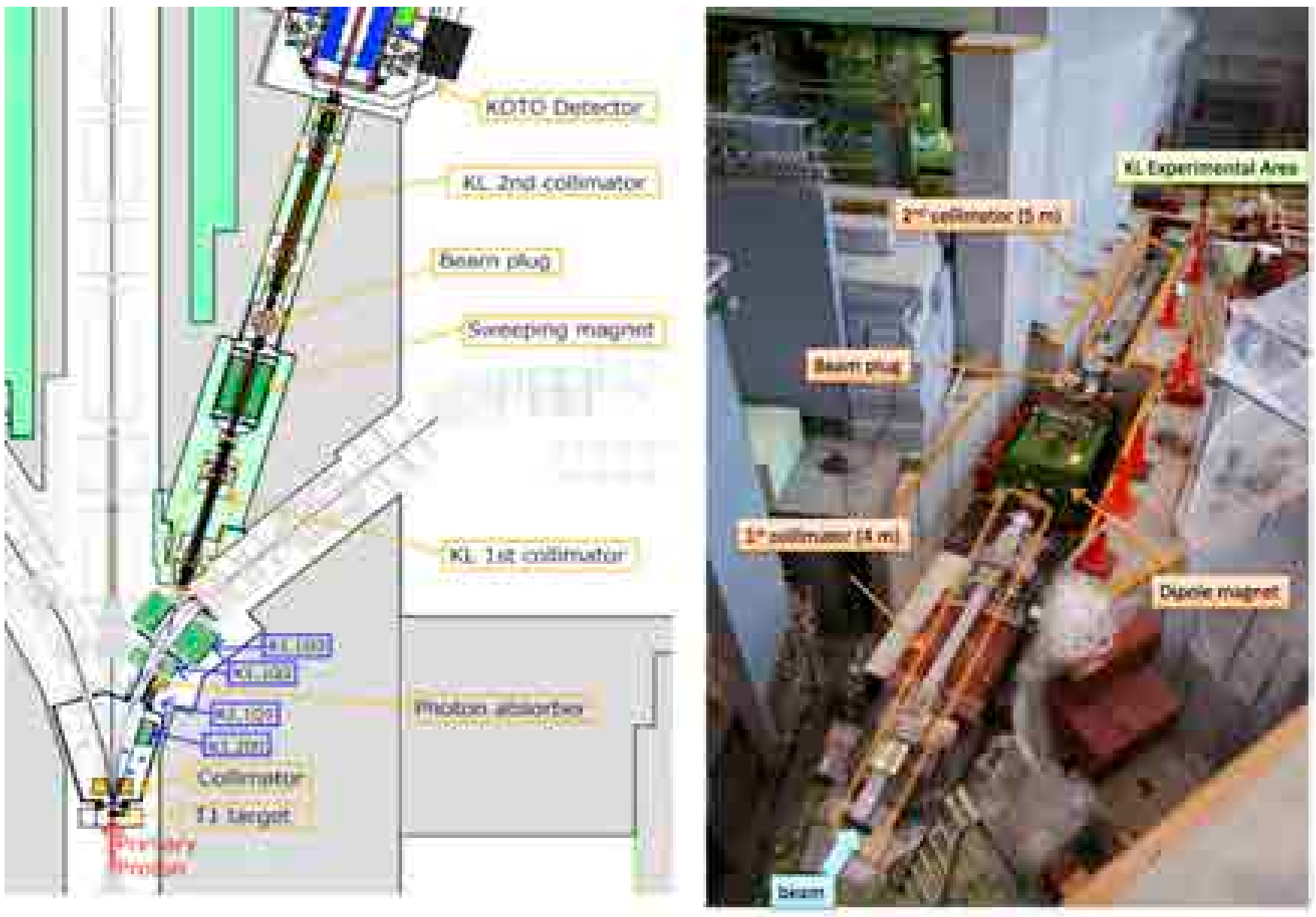

Figure 3: Layout of the neutral beam line for the KOTO experiment in the current HEF hall (left); photo of the beam line under construction (right).

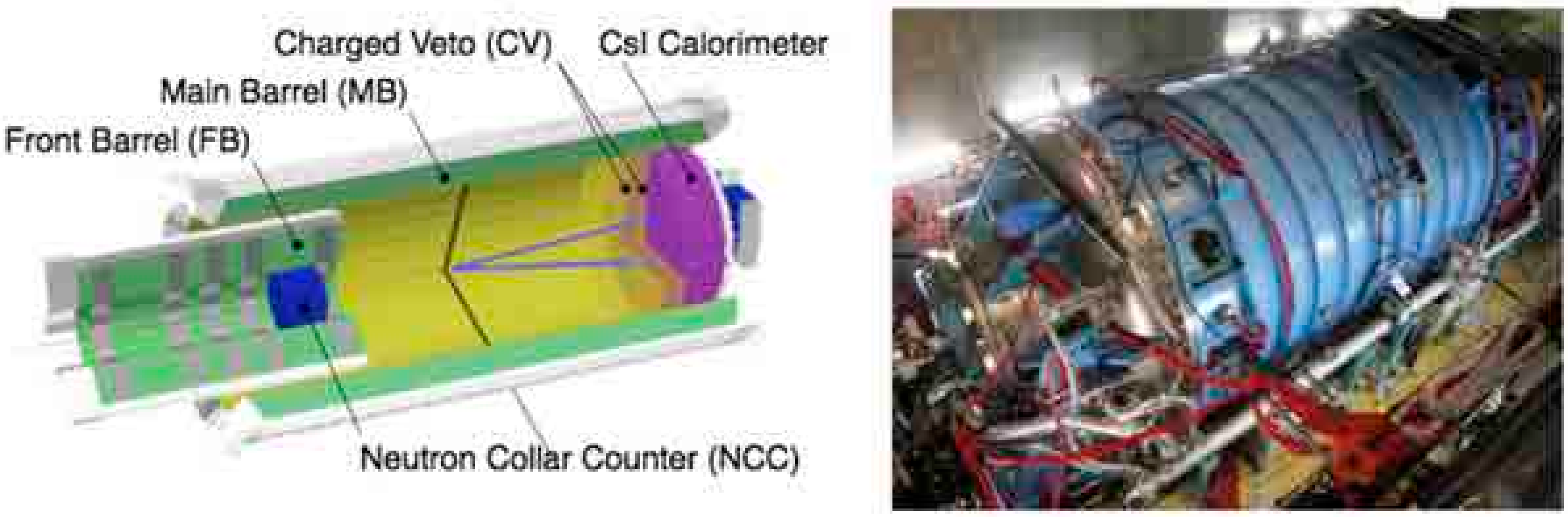

Figure 4: Schematic view of the KOTO detector in the current HEF hall (left); photo of the detector (right).

In the project for the extended Hadron Experimental Facility, the new beam line and experiment for $K_L \to \pi^0 \nu \overline{\nu}$ are regarded to be in the next phase of the KOTO experiment called the KOTO Step2. The purpose of the new experiment is to do a precise measurement of the $K_L \to \pi^0 \nu \overline{\nu}$ branching fraction.

# 3 beam line

The extraction angle of the new KL beam line will be reduced from the current 16 degrees, which was due to the limitation from the shields for the primary beam line in the current hall, to 5 degrees to increase the $K_L$ flux. The number of neutrons in the beam would also increase, but the ratio of the numbers of neutrons to kaons will be minimized to 30 at the entrance of the detector.

The momentum spectrum of the particles in the neutral beam will shift to be higher [2] and the $\Lambda$ decays in the beam could be a background source; the beam line will be longer so that $\Lambda$ would decay out in the upstream. In the layout, the neutral particles produced in a new production target located in the extended hall will be extracted and transported for 51.6 m to the new experimental area located behind the beam dump (Fig. 5). Since the downstream part of the beam line would be in the beam dump, new iron shielding around it will be made. The new area will be build in the ground as an anex of the extended hall.

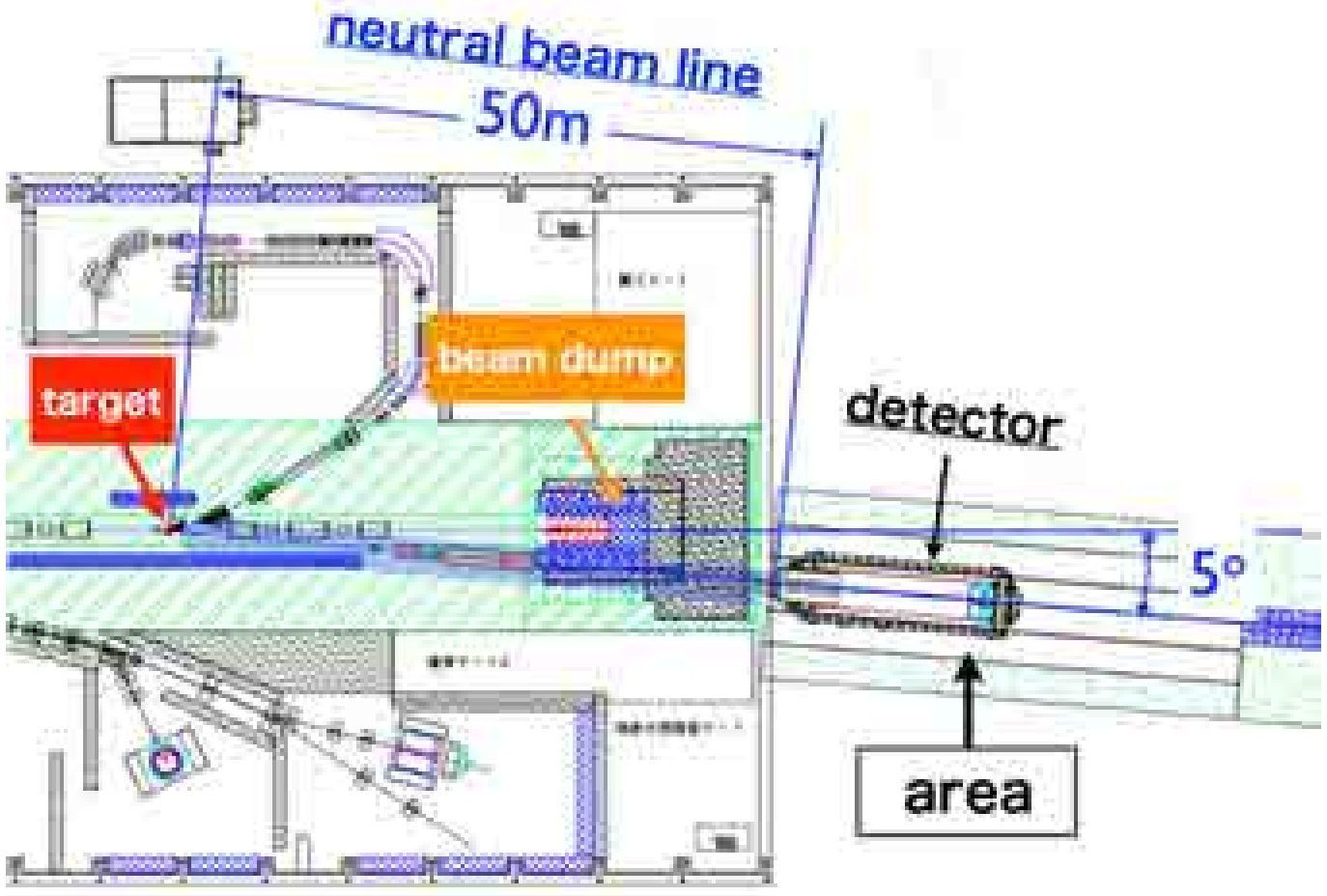

Figure 5: Plan view of the new KL beam line and experimental area at the 5-degree direction from a new production target.

# 4 detector

The concept of the new detector for $K_L \to \pi^0 \nu \overline{\nu}$ is the same as that for KOTO in the schematic view in Fig. 4; two photons from $\pi^0$ are measured in the calorimeter, and the backgrounds are identified and removed by the Veto counters surrounding the decay region. The new experiment aims to reach the single-event sensitivity at $3 \times 10^{-13}$ and observe 100 signal events, i.e. 10%-precision measurement of the branching fraction, predicted by the Standard Model. A large detector, with a longer decay region (from 2 m to 11 m) and a wider calorimeter (from 2 m to 3 m in the diameter), will be constructed in order to increase the acceptance. Since the average $K_L$ momentum will be larger than that in KOTO, and the energies of photons and other secondary particles from $K_L$ decay will be higher, the energy and timing resolutions of the calorimeter and veto counters should be better to improve their background rejection powers.

[2]The average $K_L$ momentum would be from 2.1 GeV/$c$ to 5.2 GeV/$c$.

## 5 internatonal competition

At CERN in Europe, the NA62 experiment to measure the branching fraction of the charged-kaon decay $K^+ \to \pi^+\nu\overline{\nu}$ using the decay-in-flight technique is being conducted; NA62 started the physics data taking from 2016 and the first results were published in February 2019 [11]. NA62 has a plan to proceed to the $K_L \to \pi^0\nu\overline{\nu}$ experiment in future as the KLEVER project [12]. The KLEVER detector and high intensity $K^0$-beam can in practice only be implemented in the NA62 hall, and KLEVER phasing after NA62 will depend on NA62 $K^+$ results [13].

In the accelerator facilities in USA there is no plan for the $K_L \to \pi^0\nu\overline{\nu}$ experiment.

## References

[1] A.J. Buras *et al.*, JHEP **1511**, 033 (2015).

[2] J. Brod *et al.*, Phys. Rev. **D 83**, 034030 (2011).

[3] A.J. Buras *et al.*, JHEP **1411**, 121 (2014).

[4] F. Mescia and C. Smith, *$K \to \pi\nu\nu$ decay in the Standard Model*, `http://www.lnf.infn.it/wg/vus/content/Krare.html` .

[5] T.T. Komatsubara, Prog. Part. Nucl. Phys. **67**, 995 (2012).

[6] K. Fuyuto *et al.*, Phys. Rev. Lett. **114**, 171802 (2015).

[7] J.K. Ahn *et al.* (E391a Collaboration), Phys. Rev. **D 81**, 072004 (2010).

[8] T. Yamanaka (for the KOTO Collaboration), PTEP **2012** 02B006 (2012).

[9] J-PARC KOTO Collaboration, J.K. Ahn *et al.*, PTEP **2017** 021C21 (2017).

[10] J.K. Ahn *et al.* (KOTO Collaboration), Phys. Rev. Lett. **122**, 021802 (2019).

[11] The NA62 Collaboration, Phys. Lett. **B 791**, 156 (2019).

[12] The KLEVER Project,
`https://indico.cern.ch/event/765096/contributions/3295999/` ,
input to the 2020 update of the European Strategy for Particle Physics.

[13] Jörg Jäckel, Mike Lamont and Claude Vallée,
PBC coordinators and contacts to the European Strategy Group,
`https://indico.cern.ch/event/765096/contributions/3295606/` ,
input to the 2020 update of the European Strategy for Particle Physics.